%% file: Wild_Bootstrap_Counting_Processes.tex
\documentclass[twoside,12pt]{article}
\usepackage[dvipsnames,table,xcdraw]{xcolor}
\definecolor{awesome}{rgb}{1.0, 0.13, 0.32}
\definecolor{dodgerblue}{rgb}{0.12, 0.56, 1.0}
\definecolor{trueblue}{rgb}{0.0, 0.45, 0.81}
\definecolor{amaranth}{rgb}{0.9, 0.17, 0.31}
\usepackage[T1]{fontenc}

\usepackage{comment}
\usepackage[top=1in, bottom=1in, left=1.5in, right=1in]{geometry}
\usepackage{rotating}
\usepackage{amsmath}
\usepackage{amsfonts}
\usepackage{algorithm}
\usepackage{algorithmic}
\usepackage{multirow}
\usepackage{colortbl}
\usepackage{color}
\usepackage{epigraph}
\usepackage{graphicx}
\usepackage{caption}
\usepackage{subcaption}
\usepackage{tabularx}
\usepackage{float}
\usepackage{longtable}
\usepackage{pdfpages}
\usepackage{pdflscape}
\usepackage[acronym,toc]{glossaries}
\usepackage{setspace}
\usepackage{pifont}
\usepackage{relsize}
\usepackage{bm}
\usepackage{amssymb}
\usepackage{siunitx}
\usepackage{url}
\usepackage{setspace}
\usepackage{amsmath}
\usepackage{amsthm}
\usepackage{amssymb}
\usepackage{mathrsfs} 
\usepackage{mathtools}
\usepackage{rotating}
\usepackage{adjustbox}
\usepackage{upgreek}
\usepackage{tikz}
\usetikzlibrary{arrows,calc,shapes,decorations.pathreplacing}
\usepackage{pgfplots}

\theoremstyle{plain}
 
\theoremstyle{definition}
\usepackage{titlesec}
\usepackage{pdfpages}
\usepackage{changepage}
\usepackage{realboxes}

\usepackage{hyperref}

\newcommand{\headrulewidth}{0pt}

\newcommand{\bs}{\boldsymbol}

\usepackage{amsmath}
\usepackage{bbm}
\usepackage{amssymb}
\usepackage{color}
\usepackage{hyperref}
\usepackage[capitalize]{cleveref}
\usepackage{xr}
\usepackage{graphicx}

\usepackage{rotating}

\usepackage{ulem}
\usepackage{amsthm}
\usepackage{mathtools}
\usepackage{thmtools}
\usepackage{enumitem}
\usepackage[round]{natbib}
\usepackage{mathrsfs}
\usepackage{graphicx}
\usepackage{subcaption}
\usepackage{float}
\usepackage{booktabs,amsfonts,dcolumn}
\newcolumntype{d}[1]{D..{#1}}
\usepackage[T1]{fontenc}
\usepackage{parskip}
\usepackage[tracking=true]{microtype}
\usepackage{titlesec}

\newcommand{\nc}{\newcommand}
\nc{\red}{\color{red}}
\nc{\green}{\color{green}}
\nc{\vio}{\color{violet}}
\renewcommand{\qedsymbol}{$\blacksquare$}

\declaretheoremstyle[bodyfont=\slshape]{slshape}
\declaretheorem[style=slshape,name=Theorem,numberwithin=section]{thm}
\declaretheorem[style=slshape,name=Lemma,numberlike=thm]{lem}

\declaretheorem[style=slshape,name=Assumption,numberlike=thm]{ass}
\declaretheorem[style=slshape,name=Remark,numberlike=thm]{rem}
\declaretheorem[style=slshape,name=Example,numberlike=thm]{ex}
\declaretheorem[style=slshape,name=Corollary,numberlike=thm]{cor}
\declaretheorem[style=slshape,name=Replacement,numberlike=thm]{repl}

\newlist{thmlist}{enumerate}{1}
\setlist[thmlist]{label=(\roman{thmlisti}),noitemsep}

\Crefname{thm}{Theorem}{Theorems}
\Crefname{lem}{Lemma}{Lemmas}
\Crefname{ass}{Assumption}{Assumptions}
\Crefname{defi}{Definition}{Definitions}
\Crefname{rem}{Remark}{Remarks}
\Crefname{ex}{Example}{Examples}
\Crefname{cor}{Corollary}{Corollary}
\Crefname{repl}{Replacement}{Replacement}
\Crefname{scheme}{Scheme}{Scheme}

\Crefname{listthm}{Theorem}{Theorems}
\Crefname{listlem}{Lemma}{Lemmas}
\Crefname{listass}{Assumption}{Assumptions}
\Crefname{listdefi}{Definition}{Definitions}
\Crefname{listrem}{Remark}{Remarks}
\Crefname{listex}{Example}{Examples}
\Crefname{listcor}{Corollary}{Corollary}
\Crefname{listrepl}{Replacement}{Replacement}
\Crefname{listscheme}{Scheme}{Scheme}

\addtotheorempostheadhook[thm]{\crefalias{thmlisti}{listthm}}
\addtotheorempostheadhook[lem]{\crefalias{thmlisti}{listlem}}
\addtotheorempostheadhook[ass]{\crefalias{thmlisti}{listass}}
\addtotheorempostheadhook[defi]{\crefalias{thmlisti}{listdefi}}
\addtotheorempostheadhook[rem]{\crefalias{thmlisti}{listrem}}
\addtotheorempostheadhook[ex]{\crefalias{thmlisti}{listex}}
\addtotheorempostheadhook[cor]{\crefalias{thmlisti}{listcor}}
\addtotheorempostheadhook[repl]{\crefalias{thmlisti}{listrepl}}
\addtotheorempostheadhook[scheme]{\crefalias{thmlisti}{listscheme}}

\creflabelformat{listthm}{#2\thethm.#1#3}
\creflabelformat{listlem}{#2\thethm.#1#3}
\creflabelformat{listass}{#2\thethm.#1#3}
\creflabelformat{listdefi}{#2\thethm.#1#3}
\creflabelformat{listrem}{#2\thethm.#1#3}
\creflabelformat{listex}{#2\thethm.#1#3}
\creflabelformat{listcor}{#2\thethm.#1#3}
\creflabelformat{listrepl}{#2\thethm.#1#3}
\creflabelformat{listscheme}{#2\thethm.#1#3}

\allowdisplaybreaks
\title{\mbox{Wild Bootstrap for Counting Process-Based Statistics}}

\author{Marina T.\ Dietrich\footnote{Department of Mathematics\\\phantom{X}\quad Faculty of Exact Science\\\phantom{X}\quad Vrije Universiteit Amsterdam\\\phantom{X}\quad De Boelelaan 1111\\\phantom{X}\quad 1081 HV Amsterdam\\\phantom{X}\quad The Netherlands} \!  \ Dennis Dobler$^*$, \ Mathisca C.\ M.\ de Gunst$^*$}

\date{\today}

	\renewcommand*{\thesection}{I.\arabic{section}}
\renewcommand*{\thesubsection}{I.\arabic{section}.\arabic{subsection}}
	\renewcommand*{\theequation}{I.\arabic{equation}}
	\renewcommand*{\thetable}{I.\arabic{table}}
 \renewcommand*{\thefigure}{I.\arabic{figure}}

\begin{document}
%\includepdf[pagecommand={\begin{tikzpicture}[remember picture, overlay];\end{tikzpicture}}]{BirgitSollie-Cover-Gekozen-BackRugCover}

\maketitle

\begin{abstract}
   \noindent The wild bootstrap is a popular resampling method in the context of time-to-event data analyses. Previous works established the large sample properties of it for applications to different estimators and test statistics. It can be used to justify the accuracy of inference procedures such as hypothesis tests or time-simultaneous confidence bands.
   This paper consists of two parts:
   in Part~I, a general framework is developed in which the large sample properties are established in a unified way by using martingale structures. The framework includes most of the well-known non- and semiparametric statistical methods in time-to-event analysis and parametric approaches.
   In Part II, the Fine-Gray proportional sub-hazards model exemplifies the theory for inference on cumulative incidence functions given the covariates. The model falls within the framework if the data are censoring-complete.
   A simulation study demonstrates the reliability of the method and an application to a data set about hospital-acquired infections illustrates the statistical procedure.
   %However, not all censoring times are known in most real-life applications. Hence, the wild bootstrap is additionally combined with a multiple imputation of the required yet unknown censoring times. Simulation results are shown and an application to a data set about hospital-acquired infections is illustrated.
\end{abstract}


%\newpage

\setstretch{1.3}
\newgeometry{left=3cm, right=2cm,bottom=3.5cm}

%\newpage
%\include{Acknowledgement/acknowledgement}

\renewcommand{\headrulewidth}{0.5pt}

%promotor:           prof.dr. M.C.M. de Gunst

%promotor:        prof.dr. M.R.H. Mandjes 

%%%%%
%\newpage
%\vspace{100mm}
%\begin{center}
%Give thanks to the Lord, for he is good; \\
%his love endures forever. \\
%Psalm 107:1, NIV
%\end{center}
%%%%%

%\newpage\null\thispagestyle{empty}

% set the number of sectioning levels that get number and appear in the contents
%\setcounter{secnumdepth}{3}
%\setcounter{tocdepth}{1}

%\include{Dedication/dedication}
%\include{Abstract/abstract}

%\fancyfoot[C]{\thepage}

%\fancyhead[LE,RO]{\thepage}

%\tableofcontents
%\thispagestyle{empty}

%\printglossary[title=List of Acronyms,type=\acronymtype]
%\printglossary  % Print the nomenclature 
%\addcontentsline{toc}{chapter}{Nomenclature}

\include{Chapter1}

	\setcounter{section}{0}
	\setcounter{equation}{0}
	\renewcommand*{\thesection}{II.\arabic{section}}
\renewcommand*{\thesubsection}{II.\arabic{section}.\arabic{subsection}}
	\renewcommand*{\theequation}{II.\arabic{equation}}
	\renewcommand*{\thetable}{II.\arabic{table}}
 \renewcommand*{\thefigure}{II.\arabic{figure}}

\include{Chapter2}

%\include{Chapter4/Chapter3}

%\include{Appendix1/appendix1}

%\addcontentsline{toc}{chapter}{\protect \textbf{References}}
%\include{Bibliography/bibliography}

%\rhead{}
%\lhead{}
%\fancyhead[LE,RO]{\thepage}

%\clearpage
%\phantomsection
%\label{bibliography}
\bibliographystyle{plainnat}
\bibliography{references}
%\markboth{}{}

%\addcontentsline{toc}{chapter}{\protect \textbf{Abstract}}
%\include{Abstract/abstract}
%\addcontentsline{toc}{chapter}{\protect \textbf{References}}
%\include{Bibliography/bibliography}
%\cleardoublepage

%\clearpage
% \cleardoublepage
% \phantomsection
% \addcontentsline{toc}{chapter}{\protect \textbf{Summary}}
%\include{Summary/summary_thesis}
%\cleardoublepage
%\clearpage
%\phantomsection
%\addcontentsline{toc}{chapter}{\protect \textbf{Samenvatting}}
%\include{Samenvatting/samenvatting}
%\addcontentsline{toc}{chapter}{\protect \textbf{Acknowledgement}}
%\include{Acknowledgement/acknowledgement}
%\appendix
%\include{Appendix1/appendix1}
%\include{Appendix2/appendix2}

\end{document}

%% file: Chapter1.tex
%\noindent
\textbf{Part I: A Martingale Theory Approach}%\label{chap:martingale}

%
%%
%%%
%%%%
%%%%%
%%%%%%
%%%%%%%
\section{Introduction}
%%%%%%%
%%%%%%
%%%%%
%%%%
%%%
%%
%

%-refer to its origin (bootstrap / first wild bootstrap)\\
%- references: wild bootstrap for specific counting process models

In medical studies about, say, the 5-year survival chances of patients who underwent a novel treatment, not only the point estimate after five years is of interest, but also
a confidence interval which quantifies the estimation uncertainty.
Furthermore, it makes an essential difference for the patient whether the survival chances fall rather swiftly or slowly towards the 5-year survival chance, because the rate of decrease  of the survival chance  affects, for instance, the expected remaining lifetime.
For this reason, it is more instructive to inspect confidence \textit{regions} for the entire run of the survival curve, such as \textit{time-simultaneous bands},  than confidence intervals for the survival chances at single time points.

In order to construct confidence regions, naturally information about the uncertainty of the estimation along the entire trajectory is required. Thus, one is interested in the distribution of the estimator around the target quantity  as a function in time.
Likewise, in the context of statistical testing, %in the sense that 
the distribution of the test statistic under the null hypothesis has to be determined. 
In both cases, because of the complex nature of the involved stochastic processes, the exact distribution of the estimator or the test statistic is generally unknown and needs to be approximated.

A solution to the problem of assessing the distribution of a time-dependent statistic or the null distribution of an intricate test statistic is given by resampling techniques like
%If the critical values are found this way and are as a consequence data-dependent, one speaks of a conditional test.
 random permutation, algebraic group-based re-randomization \citep{dobler21}, the bootstrap \citep{efron79} or many variants thereof such as the wild bootstrap \citep{wu86}.
Certain variants of these techniques were also proposed in survival analysis contexts where time-to-event data could be incomplete due to, e.g.,  independent left-truncation or right-censoring. Early references are \cite{efron81} and \cite{akritas86} for the classical bootstrap (drawing with replacement from the individual data points), \cite{neuhaus93} for random permutation (of the censoring indicators), and \cite{lin93} for the wild bootstrap (mimicking martingale increments related to counting processes).

Because of its popularity, elegance, and flexibility,  in this Part~I we
focus on the wild bootstrap as the method of choice in the context of survival and event history analysis.
Indeed, the wild bootstrap has been used frequently and in various models, though most often with normally distributed multipliers---an unnecessary restriction. For example, 
in \cite{lin94} and  \cite{dobler19} the wild bootstrap is applied to Cox models, and in \cite{lin97}, \cite{beyersmann13}, and \cite{dobler17} the wild bootstrap is applied to cumulative incidence functions in competing risks models. In contrast to the pioneer papers of  Lin~(et al.), in the publications of Dobler et al.\ and Beyersmann et al.\ it has been allowed for generally distributed and data-dependent multipliers, respectively.
Furthermore, in \cite{spiekerman98} multivariate failure time models are  considered, in \cite{Fine-Gray} proportional subdistribution hazard models, in \cite{lin00}  means in semiparametric models, and in \cite{scheike03} Cox-Aalen models are studied.
More recently, Bluhmki and colleagues analyzed Aalen-Johansen estimators in general Markovian multi-state models (\cite{bluhmki18}) and  general Nelson-Aalen estimators (\cite{bluhmki19}), and Feifel and Dobler treated nested case-control design models (\cite{feifel21}).
%See \cite{martinussen06} for a textbook that advocates the use of the wild bootstrap in survival models.
%Note that the above list is not exhaustive.

In this Part~I, we develop a rigorous theory to justify the use of the wild bootstrap under various survival analysis models.
As in the above-mentioned articles, we employ the wild bootstrap for mimicking the martingale processes related to individual counting processes. 
We allow the individual counting processes to have multiple jumps each. 
Nonparametric models, parametric models and semiparametric (regression) models are covered in a unified approach.
In this sense, the present Part~I provides an umbrella theory for a large variety of specific applications of the wild bootstrap in the context of counting processes.
In particular, we show that the asymptotic distribution of the resampled process coincides with that of the statistic of interest. In this way we verify the asymptotic validity of the wild bootstrap as an approximation procedure.
Our proofs rely on weak regularity conditions and, differently from those in the above-mentioned articles, are developed in a novel way based on the martingale theory for counting processes as given in Rebolledo's  original paper \cite{Rebolledo}.
In particular, our approach solves an open problem of handling the Lindeberg condition in a suitable way. 
We also  illustrate our approach for a couple of frequently used models.

%{\vio [This perhaps in Discussion or shorter?]} 

%In Part~II
%or: follow-up paper  
%we concentrate on the wild bootstrap for the Fine-Gray model under censoring-complete data.
%In particular, we apply the wild bootstrap to the estimators involved in the Fine-Gray model
%and use the corresponding asymptotic results in order to verify the wild bootstrap with generally distributed multipliers as an appropriate procedure to approximate the unknown distribution of the aforesaid estimators. 
%Moreover, we  analyze the small sample performance of the wild bootstrap confidence bands for the cumulative incidence function by means of simulation. 
% and we will apply it to a real data set.  

The present Part~I is organized as follows. In \cref{sec:Notation} we introduce the general set-up, the precise form of the counting process-based statistic, and derive its asymptotic distribution. In \cref{sec:generalBootstrap} we define the wild bootstrap counterpart of the statistic under consideration and study its asymptotic distribution. Furthermore, we illustrate our findings with some examples in \cref{sec:examples}. Finally, in \cref{sec:discussion} we provide a discussion. All proofs are presented in the appendix.

%
%%
%%%
%%%%
%%%%%
%%%%%%
%%%%%%%
\section{General Set-Up and a Weak Convergence Result for Counting Process-Based Estimators}\label{sec:Notation}
% Weak convergence of wild bootstrap estimators based on general martingale arguments
%%%%%%%
%%%%%%
%%%%%
%%%%
%%%
%%
%

\noindent Let $N_1(t), \dots, N_n(t)$, $t\in\mathcal{T}$, be independent and identically distributed counting processes, where each individual counting process $N_i$, $i=1, \dots, n$, has in total $n_i$ jumps of size 1 at the observed event times $T_{i,1},\ldots,T_{i,n_i}$. Here, $\mathcal{T} = [0,\tau]$ is a finite time window. The multivariate counting process $(N_1,\ldots,N_n)$ containing all $n$ individual counting processes is denoted by $\textbf N(t)$, $t\in\mathcal{T}$, and it is assumed that no two counting processes $N_i$ jump simultaneously. 
The corresponding at-risk indicator for individual $i$ is denoted by $Y_i(t)$, $t\in\mathcal{T}$, $i=1, \dots, n$. The multivariate at-risk indicator $(Y_1,\ldots,Y_n)$ is denoted by $\textbf Y(t)$, $t\in\mathcal{T}$. Additionally, an individual $d$-variate covariate vector $\tilde{\textbf{Z}}_i(t)$, $t\in\mathcal{T}$, possibly time-dependent, may also be available for individuals $i=1, \dots, n$. In general, $\tilde{\textbf {Z}}_i$ is available only as long as $Y_i = 1$. The observable vector of covariates $\tilde{\textbf {Z}}_i Y_i$ is denoted by $\textbf Z_i(t)$, $t\in\mathcal{T}$, $i=1,\ldots,n$. 
The list of all $n$ observable covariate vectors each of dimension $d$ is denoted by $\textbf Z(t)$, $t\in\mathcal{T}$. 
%MdG
We  assume a parametric model for  the data $(\textbf N(t) ,\textbf Y(t) ,\textbf Z (t), t\in\mathcal{T})$, but our approach is suitable for  nonparametric or semiparametric models as well. In the case of a parametric regression model, a parameter coefficient $\bs{\beta} \in \mathbb{R}^q$ with $q\geq d$ contains the $d$-dimensional parameter coefficient that specifies the influence of the covariates $\textbf Z $ on the jump times of $\textbf N$, but additional parameters may be included in $\bs{\beta}$. If a nonparametric or semiparametric regression model is preferred, the set-up changes accordingly, cf.\ Examples~\ref{ex:application1} and \ref{ex:application4}. 
%MdG
Finally, $(\Omega,\mathcal{A},\mathbb{P})$ denotes the underlying probability space, and $\stackrel{\mathbb{P}}{\longrightarrow}$, $\stackrel{\mathcal{L}}{\longrightarrow}$ denote  convergence in probability and  convergence in law, respectively. 
%We omit the $\omega\in\Omega$ in the notation of the stochastic processes.
We usually write multivariate quantities in bold type and
  when we specify a stochastic quantity as finite, this is always to be understood as almost surely finite. 

In the present context, one is often interested in the estimation of a vector-valued stochastic function $\textbf X(t)$, $t\in\mathcal{T}$, of dimension $p$ by a counting process-based statistic of the form
\begin{align} \label{eq:Xn}
    \textbf X_n(t) =\frac1n \sum_{i=1}^n \int_0^t \textbf k_{n, i}(u,\hat {\bs{\beta}}_n) d N_i(u), \quad t\in\mathcal{T}, 
\end{align}
where the $p$-dimensional integrands $\textbf k_{n, i}(t, {\bs{\beta}})$ defined on $ \mathcal{T}\times\mathbb{R}^q$ are stochastic processes that are not necessarily independent, with $\textbf k_{n, i}(\cdot, {\bs{\beta}})$ locally bounded and predictable for ${\bs{\beta}} = {\bs{\beta}_0}$, and $\textbf k_{n, i}(t,\cdot)$ almost surely continuously differentiable in $\bs{\beta}$, $i=1,\ldots,n$. We assume that $\hat{\bs{\beta}}_n$ is a consistent estimator of the true model parameter $\bs{\beta}_0$ with 
\begin{align}
\label{eq:beta-Op}
    \hat {\bs{\beta}}_n - \bs{\beta}_0 = O_p(n^{-1/2}).
\end{align}
Additionally, we impose an assumption on the asymptotic representation of $\sqrt{n}(\hat {\bs{\beta}}_n-\bs{\beta}_0)$ for $n\to\infty$, which  will be specified later in this section. In other contexts, one may be interested in employing univariate test statistics of the form \eqref{eq:Xn} %$ \check{\bs X}_n(t) =\frac1n \sum_{i=1}^n \int_0^t \check{\bs k}_{n, \hat {\bs{\beta}}, i}(u) d N_i(u) $
to test a null hypothesis $H$ against an alternative hypothesis $K$. 
Obviously, useful estimation of the process $\textbf X$ is only achievable if the distribution of $\textbf X_n - \textbf X$ is appropriately analyzed, and approximated if necessary. Likewise for the null distribution of a test statistic $ X_n$ in the case of testing.

In the following, we focus on estimation in the situation in which the exact distribution of $\textbf X_n - \textbf X$ is unknown. Thus, the goal of this section is to determine the asymptotic distribution of the stochastic process $\sqrt{n}\big(\textbf X_n - \textbf X\big)$ for $n \to \infty$, which will be used in \cref{sec:generalBootstrap} to identify the wild bootstrap as a suitable approximation procedure.
A special feature of such counting process-based statistics is that they have a strong connection to martingales, and martingale theory can be used to analyze the asymptotic distribution. 
The connection to martingale theory is established by means of the Doob-Meyer decomposition, which links the counting process $N_i$ uniquely to the process  
\begin{equation} \label{eq:Doob}
M_i(t) = N_i(t) - \Lambda_i(t,\bs {\beta}_0),\quad t\in\mathcal{T},
\end{equation}
%in $\bs{\beta} = \bs{\beta}_0$ is a zero-mean martingale 
which is a martingale with respect to the filtration 
\[\mathcal{F}_1(t) = \sigma\{ N_i(u), Y_i(u), \textbf Z_{i}(u), 0\leq u\leq t, i=1,\ldots,n\}, \quad t \in \mathcal{T}.\]
The cumulative intensity process $\Lambda_i(t, \bs {\beta}_0)$ as introduced in \eqref{eq:Doob} is the compensator of $N_i(t)$, $t \in \mathcal{T}$; it is a non-decreasing predictable function in $t$ with $\Lambda_i(0, \bs {\beta}_0) = 0$, 
%and $\Lambda_i(\tau, \bs {\beta}_0) < \infty$, 
$i=1,\ldots,n$. 
%\cite{Andersen}[pp.73-74]. 
Additionally, we assume  $\Lambda_i(t, \bs {\beta}_0)$ to be absolutely continuous with rate process 
$\lambda_i = {\displaystyle \frac{d}{dt}}\Lambda_i$ and expected value  $E(\Lambda_i(\tau, \bs {\beta}_0)) < \infty$. %In other terms, \eqref{eq:Doob} can be written as $E( d N_i(t) | \mathcal{F}_1(t-) ) = d \Lambda_i(t,\bs{\beta}_0)$ for $t\in\mathcal{T}$ with $d \Lambda_i(t,\bs{\beta}_0)=  \lambda_i(t,\bs{\beta}_0)dt$, $i=1,\ldots,n$.
Furthermore, some event times may be unobservable due to independent right-censoring, left-truncation, or more general incomplete data patterns such as independent censoring on intervals.
% in the sense of \cite[Chapter~III]{Andersen}. 
These censoring mechanisms are captured by the at-risk function $Y_i$, $i=1,\ldots,n,$ and incorporated in the structure of the rate process by assuming that the individual counting process $N_i$ satisfies the multiplicative intensity model.
% cf.\ \cite[Section~III.1.3]{Andersen}. 
In particular, we assume for $i=1,\ldots,n,$
\begin{align*}
    \lambda_i(t,\bs{\beta}_0) = Y_i(t) \alpha_i(t,\bs{\beta}_0),\quad t\in\mathcal{T},
\end{align*}
where $\alpha_i ( \cdot,\bs{\beta}_0)$ is the hazard rate related to the events registered by the counting process $N_i$, and does not depend on the censoring or the truncation. In the case of a parametric or semiparametric model the hazard rate $\alpha_i ( t,\bs{\beta}_0)$ takes the form
 $\alpha_0(t,\bs{\beta}_{1;0} ) r(\bs{\beta}_{2;0}^\top\textbf Z_i(t))$ or $\alpha_0(t ) r(\bs{\beta}_0^\top\textbf Z_i(t))$, $t\in\mathcal{T}$, respectively, with $\bs{\beta}_0 = (\bs{\beta}_{1;0},\bs{\beta}_{2;0})$. 
Here, $r(\cdot)$ is some relative risk function and $\alpha_0(\cdot,\bs{\beta}_{1;0})$, respectively, $\alpha_0$ is the corresponding parametric or nonparametric baseline hazard function.
%We also assume that the counting processes $N_i(t)$, $i=1,\dots,n$, satisfy the multiplicative intensity model, cf.\ \cite[Section~III.1.3]{Andersen}:
%\begin{align*}
%    \lambda_i(t,\bs{\beta}_0) = Y_i(t) \cdot \alpha_i(t,\bs{\beta}_0).
%\end{align*}
%Here, $\alpha_i ( t,\bs{\beta}_0)$ is the hazard rate related to the events registered by the counting process $N_i(t),i=1,\ldots,n$, and it does not depend on the censoring or the truncation.
%As mentioned before, we assume a multiplicative model for the intensity process, namely $\lambda_i(t,\bs{\beta}_0) = Y_i(t)\cdot \alpha_i (t,\bs{\beta}_0)$, where $\alpha_i ( t,\bs{\beta}_0)$ is the hazard rate related to the events registered by the counting process $N_i(t),i=1,\ldots,n$. 
For a general reference on counting processes and the ingredients of the model that we introduced above, we refer to \cite{Andersen}.

We now focus on the derivation of an asymptotic representation for $\sqrt{n}\big(\textbf X_n - \textbf X\big)$ that plays a key role in deducing the corresponding asymptotic distribution. In this regard we make a number of assumptions. In Section~\ref{sec:examples} we will illustrate with some examples that these assumptions are commonly satisfied. We start by rewriting $\sqrt{n}\big(\textbf X_n - \textbf X\big)$ in basically two steps.
%aside from algebraic considerations. 
In particular, we consecutively apply the Doob-Meyer decomposition \eqref{eq:Doob} and a Taylor expansion around  $\bs{\beta}_0$. Here, we recall that, for fixed $t\in \mathcal{T}$, the integrands $\textbf k_{n, i}(t,\cdot)$ are almost surely continuously differentiable in $\bs{\beta}$, $i=1,\ldots,n$. 
We thus find for $t\in \mathcal{T}$
% $\sqrt{n}  (\textbf X_n(t) - \textbf X(t))$ equals  %This yields
\begin{align*}
%\begin{split}
  \label{eq:Xn-X_pre}
  &{\ }{}\sqrt{n}  (\textbf X_n(t) - \textbf X(t))  \qquad\qquad\\
 &={} \sqrt{n} \Big( \frac1n \sum_{i=1}^n \int_0^t \big[ \textbf k_{n, i}(u,\hat {\bs{\beta}}_n) - \textbf k_{n, i}(u, \bs{\beta}_0) + \textbf k_{n, i}(u, \bs{\beta}_0) \big] d N_i(u) - \textbf X(t) \Big)\\
     &{} = \sqrt{n} \Big(\frac1n \sum_{i=1}^n \int_0^t \textbf k_{n, i}(u,  \bs{\beta}_0) \big(d M_i(u) + d\Lambda_i(u,{\beta}_0)\big) - \textbf X(t)\\
     &{}\qquad +\frac1n \sum_{i=1}^n \int_0^t \big[ \textbf k_{n, i}(u,\hat {\bs{\beta}}_n) - \textbf k_{n, i}(u, \bs{\beta}_0) \big] d N_i(u) \Big)\\
     &{} = \sqrt{n} \Big(\frac1n \sum_{i=1}^n \int_0^t \textbf k_{n, i}(u,  \bs{\beta}_0) d M_i(u) \stepcounter{equation}\tag{\theequation}\\
     &{}\qquad + \frac1n \sum_{i=1}^n \int_0^t \textbf k_{n, i}(u,  \bs{\beta}_0)  d\Lambda_i(u,{\beta}_0) - \textbf X(t)\\
     &{}\qquad +\big(\frac1n \sum_{i=1}^n \int_0^t \textnormal{D} \textbf k_{n,i}(u, \bs{\beta}_0) d N_i(u) \big) (\hat {\bs{\beta}}_n-\bs{\beta}_0) +  o_p(\hat{\bs{\beta}}_n - \bs{\beta}_0)   \Big),
%\end{split}
\end{align*}
%where $\nabla_{\bs{\beta}_0} \textbf k_{n,i}(t, \bs{\beta}_0)$ 
where $\textnormal D \textbf f$ denotes the Jacobian of a function $\textbf f$
%of $\textbf k_{n,i}(t, \bs{\beta})$ 
with respect to $\bs{\beta}$.
%evaluated at $\bs{\beta}_0$. 
%MdG
For the next step we make the following regularity assumption:
%assume that the quantity of interest $\textbf X(t)$ is well estimated by $\textbf X_n(t)$ in the sense that 
\begin{align}\label{eq:intLambda-X=o_p}
    \frac{1}{n} \sum_{i=1}^n \int_0^t \textbf k_{n, i}(u, \bs{\beta}_0) d \Lambda_i(u, \bs {\beta}_0) - \textbf X(t)  = o_p(n^{-1/2}) \text{ for all } t\in\mathcal{T}. 
\end{align}
%In section~\ref{sec:examples}, we illustrate for some examples that \eqref{eq:intLambda-X=o_p} and the other assumptions further down are satisfied.
%Under this assumption, w
We now continue from the right hand side of the equality labeled by \eqref{eq:Xn-X_pre}, and with \eqref{eq:beta-Op} in combination with \eqref{eq:intLambda-X=o_p} we obtain for $t\in\mathcal{T}$
%that  $\sqrt{n}  (\textbf X_n(t) - \textbf X(t))$ equals
%\begin{equation}\label{eq:Xn-X}
%\begin{array}{r@{}l}
\begin{align}
\begin{split}
    \label{eq:Xn-X}
&{\ }{}\sqrt{n}  (\textbf X_n(t) - \textbf X(t)) \qquad\qquad\\
&{}= \frac{1}{\sqrt{n}} \sum_{i=1}^n \int_0^t \textbf k_{n, i}(u,  \bs{\beta}_0) d M_i(u)\\  
 &{}\qquad +  
    % \ + \  
\big(\frac1n \sum_{i=1}^n \int_0^t \textnormal D \textbf k_{n,i}(u, \bs{\beta}_0) d N_i(u)  \big) \sqrt{n}(\hat {\bs{\beta}}_n-\bs{\beta}_0)      + o_p(1),
\end{split}
\end{align}
%\end{array}
%\end{equation}
\noindent
where we denote the $(p\times q)$-dimensional counting process integral in \eqref{eq:Xn-X} by 
\begin{align}\label{notation:Bn}
    \textbf B_n(t) = \frac1n \sum_{i=1}^n \int_0^t \textnormal D \textbf k_{n,i}(u, \bs{\beta}_0) d N_i(u), \quad t \in \mathcal{T}.
\end{align}
%We assume that $\textbf B_n(t) $, $t\in\mathcal{T}$, converges in probability to a {\green fixed completely regular continuous} $q\times p$-dimensional function $\textbf B(t)$, as $n$ goes to infinity.
%{\green  We assume that $\textbf C_n(t)$ converges in probability to a $p$-dimensional, almost surely finite quantity $\textbf{C}(t)$ for all $t\in\mathcal{T}$, as n goes to infinity. This requirement would for example be met, if $\sup_{i\in\{1,\dots,n\},t\in\mathcal{T}} \lVert \nabla_{{\bs{\beta}_0}} \textbf k_{n,i}(t, {\bs{\beta}_0}) \rVert_\infty  =  O_p(1)$ is fulfilled.  }
%Under the assumption $\sup_{i\in\{1,\dots,n\},t\in\mathcal{T}} \lVert \nabla_{{\bs{\beta}_0}} \textbf k_{n,i}(t, {\bs{\beta}_0}) \rVert_\infty  =  O_p(1)$, we know that $\textbf C_n(t)$ converges in probability to a $p$-dimensional, almost surely finite quantity $\textbf{C}(t)$ uniformly in $t\in\mathcal{T}$, as n goes to infinity. 
Moreover, we assume the following asymptotic representation:
\begin{align}
\label{eq:beta_asy_lin}
 \sqrt{n}(\hat {\bs{\beta}}_n-\bs{\beta}_0) = \textbf C_n  \frac1{\sqrt{n}} \sum_{i=1}^n \int_0^\tau \textbf g_{n, i}(u, \bs{\beta}_0) d M_i(u) + o_p(1),
\end{align}
where $\textbf C_n$ is a $(q\times b)$-dimensional random matrix that we leave unspecified and the $b$-dimensional integrands $\textbf g_{n, i}(t,\bs{\beta})$ defined on $ \mathcal{T}\times\mathbb{R}^d$ are locally bounded stochastic processes that are predictable for $\bs{\beta} = \bs{\beta}_0$, $i=1,\ldots,n$. 
In \cref{remark:MLE_beta_sec2} at the end of this section, we illustrate why \eqref{eq:beta_asy_lin} is a natural condition.
Combining \eqref{eq:Xn-X}, \eqref{notation:Bn} and \eqref{eq:beta_asy_lin} we obtain the asymptotic representation of $\sqrt{n}  (\textbf X_n - \textbf X) $ we were aiming for, i.e.,
\begin{align}
\begin{split}
    \label{eq:Xn-X_new}
&{\ }{}\sqrt{n}  (\textbf X_n(t) - \textbf X(t)) \qquad\qquad\\
&{}= \frac{1}{\sqrt{n}} \sum_{i=1}^n \int_0^t \textbf k_{n, i}(u,  \bs{\beta}_0) d M_i(u)\\  
 &{}\qquad +  
\textbf B_n(t)  
 \textbf C_n  \frac1{\sqrt{n}} \sum_{i=1}^n \int_0^\tau \textbf g_{n, i}(u, \bs{\beta}_0) d M_i(u)     + o_p(1), \quad t\in\mathcal{T}.
\end{split}
\end{align}

\noindent
In view of the similar structure of the two martingale integrals displayed in \eqref{eq:Xn-X_new},
 %In order to analyze the latter with the help of martingale theory, we work with the corresponding stochastic process $\textbf D_{n, g}(t)$, $t\in\mathcal{T}$, instead of only evaluated at $\tau$. 
we introduced the joint $(p+b)$-dimensional stochastic process $\textbf{D}_{n, h} = (\textbf{D}_{n, k}^\top,\textbf{D}_{n, g}^\top)^\top$ with
\begin{align}
\label{eq:Dn}
    \textbf D_{n, h}(t) =   \frac{1}{\sqrt{n}} \sum_{i=1}^n \int_0^t \textbf h_{n, i}(u, \bs {\beta}_0) d M_i(u),\quad t\in\mathcal{T}, 
\end{align}
where the $(p+b)$-dimensional integrands $\textbf h_{n, i}(t,\bs{\beta} ) = (\textbf k_{n, i} (t,\bs{\beta} )^\top, \textbf g_{n, i} (t,\bs{\beta} )^\top)^\top$ defined on $ \mathcal{T}\times\mathbb{R}^q$ are locally bounded stochastic processes that are predictable for $\bs {\beta} = \bs {\beta}_0$, $i=1,\ldots,n$. 
In particular, $\textbf{D}_{n, h} $ is composed of the $p$-dimensional stochastic process $\textbf{D}_{n, k}$ and the $b$-dimensional stochastic process $\textbf{D}_{n, g}$
%{\vio [But can you---or do you want to---define $D_h$  it in this way when $D_k$ has argument $t$ and$ D_g$ has argument $\tau$?]}
%To be more precise,
%We treat such martingale integrals in a general manner and introduce the multi-dimensional stochastic process 
with which we denote the first and second martingale integral on the right hand side of \eqref{eq:Xn-X_new}. 
With this notation, \eqref{eq:Xn-X_new} becomes
\begin{equation}  \label{eq:Xn-X_inD}
\sqrt{n}  (\textbf X_n(t) - \textbf X(t)) 
= \textbf{D}_{n, k}(t) +\textbf B_n(t) \textbf C_n\textbf{D}_{n, g}(\tau) + o_p(1), \quad t\in\mathcal{T}.
\end{equation}
%\add{We will slightly abuse this notation by using different functions for $h$ and below we will re-use $h$ to define a concrete function.}

In order to derive the asymptotic distribution of the right-hand side of \eqref{eq:Xn-X_inD}, 
%i.e.\ of the right-hand side of $ \textbf{D}_{n, k}(\cdot) +\textbf B_n(\cdot)\cdot \textbf C_n\cdot\textbf{D}_{n, g}(\tau) $, 
we focus on the asymptotic distribution of its components $(\textbf{D}_{n, k},\textbf{D}_{n, g})$, $\textbf B_n$, and $\textbf C_n$ first. For this, we start by analyzing the joint asymptotic distribution of $\textbf{D}_{n, h} = (\textbf{D}_{n, k}^\top,\textbf{D}_{n, g}^\top)^\top$. 
%From now on, we write $\textbf{h}_{n,i}(t,{\bs{\beta}})^\top = (\textbf{k}_{n,i}(t,{\bs{\beta}})^\top, \textbf{g}_{n,i}(t,{\bs{\beta}})^\top)$ and $\textbf{D}_{n, h}(t)^\top = (\textbf{D}_{n, k}(t)^\top,\textbf{D}_{n, g}(t)^\top)$.
According to Proposition II.4.1 of \cite{Andersen}, $\textbf D_{n, h}$ is a local square integrable martingale with respect to $\mathcal{F}_1$. By the use of this property, we will show that under regularity conditions $\textbf D_{n, h}$ converges in law to a Gaussian martingale in $(D(\mathcal{T}))^{p+b}$, as $n\rightarrow \infty$.
% = \mathbb E ( \int_0^{s\wedge t} \tilde{\textbf{k}}_{1}(u,\bs {\beta}_0)^{\otimes 2} d\Lambda_1(u))$ in $(D(\mathcal{T}))^p$, as n tends to infinity. Here, $\tilde{\textbf{k}}_{1}(t,\bs {\beta}_0)$ is the limiting function of $\textbf k_{n, 1}(t, \bs {\beta}_0)$
Here, $(D(\mathcal{T}))^{p+b}$ is the space of cadlag functions in $\mathbb R^{p+b}$ equipped with the product Skorohod topology. In the sequel, the $p\times p$ matrix $\bs v \cdot \bs v^\top$ for some $\bs v\in\mathbb R^{p}$ will be denoted by $\bs v^{\otimes 2}$, $\lVert \cdot \rVert$ will denote a norm, e.g., the Euclidean norm, and $\mathcal{B}$  a neighborhood of $\bs{\beta}_0$. Furthermore, we need the following regularity assumptions.

\begin{ass}\label{assump_general}
 For each $i\in\mathbb N$ there exists a $(p+b)$-dimensional stochastic process $\tilde{\textbf{h}}_{i}(t,\bs{\beta})$ defined on $\mathcal{T}\times\mathcal{B}$ such that
%For the stochastic process $\textbf{h}_{n,i}(t,{\bs{\beta}})$ t
%We assume that
% Ich wuerde nicht nochmal erwaehnen, dass wir das annehmen, denn das steckt ja schon im Wort Assumption.
% Ausserdem brauchen wir ja manchmal nicht alle Annahmen.
\begin{thmlist}
%Moreover, , $t\in\mathcal{T}$ and $i=1,\ldots,n$, is
%\item The cumulative intensity processes $\Lambda_i(t,\bs{\beta}_0)$ are absolutely continuous, i.e.\ $\Lambda_i(dt,\bs{\beta}_0) = \lambda_i(t,\bs{\beta}_0) dt$;\label{ass:Lambda_cont}
% \item  and integrable, i.e. $\mathbb E(\Lambda_i(t,\bs{\beta}_0)) < \infty$; \label{item:ass_Bn_4}
\item  $\sup_{t\in\mathcal{T},i\in \{1,\ldots,n\}}\lVert \textbf{h}_{n,i}(t,\check{\bs{\beta}}_n) - \tilde{\textbf{h}}_{i}(t,\bs{\beta}_0)\rVert \stackrel{\mathbb P}{\longrightarrow} 0,$
as $n\rightarrow \infty$,
for any consistent estimator $\check{\bs{\beta}}_n$ of $\bs{\beta}_0$; %and for some
%There exist
%$(p+b)$-dimensional stochastic processes $\tilde{\textbf{h}}_i(t,\bs{\beta})$ defined on $\mathcal{T}\times\mathcal{B}$ such that
%for all $i\in\mathbb{N}$ such that; {\green h at beta-hat or beta-0? }
\label{assump_general1}
\item $\tilde{\textbf{h}}_{i}(t,\cdot)$ is a continuous function in $\bs{\beta} \in \mathcal{B}$ %uniformly in $t\in\mathcal{T}$ {\green double check}
and bounded on $\mathcal{T}\times\mathcal{B}$; %for all $i\in\mathbb{N}$;
\label{assump_general2}

%\item \add{\sout{$\textbf V_{ h}(t) = \int_0^t\mathbb E (  \tilde{\textbf{h}}_{1}(u,\bs {\beta}_0)^{\otimes 2} \lambda_1(u,\bs {\beta}_0))du$ is a $(p+b)\times (p+b)$-dimensional function with positive semi-definite increments;}} {\vio [explanation: see added sentence below Lemma 2.3.]}\label{assump_general4} 
%Furthermore, it holds that
\item the $(p+b+1)$-tuples $(\tilde{\textbf{h}}_{i}(t,\bs{\beta}_0),\lambda_i(t,\bs{\beta}_0))$, $i=1,\ldots,n$, are pairwise independent and identically distributed for all $t\in\mathcal{T}$.\label{assump_general3}
\end{thmlist}
\end{ass}

\noindent
We are now ready to formulate the following result on the limit in distribution of $\textbf{D}_{n,h}$.
\begin{lem}\label{lem:Dn}
If \cref{assump_general} holds, then
\[\textbf{D}_{n, h} \stackrel{\mathcal{L}}{\longrightarrow} \textbf{D}_{ \tilde{h}},\quad\text{in}\; (D(\mathcal{T}))^{p+b}, \text{ as } n\rightarrow \infty,\]
% hier habe ich wieder "a" statt "the" martingale geschrieben, da es mehrere gibt.
where $\textbf D_{\tilde{h}} = (\textbf D_{\tilde{k}}^\top,\textbf D_{\tilde{g}}^\top)^\top$ is a continuous zero-mean Gaussian $(p+b)$-dimensional vector martingale with  $\langle \textbf D_{\tilde{h}} \rangle (t) = \textbf V_{ \tilde{h}}( t) =\int_0^t \mathbb E (  \tilde{\textbf{h}}_{1}(u,\bs {\beta}_0)^{\otimes 2} \lambda_1(u,\bs {\beta}_0))du$, $t\in \mathcal{T}$. In particular, 
\[
\textbf V_{\tilde h}
=
\begin{pmatrix}
    \textbf V_{\tilde k} & \textbf V_{\tilde k , \tilde g} \\
    \textbf V_{\tilde g , \tilde k} & \textbf V_{\tilde g}
 \end{pmatrix},
\]
with 
$$  \textbf V_{\tilde k}(t) =  \langle \textbf D_{\tilde{k}} \rangle (t) = \int_0^{ t}\mathbb E (  \tilde{\textbf{k}}_{1}(u,\bs {\beta}_0)^{\otimes 2} \lambda_1(u,\bs {\beta}_0))du, \quad t\in\mathcal{T},$$ 
$$\textbf V_{\tilde g}(t) = \langle \textbf D_{\tilde{g}} \rangle (t) =  \int_0^{t}\mathbb E (  \tilde{\textbf{g}}_{1}(u,\bs {\beta}_0)^{\otimes 2} \lambda_1(u,\bs {\beta}_0))du, \quad t\in\mathcal{T},$$
and cross-covariance 
$$\textbf V_{ \tilde{k},\tilde{g}}( t) = \textbf V_{ \tilde{g},\tilde{k}}( t)^\top =  \langle \textbf D_{\tilde{k}}, \textbf D_{\tilde{g}} \rangle (t) =  \int_0^t\mathbb E (  \tilde{\textbf{k}}_{1}(u,\bs {\beta}_0) \tilde{\textbf{g}}_{1}(u,\bs {\beta}_0)^\top \lambda_1(u,\bs {\beta}_0))du , \quad t\in\mathcal{T}.$$ 
%{\vio [explanation: we changes the order of integration, i.e. integral and expectation, such that it becomes clear that $V_h$ is continuous by construction.]}
\begin{proof}
See Appendix. %\hfill\qedsymbol %\ref{proof:Dn}. 
\end{proof}
\end{lem}

\noindent
We note that $\textbf V_{\tilde h}(t)$, $t\in\mathcal{T}$, in \cref{lem:Dn} is by construction a continuous, deterministic and positive semidefinite matrix-valued function with $\textbf V_{\tilde h}(0)=0$.

Next, we study the limiting behaviour of the counting process integral $\textbf{B}_n$, and characterize the limit in probability of the random matrix $\textbf{C}_n$.
The following assumptions are required.
\begin{ass}\label{ass:Bn_Cn}
For each $i\in\mathbb N$ there exists a $(p\times q)$-dimensional stochastic process $\tilde{\textbf K}_{i}(t,\bs{\beta})$ defined on $\mathcal{T}\times\mathcal{B}$ 
%and there exists a $(q\times b)$-dimensional matrix $\textbf C$ 
such that 
\begin{thmlist}
    \item $\sup_{ t\in\mathcal{T},i\in\{1,\ldots,n\}} \lVert \textnormal D \textbf k_{n,i}(t,\check{\bs{\beta}}_n) -  \tilde{\textbf K}_{i}(t,\bs{\beta}_0)\rVert \stackrel{\mathbb{P}}{\longrightarrow} 0$, as $n\rightarrow \infty$, for any consistent estimator $\check{\bs{\beta}}_n$ of $\bs{\beta}_0$; \label{item:ass_Bn_1}
    \item  $\tilde{\textbf K}_{i} (\cdot,\bs{\beta}_0)$ is predictable w.r.t.\  $\mathcal{F}_1$ and bounded on $\mathcal{T}$;
%{\blue \sout{bounded on $\mathcal{T}\times\mathcal{B}$}} 
%for all $i\in\mathbb N $;
\label{item:ass_Bn_2}
    \item the $(p+q+1)$-tuples $(\text{vec}(\tilde{\textbf{K}}_{i}(t,\bs{\beta}_0)),\lambda_i(t,\bs{\beta}_0))$, $i=1,\ldots,n$, are pairwise independent and identically distributed for all $t\in\mathcal{T}$.\label{item:ass_Bn_3}
%    \item $\lVert \textbf C_n - \textbf C\rVert \stackrel{\mathbb P}{\longrightarrow} 0, \text{ as } n\rightarrow \infty$, and $\textbf C$ is deterministic.\label{item:Cn-C}
\end{thmlist}
\end{ass}

\noindent
The next lemma describes the  limiting behaviour of  $\textbf{B}_n$.
\begin{lem}\label{lem:Cn-C}
If \cref{ass:Bn_Cn}
%~\ref{item:ass_Bn_1}~-~\ref{item:ass_Bn_3} 
holds, then
\[\sup_{t\in\mathcal{T}} \lVert \textbf B_n(t) - \textbf B(t) \rVert \stackrel{\mathbb{P}}{\longrightarrow} 0, \text{ as } n\rightarrow\infty,\]
where $\textbf B(t)=\int_0^t \mathbb E( \tilde{\textbf K}_{1} (u,\bs{\beta}_0)  \lambda_1(u,\bs{\beta}_0) ) du$, $t\in\mathcal{T}$, is a $(p\times q)$-dimensional continuous, deterministic function.
\begin{proof}
See Appendix. %\hfill\qedsymbol %\ref{proof:Dn}. 
\end{proof}
\end{lem}

\noindent
With respect to the limiting behaviour of  $\textbf{C}_n$, we require the following.
\begin{ass} \label{ass:Cn-C}
There exists a $(q\times b)$-dimensional matrix $\textbf C$ such that
$$\lVert \textbf C_n - \textbf C\rVert \stackrel{\mathbb P}{\longrightarrow} 0, \text{ as } n\rightarrow \infty,$$ 
where $\textbf C$ is deterministic.
\end{ass}

Finally, we can state the limit in distribution of $\sqrt{n}  (\textbf X_n - \textbf X) $. For this, we combine the results we have obtained on the weak limits of $\textbf{D}_{n, h}$, and
$\textbf B_n$ with our assumption on that of  $\textbf C_n$. 
%In particular, we derive the asymptotic distribution of $ \textbf{D}_{n, k}(\cdot) +\textbf B_n(\cdot)\cdot \textbf C_n\cdot\textbf{D}_{n, g}(\tau) $.

\begin{thm}\label{thm:Xn-X_convergence}
If the asymptotic representation \eqref{eq:Xn-X_inD} %and \eqref{eq:beta_asy_lin} are
is fulfilled, and %\cref{assump_general} and \cref{ass:Bn_Cn} 
Assumptions~\ref{assump_general}, \ref{ass:Bn_Cn}, and \ref{ass:Cn-C}
hold, then,
\[\sqrt{n} \big(\textbf{X}_{n} - \textbf{X}\big)= \textbf D_{n, k} + \textbf B_n \textbf C_n  \textbf D_{n, g}(\tau) +o_p(1) \stackrel{\mathcal{L}}{\longrightarrow} \textbf{D}_{\tilde  k} + \textbf B  \textbf C  \textbf D_{\tilde g}(\tau), \text{ in } (D(\mathcal{T}))^p,% \text{ as } n\rightarrow \infty, 
\]
as $n\rightarrow \infty$, with $\textbf D_{ \tilde k}$ and $\textbf D_{\tilde  g}$ as in \cref{lem:Dn}, and $\textbf B$ as in \cref{lem:Cn-C}. 
Moreover, the matrix-valued variance function  
%\sout{predictable covariation process $\langle \textbf D_{\tilde k} + \textbf B  \cdot \textbf C\cdot \textbf D_{\tilde g}(\tau)\rangle (t) $}} 
%{\vio [explanation: the whole expression is in general not a martingale. Thus, we must not talk about predictable variation processes.]}
of $\textbf D_{\tilde k} + \textbf B  \textbf C \textbf D_{\tilde g}(\tau)$ is given as
\[
   t \mapsto \textbf V_{\tilde k}(t) + \textbf B(t)  \textbf C\textbf V_{\tilde g}(\tau) \textbf C^\top  \textbf B(t)^\top  +\textbf V_{\tilde k,\tilde g}(t) \textbf C^\top\textbf B( t)^\top  + \textbf B(t)  \textbf C \textbf V_{\tilde g, \tilde k}(t).
\]
%{\vio [I do not understand where the $\tau$ sits in $\textbf V_{\tilde g, \tilde k}(t)$: I see $t$ for $\tilde g$.]}
%are continuous zero-mean Gaussian $p$-dimensional and $b$-dimensional vector martingales with covariance functions $\bs\Sigma_{\tilde  k}(s,t)=\textbf V_{\tilde  k}(s\wedge t) = \mathbb E ( \int_0^{s\wedge t} \tilde{\textbf{k}}_{1}(u,\bs {\beta}_0)^{\otimes 2} d\Lambda_1(u,\bs {\beta}_0))$ and $\bs\Sigma_{\tilde  g}(s,t)=\textbf V_{\tilde  g}(s\wedge t) = \mathbb E ( \int_0^{s\wedge t} \tilde{\textbf{g}}_{1}(u,\bs {\beta}_0)^{\otimes 2} d\Lambda_1(u,\bs {\beta}_0))$, respectively,
%and cross-covariance function $\bs\Sigma_{\tilde k,\tilde g}(s,t)=\textbf V_{\tilde k, \tilde g}(s\wedge t) = \mathbb E ( \int_0^{s\wedge t} \tilde{\textbf{k}}_{1}(u,\bs {\beta}_0) \cdot \tilde{\textbf{g}}_{1}(u,\bs {\beta}_0)^\top d\Lambda_1(u,\bs {\beta}_0))$, for any $(s,t)\in \mathcal{T}^2$.
\begin{proof}
See the appendix. %\hfill\qedsymbol
\end{proof}
\end{thm}
\noindent
The proof of \cref{thm:Xn-X_convergence} is based on martingale theory which we will also use in Section~\ref{sec:generalBootstrap}.
For this we make
%In the upcoming section we will make 
use of the following  notation. Given a multi-dimensional vector of local square integrable martingales $\textbf H_n(t), t \in \mathcal{T}$, its predictable covariation process and its optional covariation process are denoted by $\langle \textbf H_n \rangle (t)$ and $[\textbf H_n](t)$, respectively. %In the Lindeberg condition  it is referred to the quantity $\sigma^\epsilon[H_n](t)$, which denotes the sum over all squared jumps of $H_n$ of size greater in absolute value than $\epsilon$ until time t, i.e. 
%\[\sigma^\epsilon[H_n](t)=\sum_{s\leq t} \left\vert\Delta H_n(s)\right\vert^2 \mathbbm{1}\{\left\vert\Delta H_n(s)\right\vert > \epsilon\},\, t\in \mathcal{T}.\]
Moreover, $\mathcal{L}(\textbf H_n)$ and $\mathcal{L}(\textbf H_n|\cdot)$ denote the law and the conditional law of $\textbf H_n$, respectively. Additionally, $d[\cdot,\cdot]$ is an appropriate distance measure between probability distributions, for example the Prohorov distance.

\begin{rem}\label{remark:MLE_beta_sec2}
To illustrate that   \eqref{eq:beta_asy_lin} is a a natural condition, we note that for
parametric models it is common practice to take the maximum likelihood estimator as the estimator  $\hat{\bs{\beta}}_n$ for estimating
 the true parameter $\bs{\beta}_0$. 
%In this context, the estimator $\hat{\bs{\beta}}_n$ is the solution of the likelihood equations. 
In \cite{Borgan} parametric survival models are  considered, where for $n$-variate counting processes $(N_1, \dots, N_n)$ the likelihood equations take the form
\begin{align*}
\sum_{i=1}^n \int_0^\tau \nabla\alpha_i(u,\bs{\beta}) \alpha_i(u,\bs{\beta})^{-1} dN_i(u) - \sum_{i=1}^n \int_0^\tau \nabla \alpha_i(u,\bs{\beta}) Y_i(u) du = 0,
\end{align*}
for some parametric functions $\alpha_i$, $i=1,\dots, n$, where $\nabla\alpha_i$ denotes the gradient of $\alpha_i$ with respect to $\bs {\beta}$.
Let us denote the left-hand side of the likelihood equations above by $\textbf U_n(\bs{\beta},\tau )$. 
Then  $ \textbf U_n(\bs{\beta},\cdot )$ evaluated at $\bs {\beta} = \bs{\beta}_0$ is a local square integrable martingale. In particular,
\[\textbf U_n(\bs{\beta}_0,\tau ) = \sum_{i=1}^n \int_0^\tau \frac{\nabla\alpha_i(u,\bs{\beta}_0)}{\alpha_i(u,\bs{\beta}_0)} dM_i(u),\]
as $\alpha_i(t,\bs{\beta}_0) Y_i(t) dt = d\Lambda_i(t,\bs{\beta}_0)$ is the compensator of $dN_i(t)$. %Moreover, we consider $\textbf U(\bs{\beta},\tau )$ as the derivative with respect to $\bs{\beta}$ of 
%\[
%C_n(\bs{\beta},\tau) = \sum_{i=1}^n \int_0^\tau \log(\alpha_i(u,\bs{\beta})) dN_i(u) - \sum_{i=1}^n \int_0^\tau \alpha_i(u,\bs{\beta}) Y_i(u) du. {\green Notation!}  
%\]
Under regularity conditions a Taylor expansion of $ \textbf U_n(\hat{\bs{\beta}}_n,\tau )$ around $\bs{\beta}_0$ yields 
\[
\sqrt{n}(\hat{\bs{\beta}}_n - \bs{\beta}_0 ) = -\Big(\frac{1}{n} D\textbf  U_n(\bs {\beta}_0,\tau)\Big)^{-1} \frac{1}{\sqrt{n}}\textbf U_n(\bs {\beta}_0,\tau ) + o_p(1).
\]
Thus, \eqref{eq:beta_asy_lin} holds with $\textbf g_{n, i}(u, \bs{\beta}_0) = \nabla\alpha_i(u,\bs{\beta}_0) \alpha_i(u,\bs{\beta}_0)^{-1}$ and $\textbf C_n = -\big(\frac{1}{n} D\textbf  U_n(\bs{\beta}_0,\tau)\big)^{-1}$, where
$$ D\textbf  U_n(\bs{\beta}_0,\tau) =   \sum_{i=1}^n \int_0^\tau \nabla^2\log (\alpha_i(u,\bs{\beta}_0)) dN_i(u) 
- \sum_{i=1}^n \int_0^\tau \nabla^2 \alpha_i(u,\bs{\beta}_0) Y_i(u) d u. $$
%{\blue \sout{can be identified with the optional covariation process of $U_n(\bs{{\beta}}_0, \cdot )$ at $\tau$. }}
Note that $-\frac1n D\textbf  U_n(\bs{\beta}_0,\tau)$ is asymptotically equivalent to the optional covariation process $- \frac1n [\textbf U_n(\bs{{\beta}}_0, \cdot )]$ of $-\frac{1}{\sqrt{n}}\textbf U_n(\bs{{\beta}}_0, \cdot )$ at $\tau$, which will be of use in \cref{remark:MLE_beta_sec3}. 
%{\vio [explanation: identified was the wrong word.]}
%Under the regularity assumptions stated in \cite{Borgan},  $\textbf C_n \stackrel{\mathbb P}{\longrightarrow} \textbf C$, where $\textbf C$ the covariance bf C
\end{rem}

%
%%
%%%
%%%%
%%%%%
%%%%%%
%%%%%%%
\section{The Wild Bootstrap for Counting Process-Based Estimators and a Weak Convergence Result}\label{sec:generalBootstrap}
%based on general martingale arguments}

%%%%%%%
%%%%%%
%%%%%
%%%%
%%%
%%
%

\noindent
In \cref{sec:Notation} we have introduced the counting process-based statistic $\textbf X_n$ given in \eqref{eq:Xn} as an estimator of the multidimensional function $\textbf X$. In the current section we use the wild bootstrap as an approximation procedure to recover the unknown distribution of $\textbf X_n - \textbf X$. The wild bootstrap counterpart of $\textbf X_n$ will be denoted by $\textbf X_n^*$. In order to verify the validity of the approximation procedure, we will prove that under regularity conditions the distributions of $\sqrt{n}(\textbf X_n - \textbf X)$ and $\sqrt{n}(\textbf X_n^* - \textbf X_n)$ are asymptotically equivalent. %In order to verify the validity of this approach, we show with \cref{thm:asyEquivalence} that the statistic $\textbf D_n(t)$ and its wild bootstrap counterpart $\textbf D^*_n(t)$ converge in law to the same Gaussian martingale, as n goes to infinity.
For this we will discover that  $\sqrt{n}(\textbf X_n^* - \textbf X_n)$ can be represented by an expression with the same structure as  
$\sqrt{n}(\textbf X_n - \textbf X) =  \textbf{D}_{n, k} +\textbf B_n \textbf C_n\textbf{D}_{n, g}(\tau) +o_p(1)$. 
Additionally, we will show with the proof of \cref{thm:asyEquivalence} that the joint distribution of the components involved in the representation of $\sqrt{n}(\textbf X_n^* - \textbf X_n)$ converges to the same asymptotic distribution as the joint distribution of the components of $\sqrt{n}(\textbf X_n - \textbf X)$. With the help of the continuous mapping theorem we then obtain the asymptotic equivalence of the distributions of $\sqrt{n}(\textbf X_n - \textbf X)$ and $\sqrt{n}(\textbf X_n^* - \textbf X_n)$.

In order to define the wild bootstrap estimator $\textbf X_n^*$, we first introduce the core idea of the wild bootstrap. Naturally, the realisations of $\textbf X_n$ vary with the underlying data sets. If we would have many data sets and thus many estimates, we could draw conclusions about the distribution of the estimator. The wild bootstrap provides for this: the variation immanent in the estimates arising from different data sets is produced by so-called random multipliers such that for this procedure only the one available data set $\{\textbf N(t) ,\textbf Y (t) ,\textbf Z (t), t\in\mathcal{T}\}$ is needed. In particular, the estimate calculated based on that data set is perturbed by random multipliers such that for each random multiplier a new estimate is created. Based on these so-called wild bootstrap estimates the distribution of the estimator can be inferred. Thus, the multiplier processes, denoted by $G_i(t)$, $t\in\mathcal{T}$, with $E(G_i) = 0$ and $E(G_i^2)=1$, $i=1,\dots,n$, lie at the heart of the wild bootstrap. They are random piecewise constant functions that we consider in further detail below. 
The construction of the wild bootstrap counterpart $\textbf X^*_n$ of $\textbf X_n$, $\textbf B^*_n$ of $\textbf B_n$, $\textbf C^*_n$ of $\textbf C_n$,  $\textbf D^*_{n,h}$ of $\textbf D_{n,h}$, or of any of the quantities that arise in this context, can be attributed to the following replacements: 
\begin{repl}\label{WB_replacement}\quad
%\newline
%\phantom{X}\\
\begin{thmlist}
    \item The square integrable martingale increment $dM_i(t)$ is replaced by the randomly perturbed counting process increment $G_i(t)dN_i(t)$, $i=1,\dots, n$;\label{item:WB_step1}
    \item the unknown increment of the cumulative intensity process $\Lambda_i(dt,\bs{\beta}_0)$ is replaced by 
%an estimator $d\hat{\Lambda}_i(dt,\hat{\bs{\beta}}_n)$
the estimator $dN_i(t)$, $i=1,\ldots ,n$;\label{item:WB_step2}
    \item the unknown parameter coefficient $\bs{\beta}_0$ is replaced by the estimator $\hat{\bs{\beta}}_n$;\label{item:WB_step3}
    \item we set all $o_p(1)$ terms in asymptotic representations to 0.\label{item:WB_step4.2}
\end{thmlist}
\end{repl}
Note that the substitution $ G_i(t) dN_i(t)$ of $dM_i(t)$, $t\in\mathcal{T}$, in \cref{WB_replacement}~\ref{item:WB_step1} is a square integrable martingale increment itself, given the data set, cf.\ \cref{lemma:mgale}. Moreover, for wider applicability  we chose in \cref{WB_replacement}~\ref{item:WB_step2}  the nonparametric estimator $dN_i(t)$  rather than a semiparametric estimator $\hat{\Lambda}_i(dt,\hat{\bs{\beta}}_n)$, $t\in\mathcal{T}$. As a consequence of \cref{WB_replacement}, we also replace the counting process increments $dN_i(t)$ in two steps. First, it is decomposed into $dM_i(t) + d\Lambda_i(t,\bs{\beta}_0)$ according to the Doob-Meyer decomposition given in \eqref{eq:Doob}. Second, \cref{WB_replacement}~\ref{item:WB_step1} and \ref{item:WB_step2} are applied. Step one and two combined yield 
$$\big(G_i(t) + 1 \big) dN_i(t), \quad t\in\mathcal{T}$$ 
as the replacement for $dN_i$. 
Furthermore, we obtain a wild bootstrap counterpart of $\hat{\bs{\beta}}_n $ via its asymptotic representation given in \eqref{eq:beta_asy_lin}. According to that equation we have
\begin{align}\label{eq:beta_4_beta_star}
    \hat{\bs{\beta}}_n = \bs{\beta}_0 +  \textbf C_n\frac{1}{{n}}\sum_{i=1}^n \int_0^\tau \textbf{g}_{n,i}(u,{\bs{\beta}}_0) dM_i(u) \ + \ o_p(1). 
\end{align}
In order to define the wild bootstrap counterpart $\hat{\bs{\beta}}^*_n$ of $\hat{\bs{\beta}}_n $, we replace $\textbf C_n$ by some $(q\times b)$-dimensional random matrix $\textbf C_n^*$ which is a wild bootstrap counterpart of $\textbf C_n$, and apply \cref{WB_replacement} to the other terms on the right hand side of \eqref{eq:beta_4_beta_star}. 
This yields
%definition of the replacement $\hat{\bs{\beta}}^*_n$ for the estimator $\hat{\bs{\beta}_n}$ by inserting in  \eqref{eq:beta_asy_lin}   $\hat{\bs{\beta}}^*_n$ for  $\hat{\bs{\beta}_n}$, and next  applying \cref{WB_replacement} to the other terms of this equation. This yields
% \begin{align} \label{eq:beta*}
%     \sqrt{n} (\hat{\bs{\beta}}^*_n - \hat{\bs{\beta}}_n ) = \textbf C_n^* \frac{1}{\sqrt{n}} \sum_{i=1}^n \int_0^\tau \textbf{g}_{n,i}(u,\hat{\bs{\beta}}_n) G_i(u)dN_i(u) \ + \ 0,
%\end{align}
%with some $(q\times b)$-dimensional random matrix $\textbf C_n^*$ which is a wild bootstrap counterpart of $\textbf C_n$. Thus, we use 
\begin{align}\label{eq:beta*}
    \hat{\bs{\beta}}^*_n  =   \hat{\bs{\beta}}_n + \textbf C_n^* \frac{1}{{n}} \sum_{i=1}^n \int_0^\tau \textbf{g}_{n,i}(u,\hat{\bs{\beta}}_n) G_i(u)dN_i(u).
\end{align}
%as the replacement for $\hat{\bs{\beta}}_n$.
Note that $\textbf C_n^*$ could take many different forms as long as it is asymptotically equivalent to $\textbf C_n$, 
i.e., as long as  $\lVert \textbf C_n^* - \textbf C_n \rVert = o_p(1)$ holds for $n\rightarrow\infty$, cf.\ \cref{ass:Cn_star-Cn}. When working with a particular model a natural choice for $\textbf C_n^*$ might be apparent as we shall demonstrate in \cref{remark:MLE_beta_sec3}.

We now consider the multiplier processes $G_i(t)$, $t\in\mathcal{T}, i=1,\ldots,n,$ in more detail. We define $G_i$ as a random piecewise constant function with jump time points identical to those of the counting process $N_i$,
i.e.,  at
%. The set of jump time points of $G_i(t)$ is denoted by 
\begin{align}\label{eq:jumpTimes}
    \mathcal{T}^\Delta_{n,i} = \{t \in \mathcal{T}: \Delta N_i(t) = 1\} = \{T_{i,1},\ldots,T_{i,n_i}\}.
\end{align} 
%see \cref{sec:Notation} on notation and general set-up.
We note that the number of jumps for the $i$-th process is the random number $n_i = N_i(\tau) \geq 0$.
Moreover, the multiplier processes $G_i$ are constructed such that at the jump time points $T_{i,j} \in \mathcal{T}^\Delta_{n,i}$ 
they take the values of 
%is equivalent to a sequence of 
i.i.d.\ random variables $G_{i,j}$, $j=1,2,\ldots$, that have  mean zero, unit variance and  finite fourth moment, and that are  independent of $\mathcal{F}_1(\tau)$.
%{\vio [can they be independent of $\mathcal{F}_1(\tau)$ while their number is $n_i=N_i(\tau$)?]}
In particular, $G_i(t)= 0 $ for $t < T_{i,1}$ and $G_i(t)= G_{i,j} $ for $T_{i,j} \leq t < T_{i,j+1}$, where $T_{i, n_i+1} = \infty$. 
Furthermore, the multiplier processes $G_1(t),\ldots,G_n(t)$, $t\in\mathcal{T}$, are pairwise independent and identically distributed.
Conditionally on $\mathcal{F}_1(\tau)$, however, their jump times are fixed and the identical distribution is lost.
See \cite{bluhmki18, bluhmki19} for similar approaches. 

Let us revisit \cref{WB_replacement} and the direct consequences of its application to $N_i$ and $\hat{\bs{\beta}}_n$. Due to the construction of the multiplier processes $G_i$, $i=1,\ldots,n$, the wild bootstrap replacement $\big(G_i + 1 \big) N_i$ varies vertically around $N_i$, i.e., the jump size deviates from 1, while the jump time points are fixed. A similar behaviour holds for the wild bootstrap estimator $\hat{\bs{\beta}}^*_n $ around $ \hat{\bs{\beta}}_n $, as we will see in \cref{lemma:mgale} that the integral on the right-hand side of \eqref{eq:beta*} is a zero-mean martingale evaluated at $t=\tau$. Finally, we obtain the wild bootstrap counterpart $\textbf{X}^*_n $ of $\textbf X_n$ by applying \cref{WB_replacement} to \eqref{eq:Xn} which results in the following definition
\begin{align} \label{eq:Xn*}
  \textbf{X}^*_n (t) = \frac{1}{n} \sum_{i=1}^n \int_0^t \textbf{k}_{n,i}(u,\hat{\bs{\beta}}^*_n)\big(G_i(u)+1 \big) dN_i(u), \quad t\in \mathcal{T}.
\end{align}
%\begin{align} \label{eq:beta*}
%  \sqrt{n} (\hat{\bs{\beta}}^* - \hat{\bs{\beta}} ) = \textbf C_n^*\cdot \frac{1}{\sqrt{n}} \sum_{i=1}^n \int_0^\tau \textbf{g}_{n,i}(u,\hat{\bs{\beta}}) G_i(u)dN_i(u) ,
%\end{align}
% =  \big( \frac{1}{n}\sum_{i=1}^n \int_0^\tau  \Check{\textbf g}_{n,i}(u,\hat{\bs{\beta}})  {\green \big( G_i(u) + 1\big ) dN_i(u)  / G_i(u)dN_i(u)} \big)^{-1}$. 
%where $\textbf X_n(t)$ is given in \eqref{eq:Xn}. 
Recall that the replacement of $\hat{\bs{\beta}}_n$ by $\hat{\bs{\beta}}^*_n$ can be traced back to \cref{WB_replacement} by first substituting $\hat{\bs{\beta}}_n$ in \eqref{eq:Xn*} by the right-hand side of \eqref{eq:beta_4_beta_star} and then applying \cref{WB_replacement} to the corresponding components.
Moreover, we point out that due to the fluctuation of $\big(G_i + 1 \big) N_i$ around $N_i$ and $\hat{\bs{\beta}}^*_n $ around $ \hat{\bs{\beta}}_n $, a reasonable amount of variation of the wild bootstrap estimator $\textbf{X}^*_n $ around $\textbf{X}_n$ is induced.
The remaining part of this section concerns the asymptotic behaviour of the wild bootstrap estimator $\textbf{X}^*_n $ around $\textbf X_n$.
% and is not directly relevant for the practical implementation of this approximation procedure. 

In order to study the asymptotic distribution of $\sqrt{n}\big(\textbf X_n^* - \textbf X_n \big)$, we start by deriving a representation of $\sqrt{n}\big(\textbf X_n^* - \textbf X_n \big)$ similar to the one stated in \eqref{eq:Xn-X_inD}. 
%{\red \eqref{eq:Xn} and \eqref{eq:Xn*}, respectively.}
For this, we rewrite $\sqrt{n}\big(\textbf X_n^* - \textbf X_n \big)$ as follows, i.e., for $t\in\mathcal{T}$ we have 
%\begin{equation}\label{eq:X*-Xn_1}
%\begin{array}{r@{}l}
\begin{align}
\begin{split}
    \label{eq:X*-Xn_1}
 \hspace{-0.4cm}&{\ }{}\sqrt{n}\big(\textbf X_n^*(t) - \textbf X_n(t) \big) \qquad\qquad\\
   \hspace{-0.4cm} & = \sqrt{n}\Big( \frac{1}{n} \sum_{i=1}^n \int_0^t \big [\textbf{k}_{n,i}(u,\hat{\bs{\beta}}^*_n) -  \textbf{k}_{n,i}(u,\hat{\bs{\beta}}_n) +  \textbf{k}_{n,i}(u,\hat{\bs{\beta}}_n) \big ]\big( G_i(u) + 1 \big) dN_i(u)\\
    \hspace{-0.4cm} &{} \quad -  \frac{1}{n} \sum_{i=1}^n \int_0^t \textbf{k}_{n,i}(u,\hat{\bs{\beta}}_n)dN_i(u)\Big)\\
     \hspace{-0.4cm} &{}= \sqrt{n}\Big( \frac{1}{n} \sum_{i=1}^n \int_0^t \big[ \textbf{k}_{n,i}(u,\hat{\bs{\beta}}_n) (G_i(u)+1) - \textbf{k}_{n,i}(u,\hat{\bs{\beta}}_n)\big] dN_i(u)\\
     \hspace{-0.4cm} &{} \quad +  \frac{1}{n} \sum_{i=1}^n  \int_0^t \big [ \textbf{k}_{n,i}(u,\hat{\bs{\beta}}^*_n) - \textbf{k}_{n,i}(u,\hat{\bs{\beta}}_n)\big ] (G_i(u)+1)dN_i(u)\Big)\\
    \hspace{-0.4cm} &{}=  \sqrt{n}\Big( \frac{1}{n} \sum_{i=1}^n \int_0^t  \textbf{k}_{n,i}(u,\hat{\bs{\beta}}_n) G_i(u) dN_i(u)\\
     \hspace{-0.4cm} &{} \quad + \frac{1}{n} \sum_{i=1}^n  \int_0^t \big [ \textbf{k}_{n,i}(u,\hat{\bs{\beta}}^*_n) - \textbf{k}_{n,i}(u,\hat{\bs{\beta}}_n)\big ] (G_i(u)+1)dN_i(u) \Big).
\end{split}
\end{align}
%     \end{array}
%\end{equation}
%\add{where $\textbf{k}_{n,i}(u,\hat{\bs{\beta}})$ has been added and subtracted as an integrand to the first term on the left-hand side of the first equality of \eqref{eq:X*-Xn_1}.}% we obtain 
%\begin{equation*}\label{eq:X*-Xn_2}
%\begin{array}{r@{}l}
Next, we apply a Taylor expansion  around $\hat{\bs{\beta}}_n$ to the second term on the right-hand side of the last equality of \eqref{eq:X*-Xn_1}. Here, we recall that, for fixed $t\in \mathcal{T}$, the $\textbf k_{n, i}(t,\cdot)$ are almost surely continuously differentiable in $\bs{\beta}$, $i=1,\ldots,n$.

The Taylor expansion yields
%\begin{equation}\label{eq:X*-Xn_3}
%\begin{array}{r@{}l}
\begin{align}
\begin{split}
\label{eq:X*-Xn_3}
         &{\ }{}\sqrt{n}\big(\textbf X_n^*(t) - \textbf X_n(t) \big) \qquad\qquad\\
         &{}= \sqrt{n}\big(\frac{1}{n} \sum_{i=1}^n  \int_0^t \textbf{k}_{n,i}(u,\hat{\bs{\beta}}_n)G_i(u)dN_i(u)\\
     &{} \quad + \big( \frac{1}{n} \sum_{i=1}^n  \int_0^t \textnormal D \textbf{k}_{n,i}(u,\hat{\bs{\beta}}_n)\big(G_i(u) + 1\big) dN_i(u) \big)  (\hat{\bs{\beta}}^*_n - \hat{\bs{\beta}}_n)+ o_p(\hat{\bs{\beta}}^*_n - \hat{\bs{\beta}}_n)\big) \\
  &{}= \sqrt{n}\big(\frac{1}{n} \sum_{i=1}^n  \int_0^t \textbf{k}_{n,i}(u,\hat{\bs{\beta}}_n)G_i(u)dN_i(u)
    +   \textbf B_n^*(t)(\hat{\bs{\beta}}^*_n - \hat{\bs{\beta}}_n)+ o_p(\hat{\bs{\beta}}^*_n - \hat{\bs{\beta}}_n)\big), 
\end{split}
\end{align}
%\end{array}
%\end{equation}
%where $\nabla_{\hat{\bs{\beta}}} \textbf{k}_{n,i}(u,\hat{\bs{\beta}})$ denotes the Jacobian matrix of $\textbf{k}_{n,i}(u,\bs{\beta})$ with respect to $\bs{\beta}$ evaluated at $\hat{\bs{\beta}}$. 
where %we denote the integral contained in the second term of \eqref{eq:X*-Xn_3} as
%we denote the following $p\times q$-dimensional stochastic process of the second term of \eqref{eq:X*-Xn_3} by $\textbf B_n^*(t)$ 
\begin{align}\label{eq:C*}
    \textbf B_n^*(t) =\frac1n \sum_{i=1}^n \int_0^t \textnormal D \textbf k_{n,i}(u, \hat{\bs{\beta}}_n) ( G_i(u) + 1 ) dN_i(u),\quad t\in \mathcal{T}. 
\end{align}
We thus retrieved $\textbf B_n^*$ as the wild bootstrap version of $\textbf B_n (t) = \frac1n \sum_{i=1}^n \int_0^t \textnormal D \textbf k_{n,i}(u, \bs{\beta}_0) d N_i(u)$, $t\in\mathcal{T}$,
%given in the second term of \eqref{eq:Xn-X}, as $\textbf B_n^*(t)$ can be derived from $\textbf B_n(t)$ 
%by following 
as if we had applied \cref{WB_replacement} directly to $\textbf B_n$. 
Finally, combining \eqref{eq:beta*} and \eqref{eq:X*-Xn_3},
%with \eqref{eq:X*-Xn_1}, \eqref{eq:X*-Xn_3}, and \eqref{eq:C*}, 
we obtain the following representation of $\sqrt{n}\big(\textbf X_n^* - \textbf X_n \big)$:
%\begin{equation}\label{eq:X*-Xn_4}
%\begin{array}{r@{}l}
\begin{align}
\begin{split}
    \label{eq:X*-Xn_4}
&{\ }{}\sqrt{n}\big(\textbf X_n^*(t) - \textbf X_n(t) \big) \qquad\qquad\\
     &{} = \frac{1}{\sqrt{n}} \sum_{i=1}^n  \int_0^t \textbf{k}_{n,i}(u,\hat{\bs{\beta}}_n)G_i(u)dN_i(u) \\
     &{} \quad + \textbf B_n^*(t)\textbf C_n^* \frac{1}{\sqrt{n}} \sum_{i=1}^n \int_0^\tau \textbf{g}_{n,i}(u,\hat{\bs{\beta}}_n) G_i(u)dN_i(u) + o_p(1),\quad t\in\mathcal{T}.
\end{split}
\end{align}
%     \end{array}
%\end{equation}
Indeed, as we will see later, $\hat{\bs{\beta}}^*_n - \hat{\bs{\beta}}_n = O_p(n^{-1/2})$.
Hence, $o_p(\hat{\bs{\beta}}^*_n - \hat{\bs{\beta}}_n) = o_p(1)$.
%Clearly, the term $o_p(1)$ on the right-hand side of \eqref{eq:X*-Xn_4} can be set to zero without changing the asymptotic distribution of $\sqrt{n}\big(\textbf X_n^*(t) - \textbf X_n(t) \big)$.
Additionally, we point out that the components of \eqref{eq:X*-Xn_4} are the wild bootstrap counterparts of the components specified in \eqref{eq:Xn-X_new}. In particular, the first term of \eqref{eq:X*-Xn_4} is the wild bootstrap counterpart of $\textbf{D}_{n,k} $ and the second term of \eqref{eq:X*-Xn_4} contains the wild bootstrap counterpart of 
$\textbf{D}_{n,g} $, both of which could also have been obtained by applying \cref{WB_replacement} directly to $\textbf{D}_{n,k} $ respectively $\textbf{D}_{n,}$. This leads us to the definition of the wild bootstrap counterpart $\textbf{D}^*_{n,h} = (\textbf{D}^{*\top}_{n,k},\textbf{D}^{*\top}_{n,g})^\top$ of $\textbf{D}_{n,h} $,
\begin{align} \label{eq:Dn*}
  \textbf{D}^*_{n,h} (t) = \frac{1}{\sqrt{n}}\sum_{i=1}^n \int_0^t \textbf{h}_{n,i}(u,\hat{\bs{\beta}}_n) G_i(u) dN_i(u),\quad t\in \mathcal{T},
\end{align}
where, as before,  $\textbf h_{n,i} = (\textbf k_{n,i}^\top, \textbf g_{n,i}^\top)^\top$. 
We assume that $\textbf h_{n,i}(t,\hat{\bs{\beta}}_n)$, $t\in\mathcal{T}$, 
%, evaluated at the estimate $\bs{\beta}=\hat{\bs{\beta}}_n$
is a known, $\mathcal{F}_1(\tau)$-measurable multi-dimensional function.
%that is completely specified by the data set $\{N(t) ,Y(t) ,\textbf Z (t), t\in\mathcal{T}\}$, and the chosen estimators $\textbf X_n(t)$ of $\textbf X(t)$ and $\hat{\bs{\beta}}$ of $\bs{\beta}_0$. 
We still need to specify a filtration that reflects the available information: (i) at time zero, all data are available from the resampling-point of view, i.e., $\mathcal{F}_1(\tau)$; (ii) during the course of time $t\in \mathcal{T}$, the wild bootstrap multiplier processes $G_i$ evolve.
Hence, the following filtration is a sensible choice:
%With respect to $\mathcal F_2$, the information on $N(t) ,Y(t)$ and $\textbf Z (t)$ for the entire time window $\mathcal{T}$ is available at time zero, whereas the multiplier process $G_i(t)$, $t\in\mathcal{T}$, is realized only in the course of time.
%Thus, the only unknown at time zero about $\textbf{D}^*_{n,h}$ is $G_i$. 
%Formally, in terms of the wild bootstrap, the available information at time $t\in\mathcal{T}$ is given by the filtration 
\[\mathcal{F}_2(t) =\sigma\{G_i(s), N_i(u), Y_i(u),  \textbf Z_{i}(u), 0<s\leq t, u\in\mathcal{T},  i=1,\ldots,n\}, \quad t\in \mathcal{T}.\]
%and $\textbf h_{n,i}(t,\hat{\bs{\beta}})$ is to be understood as $\mathcal{F}_2(0)$-measurable.
Note that $\mathcal{F}_2(0) = \mathcal{F}_1(\tau)$ represents the available data. From now on, the underlying filtered probability space is $(\Omega,\mathcal{A},\mathbb{P},\mathcal{F}_2)$. In the following lemma, we identify $\textbf{D}^*_{n,h} $ as a square integrable martingale with respect to the proposed filtration and state its predictable and optional variation process.
\begin{lem}\label{lemma:mgale}
%Given the data set $\{N(t) ,Y(t) ,\textbf Z (t), t\in\mathcal{T}\}$, the stochastic process $\textbf{D}_{n,h}^*(t)$, $t\in \mathcal{T}$,
$\textbf{D}_{n,h}^*$ is a 
%$p$-dimensional vector of 
square integrable martingale with respect to $\mathcal{F}_2$. \\
Moreover, its predictable and optional covariation processes are 
%of $\textbf D_{n,h}^*(t)$ is given as
\begin{align*}
   \langle \textbf{D}^*_{n,h} \rangle (t)= \frac{1}{n}\sum_{i=1}^n\int_0^t \textbf{h}_{n,i}(u,\hat{\bs{\beta}}_n)^{\otimes 2}\, dN_i(u), \ t\in\mathcal{T}, 
\end{align*}
and
\begin{align*}
   [\textbf{D}^*_{n,h}] (t)= \frac{1}{n}\sum_{i=1}^n\int_0^t \textbf{h}_{n,i}(u,\hat{\bs{\beta}}_n)^{\otimes 2} G^2_i(u) \, dN_i(u), \ t\in\mathcal{T},
\end{align*}
respectively.
\begin{proof}
See Appendix. %\hfill\qedsymbol
\end{proof}
\end{lem}

Next, we aim at deriving the asymptotic distribution of $\textbf{D}_{n,h}^* $ by making use of martingale theory. Recall that  $\textbf{D}_{n,h}^* $ is the wild bootstrap counterpart of $\textbf{D}_{n,h} $ defined in \eqref{eq:Dn}. In particular, $\textbf{D}_{n,h}$ is an integral with respect to a counting process martingale. To prove the convergence in distribution of $\textbf{D}_{n,h} $ in \cref{lem:Dn}, we used Rebolledo's martingale central limit theorem 
as stated in Theorem~II.5.1 of \cite{Andersen} for counting process martingales (see Appendix). Although it is tempting to apply this theorem  to $\textbf{D}_{n,h}^* $ as well, this does not work for the following reason.
In Theorem~II.5.1 of \cite{Andersen} the predictable covariation process of the process which contains all the jumps of the martingales that exceed in absolute value some $\epsilon>0$ is considered. Let us call this process the $\epsilon$-jump process. As we will see in \cref{ex:epsilon-jump-process}, the $\epsilon$-jump process of the wild bootstrap counterpart $\textbf{D}^*_{n,h}$ of $\textbf{D}_{n,h}$ is in general not a martingale.
Hence, it does not make sense to speak of  its predictable covariation process.
Consequently, the above-mentioned variant of Rebolledo's theorem cannot be used to analyze the asymptotic behaviour of the martingale $\textbf{D}_{n,h}^* 
$.

\begin{ex}
\label{ex:epsilon-jump-process}
Let us consider the case where $N_i \leq 1$ and the square integrable martingale ${D}_{n,h}^* $ with integrand $ h_{n, i}(t, \hat{{\beta}})\equiv 1 $, i.e., $D_{n,h}^*(t) = \frac{1}{\sqrt{n}} \sum_{i=1}^n  \int_0^t 1\cdot G_i dN_i(u)$, $t\in \mathcal{T}$,
and $G_i$ may be considered time-constant. 
Then, for the $\epsilon$-jump process $D_{n,h}^{\epsilon,*}(t) = \int_0^t \mathbbm{1}\{\lvert \Delta D_{n,h}^*(u)\rvert \geq \epsilon \} \cdot  D_{n,h}^*(du)$, $t\in\mathcal{T}$, we have
\begin{align*}
    & \mathbb E (D_{n,h}^{\epsilon,*}(t)|\mathcal{F}_2(s)) = \mathbb E \Big( \frac{1}{\sqrt{n}} \sum_{i=1}^n  \int_0^t  \mathbbm{1}\Big\{\Big\lvert \frac{1}{\sqrt{n}}\sum_{i=1}^n G_i \Delta N_i(u) \Big\rvert \geq \epsilon \Big\}  G_i dN_i(u)\Big|\mathcal{F}_2(s)\Big)\\
    & = D_{n,h}^{\epsilon,*}(s) + \frac{1}{\sqrt{n}} \sum_{i=1}^n  \int_s^t  \mathbb E \Big( \mathbbm{1}\Big\{\Big\lvert \frac{1}{\sqrt{n}}\sum_{i=1}^n G_i \Delta N_i(u)\Big \rvert \geq \epsilon \Big\}  G_i \Big|\mathcal{F}_2(s)\Big)dN_i(u)\\
    & = D_{n,h}^{\epsilon,*}(s) + \frac{1}{\sqrt{n}} \sum_{i=1}^n    \mathbb E \Big( \mathbbm{1}\Big\{\Big\lvert \frac{1}{\sqrt{n}} G_i \Big \rvert \geq \epsilon \Big\}  G_i\Big) (N_i(t) - N_i(s)) ,
\end{align*}
which is in general not equal to $D_{n,h}^{\epsilon,*}(s)$ if the zero mean random variables $G_1,\ldots,G_n$ follow an asymmetric distribution. Hence, $D_{n,h}^{\epsilon,*}(t)$, $t\in\mathcal{T}$, does not fulfill the martingale property for the multiplier processes $G_1,\ldots, G_n$ as defined above. 
\end{ex}

The non-applicability of the mentioned version of Rebolledo's theorem constitutes a gap in the literature that needs to be filled.
Even though one may argue in a different way why the $\epsilon$-jump process is asymptotically negligible and then draw conclusions for the convergence in law of a wild bootstrap-based martingale \citep{bluhmki19, dobler19}, it is of general interest to have a broadly applicable solution that makes ad hoc workarounds superfluous. As a solution, we revisit Rebolledo's original paper \cite{Rebolledo} to examine his Lindeberg condition which requires the squared $\epsilon$-jump process to converge to zero in $\text{L}_1$, as $n\rightarrow\infty$. We combine this easily accessible Lindeberg condition with Rebolledo's theorem for square integrable martingales by using the Lindeberg condition as a replacement for the rather technical ARJ(2) condition of that theorem; see also Proposition 1.5 of the same reference. For the sake of
completeness we now state this version of Rebolledo's theorem. %

\begin{thm}[Rebolledo's martingale central limit theorem, Theorem~V.1 of \cite{Rebolledo}]
\label{thm:rebolledo}
 Let $H_n$ be a locally square integrable zero-mean martingale which satisfies the Lindeberg condition, i.e.,  for each $\epsilon >0$ and $t\in \mathcal T$,
 \begin{align}
 \label{L-condition}
     \mathbb{E}(\sigma^\epsilon[H_n](t)) = \mathbb E\Big( \sum_{s\leq t} (\Delta H_n(s))^2 \mathbbm{1}\{|\Delta H_n (s)| > \epsilon\}\Big)\rightarrow 0, \quad \text{as} \; n\rightarrow\infty.
 \end{align}
  Consider the two following relations.
 \begin{enumerate}
     \item $\langle H_n \rangle (t) \stackrel{\mathbb{P}}{\longrightarrow} V(t) $, as $n\rightarrow\infty$, for all $t \in \mathcal{T}$,\label{thm:rebolledo_1}
     \item $[ H_n ](t)\stackrel{\mathbb{P}}{\longrightarrow} V(t) $, as $n\rightarrow\infty$, for all $t\in \mathcal{T}$.\label{thm:rebolledo_2}
 \end{enumerate}
 If \ref{thm:rebolledo_1} (respectively \ref{thm:rebolledo_2}) holds, then relation \ref{thm:rebolledo_2} (respectively \ref{thm:rebolledo_1}) is also valid and 
 \begin{align*}
     H_n \stackrel{\mathcal{L}}{\longrightarrow} H, \text{ in } D(\mathcal T), \text{ as } n\rightarrow \infty.
 \end{align*}
Here, $H$ denotes the $1$-dimensional Gaussian centered continuous martingale with covariance function $\Sigma(s,t) = V(s \wedge t)$, $(s,t)\in \mathcal{T}^2$, where $V(t) = \langle H \rangle (t)$ is a continuous increasing real function with $V(0) = 0$.
 \end{thm}

\noindent
We remark that Rebolledo considers one-dimensional martingales in the aforementioned paper. In contrast, we consider multi-dimensional martingales. 
To bridge this gap, we will make use of the Cram\'er-Wold theorem.
%For the bridge from the one-dimensional case to the p-dimensional case,
%we refer to Satz 3.19 of \cite{Markus_Dis}.
%\\~[DD: describe statement. Also, this is only relevant for the wild bootstrap.]\\
%Besides the condition on the jumps of the considered martingale, Rebolledo's martingale central limit theorem requires the predictable covariation process $\langle H_n\rangle (t)$ or the optional covariation process $[H_n](t)$ of the local square integrable martingale $H_n$ to converge in probability to a function $A(t)$ introduced at the end of \cref{sec:Notation}, as n goes to infinity. 
%\\~[DD: der vorige Satz ist zu lang. Auch der folgende Satz ist wohl nicht noetig, das sieht man ja direkt im Lemma. Ggf kuerzer umschreiben.]\\
%For the square integrable martingale $\textbf{D}_{n,h}^* $ we show in the following lemma under which assumptions the predictable covariation process $\langle \textbf{D}_{n,h}^*\rangle (t)$ of $\textbf{D}_{n,h}^* (t)$ converges in probability, as $n$ tends to infinity, to the matrix-valued function $\textbf V_h(t)$, which is the multidimensional equivalent of $A(t)$. 

The following lemma takes care of the convergence of the predictable covariation process of $\textbf{D}_{n,h}^* $, as required in Condition~\ref{thm:rebolledo_1} of \cref{thm:rebolledo}.

%To do: Dreieckschema LLN
\begin{lem}\label{lem:predCov_D*}
If \cref{assump_general} holds, then, conditionally  on $\mathcal{F}_2(0)$, 
\[\langle\textbf{D}_{n,h}^*\rangle (t) \stackrel{\mathbb{P}}{\longrightarrow} \textbf V_{\tilde h}(t),\text{ as } n\rightarrow \infty,\text{ for all } t\in\mathcal{T},\]
with $\textbf V_{\tilde h}$ as defined in \cref{lem:Dn}.
\begin{proof}
See Appendix. %\hfill\qedsymbol
\end{proof}
\end{lem}

\noindent
Based on the discussed theory, we study the convergence in law of the process $\textbf D^*_{n,h}$ in the proof of the upcoming \cref{lem:D*->D}.
 From Lemmas \ref{lem:Dn} and  \ref{lem:predCov_D*} it follows that the predictable variation process $\langle \textbf D^*_{n,h} \rangle  $ of $\textbf D^*_{n,h}$ converges to the same matrix-valued function $\textbf V_{\tilde h}$ as the predictable variation process $\langle \textbf D_{n,h} \rangle  $ of $\textbf D_{n,h}$. This gives rise to the supposition that those two processes converge in distribution to the same Gaussian martingale. In fact, we show that the conditional distribution of $\textbf D_{n,h}^*$ asymptotically coincides with the distribution of $\textbf D_{n,h}$.

\begin{lem} \label{lem:D*->D}
If \cref{assump_general} holds, then, conditionally on $\mathcal{F}_2(0)$, 
\[\textbf{D}_{n,h}^* \stackrel{\mathcal{L}}{\longrightarrow} \textbf{D}_{\tilde h},\quad\text{in}\; (D(\mathcal{T}))^{p+b}, \text{ as } n\rightarrow \infty \]
in probability, with $\textbf{D}_{\tilde h} = (\textbf{D}_{\tilde k},\textbf{D}_{\tilde g})$ as given in \cref{lem:Dn}. 
\begin{proof}
See Appendix. %\hfill\qedsymbol
\end{proof}
\end{lem}

\noindent
In the proof of \cref{lem:D*->D} in the appendix one can see that under \cref{assump_general}
%\remove{not only the predictable covariation process $\langle \textbf{D}_{n,h}^*\rangle $ of $ \textbf{D}_{n,h}^*$ converges to matrix-valued function $\textbf V_h$, as stated by} {\red \cref{lem:predCov_D*}}. 
%\remove{But additionally,} 
the stochastic process $\textbf{D}_{n,h}^* $ fulfills the Lindeberg condition.
%\remove{as required in }%, which implies the ARJ(2) condition required by 
%{\red \cref{thm:rebolledo}}. 
Thus, \cref{cor:optCov_D*} below is a direct consequence of \cref{thm:rebolledo} and \cref{lem:predCov_D*}. 
However, instead of employing \cref{thm:rebolledo} we provide an alternative proof of \cref{cor:optCov_D*} in the appendix based on Lenglart's inequality. 
%using Rebolledo's martingale central limit theorem, we also provide an alternative proof of \cref{cor:optCov_D*} in the appendix based on Lenglart's inequality.  

\begin{cor}\label{cor:optCov_D*}
If \cref{assump_general} holds, then, conditionally  on $\mathcal{F}_2(0)$,
\[[ \textbf{D}_{n,h}^*] (t) \stackrel{\mathbb{P}}{\longrightarrow} \textbf V_{\tilde h}(t),\text{ as } n\rightarrow \infty,\text{ for all } t\in\mathcal{T},\]
with $\textbf V_{\tilde h}$ as defined in \cref{lem:Dn}.
\begin{proof}
See Appendix. 
\end{proof}
\end{cor}

After having assessed the joint convergence in distribution of $\textbf D^*_{n,h} = (\textbf D^*_{n,k},\textbf D^*_{n,g})$ by means of \cref{lem:D*->D}, we focus again on the representation of $\sqrt{n}(\textbf X_n^* - \textbf X_n) = \textbf D^*_{n,k} + \textbf B^*_n \textbf C^*_n \textbf D^*_{n,g}(\tau) +o_p(1) $ given in \eqref{eq:X*-Xn_4} together with \eqref{eq:Dn*}. 
%It is basically the sum of two terms. The first term is $\textbf D^*_{n,k}$. The second term equals $\textbf B^*_n\cdot \textbf C^*_n\cdot \textbf D^*_{n,g}(\tau)$. 
%We assume that $\textbf B^*_n$ converges in probability to $\textbf B$, as $n$ tends to infinity. 
We first address the convergence of the components $\bs B_n^*$ and $\textbf C_n^*$ before we eventually consider the representation as a whole.
%we show that $\textbf B^*_n(t)$ converges in probability to the limiting function $\textbf B(t)$.

\begin{lem}\label{lem:Cn*-->C}
If \cref{assump_general}~\ref{assump_general3} and \cref{ass:Bn_Cn} %~\ref{item:ass_Bn_1}~-~\ref{item:ass_Bn_3}
 hold, then, conditionally on $\mathcal{F}_2(0)$,
\[\sup_{t\in\mathcal{T}} \lVert \textbf B^*_n(t) - \textbf B(t) \rVert \stackrel{\mathbb{P}}{\longrightarrow} 0, \text{ as } n\rightarrow\infty\]
with $\textbf B$ as in \cref{lem:Cn-C}.
%$=\int_0^t \mathbb E( \tilde{\textbf K}_{1} (u,\bs{\beta}_0)  \lambda_1(u,\bs{\beta}_0) ) du$, $t\in\mathcal{T}$, is a $(p\times q)$-dimensional continuous, deterministic function.
%there exists a $(p \times q)$-dimensional, almost surely finite {\red (or deterministic / the same as in Section 2? possible refer there directly)} matrix $\textbf B(t)$, such that
\begin{proof}
See Appendix.
\end{proof}
\end{lem}

\begin{ass}\label{ass:Cn_star-Cn}
Under \cref{ass:Cn-C} we further assume that the $(q\times b)$-dimensional random matrices $\textbf C_n$ and $\textbf C_n^*$ are asymptotically equivalent,
$$\lVert \textbf C_n^* - \textbf C_n \rVert \stackrel{\mathbb P}{\longrightarrow} 0, \quad n \rightarrow\infty.$$
\end{ass}

Finally, we are ready to derive the asymptotic distribution of $\sqrt{n}(\textbf X_n^* - \textbf X_n)$.

\begin{thm}
\label{thm:asyEquivalence}
If the representation \eqref{eq:X*-Xn_4} is fulfilled, and Assumptions \ref{assump_general},  \ref{ass:Bn_Cn}, \ref{ass:Cn-C}, and \ref{ass:Cn_star-Cn}
%~\ref{item:ass_Bn_1}~-~\ref{item:ass_Bn_3} 
hold, 
%and if $\lVert \textbf C^*_n - \textbf{\vio C} \rVert \stackrel{\mathbb P}{\longrightarrow} 0, \text{ as } n\rightarrow \infty$, 
then, conditionally on $\mathcal{F}_2(0)$,
\[\sqrt{n} \big(\textbf{X}^*_{n} - \textbf{X}_n\big)= \textbf D^*_{n, k} + \textbf B^*_n \textbf C^*_n  \textbf D^*_{n, g}(\tau) +o_p(1) \stackrel{\mathcal{L}}{\longrightarrow} \textbf{D}_{\tilde k} + \textbf B  \textbf C  \textbf D_{\tilde g}(\tau), \text{ in } (D(\mathcal{T}))^p,\]
in probability, as $n\rightarrow \infty$, with $\textbf{D}_{\tilde k}, \textbf D_{\tilde g}$, and $\textbf B$ as stated in \cref{lem:Dn} and \cref{lem:Cn-C}, respectively. % where $\textbf D_{ \tilde k}$ and $\textbf D_{\tilde g}$ as in \cref{thm:Xn-X_convergence}. %are continuous zero-mean Gaussian $p$-dimensional and $b$-dimensional vector martingales with covariance function $\bs\Sigma_{\tilde k}(s,t)=\textbf V_{ \tilde k}(s\wedge t)  = \textbf V_{ k}(s\wedge t) = \mathbb E ( \int_0^{s\wedge t} \tilde{\textbf{k}}_{1}(u,\bs {\beta}_0)^{\otimes 2} d\Lambda_1(u,\bs {\beta}_0))$ and $\bs\Sigma_{ g}(s,t)=\textbf V_{ g}(s\wedge t),\,(s,t)\in \mathcal{T}^2,$ with $\textbf V_{ g}(t) = \mathbb E ( \int_0^t \tilde{\textbf{g}}_{1}(u,\bs {\beta}_0)^{\otimes 2} d\Lambda_1(u,\bs {\beta}_0))$, respectively.
%
%If additionally \cref{ass:Cn-C} holds, then
If additionally \eqref{eq:Xn-X_inD} is satisfied, we have
\[d[\mathcal{L}(\sqrt{n}(\textbf{X}_{n}^*-\textbf{X}_{n})|\mathcal{F}_2(0)),\mathcal{L}(\sqrt{n}(\textbf{X}_{n}-\textbf{X}))]\stackrel{\mathbb P}{\longrightarrow} 0, \text{ as } n\rightarrow \infty. \]

%, \text{ on } \mathcal{T}.\]
\begin{proof}
See Appendix. 
\end{proof}
\end{thm}

\noindent
In conclusion, with \cref{thm:asyEquivalence} we verify the asymptotic validity of the wild bootstrap as an appropriate approximation procedure for counting process-based statistics of the form given in \eqref{eq:Xn}.

 \begin{rem}\label{remark:MLE_beta_sec3}
We continue \cref{remark:MLE_beta_sec2} in order to illustrate how to choose the wild bootstrap counterpart $\textbf C_n^*$ of $\textbf C_n$ in parametric survival models such that \ref{ass:Cn_star-Cn} holds. In this way, we underline the wild bootstrap as an alternative to the parametric bootstrap. As stated in \cref{remark:MLE_beta_sec2}, $\textbf C_n$ is asymptotically related to the optional covariation process $\frac{1}{n}[\textbf U_n(\bs {\beta}_0, \cdot)]$ of $ \frac{1}{\sqrt{n}}\textbf U_n(\bs {\beta}_0, \cdot)$.
Hence, we propose to choose $\textbf C_n^*$ similarly based on the optional covariation process $\frac{1}{n}[\textbf U_n^*(\hat {\bs {\beta}}_n, \cdot)]$ of the wild bootstrap version $\frac{1}{\sqrt{n}}\textbf U_n^*(\hat {\bs {\beta}}_n, \cdot)$ of the martingale $\frac{1}{\sqrt{n}}\textbf U_n(\bs {\beta}_0, \cdot)$. Application of \cref{WB_replacement} to  $\frac{1}{\sqrt{n}}\textbf U_n(\bs {\beta}_0, \cdot)$ yields
$$ \textbf D_{n,g}^*(\tau) = \frac{1}{\sqrt{n}}\textbf U_n^*(\hat {\bs {\beta}}_n, \tau) =  \frac{1}{\sqrt{n}}\sum_{i=1}^n \int_0^\tau \frac{\nabla\alpha_i(u,\hat{\bs{\beta}}_n)}{\alpha_i(u,\hat{\bs{\beta}}_n)}G_i(u) dN_i(u).$$
According to \cref{lemma:mgale} we obtain the following structure:
$$ \textbf C_n^* = \big(-\frac{1}{n} [\textbf U_n^*(\hat {\bs {\beta}}_n, \cdot)](\tau)\big)^{-1} = -\Big(\frac{1}{n} \sum_{i=1}^n \int_0^\tau \frac{(\nabla \alpha_{i}(u,\hat{\bs{\beta}}_n))^{\otimes 2}}{\alpha_{i}(u,\hat{\bs{\beta}}_n)^2} G_i^2(u)dN_i(u) \Big)^{-1}.$$
This is a natural choice for $\textbf C_n^*$ in the present context, because under regularity conditions the (conditional) distributions of $\textbf D_{n,g}^*$ and $\textbf D_{n,g}=\frac{1}{\sqrt{n}}\textbf U_n(\bs {\beta}_0, \cdot)$ are asymptotically equivalent and the same holds for their optional covariation processes, cf.\ \cref{lem:Dn} and \cref{lem:D*->D} in combination with \cref{thm:rebolledo}.
\end{rem}   

%{\vio [In this section I see everywhere  in lemmas etc, `conditionally on $\mathcal{F}_2(0)$' . Two questions about this:\\
%i) $\mathcal{F}_2(0)$    is equal to $\mathcal{F}_1(\tau)$, so one is inclined to think: why is $\mathcal{F}_2(0)$ needed? 
%I do see $\mathcal{F}_2(t)$,  for $t$ other than 0, in the proofs, so it does seem to be needed, but why then does it occur in the lemma's/theorems only as $\mathcal{F}_2(0)$ and not as $\mathcal{F}_2(t)$? \\
%ii) shouldn't then also in the lemmas etc of Section 2, be added  `conditionally on $\mathcal{F}_1(t?)$' ?]}

%
%%
%%%
%%%%
%%%%%
%%%%%%
%%%%%%%
\section{Examples}
%%%%%%%
%%%%%%
%%%%%
%%%%
%%%
%%
%
\label{sec:examples}

We will now present a series of examples,  which is by no means exhaustive, of specific cases of the general set-up described in Sections \ref{sec:Notation} and \ref{sec:generalBootstrap}. In particular, it is briefly outlined how the theory developed in this Part~I can be applied to these models.
%In the examples we will use the notation that is common in that particular setting.
In Part~II we apply the present approach to the Fine-Gray model under censoring-complete data and  work out the details of the wild bootstrap for this specific model.
\begin{ex}\label{ex:application1}
 \textnormal{(Nelson-Aalen estimator)}
Let $X(t)= A(t) = \int_0^t \alpha(u) du$, $t \in \mathcal{T}$, be the cumulative hazard function of a continuous survival time $T$, i.e., $\alpha(u)du = \mathbb{P}(T \in [u,u + du] | T \geq u)$.
Let $N_1(t), \dots, N_n(t)$, $t\in\mathcal{T}$, be the counting processes that are related to $n$ independent copies of $T$ which possibly involve right-censoring.
For  $\hat X_n(t)$, $t \in \mathcal{T}$, we take the Nelson-Aalen estimator $\hat A_n(t) = {\displaystyle \sum_{i=1}^n \int_0^t } {\displaystyle \frac{J(u) }{Y(u)}}d N_i(u)$, $t\in\mathcal{T}$, \cite{aalen78}, where $Y_i(t)$ is the at-risk indicator for individual $i$ at time $t$, $Y(t)= \sum_{i=1}^n Y_i(t)$, and $J(t) = \mathbbm{1}\{Y(t) > 0 \}$. Thus, the counting process-based estimator $\hat A_n$ exhibits the general structure stated in \eqref{eq:Xn} with $k_{n}(t) = \tfrac{n  J(t)}{Y(t)}$, $t\in\mathcal{T}$. Furthermore, we have for $t\in\mathcal{T}$,
\begin{align}\label{eq:exampl_NA_1}
    \sqrt{n}(\hat A_n(t) - A(t)) = \sqrt{n}\sum_{i=1}^n \int_0^t \frac{J(u) }{Y(u)}(d N_i(u) - d \Lambda_i(u)) + \sqrt{n}\int_0^t (J(u) -1 ) d A(u),
\end{align}
where $d\Lambda_i = Y_i d A $. As the integrand $k_{n} = \tfrac{n  J}{Y}$ %{\green index i redundant and it is also not a fct in beta} 
is bounded by $J$ and predictable due to the predictability of $Y$, the first term on the right-hand side of \eqref{eq:exampl_NA_1} is a local square integrable martingale. This martingale refers to $D_{n,k}$, cf.\ \eqref{eq:Dn}. The second term on the right-hand side of \eqref{eq:exampl_NA_1} is asymptotically negligible as $n \to \infty$, because $J(t) \stackrel{\mathbb P}{\longrightarrow} 1 $ as $n \to \infty$, $t\in\mathcal{T}$. Hence, \eqref{eq:intLambda-X=o_p} is satisfied.
Furthermore, we make the natural assumption that there exists a deterministic function $y$, which is bounded away from zero on $\mathcal{T}$ and such that
\begin{align}\label{eq:ass_risk_set}
        \sup_{t\in\mathcal{T}}\big\lvert \frac{Y(t)}{n} - y(t)\big\rvert = o_p(1).
\end{align}
This weak assumption implies \cref{assump_general}. 
Moreover, we deal with a nonparametric model and as such we have for $t \in \mathcal{T}$,
$D k_{n}(t) \equiv 0 $. This implies that \cref{ass:Bn_Cn} is trivially satisfied and that $\textbf B_n \equiv 0$. Additionally, due to the nonparametric model, the assumption on the asymptotic representation of the parameter estimator stated in \eqref{eq:beta_asy_lin} is superfluous and we set $ \textbf C_n=0 $ and $ \textbf D_{n,g}(\tau )= 0$. Therefore, also Assumptions \ref{ass:Cn-C} and \ref{ass:Cn_star-Cn} are redundant. In conclusion, we point out that for the normalized Nelson-Aalen process $\sqrt{n}(\hat A_n - A)$ stated in \eqref{eq:exampl_NA_1} the asymptotic representation \eqref{eq:Xn-X_inD} holds with $\textbf B_n  \textbf C_n  \textbf D_{n,g}(\tau )\equiv 0$, i.e., $\sqrt{n}(\hat A_n - A) = D_{n,k} + o_p(1)$.
%$D k_{n, i}(t, \bs {\beta}_0) \equiv 0 $, 
%and thus, $\textbf B_n  \textbf C_n  \textbf D_{n,g}(\tau )\equiv 0$. Thus, Assumptions \ref{ass:Bn_Cn}, \ref{ass:Cn-C}, \ref{ass:Cn_star-Cn}, and \eqref{eq:beta_asy_lin} 
%$D k_{n, i}(t, \check{\bs {\beta}})$ 
%are trivially satisfied. 
According to \cref{WB_replacement}, the 
wild bootstrap version of the normalized Nelson-Aalen process is
\begin{align*}
   \sqrt{n}(\hat A^*_n(t) - \hat A_n(t)) &= \sqrt{n} \big(\sum_{i=1}^n  \int_0^t \frac{J(u)}{Y(u)} (G_i + 1)dN_i(u) - \sum_{i=1}^n \int_0^t \frac{J(u)}{Y(u)} dN_i(u)\big) \\
   &= \sqrt{n} \sum_{i=1}^n  \int_0^t \frac{J(u)}{Y(u)} G_i dN_i(u) , \quad  t \in \mathcal{T},
\end{align*}
where the term on the right-hand side of the second equality of the equation above refers to $D_{n,k}^*$, cf.\ \eqref{eq:Dn*}. Thus, also \eqref{eq:X*-Xn_4} holds with $\textbf B^*_n  \textbf C^*_n  \textbf D^*_{n,g}(\tau )\equiv 0$ and $o_p(1)$ set to zero, i.e., $\sqrt{n}(\hat A^*_n - \hat A_n) =D_{n,k}^*$. Note, that the multipliers $G_i$ can be chosen time-independent, $i=1,\ldots , n$. Finally, \cref{thm:asyEquivalence} can be used to justify the wild bootstrap as a suitable resampling method for the Nelson-Aalen estimator. In particular, the (conditional) distributions of $\sqrt{n}(\hat A_n(t) - A(t))$ and $\sqrt{n}(\hat A^*_n(t) - \hat A_n(t))$ are asymptotically equivalent.
Furthermore, similar structures hold for more general multivariate Nelson-Aalen estimators in not necessarily survival set-ups, except that the multiplier processes might be time-dependent \citep{bluhmki19}.
\end{ex}

\begin{ex}\label{ex:application3} 
\textnormal{(Weighted logrank test)}
    The two-sample weighted logrank statistic is 
    \begin{align}\label{eq:example_log_rank_1}
    \begin{split}
       T_{n_1,n_2}(w) &= \sqrt{\frac{n_1+ n_2}{n_1 n_2}} \int_0^\infty w(\hat S_n(t-)) \frac{Y^{(1)}(t) Y^{(2)}(t)}{Y(t)} (d\hat A^{(1)}_{n}(t) - d \hat A^{(2)}_{n}(t))\\
    &= \frac{1}{\sqrt{n_1}} \sum_{i=1}^{n_1}\int_0^\infty \sqrt{\frac{n_1+ n_2}{n_2}} w(\hat S_n(t-)) \frac{Y^{(2)}(t) }{Y(t)}dN_i^{(1)}(t) \\
    & \quad - \frac{1}{\sqrt{n_2}} \sum_{i=1}^{n_2}\int_0^\infty \sqrt{\frac{n_1+ n_2}{n_1}} w(\hat S_n(t-)) \frac{Y^{(1)}(t) }{Y(t)}dN_i^{(2)}(t) ,
    \end{split}
    \end{align}
    where $\hat A^{(j)}_{n}$ are the Nelson-Aalen estimators, $N^{(j)}_i$, $i=1,\ldots , n$, the counting processes, and $Y^{(j)}$ the at-risk counters in samples $j=1,2$, $n_1,n_2$ are the sample sizes, $Y=Y^{(1)} + Y^{(2)}$, $w$ is a positive weight function, and $\hat S_n$ is the Kaplan-Meier estimator \citep{kaplan58} in the pooled sample, cf., e.g., \cite{ditzhaus20} who conducted weighted logrank tests as permutation tests and \cite{ditzhaus19} who used the wild bootstrap. Hence, $T_{n_1,n_2}(w) $ is the sum of two counting process-based statistics, say, $X^{(1)}_{n_1,n_2}(\infty )$ and $ X^{(2)}_{n_1,n_2}(\infty )$ of a form similar to the one given in \eqref{eq:Xn} evaluated at the upper integration bound $\infty$, where the integrand of the statistic $X^{(1)}_{n_1,n_2}(\infty )$ equals $k^{(1)}_{n_1,n_2}(t) = { \displaystyle \sqrt{\frac{n_1+n_2}{n_2}} w(\hat S_n(t-)) \frac{Y^{(2)}(t)}{Y(t)}}$ and the integrand of the statistic $X^{(2)}_{n_1,n_2}(\infty )$ equals $ k^{(2)}_{n_1,n_2}(t) ={ \displaystyle - \sqrt{\frac{n_1+n_2}{n_1}} w(\hat S_n(t-)) \frac{Y^{(1)}(t)}{Y(t)}}$, $t\geq 0$.

    Under the null hypothesis of equal hazards or, equivalently, equal survival functions, $H_0: A^{(1)} = A^{(2)}$, we have 
    \begin{align}\label{eq:example_logRank_2_mgale}
    \begin{split}
    &Y^{(2)}\sum_{i=1}^{n_1}dN_i^{(1)} - Y^{(1)}\sum_{i=1}^{n_2}dN_i^{(2)} \\
    &= Y^{(2)}\big(\sum_{i=1}^{n_1}dM_i^{(1)}+Y^{(1)}dA^{(1)}\big) - Y^{(1)}\big(\sum_{i=1}^{n_2}dM_i^{(2)}+Y^{(2)}dA^{(2)}\big)\\
    &\stackrel{H_0}{=} Y^{(2)}\sum_{i=1}^{n_1}dM_i^{(1)} - Y^{(1)}\sum_{i=1}^{n_2}dM_i^{(2)},
    \end{split}
\end{align} 
where we have applied the Doob-Meyer decomposition in the first step of \eqref{eq:example_logRank_2_mgale} (cf.\ \eqref{eq:Doob}), and $M^{(j)}_{i}$, $i=1,\dots, n_j$, are the sample $j$-specific counting process martingales.
    %$d\hat A^{(1)}_{n}(t) - d \hat A^{(2)}_{n}(t) = \frac{J^{(1)}(t)}{Y^{(1)}}\sum_{i=1}^{n_1}dM^{(1)}_i - \frac{J^{(2)}(t)}{Y^{(2)}}\sum_{i=1}^{n_2}dM^{(2)}_i  + (J^{(1)}(t) - J^{(2)}(t))dA(t)$ and $(J^{(1)}(t) - J^{(2)}(t))dA(t) = o_p(1)$, where $J^{(j)}(t) = \mathbbm{1}\{Y^{(j)}(t)>0\}$, $j=1,2$.
    Due to \eqref{eq:example_logRank_2_mgale}, the test statistic $ T_{n_1,n_2}(w)$ has the following form %asymptotically equivalent 
    %martingale representation 
    under the null hypothesis:
    \begin{align}\label{eq:xample_logRank_mgale}
    \begin{split}
       T_{n_1,n_2}(w) &\stackrel{H_0}{=} %\sum_{j=1}^2 (-1)^{j+1}
       \frac{1}{\sqrt{n_1} }  \sum_{i=1}^{ n_1 } \int_0^\infty \sqrt{\frac{n_1 + n_2}{n_2}}w(\hat S_n(t-)) \frac{Y^{(2)}(t)}{Y(t)}  d M^{(1)}_{i}(t) \\
       &\quad - \frac{1}{\sqrt{n_2} }  \sum_{i=1}^{ n_2 } \int_0^\infty \sqrt{\frac{n_1 + n_2}{n_1}}w(\hat S_n(t-)) \frac{Y^{(1)}(t)}{Y(t)}  d M^{(2)}_{i}(t).
    \end{split}
    \end{align}
    %In this case, we have that $k_{n,j}(t)$ equals either ${ \displaystyle \sqrt{\frac{n_1+n_2}{n_1}} w(\hat S_n(t-)) \frac{Y^{(2)}(t)}{Y(t)}}$ ($j=1$)  or $ { \displaystyle - \sqrt{\frac{n_1+n_2}{n_1}} w(\hat S_n(t-)) \frac{Y^{(1)}(t)}{Y(t)}}$ ($j=2$).  
Under regularity conditions on the weight function and the sample sizes
{\rm (}$ { \displaystyle \frac{n_j}{n_1+n_2}\rightarrow\nu_j }$ as $\min(n_1,n_2) \rightarrow\infty$,  with $\nu_j \in (0,1)$, $j=1,2${\rm )}, 
the stochastic processes $k^{(j)}_{n_1,n_2}$, $j=1,2$, are uniformly bounded on any interval $\mathcal T = [0,\tau]$. Clearly, they are also predictable. Thus, under $H_0$, the test statistic can be written as the sum of two local square integrable martingales of a form similar to the one given in \eqref{eq:Dn} evaluated at the upper integration bound $\infty$, i.e., $T_{n_1,n_2}(w) \stackrel{H_0}{=} D_{n_1,n_2,k^{(1)}}(\infty) + D_{n_1,n_2,k^{(2)}}(\infty) $, where the local square integrable martingale $D_{n_1,n_2,k^{(1)}}(t)$, $t\geq 0$, relates to the first term on the right-hand side of \eqref{eq:xample_logRank_mgale} and the local square integrable martingale $D_{n_1,n_2,k^{(2)}}(t)$, $t\geq 0$, relates to the second term on the right-hand side of \eqref{eq:xample_logRank_mgale}.
In order to obtain a similar structure for $T_{n_1,n_2}(w)$ as given in \eqref{eq:Xn-X_inD}, we consider the $2$-dimensional vectors $\textbf M_{n_1,n_2}^\top = (\frac{1}{\sqrt{n_1}}\sum_{i=1}^{ n_1 } M_i^{(1)}, \frac{1}{\sqrt{n_2}}\sum_{i=1}^{ n_2 } M_i^{(2)})^\top$ and  $\textbf{k}_{n_1,n_2}^\top = ({k}^{(1)}_{n_1,n_2},{k}^{(2)}_{n_1,n_2})^\top$, $t\geq 0$. With this notation we get 
\begin{align}\label{eq:logrank_like_eq_11}
    T_{n_1,n_2}(w) \stackrel{H_0}{=} \int_0^\infty \textbf{k}_{n_1,n_2}(t)^\top d\textbf M_{n_1,n_2}(t),
\end{align} 
where the right-hand side of \eqref{eq:logrank_like_eq_11} is the multidimensional martingale counterpart of the first term on the right-hand side of \eqref{eq:Xn-X_inD}.
With \eqref{eq:logrank_like_eq_11} we thus obtained a similar structure for $T_{n_1,n_2}(w)$ as in \eqref{eq:Xn-X_inD} with the second term on the right-hand side of \eqref{eq:Xn-X_inD} set to zero due to the nonparametric setting. 
%As the weighted logrank test $T_{n_1,n_2}(w)$ is nonparametric, there is no counterpart of the second term on the right hand-side of \eqref{eq:Xn-X_inD} in \eqref{eq:logrank_like_eq_11}. 
The wild bootstrap version $T^*_{n_1,n_2}(w)$ of $T_{n_1,n_2}(w)$ under $H_0$ is obtained by applying \cref{WB_replacement} to \eqref{eq:logrank_like_eq_11}:
\begin{align}\label{eq:logrank_like_eq_19}
    T^*_{n_1,n_2}(w) \stackrel{H_0}{=} \int_0^\infty \textbf{k}^*_{n_1,n_2}(t)^\top d\textbf M^*_{n_1,n_2}(t),
\end{align}
where $\textbf M^{*\top}_{n_1,n_2} = (\frac{1}{\sqrt{n_1}}\sum_{i=1}^{ n_1 } G^{(1)}_i N_i^{(1)}, \frac{1}{\sqrt{n_2}}\sum_{i=1}^{ n_2 } G^{(2)}_i N_i^{(2)})^\top$ is the wild bootstrap counterpart of $\textbf M_{n_1,n_2}$, and $\textbf{k}^{*\top}_{n_1,n_2} = ({k}^{*(1)}_{n_1,n_2},{k}^{*(2)}_{n_1,n_2})^\top$ with
$$ {k}^{*(j)}_{n_1,n_2}(t) = (-1)^{j+1} \sqrt{\frac{n_1 + n_2}{n_{3-j}}} w(\hat S_n^*(t-)) \frac{Y^{(3-j)}(t)}{Y(t)}, \quad t\geq 0, j=1,2,$$ 
is the wild bootstrap counterpart of $\textbf{k}_{n_1,n_2}$. Here, the multiplier processes $G^{(1)}_1,\ldots ,G^{(1)}_{n_1},$\\
$G^{(2)}_1,\ldots , G^{(2)}_{n_2} $ are pairwise independent and identically distributed.
Note that this definition of $T^*_{n_1,n_2}(w)$ deviates slightly from the corresponding definition given in \cite{ditzhaus19} as it contains the wild bootstrap counterpart $\hat S_n^*$ of the pooled Kaplan-Meier estimator $\hat S_n$. 
%Additionally, $\hat S^*_n$ is a wild bootstrap version of the Kaplan-Meier estimator. 
In Part~II we will give an idea  of how such a reampling version may be constructed based on a functional relationship between the estimator of interest and Nelson-Aalen estimators; we will exemplify this by means of cumulative incidence functions in semiparametric models.
With \eqref{eq:logrank_like_eq_19} we thus obtained a similar structure for $T^*_{n_1,n_2}(w)$ as stated in \eqref{eq:X*-Xn_4} with 
%the second term on the right-hand side of \eqref{eq:X*-Xn_4} set to zero 
$\textbf B^*_n  \textbf C^*_n  \textbf D^*_{n,g}(\tau )\equiv 0$ due to the nonparametric setting and $o_p(1)$ set to zero.  
It is left to show that a result as stated in \cref{thm:asyEquivalence} holds for $T_{n_1,n_2}(w)$ and $T^*_{n_1,n_2}(w)$ under the null hypothesis. For this, one may first argue with respect to any finite upper bound of integration $\tau$. 
With one additional argument, the remaining integral from $\tau$ to $\infty$ can be shown to be asymptotically negligible for $n \to \infty $ followed by $\tau \to \infty$; use for instance Theorem~3.2 in \cite{billingsley99}. In this way, one obtains a justification of the wild bootstrap for the weighted logrank test within a multidimensional martingale framework which can be seen as an extension of the setting presented in this Part~I.
\end{ex}

\begin{ex}\label{ex:application4} 
\textnormal{(Cox model)} Given the $d$-variate predictable covariate vectors $\textbf Z_i(t)$, $t\in \mathcal{T}$,
% $i=1,\dots,n$, $t\geq 0$, let 
the intensity process of the counting process $N_i$ is 
    $E(d N_i(t) | \textbf Z_i(t)) = \lambda_i(t,\textbf Z_i(t),\bs{\beta}_0) dt = Y_i(t) \exp(\textbf Z_i^\top(t) \bs{\beta}_0) \alpha_0(t) dt $,
$t\in \mathcal{T}$,  $i=1,\dots,n$.
    Here, $\alpha_0$ is the so-called baseline hazard rate for an individual with the zero covariate vector.
    In this case the processes $M_i(t) = N_i(t) - \Lambda_i(t,\textbf Z_i(t),\bs{\beta}_0)$, $t\in \mathcal{T}$, are martingales, where $\Lambda_i(t,\textbf Z_i(t),\bs{\beta}) = \int_0^t \lambda_i(u,\textbf Z_i(t),\bs{\beta}) du$.
    The Breslow estimator for the cumulative baseline hazard function $X(t)=A_0(t) = \int_0^t \alpha_0(u) du$, $t\in \mathcal{T}$, is given by
    $$ \hat X_n(t)=\hat A_{0,n}(t, \hat {\bs{\beta}}_n) = \sum_{i=1}^n \int_0^t \frac{J(u)}{S^{(0)}_n(u,\hat {\bs{\beta}}_n)} dN_i(u),\quad t\in \mathcal{T},$$
    where $\hat {\bs{\beta}}_n$ is the solution to the score equation $$\sum_{i=1}^n \int_0^\tau \Big(\textbf Z_i(u) - \frac{\textbf S^{(1)}_n(u,\bs{\beta})}{S^{(0)}_n(u, \bs{\beta})}\Big) dN_i(u)= 0,$$ 
    $\tau >0$ is the terminal evaluation time,
    and $S^{(0)}_n(t,\bs{\beta}) = \sum_{i=1}^n Y_i(t) \exp(\textbf Z_i^\top(t) \bs{\beta}) $, 
     $\textbf S^{(1)}_n(t, \bs{\beta}) = \sum_{i=1}^n Y_i(t) \textbf Z_i(t) \exp(\textbf Z_i^\top(t) \bs{\beta}) $, 
     $\textbf S^{(2)}_n(t, \bs{\beta}) = \sum_{i=1}^n Y_i(t) \textbf Z_i(t)^{\otimes 2} \exp(\textbf Z_i^\top(t) \bs{\beta}) $, $t\in \mathcal{T}$.
In particular, $\hat A_{0,n}(\cdot, \hat {\bs{\beta}}_n)$ follows the general counting process-based structure stated in \eqref{eq:Xn} with $k_{n}(t,\bs{\beta}_0) = {\displaystyle {\frac{n  J(t)}{S^{(0)}_n(t,\bs{\beta}_0)}}}$, $t\in \mathcal{T}$.  
     For the Breslow estimator it is well-known that for $t\in \mathcal{T}$
     \begin{align}
        \begin{split}
\label{eq:Cox}
         \sqrt{n} (\hat A_{0,n}(t, \hat {\bs{\beta}}_n) - A_0(t)) &= \sqrt{n} \sum_{i=1}^n \int_0^t \frac{J(u)}{S^{(0)}_n(u,\bs{\beta}_0)} d M_i(u)\\
          &\quad - \int_0^t  \frac{ J(u)  \textbf S^{(1)}_n(u, \bs{\beta}_0)}{S^{(0)}_n(u,\bs{\beta}_0)^2} dN_i(u)\\
          &\quad\cdot \textbf C_n\frac1{\sqrt{n}} \Big(\sum_{i=1}^n \int_0^\tau \Big(\textbf Z_i(u) - \frac{\textbf S^{(1)}_n(u, \bs{\beta}_0)}{S^{(0)}_n(u, \bs{\beta}_0)}\Big) dM_i(u)\Big) +o_p(1) ,
  %       & = \sqrt{n} \sum_{i=1}^n \int_0^t \frac{J(u) }{S^{(0)}_n(u,\bs{\beta}_0)}d M_i(u)\\
  %       &\quad - \sum_{i=1}^n\int_0^t \frac{ J(t)  \textbf S^{(1)}_n(t, \bs{\beta}_0)}{S^{(0)}_n(t,\bs{\beta}_0)^2}   dN_i(u) \cdot \textbf C_n^\top 
  %       \frac1{\sqrt{n}} \Big(\sum_{i=1}^n \int_0^\tau \Big(\textbf Z_i(u) - \frac{\textbf S^{(1)}_n(u, \bs{\beta}_0)}{S^{(0)}_n(u, \bs{\beta}_0)}\Big) dM_i(u)\Big)^\top +o_p(1),
    \end{split}
 \end{align}
where $\textbf C_n$ is a certain (random) $d\times d$ matrix. %{\green and $\textbf e$ is the uniform limit in probability of $\textbf S^{(1)}_n/S^{(0)}_n$, see e.g.\ \cite{dobler19}.} 
Note that in \eqref{eq:Cox} it has been used that \eqref{eq:intLambda-X=o_p} and \eqref{eq:beta_asy_lin} are satisfied, i.e.,
%as part of the $o_p(1)$-term in the equation above, we have 
$$\sqrt{n}\big( \frac1n \sum_{i=1}^n \int_0^t k_{n}(u,\bs{\beta}_0) d \Lambda_i (u,\bs{\beta}_0) - A_0(t)\big ) = \sqrt{n} \int_0^t (J(u) - 1) d A_0(u) = o_p(1),\quad t\in\mathcal{T},$$
and 
\begin{align*}
    \sqrt{n}(\hat{\bs{\beta}}_n - \bs{\beta}_0) = \textbf C_n\frac1{\sqrt{n}} \Big(\sum_{i=1}^n \int_0^\tau \Big(\textbf Z_i(u) - \frac{\textbf S^{(1)}_n(u, \bs{\beta}_0)}{S^{(0)}_n(u, \bs{\beta}_0)}\Big) dM_i(u)\Big) + o_p(1).
\end{align*}
Additionally, we have $D k_{n}(t,\bs{\beta}_0) =  {\displaystyle - \frac{n  J(t) \textbf S^{(1)}_n(t, \bs{\beta}_0)}{S^{(0)}_n(t,\bs{\beta}_0)^2}}$ and $\textbf g_{n,i}(t,\bs{\beta}_0) =  {\displaystyle { \textbf Z_i(t) - \frac{\textbf S^{(1)}_n(t,\bs{\beta}_0)}{S^{(0)}_n(t,\bs{\beta}_0)} }}$, $t\in \mathcal{T}$.
As a result of the boundedness of the covariates and the boundedness of $J S^{(0)}_n$ away from zero on $\mathcal T$,   $k_{n}, D k_{n}, $ and $ \textbf g_{n,i}$ as functions in $t$ are bounded on $\mathcal T$. Additionally, they are predictable due to the predictability of the covariates. Thus, the first term and the martingale integral in the second term of the form \eqref{eq:Dn} on the right-hand side of \eqref{eq:Cox} are local square integrable martingales. 
     %Hence, \eqref{eq:intLambda-X=o_p} is satisfied.
In conclusion, with \eqref{eq:Cox} we retrieve the asymptotic representation \eqref{eq:Xn-X_inD}, i.e., $\sqrt{n} (\hat A_{0,n}(\cdot, \hat {\bs{\beta}}_n) - A_0) = D_{n,k} + \textbf{B}_n\textbf C \textbf D_{n,g}(\tau) +o_p(1)$. The uniform limits in probability of $k_{n}$ and $\textbf g_{n,i}$ are $\tilde k ={\displaystyle \frac{1}{s^{(0)}}}$ and $\tilde{\textbf g}_i = \textbf Z_i - {\displaystyle \frac{s^{(1)}}{s^{(0)}}} $, respectively, where $s^{(j)}$ are the uniform deterministic limits in probability of $n^{-1} S_n^{(j)}$, $j=0,1$. 
Under the typically made assumptions (Condition VII.2.1 of \citealt{Andersen}) and under the assumption that the covariate vectors $\textbf Z_i$, $i=1,\ldots ,n$, are pairwise independent and identically distributed, \cref{assump_general} is  fulfilled.
Similarly, the uniform limit in probability of $ D k_{n}$ is $\tilde{K} =  {\displaystyle\frac{s^{(1)}}{(s^{(0)})^2}}$. 
     Again, under Condition VII.2.1 and $(7.2.28)$ of \cite{Andersen}, Assumptions \ref{ass:Bn_Cn} and \ref{ass:Cn-C} are valid. In particular, $\textbf C_n$ in Assumption~\ref{ass:Cn-C} takes the form
     $$\Big[\frac{1}{n}\sum_{i=1}^n \int_0^\tau \Big( \frac{S^{(2)}_n(u,\bs{\beta}_0)}{S^{(0)}_n(u,\bs{\beta}_0)} - \Big( \frac{S^{(1)}_n(u,\bs{\beta}_0)}{S^{(0)}_n(u,\bs{\beta}_0)}\Big)^{\otimes 2} \Big)dN_i(u)\Big]^{-1}.$$
    Eventually, the wild bootstrap counterpart $\sqrt{n} (\hat A^*_{0,n}(\cdot, \hat {\bs{\beta}}^*_n) - \hat A_{0,n}(\cdot, \hat {\bs{\beta}}_n))$ of $\sqrt{n} (\hat A_{0,n}(\cdot, \hat {\bs{\beta}}_n) - A_0)$ can be formulated by applying \cref{WB_replacement} to \eqref{eq:Cox}. This yields for $t\in\mathcal{T}$
     \begin{align}\label{eq:cox_wb_mgale}
     \begin{split}
         \sqrt{n} (\hat A^*_{0,n}(t, \hat {\bs{\beta}}^*_n) - \hat A_{0,n}(t, \hat {\bs{\beta}}_n)) &= \sqrt{n} \sum_{i=1}^n \int_0^t \frac{J(u) }{S^{(0)}_n(u,\hat{\bs{\beta}})}G_i\,d N_i(u)\\
     &\quad - \sum_{i=1}^n\int_0^t \frac{ J(u)  \textbf S^{(1)}_n(u, \hat{\bs{\beta}})}{S^{(0)}_n(u,\hat{\bs{\beta}})^2}(G_i + 1) dN_i(u)\\
     &\quad\cdot \textbf C^{*}_n  \frac1{\sqrt{n}} \Big(\sum_{i=1}^n \int_0^\tau \Big(\textbf Z_i(u) - \frac{\textbf S^{(1)}_n(u, \hat{\bs{\beta}})}{S^{(0)}_n(u, \hat{\bs{\beta}})}\Big) G_i\,dN_i(u)\Big) .
     \end{split}
     \end{align}
   Here $\textbf C^*_n $ as given in \cref{remark:MLE_beta_sec3} simplifies for the Cox model to $$\textbf C^*_n  = \Big[\frac{1}{n}\sum_{i=1}^n \int_0^\tau \Big( \frac{S^{(2)}_n(u,\hat{\bs{\beta}})}{S^{(0)}_n(u,\hat{\bs{\beta}})} - \Big( \frac{S^{(1)}_n(u,\hat{\bs{\beta}})}{S^{(0)}_n(u,\hat{\bs{\beta}})}\Big)^{\otimes 2} \Big) G_i^2 dN_i(u)\Big]^{-1}.$$
Additionally, \cref{ass:Cn_star-Cn} is satisfied as argued in \cref{remark:MLE_beta_sec3}. In conclusion, \eqref{eq:cox_wb_mgale} implies that \eqref{eq:X*-Xn_4} holds with $o_p(1)$ set to zero, i.e., $\sqrt{n} (\hat A^*_{0,n}(\cdot, \hat {\bs{\beta}}^*_n) - \hat A_{0,n}(\cdot, \hat {\bs{\beta}}_n)) = D^*_{n,k} + \textbf{B}^*_n\textbf C^* \textbf D^*_{n,g}(\tau) $. Finally, \cref{thm:asyEquivalence} can be applied to verify the asymptotic validity of the wild bootstrap for statistical inference on the Breslow estimator. Note that all expressions used in this example are similar to the ones in \cite{dobler19}.
\end{ex}

%
%%
%%%
%%%%
%%%%%
%%%%%%
%%%%%%%
\section{Discussion}
%%%%%%%
%%%%%%
%%%%%
%%%%
%%%
%%
%
\label{sec:discussion}

We have proposed and validated a widely applicable wild bootstrap procedure for general nonparametric and (semi-)parametric counting process-based statistics. We gave a step by step description of how to construct the wild bootstrap counterpart of the statistic. In particular, it is crucial to match each individual with one multiplier process.
In order to justify the validity of the wild bootstrap, we have studied the asymptotic distributions of the statistic of interest and of the wild bootstrap counterpart which turned out to  coincide. We have found the wild bootstrapped martingales to be martingales as well. Thus, in the corresponding proof, we made use of a carefully chosen variant of Rebolledo's martingale central limit theorem. We illustrated the method for several main models in survival analysis. 

As we have seen in Examples~\ref{ex:application1}-\ref{ex:application4}, the assumptions we have made throughout the Part~I are rather weak: they are satisfied under very natural regularity conditions.
However, \cref{assump_general}~\ref{assump_general3} is, for example, not satisfied in shared frailty models, because in these models it is assumed that common unobserved variables influence the intensity processes of multiple individuals.

For the construction of the wild bootstrap counterpart of a given counting process-based statistic we have chosen the  nonparametric estimator $G_i dN_i$ for the martingale increment $dM_i$, cf.\ \cref{WB_replacement}~\ref{item:WB_step1}. This choice guarantees a more general applicability of the proposed wild bootstrap resampling procedure, because no specifications on the form of the cumulative hazard rate have to be made. In contrast, Spiekerman and Lin proposed a semiparametric approach by choosing $G_i [dN_i - d\hat\Lambda_i(\cdot,\hat{\bs{\beta}}_n)]$ as the replacement for the martingale increment (\cite{spiekerman98}). Under this semiparametric estimator the information encoded in the parameter $\bs{\beta}$ is incorporated in the wild bootstrap estimators, which could potentially lead to more accurate results. However, their  approach is not as widely applicable as the nonparametric one that we decided to employ. Moreover, in the context of Cox models, in \cite{dobler19} %Dobler et al {\green ref to the paper}
it is revealed by means of a substantial simulation study that the difference between the results of the two methods is not significant.
% {\vio [explanation: I was confusing two different things and clarified it now.]}

In conclusion, the wild bootstrap procedure as proposed in this Part~I is applicable to a wide range of models and simple to implement. By means of this method, one may easily approximate the unknown distribution of a counting process-based statistic around the target quantity. Aside from the theoretical justification of this resampling procedure, in Part~II we present an extensive simulation study  based on which we explore the small sample performance of the method. That Part~I concentrates on Fine-Gray models for censoring-complete data. In particular, we explain on the basis of the cumulative incidence function how to obtain wild bootstrap confidence bands for a functional applied to a vector of two statistics of the form considered in the present Part~I.
%We conclude the second part of this paper with additional statistical analyses within Fine-Gray models.

%
%%
%%%
%%%%
%%%%%
%%%%%%
%%%%%%%
\section*{Appendix A: Proofs}
%\appendix
%%%%%%%
%%%%%%
%%%%%
%%%%
%%%
%%
%

%\subsection{Proofs}

For the proofs we introduce some additional notation: we write  $\lVert \cdot \rVert_\infty$ for the maximum norm of a vector $\bs v\in\mathbb{R}^p$ or a matrix $\textbf G\in \mathbb{R}^{p\times p}$, which denotes the largest element in absolute value of $\bs v$ and $\textbf G$, respectively. Moreover, $\mathcal{C}[0,\tau]^{m}$ denotes the set of all continuous functions with values from $[0,\tau]$ to $\mathbb{R}^{m}$ for any $m\in\mathbb N$.

%%%%%%%%%%%%%%%%%%%%
%%%%%%%%%%%%%%%%%%%%
\subsection*{A.1 Proofs of \cref{sec:Notation}}
%\begin{proof}{of \cref{lem:Dn}}

\noindent
\textbf{Proof of \cref{lem:Dn}.}\label{proof:Dn}\\
As explained in \cref{sec:Notation} below \eqref{eq:Dn}, $\textbf D_{n,h}$ is a local square integrable counting process martingale. Thus, we can apply Rebolledo's martingale central limit theorem as stated in Theorem~II.5.1 of \cite{Andersen}. It follows that we have to show two conditions. The predictable covariation process $\langle\textbf D_{n,h} \rangle (t)$ or the optional covariation process $[ \textbf D_{n,h} ] (t)$ of $\textbf D_{n,h}$ must converges in probability, as $n\rightarrow\infty$, to a continuous, deterministic and positive semidefinite matrix-valued function on $\mathcal{T}$ with $\textbf V_{\tilde h}(0)=0$. Additionally, condition (2.5.3) of \cite{Andersen} on the jumps of $\textbf D_{n,h}$ must hold. 

We first show the convergence in probability of the predictable covariation process $\langle\textbf D_{n,h} \rangle (t)$ to the matrix-valued function $\textbf V_{\tilde h}(t)$ for all $t\in\mathcal{T}$, as $n \rightarrow\infty$.  According to Proposition~II.4.1 of \cite{Andersen} together with \eqref{eq:Dn}, we have 
%\begin{equation}\label{eq:cov_Dn}
    %\begin{array}{r@{}l}
\begin{align}
\begin{split}
    \label{eq:cov_Dn}
    \langle\textbf D_{n,h} \rangle (t) &{}= \frac{1}{n} \sum_{i=1}^n \int_0^t \textbf h_{n, i}(u, \bs {\beta}_0)^{\otimes 2} d\Lambda_i(u,\bs {\beta}_0)\\
    &{}= \frac{1}{n} \sum_{i=1}^n \int_0^t [ \textbf h_{n, i}(u, \bs {\beta}_0)^{\otimes 2} - \tilde{\textbf{h}}_{ i}(u,\bs {\beta}_0)^{\otimes 2} ] d\Lambda_i(u,\bs {\beta}_0) \\
    &{} \quad + \frac{1}{n} \sum_{i=1}^n \int_0^t \tilde{\textbf{h}}_{ i}(u,\bs {\beta}_0)^{\otimes 2} d\Lambda_i(u,\bs {\beta}_0).
\end{split}
\end{align}
%    \end{array}
%\end{equation}
We start with focusing on the first term of the second step of \eqref{eq:cov_Dn}. We want to show that 
\begin{align}\label{eq:cov_Dn_term1}
    \frac{1}{n} \sum_{i=1}^n \int_0^t [\textbf h_{n, i}(u, \bs {\beta}_0)^{\otimes 2} -\tilde{\textbf{h}}_{ i}(u,\bs {\beta}_0)^{\otimes 2}] d\Lambda_i(u,\bs {\beta}_0)= o_p(1),\text{ for all } t\in\mathcal{T}, \text{ as } n\rightarrow\infty.
\end{align}
For this it suffices to bound its largest component: 
\begin{align}\label{eq:cov_Dn_term1.2}
\begin{split}
    & \frac{1}{n} \sum_{i=1}^n \int_0^t \lVert \textbf h_{n, i}(u, \bs {\beta}_0)^{\otimes 2} -\tilde{\textbf{h}}_{ i}(u,\bs {\beta}_0)^{\otimes 2} \rVert_\infty  d\Lambda_i(u,\bs {\beta}_0)\\
    & \leq \sup_{i\in\{1,\dots,n\},t\in \mathcal{T}} \lVert \textbf h_{n, i}(t, \bs {\beta}_0)^{\otimes 2} -\tilde{\textbf{h}}_{ i}(t,\bs {\beta}_0)^{\otimes 2} \rVert_\infty  \frac{1}{n} \sum_{i=1}^n \Lambda_i(\tau,\bs {\beta}_0) \\
    & \leq \Big(\sup_{i\in\{1,\dots,n\},t\in \mathcal{T}} \lVert ( \textbf h_{n, i}(t, \bs {\beta}_0) -\tilde{\textbf{h}}_{ i}(t,\bs {\beta}_0) ) \textbf h_{n, i}(t, \bs {\beta}_0)^\top \rVert_\infty \\
    & \qquad + \sup_{i\in\{1,\dots,n\},t\in \mathcal{T}} \lVert \tilde{\textbf{h}}_{ i}(t,\bs {\beta}_0) (\textbf h_{n, i}(t, \bs {\beta}_0) - \tilde{\textbf{h}}_{ i}(t,\bs {\beta}_0))^\top \rVert_\infty \Big)  \frac{1}{n} \sum_{i=1}^n \Lambda_i(\tau,\bs {\beta}_0) 
\end{split}
\end{align}
where the last step is due to the triangle inequality and $\bs a^{\otimes2} - \bs b^{\otimes2} = (\bs a - \bs b) \bs a^\top + \bs b (\bs a - \bs b)^\top $ for two vectors $\bs a, \bs b$. Both terms in brackets converge to zero in probability, as $n\rightarrow \infty$, according to \cref{assump_general}~\ref{assump_general1},~\ref{assump_general2}, and since $\textbf h_{n, i}(t, \bs {\beta}_0)$ is locally bounded for $i=1,\ldots , n$. Note that \cref{assump_general}~\ref{assump_general1} holds for any consistent estimator $\check{\bs{\beta}}_n$, in particular for $\bs {\beta}_0$ itself.   
%the continuous mapping theorem. 
From \cref{assump_general}~\ref{assump_general3} in combination with the integrability of the cumulative intensities and the law of large numbers, we get $\frac{1}{n} \sum_{i=1}^n \Lambda_i(\tau,\bs {\beta}_0) \stackrel{\mathbb P}{\longrightarrow} \mathbb E (\Lambda_1(\tau,\bs {\beta}_0))$, as $n \rightarrow \infty$. Hence, the whole expression
converges to zero in probability, as $n\rightarrow\infty$, and we conclude that \eqref{eq:cov_Dn_term1} holds. 

The subsequent considerations relate to the second term of the second step of \eqref{eq:cov_Dn}. According to \cref{assump_general}~\ref{assump_general2} it holds that $\sup_{t\in\mathcal{T}}\lVert\tilde{\textbf{h}}_{1}(t,\bs{\beta}_0)\rVert_\infty$ is bounded. Moreover, we have $ \mathbb{E}(\Lambda_1(\tau,\bs{\beta}_0)) < \infty$ by assumption. %\cref{assump_general}~\ref{item:ass_Bn_4}.
These two statements combined yield for all $t\in\mathcal{T}$,
\begin{align}\label{eq:cov_Dn_exp}
    \mathbb{E} \Big(\int_0^t  \lVert\tilde{\textbf{h}}_{1}(u,\bs{\beta}_0)^{\otimes 2}\rVert_\infty \,d\Lambda_1(u,\bs{\beta}_0)\Big)\leq   \mathbb{E}\Big(\sup_{t\in\mathcal{T}}\lVert\tilde{\textbf{h}}_{1}(t,\bs{\beta}_0)^{\otimes 2}\rVert_\infty\Lambda_1(t,\bs{\beta}_0)\Big) <\infty.
\end{align}
On the basis of \eqref{eq:cov_Dn_exp} and \cref{assump_general}~\ref{assump_general3}, we make use of the law of large numbers and get for the second term of the second step of \eqref{eq:cov_Dn}
\begin{align*}
    \frac{1}{n} \sum_{i=1}^n \int_0^t \tilde{\textbf{h}}_{ i}(u,\bs {\beta}_0)^{\otimes 2} d\Lambda_i(u,\bs {\beta}_0) \stackrel{\mathbb P}{\longrightarrow} \mathbb E ( \int_0^t \tilde{\textbf{h}}_{1}(u,\bs {\beta}_0)^{\otimes 2} d\Lambda_1(u,\bs {\beta}_0)),\quad n\rightarrow\infty, 
\end{align*}
for any fixed $t\in\mathcal{T}$. Note that the integrability of the intensity process $\lambda_1(t,\bs{\beta}_0)$ follows from the integrability of the cumulative intensity process $\Lambda_1(t,\bs{\beta}_0)$. Thus,
due to the integrability of the cumulative intensities and
%under \cref{assump_general}~\ref{item:ass_Bn_4} and 
\cref{assump_general}~\ref{assump_general2}, we can make use of Fubini's theorem, due to which we can exchange the order of integration. Thus, we have 
\begin{align}\label{eq:cov_Dn_3}
   \frac{1}{n} \sum_{i=1}^n \int_0^t \tilde{\textbf{h}}_{ i}(u,\bs {\beta}_0)^{\otimes 2} d\Lambda_i(u,\bs {\beta}_0)  \stackrel{\mathbb P}{\longrightarrow} \int_0^t \mathbb E( \tilde{\textbf{h}}_{1} (u,\bs{\beta}_0)^{\otimes 2}  \lambda_1(u,\bs{\beta}_0) ) du , 
\end{align}
for all $t\in\mathcal{T}$, as $n\rightarrow\infty$. Finally, combining \eqref{eq:cov_Dn} with \eqref{eq:cov_Dn_term1} and \eqref{eq:cov_Dn_3} yields
\begin{align*}
    \langle\textbf D_{n,h} \rangle (t)\stackrel{\mathbb P}{\longrightarrow} \int_0^t \mathbb E( \tilde{\textbf{h}}_{1} (u,\bs{\beta}_0)^{\otimes 2} \lambda_1(u,\bs{\beta}_0) ) du =\textbf V_{\tilde h}(t),\text{ for all } t\in\mathcal{T}, \text{ as }  n\rightarrow\infty.
\end{align*}
%We identify the limit in probability $\textbf V_{ h}(t)$ of $\langle\textbf D_{n,h} \rangle (t)$ with $\mathbb E ( \int_0^t \tilde{\textbf{h}}_{1}(u,\bs {\beta}_0)^{\otimes 2} d\Lambda_1(u,\bs {\beta}_0))$, $t\in\mathcal{T}$. The properties of $\textbf V_{ h}(t)$ are given in 
When taking into consideration that we have $\tilde{\textbf h} = (\tilde{\textbf k},\tilde{\textbf g}) $, we can write the covariance matrix in block form
\[
\textbf V_{\tilde h} = \textbf V_{(\tilde k,\tilde g )}
=
\begin{pmatrix}
    \textbf V_{\tilde k} & 
    \textbf V_{\tilde k , \tilde g} \\
    \textbf V_{\tilde g , \tilde k} & \textbf V_{\tilde g}
 \end{pmatrix},
\]
where for $t \in \mathcal{T}$, 
$$ \textbf V_{\tilde k}(t)  = \langle \textbf D_{\tilde k} \rangle (t) =\int_0^{ t} \mathbb E (  \tilde{\textbf{k}}_{1}(u,\bs {\beta}_0)^{\otimes 2} \lambda_1(u,\bs {\beta}_0))du,$$
$$ \textbf V_{\tilde g}(t) = \langle \textbf D_{\tilde g} \rangle (t) =\int_0^{ t} \mathbb E (  \tilde{\textbf{g}}_{1}(u,\bs {\beta}_0)^{\otimes 2} \lambda_1(u,\bs {\beta}_0))du,$$
%and 
 $$\textbf V_{\tilde k, \tilde g}(t) =\textbf V_{\tilde g, \tilde k}(t) =\langle \textbf D_{\tilde k} , \textbf D_{\tilde g}\rangle (t)
=\int_0^{t} \mathbb E (  \tilde{\textbf{k}}_{1}(u,\bs {\beta}_0)\cdot \tilde{\textbf{g}}_{1}(u,\bs {\beta}_0)^\top \lambda_1(u,\bs {\beta}_0))du.$$

Second, we verify condition (2.5.3) of Rebolledo's theorem of \cite{Andersen}. For this we introduce the stochastic process $\textbf D_{n,h}^\epsilon$ given by
\begin{align}\label{eq:Dn_epsilon}
    \textbf D_{n,h}^\epsilon(t) &{}= \int_0^t\mathbbm{1}\{\vert \Delta \textbf D_{n,h}(u)\rvert > \epsilon \} \textbf D_{n,h}(du),\quad t\in \mathcal{T},
\end{align}
which we refer to as the $\epsilon$-jump process of $\textbf D_{n,h}$. Here, the indicator function is to be understood vector-wise, specifying for each element $D_{n,h}^{j}(t)$ of the $p$-dimensional vector $\textbf D_{n,h} (t)= (D_{n,h}^1(t),\ldots,D_{n,h}^p(t)) $ whether the jump at time t is larger in absolute value than $\epsilon$. Note that the elements of the indicator function $\mathbbm{1}\{\vert \Delta \textbf D_{n,h}(u)\rvert \geq \epsilon \}$ may be unequal to zero only at discontinuities of $D_{n,h}^j$, which correspond to discontinuities of the martingale $M_i$. In addition, the jumps of the martingale $M_i$ occur only at event times registered by the counting processes $N_i$, because we assumed the cumulative intensity process $\Lambda_i(\cdot,\bs{\beta}_0)$ to be absolutely continuous. % (cf. \cref{assump_general}~\ref{ass:Lambda_cont}).
This means that the $\epsilon$-jump process $\textbf D_{n,h}^\epsilon$ accumulates all the jumps of components of $\textbf D_{n,h}$ that are larger in absolute value than $\epsilon$. Recall that no two counting processes $N_i$, $i=1,\ldots,n,$ jump simultaneously. Combining \eqref{eq:Dn_epsilon} with the above reasoning yields
\begin{align*}
    \textbf D_{n,h}^\epsilon &{}= \frac{1}{\sqrt{n}}\sum_{i=1}^n \int_0^t \textbf h_{n, i}(u, \bs {\beta}_0) \mathbbm{1}\Big\{\Big\vert \frac{1}{\sqrt{n}}\sum_{i=1}^n \textbf h_{n, i}(u, \bs {\beta}_0) \Delta M_i(u) \Big\rvert > \epsilon \Big\} dM_i(u),\\
    &{}= \frac{1}{\sqrt{n}}\sum_{i=1}^n \int_0^t \textbf h_{n, i}(u, \bs {\beta}_0) \mathbbm{1}\Big\{\Big\vert \frac{1}{\sqrt{n}} \textbf h_{n, i}(u, \bs {\beta}_0) \Delta N_i(u) \Big\rvert > \epsilon \Big\} dM_i(u).
\end{align*}
The aforementioned condition (2.5.3) is fulfilled, if the predictable covariation process $\langle \textbf D_{n,h}^\epsilon \rangle (t) $ of $\textbf D_{n,h}^\epsilon$ converges to zero in probability for all $t\in \mathcal{T}, \epsilon > 0$, as $n\rightarrow\infty$. 
Note that the predictable covariation process $\langle \textbf D_{n,h}^\epsilon \rangle (t) $ is defined as the $(p+b)\times (p+b)$-dimensional matrix of the predictable covariation processes $ \big(\langle  D_{n,h}^{\epsilon,j}, D_{n,h}^{\epsilon,l} \rangle (t)\big)_{j,l=1}^{p+b}$ of the components 
$$D_{n,h}^{\epsilon,j} = \frac{1}{\sqrt{n}}\sum_{i=1}^n \int_0^t h^j_{n, i}(u, \bs {\beta}_0) \mathbbm{1}\Big\{\Big\vert \frac{1}{\sqrt{n}} h_{n, i}^j(u, \bs {\beta}_0) \Delta N_i(u) \Big\rvert > \epsilon \Big\} dM_i(u),$$
where ${h}_{n,i}^j$ denotes the $j$-th component of the $(p+b)$-dimensional function $\textbf{h}_{n,i}$, $j=1,\ldots,p+b$. It is easy to see that the largest entry (in absolute value)  of $\langle \textbf D_{n,h}^\epsilon \rangle (t) $ is located on the diagonal and that a diagonal element takes the following form:
\begin{align*}
   & \langle  D_{n,h}^{\epsilon,j} \rangle (t) = \frac{1}{n} \sum_{i=1}^n \int_0^t h_{n,i}^j(u,\bs {\beta}_0)^{2} \mathbbm{1}\Big\{\Big|\frac{1}{\sqrt{n}}  h_{n, i}^j(u,\bs {\beta}_0)\Delta N_i(u)\Big|> \epsilon\Big\} d\Lambda_i(u),
\end{align*}
$j=1,\ldots,p+b$. Thus, it suffices to show that the diagonal elements $\langle  D_{n,h}^{\epsilon,j} \rangle (t)$ of $\langle \textbf D_{n,h}^{\epsilon} \rangle (t) $ converge to 0 in probability as $n\to\infty$ for each $t\in \mathcal{T}$, $j=1,\dots,p+b$. That is, for every $\delta > 0$ the probability $\mathbb P(\langle  D_{n,h}^{\epsilon,j} \rangle (t) \geq \delta)$ must go to zero for all $j=1,\dots,p+b$.
For this, we bound this probability from above as follows:
\begin{align}\label{eq:lem2.2}
\begin{split}
&\mathbb{P}(\langle  D_{n,h}^{\epsilon,j}  \rangle (t) \geq \delta )\\
  & \leq \mathbb{P}\Big(\sup_{t\in\mathcal{T},i\in\{1,\ldots,n\}} \mathbbm{1}\{\frac{1}{\sqrt{n}} \lVert \textbf h_{n, i}(t,\bs {\beta}_0)\rVert_\infty > \epsilon\} \frac{1}{n}\sum_{i=1}^n \int_0^t  h_{n,i}^j(u,\bs {\beta}_0)^{2}d\Lambda_i(u) \geq \delta \Big)\\
  & \leq \mathbb{P}\Big(\sup_{t\in\mathcal{T},i\in\{1,\ldots,n\}} \mathbbm{1}\{\frac{1}{\sqrt{n}} \lVert\textbf h_{n, i}(t,\bs {\beta}_0)\rVert_\infty > \epsilon\} =1\Big) \\
  & = 1 - \mathbb{P}\Big(\text{for all } i,t: \frac{1}{\sqrt{n}} \lVert\textbf h_{n, i}(t,\bs {\beta}_0)\rVert_\infty  \leq \epsilon\Big) \\
  & = o(1) + 1 -  \mathbb{P}\Big(\text{for all } i,t: \frac{1}{\sqrt{n}} \lVert\textbf h_{n, i}(t,\bs {\beta}_0)\rVert_\infty  \leq \epsilon, \\
  & \qquad \sup_{i\in\{1,\ldots,n\}, t\in \mathcal T} \lVert\textbf h_{n, i}(t,\bs {\beta}_0) -\tilde{\textbf h}_{i} (t,\bs {\beta}_0) \rVert_\infty < \eta \Big)  
  \\
  & \leq o(1) + 1 -  \mathbb{P}\Big(\text{for all } i,t: \frac{1}{\sqrt{n}} \lVert\tilde{\textbf h}_{i}(t,\bs {\beta}_0)\rVert_\infty  + \frac{\eta}{\sqrt{n}}< \epsilon\Big).   
\end{split}
\end{align}
where the one but  last equality of \eqref{eq:lem2.2}
%{\vio [correct reference?]}
 is due to \cref{assump_general}~\ref{assump_general1} and holds for any $\eta >0$.
The inequality in the last line of \eqref{eq:lem2.2}
%{\vio [correct reference?]} 
was obtained by adding and subtracting $\tilde{\textbf h}_{i} (t,\bs {\beta}_0)$ to the norm two lines above it, 
%{\vio [correct reference?]}
namely by writing $ \lVert\textbf h_{n, i}(t,\bs {\beta}_0) \rVert_\infty = \lVert\textbf h_{n, i}(t,\bs {\beta}_0) - \tilde{\textbf h}_{i} (t,\bs {\beta}_0) + \tilde{\textbf h}_{i} (t,\bs {\beta}_0)\rVert_\infty$.

Under \cref{assump_general}~\ref{assump_general2} the probability $\mathbb{P}(\text{for all } i,t:\ {\displaystyle \frac{1}{\sqrt{n}} \lVert\tilde{\textbf h}_{i}(t,\bs {\beta}_0)\rVert_\infty  + \frac{\eta}{\sqrt{n}}}< \epsilon)$ converges to one and, hence, the initial probability $\mathbb{P}(\langle  D_{n,h}^{\epsilon,j}  \rangle (t) \geq \delta )$ to zero as $n\to\infty$ for each $t\in \mathcal{T}$ and across all components $j=1,\dots,d$. Thus, condition (2.5.3) of Rebolledo's theorem as stated in Theorem~II.5.1 of  \cite{Andersen}  is fulfilled. In conclusion, both requirements of Rebolledo's theorem have been verified and the proof of \cref{lem:Dn} is complete. \hfill \hfill\qedsymbol
%\end{proof}

\medskip

%\bigskip
%\begin{proof}[of \cref{lem:Cn-C}]
\noindent
\textbf{Proof of \cref{lem:Cn-C}.}\\
We wish to show that $$\sup_{t\in\mathcal{T}} \lVert \textbf B_n(t) - \textbf B(t) \rVert \stackrel{\mathbb{P}}{\longrightarrow} 0, \text{ as } n\rightarrow\infty,$$ 
where $\textbf B_n(t) = \frac1n \sum_{i=1}^n \int_0^t \textnormal D \textbf k_{n,i} (u, \bs{\beta}_0) d N_i(u)$ and $\textbf B(t)=\int_0^t \mathbb E( \tilde{\textbf K}_{1} (u,\bs{\beta}_0)  \lambda_1(u,\bs{\beta}_0) ) du$, $t\in\mathcal{T}$. For this we point out that the compensator of $\frac{1}{n}\sum_{i=1}^n N_i(\tau)$ is equal to $\frac{1}{n}\sum_{i=1}^n \Lambda_i(\tau,\bs{\beta}_0)$. From the integrability of the cumulative intensities, %\cref{assump_general}~\ref{item:ass_Bn_4},
\cref{ass:Bn_Cn}~\ref{item:ass_Bn_3}, and the law of large numbers, we can conclude that $\frac{1}{n}\sum_{i=1}^n \Lambda_i(\tau,\bs{\beta}_0) = O_p(1)$. Thus, we get from Lenglart's inequality that $\frac{1}{n}\sum_{i=1}^n N_i(\tau) = O_p(1)$. Combining this argument with \cref{ass:Bn_Cn}~\ref{item:ass_Bn_1} yields
%\begin{equation}\label{eq:Bn_B}
%\begin{array}{r@{}l}
\begin{align}
\begin{split}
    \label{eq:Bn_B}
    &{} \sup_{t\in\mathcal{T}} \lVert \textbf B_n(t) - \textbf B(t) \rVert \\
 &{}\leq \sup_{t\in\mathcal{T}} \Big\lVert  \frac{1}{n}\sum_{i=1}^n \int_0^t [\textnormal D\textbf k_{n,i}(u,\bs{\beta}_0) -  \tilde{\textbf K}_{i} (u,\bs{\beta}_0)] dN_i(u)\Big\rVert\\
&{} \quad  + \sup_{t\in\mathcal{T}}\Big \lVert  \frac{1}{n}\sum_{i=1}^n \int_0^t  \tilde{\textbf K}_{i} (u,\bs{\beta}_0) dN_i(u) - \int_0^t \mathbb E( \tilde{\textbf K}_{1} (u,\bs{\beta}_0)  \lambda_1(u,\bs{\beta}_0) ) du\Big\rVert \\
&{} \leq  \sup_{t\in\mathcal{T}} \Big\lVert  \frac{1}{n}\sum_{i=1}^n  \int_0^t \tilde{\textbf K}_{i} (u,\bs{\beta}_0) dM_i(u) \Big \rVert \\
&{} \quad  + \sup_{t\in\mathcal{T}} \Big\lVert \frac{1}{n}\sum_{i=1}^n  \int_0^t \tilde{\textbf K}_{i} (u,\bs{\beta}_0) d\Lambda_i(u,\bs{\beta}_0) - \int_0^t \mathbb E( \tilde{\textbf K}_{1} (u,\bs{\beta}_0)  \lambda_1(u,\bs{\beta}_0) ) du \Big\rVert + o_p(1) ,
\end{split}
\end{align}
%\end{array}
%\end{equation}
where in the last step the Doob-Meyer decomposition \eqref{eq:Doob} has been applied.
With \cref{ass:Bn_Cn}~\ref{item:ass_Bn_2} and Proposition II.4.1. of \cite{Andersen} it follows that the integral $\frac{1}{n}\sum_{i=1}^n  \int_0^t \tilde{\textbf K}_{i} (u,\bs{\beta}_0) dM_i(u)$ is a local square integrable martingale. The elements of the corresponding predictable covariation process at $\tau$ can be bounded from above by 
$$\frac{1}{n^2}\sum_{i=1}^n \int_0^\tau  \lVert \tilde{\textbf K}_{i} (u,\bs{\beta}_0)\rVert^2_\infty d\Lambda_i(u,\bs{\beta}_0).$$
According to \cref{ass:Bn_Cn}~\ref{item:ass_Bn_2}, $\sup_{i\in\{1,\ldots,n\}, t\in\mathcal{T}} \lVert  \tilde{\textbf K}_{i} (t,\bs{\beta}_0)\rVert^2_\infty $ is bounded for $i\in\mathbb{N}$, and, as stated above, it holds $\frac{1}{n}\sum_{i=1}^n \Lambda_i(\tau,\bs{\beta}_0) = O_p(1)$. Hence, the considered predictable covariation process and further, according to Lenglart's inequality, the first term of the second step on the right-hand side of \eqref{eq:Bn_B} converges to zero in probability, as $n\rightarrow\infty$.
It is only left to show that 
\begin{align}\label{eq:kvgz_2_B-1}
\sup_{t\in\mathcal{T}} \Big\lVert \frac{1}{n}\sum_{i=1}^n  \int_0^t \tilde{\textbf K}_{i} (u,\bs{\beta}_0) d\Lambda_i(u,\bs{\beta}_0) - \int_0^t \mathbb E( \tilde{\textbf K}_{1} (u,\bs{\beta}_0)  \lambda_1(u,\bs{\beta}_0) ) du \Big\rVert = o_p(1),
\end{align}
as $n\rightarrow\infty$. According to the integrability of the cumulative intensities and %\cref{assump_general}~\ref{item:ass_Bn_4} and 
\cref{ass:Bn_Cn}~\ref{item:ass_Bn_2} it follows that $\mathbb E( \int_0^t \lVert \tilde{\textbf K}_{1} (u,\bs{\beta}_0) \rVert_\infty \lambda_1(u,\bs{\beta}_0) du) < \infty$. From this argument in combination with \cref{ass:Bn_Cn}~\ref{item:ass_Bn_3} and the law of large numbers, we have that  
%\begin{align}\label{eq:kvgz_2_B}
    $\frac{1}{n}\sum_{i=1}^n  \int_0^t \tilde{\textbf K}_{i} (u,\bs{\beta}_0)  \lambda_i(u,\bs{\beta}_0) du $ converges almost surely to $ \mathbb E( \int_0^t \tilde{\textbf K}_{1} (u,\bs{\beta}_0)  \lambda_1(u,\bs{\beta}_0) du)$
%\end{align}
for any fixed $t\in\mathcal{T}$, as $n\rightarrow\infty$. Note that the integrability of the intensity process $\lambda_1(t,\bs{\beta}_0)$ follows from the integrability of the cumulative intensity process $\Lambda_1(t,\bs{\beta}_0)$. Thus,
due to the integrability of the cumulative intensities and
%under \cref{assump_general}~\ref{item:ass_Bn_4} and 
\cref{ass:Bn_Cn}~\ref{item:ass_Bn_2}, we can make use of Fubini's theorem, by which we can exchange the order of integration. We can conclude that 
\begin{align}\label{eq:kvgz_2_B-new}
    \frac{1}{n}\sum_{i=1}^n  \int_0^t \tilde{\textbf K}_{i} (u,\bs{\beta}_0) d\Lambda_i(u,\bs{\beta}_0) \stackrel{{\mathbb P}}{\longrightarrow} \int_0^t \mathbb E( \tilde{\textbf K}_{1} (u,\bs{\beta}_0)  \lambda_1(u,\bs{\beta}_0) ) du , 
\end{align}
pointwise in $t\in\mathcal{T}$, as $n\rightarrow\infty$.

Next, we show the corresponding uniform convergence in probability on $\mathcal{T}$. For this, we divide the interval $\mathcal{T} = [0,\tau]$ into $N$ equidistant subintervals $[t_l,t_{l+1}]$ with $t_0 = 0$, $t_N=\tau $, and $l\in\{0,1,\ldots,N-1\}$. 
The width of a subinterval is chosen such that 
$$\int_{t_l}^{t_{l+1}} \mathbb{E}(\lVert \tilde{\textbf K}_1(u,\bs{\beta}_0) \rVert \lambda(u,\bs{\beta}_0))du \leq \delta/2$$
for all $l\in\{0,1,\ldots,N-1\}$. For $t\in [0,\tau)$ we denote the lower and upper endpoint of the subinterval containing $t$ by $t_{l(t)} = \max_{l\in \{0,1,\ldots,N-1 \} }\{t_l : t_l \leq t\}$ and $t_{l(t)+1} = \min_{l\in\{1,\ldots,N\}}\{t_l: t_l > t\}$, respectively. For $t=\tau$ we choose $t_{l(\tau)} = t_{l(\tau)+1} = \tau$. In the following derivation we make use of \eqref{eq:kvgz_2_B-new} and get 
%\begin{equation*}
%\begin{array}{r@{}l}
\begin{align*}
&{}\sup_{t\in\mathcal{T}} \Big\lVert \frac{1}{n} \sum_{i=1}^n \int_0^t \tilde{\textbf K}_i(u,\bs{\beta}_0)\lambda_i(u,\bs{\beta}_0)du - \int_0^t \mathbb E( \tilde{\textbf K}_{1} (u,\bs{\beta}_0)  \lambda_1(u,\bs{\beta}_0) ) du \Big\lVert\\
    &{} =  \sup_{t\in \mathcal{T}} \Big\lVert \frac{1}{n} \sum_{i=1}^n \int_0^t \tilde{\textbf K}_i(u,\bs{\beta}_0)\lambda_i(u,\bs{\beta}_0)du 
    - \frac{1}{n} \sum_{i=1}^n \int_0^{t_{l(t)}} \tilde{\textbf K}_i(u,\bs{\beta}_0)\lambda_i(u,\bs{\beta}_0)du \\
    &{} \quad + \frac{1}{n} \sum_{i=1}^n \int_0^{t_{l(t)}} \tilde{\textbf K}_i(u,\bs{\beta}_0)\lambda_i(u,\bs{\beta}_0)du
        -\int_0^{t_{l(t)}} \mathbb E( \tilde{\textbf K}_{1} (u,\bs{\beta}_0)  \lambda_1(u,\bs{\beta}_0) ) du\\
    &{} \quad +\int_0^{t_{l(t)}} \mathbb E( \tilde{\textbf K}_{1} (u,\bs{\beta}_0)  \lambda_1(u,\bs{\beta}_0) ) du
    - \int_0^t \mathbb E( \tilde{\textbf K}_{1} (u,\bs{\beta}_0)  \lambda_1(u,\bs{\beta}_0) ) du \Big\lVert\\
    &{}\leq \sup_{t\in \mathcal{T}} \Big( \Big\lVert \frac{1}{n} \sum_{i=1}^n \int_{t_{l(t)}}^t \tilde{\textbf K}_i(u,\bs{\beta}_0)\lambda_i(u,\bs{\beta}_0)du - \int_{t_{l(t)}}^t \mathbb E( \tilde{\textbf K}_{1} (u,\bs{\beta}_0)  \lambda_1(u,\bs{\beta}_0) ) du \Big\lVert \Big) + o_p(1)\\
    &{} \leq \sup_{t\in \mathcal{T}} \Big(  \frac{1}{n} \sum_{i=1}^n \int_{t_{l(t)}}^t \lVert\tilde{\textbf K}_i(u,\bs{\beta}_0)\rVert\lambda_i(u,\bs{\beta}_0)du  + \int_{t_{l(t)}}^t \mathbb E( \lVert \tilde{\textbf K}_{1} (u,\bs{\beta}_0)\lVert  \lambda_1(u,\bs{\beta}_0) ) du  \Big) + o_p(1)\\
    &{} \leq  \max_{l\in\{0,\ldots,N-1\}}\Big(\frac{1}{n} \sum_{i=1}^n \int_{t_{l}}^{t_{l+1}} \lVert\tilde{\textbf K}_i(u,\bs{\beta}_0)\rVert\lambda_i(u,\bs{\beta}_0)du  \\
    &{} \quad + \int_{t_{l}}^{t_{l+1}} \mathbb E( \lVert \tilde{\textbf K}_{1} (u,\bs{\beta}_0)\lVert  \lambda_1(u,\bs{\beta}_0) ) du \Big) + o_p(1)\\
 &{}  \stackrel{\mathbb{P}}{\longrightarrow} 2\cdot \max_{l\in\{0,\ldots,N-1\}} \Big( \int_{t_{l}}^{t_{l+1}} \mathbb E( \lVert \tilde{\textbf K}_{1} (u,\bs{\beta}_0)\lVert  \lambda_1(u,\bs{\beta}_0) ) du \Big) \leq \delta,\quad n\rightarrow\infty.
\end{align*}
    %\end{array}
%\end{equation*}
%in probability, as $n\rightarrow\infty$. 
The convergence involved in the last step of the considerations above, follows from the same arguments that led to \eqref{eq:kvgz_2_B-new}. As we can choose the length of the subintervals $[t_l,t_{l+1}]$ such that $\delta > 0$ is arbitrarily small, we obtain \eqref{eq:kvgz_2_B-1}. \hfill \hfill\qedsymbol

\bigskip
\bigskip

%\end{proof}

%\begin{proof}[of \cref{thm:Xn-X_convergence}]
\noindent
\textbf{Proof of \cref{thm:Xn-X_convergence}.}\\
We aim to derive the limit in law of $\textbf D_{n,k} + \textbf B_n  \textbf C_n \textbf D_{n,g}(\tau)$, as $n\rightarrow\infty$, where $\textbf D_{n,k}$ and $\textbf D_{n,g}$ are vector-valued local square integrable martingales, $\textbf B_n$ is a matrix-valued stochastic process and $\textbf C_n$ is a random matrix. 
For this, we first show that the weak limit of $(\textbf D_{n,k}^\top, \textbf D_{n,g}^\top,\text{vec}(\textbf B_n)^\top, \text{vec}(\textbf C_n)^\top)$ is $(\textbf D_{k}^\top, \textbf D_{g}^\top,\text{vec}(\textbf B)^\top,\text{vec}(\textbf C)^\top)$, as $n\rightarrow\infty$. 
%For this, we first show that $(\textbf D_{n,k}^\top, \textbf D_{n,g}^\top,\text{vec}(\textbf B_n)^\top, \text{vec}(\textbf C_n)^\top)$ converges weakly to $(\textbf D_{k}^\top, \textbf D_{g}^\top,\text{vec}(\textbf B)^\top,\text{vec}(\textbf C)^\top)$, as $n\rightarrow\infty$. 
%In order to study the joint weak convergence of $(\textbf D_{n,k}(t), \textbf D_{n,g}(t),\text{vec}(\textbf B_n(t)),\text{vec}(\textbf C_n))$, we first determine the weak limit of $(\textbf D_{n,k}(t), \textbf D_{n,g}(t))$, as $n\rightarrow\infty$, and add the remaining elements in the upcoming steps. For this, we denote the local square integrable martingale $(\textbf D_{n,k}(t), \textbf D_{n,g}(t))$ as $\textbf D_{n,(k,g)}(t)$ and the corresponding limiting process $(\textbf D_{\tilde k}(t), \textbf D_{\tilde g}(t))$ as $\textbf D_{(\tilde k,\tilde g)}(t)$ {\green might be redundant after formulating Lemma 2.2 in terms of (k,g)}. %\[\textbf D_{n,h}(t) = \frac{1}{\sqrt{n}}\sum_{i=1}^n\int_0^t \textbf h_{n,i}(u,\bs{\beta}_0)dM_i(u)\] 
%with $\textbf h_{n,i}(u,\bs{\beta}_0) = (\textbf k_{n,i}(u,\bs{\beta}_0),\textbf g_{n,i}(u,\bs{\beta}_0) ) $.  
According to \cref{lem:Dn}, we have 
\[ (\textbf D_{n,k}^\top, \textbf D_{n,g}^\top)^\top  = \textbf{D}_{n,h} \stackrel{\mathcal{L}}{\longrightarrow} \textbf{D}_{\tilde h} = (\textbf D_{\tilde k}^\top, \textbf D_{\tilde g}^\top)^\top ,\quad\text{in}\; (D(\mathcal{T}))^{p+b}, \text{ as } n\rightarrow \infty,\]
where $\textbf D_{\tilde{h}}$ is a continuous zero-mean Gaussian $(p+b)$-dimensional vector martingale with covariance function $ \textbf V_{\tilde h}( t) = \int_0^t\mathbb E (  \tilde{\textbf{h}}_{1}(u,\bs {\beta}_0)^{\otimes 2} \lambda_1(u,\bs {\beta}_0))du$, $t\in \mathcal{T}$. 
As $\textbf D_{\tilde h}\in \mathcal{C}[0,\tau]^{p+b}$, we know that $\textbf D_{\tilde h}$ is separable. Furthermore, we have shown in \cref{lem:Cn-C} that there exists a $p\times q$-dimensional continuous, deterministic function $\textbf B(t)$, $t\in\mathcal{T}$, such that $\sup_{t\in\mathcal{T}} \lVert \textbf B_n(t) - \textbf B(t) \rVert \stackrel{\mathbb{P}}{\longrightarrow} 0$, as $n\rightarrow\infty$. In other words, the limit in law $\text{vec}(\textbf B)$ of $\text{vec}(\textbf B_n)$ is a constant of the space $\mathcal{C}[0,\tau]^{p q}$. Thus, we conclude with  Example 1.4.7 of \cite{Vaart_Wellner} that 
\[(\textbf D_{n,h}^\top,\text{vec}(\textbf B_n)^\top)\stackrel{\mathcal{L}}{\longrightarrow} (\textbf D_{\tilde h}^\top,\text{vec}(\textbf B)^\top), \text{ in } D[0,\tau]^{p+b+p q}, \text{ as } n\rightarrow\infty.\] 
As the last step of the first part of this proof we argue that 
\begin{align}\label{eq:D_B_C}
    (\textbf D_{n,h}^\top, \text{vec}(\textbf B_n)^\top,\text{vec}(\textbf C_n)^\top) \stackrel{\mathcal{L}}{\longrightarrow} (\textbf D_{\tilde h}(t)^\top,\text{vec}(\textbf B)^\top,\text{vec}(\textbf C)^\top) ,
\end{align}
in $\mathcal D[0,\tau]^{p+b+pq}\times \mathbb{R}^{pq}$, as $n\rightarrow\infty$. For this, we point out that $(\textbf D_{\tilde h}^\top, \text{vec}(\textbf B)^\top)\in \mathcal{C}[0,\tau]^{p+b+p q}$. Thus, $(\textbf D_{\tilde h}^\top,\text{vec}(\textbf B)^\top)$ is separable. Additionally, we have assumed in \cref{ass:Cn-C} that the random $q\times p$-dimensional matrix $\textbf C_n$ converges in probability to the deterministic matrix $\textbf C$, as $n\rightarrow\infty$. Because $\textbf C_n$ is asymptotically degenerate and $(\textbf D_{\tilde h}^\top,\text{vec}(\textbf B)^\top)$ is separable, we again use Example 1.4.7 of \cite{Vaart_Wellner} and infer that \eqref{eq:D_B_C} holds.
%Here, for every fixed $t\in\mathcal{T}$, $\tilde{\textbf{h}}_{1}(t,\bs {\beta}_0)^{\otimes 2}$ is the $2p \times 2p$-dimensional matrix with $\tilde{\textbf{h}}_{1}(t,\bs {\beta}_0) = (\tilde{\textbf k}_{1}(t,\bs{\beta}_0),\tilde{\textbf g}_{1}(t,\bs{\beta}_0))$ {\green we probably also need to assume that $\tilde{h} = \tilde{(k,g)} = (\tilde{k}, \tilde{g})$, or is this immediately clear? why?}. 

It only remains to apply the continuous mapping theorem to \eqref{eq:D_B_C} in order to derive the weak limit of $\textbf D_{n,k} + \textbf B_n  \textbf C_n \textbf D_{n,g}$, as $n\rightarrow\infty$. In particular, we use the following three maps
\begin{equation*}
\begin{array}{r@{}l}
&{} f_1: (\textbf D_{n,k}^\top, \textbf D_{n,g}(\tau)^\top, \text{vec}(\textbf B_n)^\top,\text{vec}(\textbf C_n)^\top) \mapsto (\textbf D_{n,k}^\top, \textbf D_{n,g}(\tau)^\top, \text{vec}(\textbf B_n \textbf C_n)^\top)\\
&{} f_2: (\textbf D_{n,k}^\top, \textbf D_{n,g}(\tau)^\top, \text{vec}(\textbf B_n \textbf C_n)^\top) \mapsto (\textbf D_{n,k}^\top, (\textbf B_n \textbf C_n  \textbf D_{n,g}(\tau))^\top)\\
&{} f_3: (\textbf D_{n,k}^\top, (\textbf B_n \textbf C_n  \textbf D_{n,g}(\tau))^\top) \mapsto (\textbf D_{n,k} + \textbf B_n \textbf C_n  \textbf D_{n,g}(\tau)).\\
\end{array}
\end{equation*}
Recall that $(\textbf D_{\tilde k}^\top, \textbf D_{\tilde g}^\top, \text{vec}(\textbf B)^\top, \text{vec}(\textbf C)^\top)\in \mathcal{C}[0,\tau]^{p+b+2pq}$. Thus, it follows successively with the continuous mapping theorem and the maps $f_1, f_2$ and $f_3$ that
\[\textbf D_{n,k} + \textbf B_n  \textbf C_n \textbf D_{n,g}(\tau)\stackrel{\mathcal{L}}{\longrightarrow}\textbf D_{\tilde k} + \textbf B  \textbf C \textbf D_{\tilde g}(\tau) \text{ in } D[0,\tau]^{p},\]
as $n\rightarrow\infty$. 
Moreover, the covariance function of $\textbf D_{\tilde k} + \textbf B  \textbf C \textbf D_{\tilde g}(\tau)$ at $t\in\mathcal{T} $ maps $t $ to  
%{\blue \sout{the predictable covariation process of $\textbf D_{\tilde k} + \textbf B \cdot \textbf C\cdot \textbf D_{\tilde g}(\tau)$ is given as}}
\begin{align*}
   &{} \textbf V_{\tilde k}(t) + \textbf B(t) \textbf C\textbf V_{\tilde g}(\tau) \textbf C^\top  \textbf B(t)^\top + [\textbf V_{\tilde k,\tilde g}(t) +  
\text{Cov}(\textbf D_{\tilde k}(t),\textbf D_{\tilde g}(\tau) - \textbf D_{\tilde g}(t)) ] \textbf C^\top\textbf B( t)^\top \\
    &{} \quad + \textbf B(t)  \textbf C [\textbf V_{\tilde g, \tilde k}(t) + \text{Cov}( \textbf D_{\tilde g}(\tau) - \textbf D_{\tilde g}(t),\textbf D_{\tilde k}(t)) ] \\
     &={} \textbf V_{\tilde k}(t) + \textbf B(t)  \textbf C\textbf V_{\tilde g}(\tau) \textbf C^\top  \textbf B(t)^\top +\textbf V_{\tilde k,\tilde g}(t) \textbf C^\top\textbf B( t)^\top  + \textbf B(t)  \textbf C \textbf V_{\tilde g, \tilde k}(t) ,
\end{align*}
where $\text{Cov}(\textbf D_{\tilde k}(t),\textbf D_{\tilde g}(\tau) - \textbf D_{\tilde g}(t))  = \text{Cov}( \textbf D_{\tilde g}(\tau) - \textbf D_{\tilde g}(t),\textbf D_{\tilde k}(t))^\top = 0$, because 
\begin{align*}
    \mathbb E (\textbf D_{\tilde k}(t)(\textbf D_{\tilde g}(\tau) - \textbf D_{\tilde g}(t))^\top) &=  \mathbb E (\mathbb E (\textbf D_{\tilde k}(t)(\textbf D_{\tilde g}(\tau) - \textbf D_{\tilde g}(t))^\top | \mathcal{F}_1(t)))\\
    &= \mathbb E (\textbf D_{\tilde k}(t) \mathbb E ((\textbf D_{\tilde g}(\tau) - \textbf D_{\tilde g}(t))^\top ))\\
    &= 0.
\end{align*}
Here the one but last step holds because $\sigma(\textbf D_{\tilde k}(t))\in\mathcal{F}_1(t)$ and $\textbf D_{\tilde g}(\tau) - \textbf D_{\tilde g}(t)$ is independent of $\mathcal{F}_1(t)$. In the last step it has been applied that $\mathbb E (\textbf D_{\tilde g}(\tau) - \textbf D_{\tilde g}(t) )=0$.
%{\blue \sout{where $\langle \textbf D_{\tilde k},\textbf D_{\tilde g}(\tau) - \textbf D_{\tilde g}\rangle (t) = \langle \textbf D_{\tilde g}(\tau) - \textbf D_{\tilde g},\textbf D_{\tilde k}\rangle (t) ^\top = 0$, because the underlying Gaussian martingale has independent increments.}} 
%{\vio [explanation: not very good/clear reasoning before.]} This completes the proof of \cref{thm:Xn-X_convergence}.
\hfill \hfill\qedsymbol

%\bigskip
%\bigskip
%\newpage
%%%%%%%%%%%%%%%%%%%%%%%
%%%%%%%%%%%%%%%%%%%%%%%
\subsection*{A.2 Proofs of \cref{sec:generalBootstrap}}

\noindent
\textbf{Proof of \cref{lemma:mgale}.}\\
%\begin{proof}[of \cref{lemma:mgale}]
In the first part of this proof, we show that, conditionally on the initial $\sigma$-algebra $\mathcal{F}_2(0)$, the stochastic process $\textbf{D}^*_{n,h}(t)= ({D}^{*,1}_{n,h}(t), \ldots,{D}^{*,p+b}_{n,h}(t) )$, $t\in\mathcal{T}$, is a $(p+b)$-dimensional vector of square integrable martingales with respect to $\mathcal{F}_2(t)$. 
Here, the j-th element ${D}^{*,j}_{n,h}$ of $\textbf{D}^*_{n,h}$, $j=1,\ldots,p+b$, is given by
$${D}^{*,j}_{n,h}(t)=\frac{1}{\sqrt{n}}\sum_{i=1}^n\int_0^t {h}^j_{n,i}(u,\hat{\bs{\beta}}_n) G_i(u) \, dN_i(u),\quad t\in\mathcal{T},$$ 
where ${h}^j_{n,i}(t,\hat{\bs{\beta}}_n)$ denotes the j-th element of the $(p+b)$-dimensional function $\textbf{h}_{n,i}(t,\hat{\bs{\beta}}_n)$. 
%Thus, we show that the stochastic process ${D}^{*,j}_{n,h}(t) $, $t\in\mathcal{T}$, is a square integrable martingale for all $j=1,\ldots,p+b$. 
For later use we write ${D}^{*,j}_{n,h}$ as the scaled sum over ${D}^{*,j}_{n,h,i} = \int_0^\cdot {h}^j_{n,i}(u,\hat{\bs{\beta}}_n) G_i(u) \, dN_i(u)$, namely $ {D}^{*,j}_{n,h}(t) = \frac{1}{\sqrt{n}}\sum_{i=1}^n {D}^{*,j}_{n,h,i}(t) $, $t\in\mathcal{T}$. Furthermore, by incorporating the jump time points $T_{i,1},\ldots , T_{i,n_i}$ of the counting process $N_i$, we can write
$${D}^{*,j}_{n,h}(t) = \frac{1}{\sqrt{n}}\sum_{i=1}^n \sum_{r:T_{i,r}\leq t} {h}^j_{n,i}(T_{i,r},\hat{\bs{\beta}}) G_i(T_{i,r}),\quad t\in\mathcal{T}.$$
Clearly, all stochastic processes ${D}^{*,j}_{n,h}(t)$, $t\in\mathcal{T}$, $j=1,\ldots,p+b$, are adapted to the filtration $\mathcal{F}_2(t)$, $t\in \mathcal{T}$. Moreover, for all $j=1,\ldots,p+b$, ${D}^{*,j}_{n,h}$ is cadlag, as the same holds for the counting processes $N_i$, $i=1,\ldots,n$. As we work with a probability space, square integrability implies integrability of a stochastic process. 
Thus, we directly show that ${D}^{*,j}_{n,h}$ is square integrable for all $j=1,\ldots, p+b$. For this we wish to show that
\[\sup_{t\in\mathcal{T}}\mathbb{E}_0( D^{*,j}_{n,h}(t)^2) = \sup_{t\in\mathcal{T}}\mathbb{E}_0\Big(\frac{1}{n}\Big(\sum_{i=1}^n D^{*,j}_{n,h,i}(t)\Big)^2\Big) < \infty, \]
where $\mathbb{E}_0$ denotes the conditional expectation $\mathbb{E}(\cdot|\mathcal{F}_2(0))$. In preparation for this, we state
\begin{align}
\begin{split}
\label{eq:squareInt_0}
  &{} \frac{1}{n}\Big(\sum_{i=1}^n D^{*,j}_{n,h,i}(t)\Big)^2 = \frac{1}{n} \sum_{i=1}^n \sum_{l=1}^n D^{*,j}_{n,h,i}(t)  D^{*,j}_{n,h,l}(t)\\
    &{} = \frac{1}{n} \sum_{i=1}^n \sum_{l=1}^n \sum_{r:T_{i,r}\leq t} \sum_{v:T_{i,v}\leq t}  {h}^j_{n,i}(T_{i,r},\hat{\bs{\beta}}_n) {h}^j_{n,l}(T_{l,v},\hat{\bs{\beta}}_n) G_i(T_{i,r}) G_l(T_{l,v}).
\end{split}
\end{align}
%\begin{equation}
%    \begin{array}{r@{}l}
    %\end{array}
%\end{equation}
In the next step we use that the functions $\textbf h_{n,i}(t,\hat{\bs{\beta}}_n)$, $i=1,\ldots,n$, are $\mathcal{F}_2(0)$-measurable. Additionally, we apply that the values of the multiplier process $G_i(t)$, $t \in \mathcal{T}^\Delta_{n,i}$,
%the sigma-algebra $\sigma(\{G_i(t)\}_{t>0})$ generated by the stochastic process $(G_i(t):t\in \mathcal{T})$ 
%
are independent of the $\sigma$-algebra $\mathcal{F}_2(0)$. 
%\newpage
Combining these assumptions with \eqref{eq:squareInt_0}, we get
\begin{align}
\begin{split}
\label{eq:squareInt_1}
 &{} \mathbb{E}_0( (D^{*,j}_{n,h}(t))^2) \\
        &{} = \frac{1}{n} \sum_{i=1}^n \sum_{l=1}^n \sum_{r:T_{i,r}\leq t} \sum_{v:T_{i,v}\leq t}  {h}^j_{n,i}(T_{i,r},\hat{\bs{\beta}}_n) {h}^j_{n,l}(T_{l,v},\hat{\bs{\beta}}_n) \mathbb{E}(G_i(T_{i,r}) G_l(T_{l,v})). 
\end{split}
\end{align}
%\begin{equation}\label{eq:squareInt_1}
%    \begin{array}{r@{}l}
    %\end{array}
%\end{equation}
By construction of the multiplier processes we have for $i\neq l$ or  $\{i= l, r\neq v\}$ 
\[\mathbb{E}(G_i(T_{i,k}) G_l(T_{l,v})) = \mathbb{E}(G_i(T_{i,k})) \mathbb{E}(G_l(T_{l,v})) = 0,\]
and for $\{i=l,r= v\}$
\[\mathbb{E}(G_i(T_{i,r}) G_l(T_{l,v})) = \mathbb{E}(G_i(T_{i,r})^2) = 1.\]
Thus, \eqref{eq:squareInt_1} simplifies to $\mathbb{E}_0( (D^{*,j}_{n,h}(t))^2) {}=  \frac{1}{n} \sum_{i=1}^n  \sum_{r:T_{i,r}\leq t}  {h}^j_{n,i}(T_{i,r},\hat{\bs{\beta}}_n)^2$. Finally, it holds that
%\begin{equation*}
%    \begin{array}{r@{}l}
    $$\sup_{t\in\mathcal{T}}\mathbb{E}_0( D^{*,j}_{n,h}(t)^2) \leq  \sup_{t\in\mathcal{T}, i\in\{1,\ldots,n\}} {h}^j_{n,i}(t,\hat{\bs{\beta}}_n)^2\cdot  \max_{i\in\{1,\ldots,n\}} N_i(\tau)< \infty,$$
    %\end{array}
%\end{equation*}
since $\textbf h_{n,i}(t,\hat{\bs{\beta}}_n)$ is a known function and hence, all components ${h}^j_{n,i}(t,\hat{\bs{\beta}}_n)$, $j=1,\ldots,p+b$, are bounded on $\mathcal{T}$. 
Moreover, the observed number of events within the time frame $\mathcal{T}=[0,\tau]$, $N_i(\tau)$, is finite for all individuals $i=1,\ldots,n$. In conclusion, ${D}^{*,j}_{n,h}(t)$, $t\in\mathcal{T}$, is square integrable for all $j=1,\ldots,p+b$, given the initial $\sigma$-algebra $\mathcal{F}_2(0)$.

Next, we consider the martingale property for the stochastic process ${D}^{*,j}_{n,h}(t)$, $t\in\mathcal{T}$. 
Due to the linearity of the conditional expectation, is suffices to verify the martingale property for the summands ${D}^{*,j}_{n,h,i}(t)$ of the scaled sum ${D}^{*,j}_{n,h}(t)$, $i=1,\ldots,n$. For this, we recall that the function $\textbf{h}_{n,i}(t,\hat{\bs{\beta}}_n)$ and the counting process $N_i(t)$ are $\mathcal{F}_2(0)\subset \mathcal{F}_2(t)$-measurable for $t\in\mathcal{T}$, respectively, $i=1,\ldots ,n$. Furthermore, for a jump at $u \leq s$, the multiplier process $G_i(u)$ is $\mathcal{F}_2(s)$-measurable, and,
%for $u>s$ 
if $u$ is greater than or equal to the earliest jump time point, say $T_{i}(s^+)$, of process $i$ in $(s,\tau]$,
the values of $G_i(u)$ and the filtration $\mathcal{F}_2(s)$ are independent, $i=1,\ldots,n$. Moreover, we use that the multiplier process $G_i(t)$, $t \in \mathcal{T}$, has mean zero. This yields for any $t>s$,
%\begin{eqnarray*}
 \begin{align*}
%\label{mgaleProp2}
%\begin{split}
  &{} \mathbb{E}[{D}^{*,j}_{n,h,i}(t)|\mathcal{F}_2(s)] \\
  &{} = \mathbb{E}\Big[\int_0^t {h}^j_{n,i}(u,\hat{\bs{\beta}}_n) G_i(u)\, dN_i(u)|\mathcal{F}_2(s)\Big] \\
  &{} = \mathbb{E}\Big[\int_0^s {h}^j_{n,i}(u,\hat{\bs{\beta}}_n)  G_i(u) \, dN_i(u)+ \int_s^t {h}^j_{n,i}(u,\hat{\bs{\beta}}_n) G_i(u) \, dN_i(u)\Big|\mathcal{F}_2(s)\Big] \,  \\
  &{} = {D}^{*,j}_{n,h,i}(s) + \int_s^t {h}^j_{n,i}(u,\hat{\bs{\beta}}_n) \, \mathbb{E}(G_i(u)| \mathcal{F}_2(s)) \, dN_i(u) \nonumber \\  &{} ={D}^{*,j}_{n,h,i}(s) + \int_{T_{i}(s^+)}^t {h}^j_{n,i}(u,\hat{\bs{\beta}}_n) \, \mathbb{E}(G_i(u)) \, dN_i(u)  \\
  &{} ={D}^{*,j}_{n,h,i}(s).
%\end{eqnarray*}
%\end{split}
\end{align*}
 Thus, we have shown that all elements ${D}^{*,j}_{n,h}$ of $\textbf{D}^{*}_{n,h}$, $j=1,\ldots, p+b$, fulfill the martingale property. In conclusion, the stochastic process $\textbf{D}^{*}_{n,h}$ is a $(p+b)$-dimensional vector of square integrable martingales with respect to $\mathcal{F}_2(t)$, $t\in\mathcal{T}$. With this the first part of \cref{lemma:mgale} has been proven.

In the second part of this proof we derive the predictable covariation process $\langle \textbf{D}^*_{n,h}\rangle $ and the optional covariation process $[ \textbf{D}^*_{n,h}] $ of $ \textbf{D}^*_{n,h}$. First, we consider the predictable covariation 
process $\langle\textbf{D}^*_{n,h}\rangle (t)$:
%\begin{equation}\label{eq:pred_1}
 %   \begin{array}{r@{}l}
\begin{align}
\begin{split}
\label{eq:pred_2}
    &{} \langle\textbf{D}^*_{n,h}\rangle
   = \frac{1}{{n}}\Big\langle\sum_{i=1}^n ( {D}^{*,1}_{n,h,i},\ldots , {D}^{*,p+b}_{n,h,i})\Big\rangle  \\ 
    &{} = \frac{1}{{n}}\Big(\Big\langle\sum_{i=1}^n {D}^{*,j}_{n,h,i},\sum_{i=1}^n {D}^{*,r}_{n,h,i}\Big\rangle  \Big)_{j,r=1}^{p+b}\\
    &{}= \frac{1}{{n}}\Big(\sum_{i=1}^n\sum_{l=1}^n\langle {D}^{*,j}_{n,h,i}, {D}^{*,r}_{n,h,l}\rangle  \Big)_{j,r=1}^{p+b}\\
    &{} = \frac{1}{{n}}\sum_{i=1}^n\sum_{l = i}\big(\langle {D}^{*,j}_{n,h,i}, {D}^{*,r}_{n,h,l}\rangle\big)_{j,r=1}^{p+b} 
    \quad + \quad \frac{1}{{n}}\sum_{i=1}^n\sum_{l\neq i}\big( \langle {D}^{*,j}_{n,h,i}, {D}^{*,r}_{n,h,l}\rangle \big)_{j,r=1}^{p+b},
\end{split}
\end{align}
where in the second step of \eqref{eq:pred_2} we used that the predictable covariation process of a vector valued martingale is the matrix of the predictable covariation processes of its components. 
In the following we consider the predictable covariation processes $\langle {D}^{*,j}_{n,h,i}, {D}^{*,r}_{n,h,l}\rangle $ for $i=l$ and $i\neq l$ separately. Recall that the functions $\textbf h_{n,i}(t,\hat{\bs{\beta}}_n)$ and the counting processes $N_i$ are $\mathcal{F}_2(0)\subset \mathcal{F}_2(t)$-measurable, respectively, and that the values of the multiplier processes $G_i(t)$, $t \in \mathcal{T}$, are independent of the $\sigma$-algebra $\mathcal{F}_2(t-)$, $i=1,\ldots,n$.  
%\newpage
We then get for $i = l$,
%\begin{equation*}
%    \begin{array}{r@{}l}
\begin{align}
\begin{split}\label{eq:i=l}
 &{} \langle D^{*,j}_{n,h,i},D^{*,r}_{n,h,i} \rangle (t) \\
&{}= \int_0^t \text{Cov}\big( d D^{*,j}_{n,h,i}(u),dD^{*,r}_{n,h,i}(u) | \mathcal{F}_2(u-) \big)\\ 
    %&=&  \int_0^t \big\{ \mathbb{E}_{u-}(dD^{*,i,j}_{n,h}(u)\cdot dD^{*,i,l}_{n,h}(u)) - \mathbb{E}_{u-}(dD^{*,i,j}_{n,h}(u))\cdot \mathbb{E}_{u-}(dD^{*,i,l}_{n,h}(u)) \big\}\nonumber \\ 
    &{}=  \int_0^t \text{Cov}\big(h^{j}_{n,i}(u,\hat{\bs{\beta}}_n)\, G_i(u)\, dN_i(u), h^{r}_{n,i}(u,\hat{\bs{\beta}}_n)\, G_i(u)\, dN_i(u) | \mathcal{F}_2(u-) \big) \\ 
    &{}=  \int_0^t h^{j}_{n,i}(u,\hat{\bs{\beta}}_n)h^{r}_{n,i}(u,\hat{\bs{\beta}}_n)\text{Var}( G_i(u) )dN_i(u)   \\ 
    &{}= \int_0^t h^{j}_{n,i}(u,\hat{\bs{\beta}}_n)h^{r}_{n,i}(u,\hat{\bs{\beta}}_n)dN_i(u)  ,
\end{split}
\end{align}
%    \end{array}
%\end{equation*}
where for the last  equation above we have used that the multiplier processes $G_i(t)$, $t\in\mathcal{T}$, have unit variance, $i=1,\ldots,n$.

For $i \neq l$ it holds that
\begin{align}
\begin{split}\label{eq:i_neq_l}
   d\langle {D}^{*,j}_{n,h,i}, {D}^{*,r}_{n,h,l}\rangle (t) &{}= 
   \text{Cov}\big( d D^{*,j}_{n,h,i}(u),d D^{*,r}_{n,h,l}(u) | \mathcal{F}_2(u-) \big)\\
   &{}= \text{Cov}\big( h_{n,i}^j(u,\hat{\bs{\beta}}_n)G_i(u) dN_i(u),h_{n,l}^r (u,\hat{\bs{\beta}}_n)G_l(u) dN_l(u) | \mathcal{F}_2(u-) \big)\\
   &{}= h_{n,i}^j(u,\hat{\bs{\beta}}_n)h_{n,l}^r (u,\hat{\bs{\beta}}_n)\text{Cov}\big(G_i(u) ,G_l(u)  \big) dN_i(u)dN_l(u)\\
   &{}=0,
   \end{split}
\end{align}
%\begin{equation*}
    %\begin{array}{r@{}l}
    %\end{array}
%\end{equation*}
where in the last step  we have applied that the multiplier processes $G_1(t),\ldots,G_n(t)$, $t\in\mathcal{T}$, are pairwise independent and no two processes jump simultaneously. Hence, $\langle {D}^{*,j}_{n,h,i}, {D}^{*,r}_{n,h,l}\rangle (t) = 0$ for $i \neq l$. Combining \eqref{eq:pred_2}, \eqref{eq:i=l}, and \eqref{eq:i_neq_l}, we can state the final form of the predictable covariation process $\langle\textbf{D}^*_{n,h}\rangle  $ of $\textbf{D}^*_{n,h} $ at $t\in \mathcal{T}$ in matrix notation
\begin{align*}
%\begin{equation*}
%    \begin{array}{r@{}l}
    \langle\textbf{D}^*_{n,h}\rangle (t) &{} = \frac{1}{n}\sum_{i=1}^n\int_0^t\big(   h^{j}_{n,i}(u,\hat{\bs{\beta}}_n) h^{r}_{n,i}(u,\hat{\bs{\beta}}_n)\big)_{j,r=1}^{p+b} dN_i(u)
    = \frac{1}{n}\sum_{i=1}^n \int_0^t\textbf{h}_{n,i}(u,\hat{\bs{\beta}}_n)^{\otimes 2}dN_i(u),
%    \end{array}
%\end{equation*}
\end{align*}
which proves the second part of \cref{lemma:mgale}.

For the optional covariation process $[\textbf{D}^*_{n,h}] $ of $\textbf{D}^*_{n,h}$ we can write analogously to \eqref{eq:pred_2}
\begin{align}
\begin{split}
\label{eq:pred_3}
    [\textbf{D}^*_{n,h}] (t) &{}= \frac{1}{{n}}\sum_{i=1}^n\sum_{l = i}\big([ {D}^{*,j}_{n,h,i}, {D}^{*,r}_{n,h,l}] (t)\big)_{j,r=1}^{p+b} 
     + \frac{1}{{n}}\sum_{i=1}^n\sum_{l\neq i}\big( [ {D}^{*,j}_{n,h,i}, {D}^{*,r}_{n,h,l}] (t)\big)_{j,r=1}^{p+b}.
\end{split}
\end{align}
%\begin{equation}
%    \begin{array}{r@{}l}
%    \end{array}
%\end{equation}
Again, we consider the optional covariation process $[ {D}^{*,j}_{n,h,i}, {D}^{*,r}_{n,h,l}]$ for $i=l$ and $i\neq l$ 
%\newpage
separately. For $i=l$ we get
%\begin{equation*}
    %\begin{array}{r@{}l}
\begin{align}
\begin{split}\label{eq:i=l_2}
[D^{*,j}_{n,h,i},D^{*,r}_{n,h,i}](t) &{}= \sum_{u\leq t} \Delta D^{*,j}_{n,h,i}(u) \Delta D^{*,r}_{n,h,i}(u)\\
&{}= \sum_{u\leq t} h^j_{n,i}(u,\hat{\bs{\beta}}_n)\, G_i(u) \, \Delta N_i(u) h^r_{n,i}(u,\hat{\bs{\beta}}_n)\, G_i(u) \, \Delta N_i(u)\\
&{}=  \int_0^t h^j_{n,i}(u,\hat{\bs{\beta}}_n) h^r_{n,i}(u,\hat{\bs{\beta}}_n) \, G^2_i(u)\, dN_i(u).
\end{split}
\end{align}
    %\end{array}
%\end{equation*}
%\newpage
For $i\neq l$ it holds that
%\begin{equation*}
    %\begin{array}{r@{}l}
\begin{align}
\begin{split}\label{eq:i_neq_l_2}
    [D^{*,j}_{n,h,i},D^{*,r}_{n,h,l}](t) &{}= \sum_{u\leq t} \Delta D^{*,j}_{n,h,i}(u) \Delta D^{*,r}_{n,h,l}(u)\\
&{}= \sum_{u\leq t}  h^j_{n,i}(u,\hat{\bs{\beta}}_n)\, G_i(u) \, \Delta N_i(u) h^r_{n,l}(u,\hat{\bs{\beta}}_n)\, G_l(u) \,  \Delta N_l(u)\\
&{}=  0, 
\end{split}
\end{align}
    %\end{array}
%\end{equation*}
where in the last step of the equation above we have used that no two counting processes jump at the same time. Combining \eqref{eq:pred_3}, \eqref{eq:i=l_2}, and \eqref{eq:i_neq_l_2}, we find for the optional covariation process $[\textbf{D}^*_{n,h}]$ of $\textbf{D}^*_{n,h}  $ at $t\in\mathcal{T}$ in matrix notation:
%\begin{equation*}
%    \begin{array}{r@{}l}
\begin{align*}
    [\textbf{D}^*_{n,h}] (t) &{} = \frac{1}{n}\sum_{i=1}^n\int_0^t\big(   h^{j}_{n,i}(u,\hat{\bs{\beta}}_n) h^{r}_{n,i}(u,\hat{\bs{\beta}}_n)\big)_{j,r=1}^{p+b} \,G_i(u)^2 dN_i(u)
\end{align*}
\begin{align*}    
    &{}= \frac{1}{n}\sum_{i=1}^n \int_0^t\textbf{h}_{n,i}(u,\hat{\bs{\beta}}_n)^{\otimes 2}\,G_i(u)^2dN_i(u),
\end{align*}
%    \end{array}
%\end{equation*}
which proves the third part of \cref{lemma:mgale} and the proof of the lemma is complete.
\hfill \hfill\qedsymbol
%\end{proof}

\bigskip

%\smallskip
%\newpage
\noindent
\textbf{Proof of \cref{lem:predCov_D*}.}\\
%\begin{proof}[of \cref{lem:predCov_D*}]
According to \cref{lemma:mgale}, $\textbf{D}_{n,h}^*$ is a vector of square integrable martingales and its predictable covariation process takes the form
\begin{align*}
\langle \textbf{D}_{n,h}^* \rangle (t) &{}= \langle \frac{1}{\sqrt{n}} \sum_{i=1}^n \int_0^\cdot \textbf{h}_{n,i}(u,\hat{\bs{\beta}}_n) G_i(u)\,dN_i(u)\rangle (t)\\
&{}= \frac{1}{n} \sum_{i=1}^n\int_0^t \textbf{h}_{n,i}(u,\hat{\bs{\beta}}_n)^{\otimes 2} \,dN_i(u)\\
&{}=\frac{1}{n} \sum_{i=1}^n\int_0^t \textbf{h}_{n,i}(u,\hat{\bs{\beta}}_n)^{\otimes 2} \,(dM_i(u)+d\Lambda_i(u,\bs{\beta}_0)),
\end{align*}
where in the third step we have used the Doob-Meyer decomposition with $M_i$ a square integrable martingale with respect to $\mathcal F_1$ and $\Lambda_i(\cdot,\bs{\beta}_0)$ its compensator. Note the similarity of the integral with respect to $\Lambda_i(t,\bs{\beta}_0)$ to that of  $\langle \textbf D_{n,h} \rangle (t)$ in \eqref{eq:cov_Dn}, the only difference being that the integrand is evaluated at $\hat{\bs{\beta}_n}$ instead of at $\bs{\beta}_0$.
%\newpage
We make use of the result about $\langle \textbf D_{n,h} \rangle (t)$ and consider
\begin{align}\label{eq:lem3.6_1}
\begin{split}
        &{} \frac{1}{n} \sum_{i=1}^n\int_0^t \textbf{h}_{n,i}(u,\hat{\bs{\beta}}_n)^{\otimes 2} \,d\Lambda_i(u,\bs{\beta}_0) - \langle \textbf D_{n,h} \rangle (t) + \langle \textbf D_{n,h} \rangle (t)\\
    &{} = \frac{1}{n} \sum_{i=1}^n\int_0^t [\textbf{h}_{n,i}(u,\hat{\bs{\beta}}_n)^{\otimes 2} - \textbf{h}_{n,i}(u,{\bs{\beta}_0})^{\otimes 2} ] \,d\Lambda_i(u,\bs{\beta}_0) +  \langle \textbf D_{n,h} \rangle (t),
\end{split}
\end{align}    
where the first term on the right-hand side can be bounded from above in the following way.
\begin{align*}
%\begin{split}
    & \frac{1}{n} \sum_{i=1}^n\int_0^t [\textbf{h}_{n,i}(u,\hat{\bs{\beta}}_n)^{\otimes 2} - \textbf{h}_{n,i}(u,{\bs{\beta}_0})^{\otimes 2} ] \,d\Lambda_i(u,\bs{\beta}_0)\\
    &\leq \sup_{i\in\{1,\ldots ,n\},t\in\mathcal{T}} \lVert \textbf{h}_{n,i}(u,\hat{\bs{\beta}}_n)^{\otimes 2} -\tilde{\textbf{h}}_i(t,\bs{\beta}_0 )^{\otimes 2} +\tilde{\textbf{h}}_i(t,\bs{\beta}_0 )^{\otimes 2}
    - \textbf{h}_{n,i}(u,{\bs{\beta}_0})^{\otimes 2}  \rVert_\infty  \frac{1}{n} \sum_{i=1}^n \Lambda_i(t,\bs{\beta}_0)\\
     & \leq \Big(\sup_{i\in\{1,\dots,n\},t\in \mathcal{T}} \lVert ( \textbf h_{n, i}(t, \hat{\bs {\beta}}_n) -\tilde{\textbf{h}}_{ i}(t,\bs {\beta}_0) ) \textbf h_{n, i}(t, \hat{\bs {\beta}}_n)^\top \rVert_\infty \\
     &\qquad + \sup_{i\in\{1,\dots,n\},t\in \mathcal{T}} \lVert \tilde{\textbf{h}}_{i}(t,\bs {\beta}_0) (\textbf h_{n, i}(t, \hat{\bs {\beta}}_n) - \tilde{\textbf{h}}_{i}(t,\bs {\beta}_0))^\top \rVert_\infty \\
     &\qquad + \sup_{i\in\{1,\dots,n\},t\in \mathcal{T}} \lVert ( \textbf h_{n, i}(t, {\bs {\beta}}_0) -\tilde{\textbf{h}}_{ i}(t,\bs {\beta}_0) ) \textbf h_{n, i}(t, {\bs {\beta}}_0)^\top \rVert_\infty \\
     &\qquad + \sup_{i\in\{1,\dots,n\},t\in \mathcal{T}} \lVert \tilde{\textbf{h}}_{i}(t,\bs {\beta}_0) (\textbf h_{n, i}(t, {\bs {\beta}}_0) - \tilde{\textbf{h}}_{ i}(t,\bs {\beta}_0))^\top \rVert_\infty  \Big)
       \frac{1}{n} \sum_{i=1}^n \Lambda_i(t,\bs {\beta}_0) .
%      \end{split}
\end{align*}
All four terms in brackets converge to zero in probability, as $n\rightarrow\infty$, according to \cref{assump_general}~\ref{assump_general1},~\ref{assump_general2}, and the fact that $\textbf h_{n, i}(t, {\bs {\beta}}_0) $ and $ \textbf h_{n, i}(t, {\hat{\bs {\beta}}}_n)$ are (locally) bounded. In the following we make use of results of the proof of \cref{lem:Dn}. For this we note that convergence in probability is equivalent to convergence in conditional probability, cf.\ Fact 1 of the supplement of \cite{dobler19}. As stated in the proof of \cref{lem:Dn}, $\frac{1}{n} \sum_{i=1}^n \Lambda_i(t,\bs {\beta}_0) = O_p(1)$, according to \cref{assump_general}~\ref{assump_general3}, the integrability of $\Lambda_i(t,\bs {\beta}_0)$ and the law of large numbers. Hence, the first term on the right-hand side of \eqref{eq:lem3.6_1} converges to zero in probability, as $n\rightarrow\infty$. Additionally, according to \cref{assump_general}~\ref{assump_general2},~\ref{assump_general3}, the integrability of $\Lambda_i(t,\bs {\beta}_0)$ and the law of large numbers, we have shown in the proof of \cref{lem:Dn} that
\begin{align*}
    \langle\textbf D_{n,h} \rangle (t)\stackrel{\mathbb P}{\longrightarrow} \int_0^t\mathbb E \Big(  \tilde{\textbf{h}}_{1}(u,\bs {\beta}_0)^{\otimes 2} \lambda_1(u,\bs {\beta}_0)\Big)du =\textbf V_{\tilde h}(t),\text{ for all } t\in\mathcal{T}, \text{ as }  n\rightarrow\infty;
\end{align*}
cf.~\cref{assump_general}~\ref{assump_general3}. In particular, $\frac{1}{n} \sum_{i=1}^n\int_0^t \textbf{h}_{n,i}(u,\hat{\bs{\beta}}_n)^{\otimes 2} \,d\Lambda_i(u,\bs{\beta}_0)$ and $\langle\textbf D_{n,h} \rangle (t)$ are asymptotically equivalent.

Next, we consider the integral with respect to the local square integrable martingale $M_i$, $i=1,\ldots,n$. As, conditionally on $\mathcal{F}_2(0)$, the integrands $\textbf{h}_{n,i}(\cdot,\hat{\bs{\beta}})^{\otimes 2}$, $i=1,\ldots,n$, are known and, hence, predictable with respect to $\mathcal{F}_2$
and locally bounded, the corresponding integral $\textbf W_{n}(t) = \frac{1}{n}\sum_{i=1}^n\int_0^t  \textbf{h}_{n,i}(u,\hat{\bs{\beta}})^{\otimes 2} \,dM_i(u) $ is a local square integrable martingale (Proposition II.4.1, \citealt[p. 78]{Andersen}). Hence, we apply Lenglart's inequality in order to show that $\textbf W_{n}(t)$ converges to zero in probability for all $t\in \mathcal{T}$, as $n\rightarrow\infty$. 
For this purpose, we consider its predictable covariation process
\begin{align}\label{eq:lem3.6_2}
\begin{split}
\langle \text{vec}(\textbf W_{n}) \rangle (\tau) &{}= \langle \frac{1}{n}\sum_{i=1}^n\int_0^\cdot  \text{vec}(\textbf{h}_{n,i}(u,\hat{\bs{\beta}})^{\otimes 2}) \,dM_i(u)\rangle (\tau)\\
 &{}= 
\frac{1}{n^2}\sum_{i=1}^n\int_0^\tau  \text{vec}(\textbf{h}_{n,i}(u,\hat{\bs{\beta}})^{\otimes 2})^{\otimes 2} \,d\Lambda_i(u,\bs{\beta}_0),\\
&{}=
\frac{1}{n^2}\sum_{i=1}^n\int_0^\tau  [\text{vec}(\textbf{h}_{n,i}(u,\hat{\bs{\beta}})^{\otimes 2})^{\otimes 2} - \text{vec}(\tilde{\textbf{h}}_{i}(u,{\bs{\beta}_0})^{\otimes 2})^{\otimes 2}] \,d\Lambda_i(u,\bs{\beta}_0)\\
&{}\quad  + 
\frac{1}{n^2}\sum_{i=1}^n\int_0^\tau   \text{vec}(\tilde{\textbf{h}}_{i}(u,{\bs{\beta}_0})^{\otimes 2})^{\otimes 2} \,d\Lambda_i(u,\bs{\beta}_0),
%&\stackrel{P}{\longrightarrow}& 0,\; n\rightarrow\infty ,
\end{split}
\end{align} 
where in the second equality it has been used that the martingales $M_1(t),\dots,M_n(t)$ are independent. We wish to show that the first term on the right-hand side of the third step 
converges to zero in probability, as $n\rightarrow\infty$. For this, it suffices to consider the largest component
\begin{align*}
&\frac{1}{n^2}\sum_{i=1}^n\int_0^\tau  \lVert \text{vec}(\textbf{h}_{n,i}(u,\hat{\bs{\beta}}_n)^{\otimes 2})^{\otimes 2} - \text{vec}(\tilde{\textbf{h}}_{i}(u,{\bs{\beta}_0})^{\otimes 2})^{\otimes 2}\rVert_\infty \,d\Lambda_i(u,\bs{\beta}_0)\\
&\leq \sup_{i\in\{1\ldots ,n\},t\in \mathcal{T}} \lVert \text{vec}(\textbf{h}_{n,i}(t,\hat{\bs{\beta}}_n)^{\otimes 2})^{\otimes 2} - \text{vec}(\tilde{\textbf{h}}_{i}(t,{\bs{\beta}_0})^{\otimes 2})^{\otimes 2} \rVert_\infty  \frac{1}{n^2}\sum_{i=1}^n \Lambda_i(\tau,\bs{\beta}_0).
\end{align*}
It holds that 
\begin{align*}
&{}\lVert\text{vec}(\textbf{h}_{n,i}(t,\hat{\bs{\beta}}_n)^{\otimes 2})^{\otimes 2} - \text{vec}(\tilde{\textbf{h}}_{i}(t,{\bs{\beta}_0})^{\otimes 2})^{\otimes 2}\rVert_\infty \\
    &{} \leq \quad   \rVert \textbf{h}_{n,i}(t,\hat{\bs{\beta}}_n) \lVert_\infty^2      
   \Big[ \lVert \textbf{h}_{n,i}(t,\hat{\bs{\beta}}_n) -
    \tilde{\textbf{h}}_{i}(t,{\bs{\beta}_0})\rVert_\infty
    \lVert \textbf{h}_{n,i}(t,\hat{\bs{\beta}}_n) \rVert_\infty \\
&{}\quad + 
    \lVert 
    \tilde{\textbf{h}}_{i}(t,{\bs{\beta}_0})\rVert_\infty 
    \lVert\textbf{h}_{n,i}(t,\hat{\bs{\beta}}_n) -
    \tilde{\textbf{h}}_{i}(t,{\bs{\beta}_0})\rVert_\infty
   \Big ]\\
&{} \quad + \lVert \tilde{\textbf{h}}_{i}(t,{\bs{\beta}_0}) \rVert_\infty^2 
  \Big[ \lVert \textbf{h}_{n,i}(t,\hat{\bs{\beta}}_n) -
    \tilde{\textbf{h}}_{i}(t,{\bs{\beta}_0})\rVert_\infty 
    \lVert \textbf{h}_{n,i}(t,\hat{\bs{\beta}}_n) \rVert_\infty \\
&{}\quad + 
    \lVert  \tilde{\textbf{h}}_{i}(t,{\bs{\beta}_0})\rVert_\infty
    \lVert \textbf{h}_{n,i}(t,\hat{\bs{\beta}}_n) -
    \tilde{\textbf{h}}_{i}(t,{\bs{\beta}_0})\rVert_\infty
   \Big ],
\end{align*}
%\newpage
where we used the triangle inequality and applied $\textbf a^{\otimes 2} - \textbf b^{\otimes 2} = (\textbf a - \textbf b)\textbf a^\top + \textbf b (\textbf a - \textbf b)^\top$ for two vectors $\textbf a , \textbf b$ twice, i.e.,
\begin{align}\label{eq:otimes_2_twice}
\begin{split}
    \text{vec}[\textbf a^{\otimes 2}]^{\otimes 2} - \text{vec}[\textbf b^{\otimes 2}]^{\otimes 2} &= \text{vec}[(\textbf a - \textbf b)\textbf a^\top + \textbf b(\textbf a - \textbf b)^\top] \text{vec}[\textbf a \textbf a^\top]^\top\\
    &\quad +  \text{vec}[\textbf b \textbf b^\top]\text{vec}[(\textbf a - \textbf b)\textbf a^\top  + \textbf b(\textbf a - \textbf b)^\top]^\top.
\end{split}
\end{align}
Hence, according to \cref{assump_general}~\ref{assump_general1},~\ref{assump_general2}, and since  $\textbf h_{n, i}(t, \hat{\bs {\beta}}_n)$ is locally bounded for $i=1,\ldots , n$, it follows that $\sup_{i\in\{1\ldots ,n\},t\in \mathcal{T}} \lVert \text{vec}(\textbf{h}_{n,i}(u,\hat{\bs{\beta}}_n)^{\otimes 2})^{\otimes 2} - \text{vec}(\tilde{\textbf{h}}_{i}(u,{\bs{\beta}_0})^{\otimes 2})^{\otimes 2} \rVert_\infty = o_p(1) $. As explained before, we have $\frac{1}{n} \sum_{i=1}^n \Lambda_i(\tau,\bs {\beta}_0) = O_p(1)$. In conclusion, the first term on the right-hand side of the third step of \eqref{eq:lem3.6_2} converges to zero in probability, as $n\rightarrow\infty$.   

We futher need to show that the corresponding second term vanishes asymptotically. For this we consider the largest component of ${\displaystyle \mathbb{E}\Big(\int_0^\tau  \text{vec}(\tilde{\textbf{h}}_{1}(u,\bs{\beta}_0)^{\otimes 2})^{\otimes 2} \,d\Lambda_1(u,\bs{\beta}_0)\Big)}$, for which it holds that
\[\mathbb{E}(\int_0^\tau  \lVert \tilde{\textbf{h}}_{1}(u,\bs{\beta}_0)\rVert_\infty^4 \,d\Lambda_1(u,\bs{\beta}_0)) = \mathbb{E}(\sup_{t\in\mathcal{T}}\lVert \tilde{\textbf{h}}_{1}(t,\bs{\beta}_0)\rVert_\infty^4  \Lambda_1(\tau,\bs{\beta}_0)) <\infty,\]
due to \cref{assump_general}~\ref{assump_general2} and the integrability of  $\Lambda_i(\tau,\bs{\beta}_0)$. Combining this with 
\cref{assump_general}~\ref{assump_general3} and the law of large numbers yields
$$\frac{1}{n}\sum_{i=1}^n\int_0^\tau  \text{vec}(\tilde{\textbf{h}}_{i}(u,\bs{\beta}_0)^{\otimes 2})^{\otimes 2} \,d\Lambda_i(u,\bs{\beta}_0)\stackrel{\mathbb P}{\longrightarrow}\mathbb{E}\Big(\int_0^\tau  \text{vec}(\tilde{\textbf{h}}_{1}(u,\bs{\beta}_0)^{\otimes 2})^{\otimes 2} \,d\Lambda_1(u,\bs{\beta}_0)\Big),$$
as $n\rightarrow\infty$. 
Finally, for the second term on the right-hand side of the third step of \eqref{eq:lem3.6_2} we have 
${\displaystyle \frac{1}{n^2}\sum_{i=1}^n\int_0^\tau   \text{vec}(\tilde{\textbf{k}}_{i}(t,{\bs{\beta}_0})^{\otimes 2})^{\otimes 2} \,d\Lambda_i(u,\bs{\beta}_0) = o(1)\cdot O_p(1)}$. 
Thus, $\textbf W_{n}(t)$  converges to zero in probability for all $t\in \mathcal{T}$, as $n\rightarrow\infty$, according to Lenglart's inequality.
In conclusion, the predictable covariation process $\langle \textbf{D}_{n,h}^* \rangle (t)$ of $\textbf{D}_{n,h}^*$ at $t$ converges to the matrix-valued function $ {\displaystyle \textbf V_{\tilde h}(t)=\int_0^t\mathbb{E}(  \tilde{\textbf{h}}_{1}(u,\bs{\beta}_0)^{\otimes 2} \,\lambda_1(u,\bs{\beta}_0))du}$ in probability, as $n \rightarrow\infty$, for all $t\in\mathcal{T}$  (cf.~\cref{assump_general}~\ref{assump_general3}).
 This completes the proof of \cref{lem:predCov_D*}.
%Note that, according to \cref{assump_general}~\ref{assump_general4}, $\textbf V_h(t)$ is a continuous and positive semi-definite function with positive semi-definite increments.
\hfill\hfill\qedsymbol

%\end{proof}

\bigskip
%\bigskip

\noindent
\textbf{Proof of \cref{lem:D*->D}.}\\
%\begin{proof}[of \cref{lem:D*->D}]
We use the modified version of Rebolledo's central limit theorem %\remove{for square integrable martingales as described in
%\cref{sec:generalBootstrap}
%} 
as stated in \cref{thm:rebolledo} to prove the weak convergence of %\remove{the sequence of martingales} 
$\textbf{D}_{n,h}^*$ to the zero-mean Gaussian martingale $\textbf{D}_{\tilde h}$. %\remove{Hence, the proof consists of two main arguments. First, the predictable covariation process of $\textbf{D}_{n,h}^*$ converges in probability to a continuous and positive semi-definite function with positive semi-definite increments, as $n\rightarrow\infty$. This condition holds according to %\cref{lem:predCov_D*}. 
%Second, we show that $\textbf D_{n,h}^*$ fulfills the Lindeberg condition.} 
For this purpose, we first consider the term $\sigma^\epsilon[\bs\lambda^\top \textbf{D}_{n,h}^*](\tau)$ for some $\bs\lambda\in {S}^{p+b-1},$ where $ {S}^{p+b-1}$ denotes the unit $(p+b-1)$-sphere. It can be seen that
\begin{align*}
&\sigma^\epsilon[\bs\lambda^\top \textbf{D}_{n,h}^*](\tau)
= \sum_{u\leq \tau} \lvert \Delta \bs\lambda^\top\textbf{D}_{n,h}^*(u)\rvert ^2 \mathbbm{1}\{\lvert \Delta \bs\lambda^\top \textbf{D}_{n,h}^*(u)\rvert > \epsilon\}\\
&= \sum_{u\leq \tau} \lvert \frac{1}{\sqrt{n}}\sum_{i=1}^n\bs\lambda^\top \textbf{h}_{n,i}(u,\hat{\bs{\beta}}_n) G_i(u)\,\Delta N_i(u)\rvert^2 \mathbbm{1}\{\lvert \frac{1}{\sqrt{n}}\sum_{i=1}^n\bs\lambda^\top \textbf{h}_{n,i}(u,\hat{\bs{\beta}}_n) G_i(u)\,\Delta N_i(u)\rvert > \epsilon\}\\
%\footnote{Dreiecks-Ungl and no two jumps}{\leq}
&\leq  \frac{1}{n}\sum_{u\leq \tau} \sum_{i=1}^n \lvert \bs\lambda^\top \textbf{h}_{n,i}(u,\hat{\bs{\beta}}_n) G_i(u)\,\Delta N_i(u) \rvert^2 \mathbbm{1}\{\lvert \frac{1}{\sqrt{n}}\sum_{i=1}^n\bs\lambda^\top \textbf{h}_{n,i}(u,\hat{\bs{\beta}}_n) G_i(u)\,\Delta N_i(u)\rvert > \epsilon\}\\
%&\stackrel{N(t)\leq 1}{\leq}&
&= \frac{1}{n}\sum_{i=1}^n \sum_{j:T_{i,j} \in \mathcal{T}_{n,i}^\Delta} (\bs\lambda^\top \textbf{h}_{n,i}(T_{i,j},\hat{\bs{\beta}}_n ))^2  G_i^2(T_{i,j}) 
 \mathbbm{1}\{\lvert \frac{1}{\sqrt{n}}\bs\lambda^\top \textbf{h}_{n,i}(T_{i,j},\hat{\bs{\beta}}_n) G_i(T_{i,j})\rvert > \epsilon\},
\end{align*}
where in the third step of the derivation above it has been used that no two counting processes jump at the same time, i.e., $\Delta N_i(t)\Delta N_j(t) = 0$, for $i\neq j$. 
From this it follows that
%\begin{eqnarray*}
%\mathbb{E}_0(\sigma^\epsilon[\bs\lambda^\top \textbf{D}_{n}^*](t)) %\stackrel{P}{\longrightarrow}  0, \; n\rightarrow\infty,
%\end{eqnarray*}
 %This condition would follow from:
\begin{align*}
%\begin{split}
%\label{eq:condLindeberg_1}
&{}\mathbb{E}_0(\sigma^\epsilon[\bs\lambda^\top \textbf{D}_{n,h}^*](\tau))\\
&{}\leq \mathbb{E}_0(\frac{1}{n}\sum_{i=1}^n \sum_{j:T_{i,j} \in \mathcal{T}_{n,i}^\Delta} (\bs\lambda^\top \textbf{h}_{n,i}(T_{i,j} ,\hat{\bs{\beta}}_n))^2  G_i^2(T_{i,j}) 
 \mathbbm{1}\{\lvert \frac{1}{\sqrt{n}}\bs\lambda^\top \textbf{h}_{n,i}(T_{i,j},\hat{\bs{\beta}}_n) G_i(T_{i,j})\rvert > \epsilon\})\\
 &{} = \frac{1}{n}\sum_{i=1}^n \sum_{j:T_{i,j} \in \mathcal{T}_{n,i}^\Delta} (\bs\lambda^\top \textbf{h}_{n,i}(T_{i,j},\hat{\bs{\beta}}_n))^2  \mathbb{E}_0( G_i^2(T_{i,j}) 
 \mathbbm{1}\{\lvert \frac{1}{\sqrt{n}}\bs\lambda^\top \textbf{h}_{n,i}(T_{i,j},\hat{\bs{\beta}}_n) G_i(T_{i,j})\rvert > \epsilon\})\\
 &{} \leq \frac{1}{n}\sum_{i=1}^n \sum_{j:T_{i,j} \in \mathcal{T}_{n,i}^\Delta} (\bs\lambda^\top \textbf{h}_{n,i}(T_{i,j},\hat{\bs{\beta}}_n))^2  \big ( \mathbb{E}( G_{1,1}^4) 
 \mathbb{P}_0(\lvert \frac{1}{\sqrt{n}}\bs\lambda^\top \textbf{h}_{n,i}(T_{i,j},\hat{\bs{\beta}}_n) G_{1,1}\rvert > \epsilon)\big )^{1/2}\\
 &{} \leq  \sup_{t\in \mathcal{T}, i\in \{1,\ldots,n\}}(\bs\lambda^\top \textbf{h}_{n,i}(t,\hat{\bs{\beta}}_n))^2  (\mathbb{E}( G_{1,1}^4))^{1/2} [ \mathbb{P}_0(\sup_{t\in \mathcal{T}, i\in \{1,\ldots,n\}}\lvert\bs\lambda^\top \textbf{h}_{n,i}(t,\hat{\bs{\beta}}_n)\rvert \lvert G_{1,1}\rvert > \epsilon \sqrt{n} )]^{1/2}  \\
 &{} \quad \cdot  \frac{1}{n}\sum_{i=1}^n N_i(\tau),
% \end{split}
\end{align*}
where $\mathbb{E}_0(\cdot) $ and $\mathbb{P}_0(\cdot) $ denote the conditional expectation $ \mathbb{E}(\cdot|\mathcal{F}_2(0))$ and the conditional probability $ \mathbb{P}(\cdot|\mathcal{F}_2(0))$, respectively, given the initial filtration $\mathcal{F}_2(0)$. In the first step of the equation above we have used that $\textbf{h}_{n,i}(t ,\hat{\bs{\beta}})\in \mathcal{F}_2(0)$. In the second step, the Cauchy-Schwarz inequality has been applied. In the same step it has additionally been used that the multiplier processes $G_i(t)$, $t\in\mathcal{T},i=1,\ldots,n,$ are i.i.d.\ and independent of $\mathcal{F}_2(0)$. As our first goal is to verify the conditional Lindeberg condition in probability, i.e., $\mathbb{E}_0(\sigma^\epsilon[\bs\lambda^\top \textbf{D}_{n,h}^*](\tau)) \stackrel{\mathbb P}{\longrightarrow}  0$, $n\rightarrow\infty$, we point out that for the terms of the last step of the equation above we have $\mathbb{E}( G_{1,1}^4)<\infty$ and $\frac{1}{n}\sum_{i=1}^n N_i(\tau) = O_p(1)$. The latter holds according to the integrability of $\Lambda_i(\tau, \mathbf{{\beta}}_0)$ %\cref{assump_general}~\ref{item:ass_Bn_4} 
and \cref{assump_general}~\ref{assump_general3}, as explained at the beginning of the proof of \cref{lem:Cn-C} in combination with Fact 1 of the supplement of \cite{dobler19}. Furthermore, the limiting function $\tilde{\textbf{h}}_{i}(t,\bs{\beta}_0)$ of $\textbf{h}_{n,i}(t,\hat{\bs{\beta}}_n)$ exists and is assumed to be bounded on $\mathcal{T}$ for all $n\in\mathbb N$, according to \cref{assump_general}~\ref{assump_general1} and \ref{assump_general2}. Therefore, $\sup_{t\in \mathcal{T}, i\in \{1,\ldots,n\}}(\bs\lambda^\top \textbf{h}_{n,i}(t,\hat{\bs{\beta}}))^2$  is stochastically bounded:
\begin{align*}
& \sup_{t\in \mathcal{T}, i\in \{1,\ldots,n\}}(\bs\lambda^\top \textbf{h}_{n,i}(t,\hat{\bs{\beta}}_n))^2 \\
& \leq (p+b)  \sup_{t\in \mathcal{T}, i\in \{1,\ldots,n\}} \sum_{j=1}^{p+b} \lambda_j^2 \textbf{h}_{n,i}^j(t,\hat{\bs{\beta}}_n))^2\\
& \leq (p+b)^2  \lVert \bs\lambda \rVert_\infty^2  \sup_{t\in \mathcal{T}, i\in \{1,\ldots,n\}} \lVert \textbf{h}_{n,i}(t,\hat{\bs{\beta}}_n)) - \tilde{\textbf{h}}_{i}(t,\bs{\beta}_0)) + \tilde{\textbf{h}}_{i}(t,\bs{\beta}_0)) \rVert_\infty^2\\
&  \leq 2 (p+b)^2 \lVert \bs\lambda \rVert_\infty^2  \big( \sup_{t\in \mathcal{T}, i\in \{1,\ldots,n\}} \lVert \textbf{h}_{n,i}(t,\hat{\bs{\beta}}_n)) - \tilde{\textbf{h}}_{i}(t,\bs{\beta}_0)) \rVert_\infty^2 + \sup_{t\in \mathcal{T}, i\in \{1,\ldots,n\}} \lVert \tilde{\textbf{h}}_{i}(t,\bs{\beta}_0)) \rVert_\infty^2 \big)\\
& = 2(p+b)^2  \lVert \bs\lambda \rVert_\infty^2 \big( o_p(1) + \sup_{t\in \mathcal{T}, i\in \{1,\ldots,n\}} \lVert \tilde{\textbf{h}}_{i}(t,\bs{\beta}_0))\rVert_\infty^2 \big) \\
& = O_p(1).
\end{align*}
Hence, it is only left to show that $\mathbb{P}(\sup_{t\in \mathcal{T}, i\in \{1,\ldots,n\}}\lvert\bs\lambda^\top \textbf{h}_{n,i}(t,\hat{\bs{\beta}}_n)\rvert \lvert G_{1,1}\rvert > \epsilon \sqrt{n} |\mathcal{F}_2(0)) = o_p(1)$. For this purpose, recall that $\mathbbm{1}\{\sup_{t\in\mathcal{T},i\in \{1,\ldots,n\}}\lVert \textbf{h}_{n,i}(t,\hat{\bs{\beta}}_n)-\tilde{\textbf{h}}_{i}(t,\bs{\beta}_0) \rVert_\infty < \delta\}$ converges to one in probability, according to \cref{assump_general}~\ref{assump_general1}.
Thus, we can proceed with the following term:
\begin{align*}
&\mathbb{P}_0(\sup_{t\in\mathcal{T},i\in \{1,\ldots,n\}}\lvert \bs\lambda^\top \textbf{h}_{n,i}(t,\hat{\bs{\beta}}_n)\rvert  \lvert G_{1,1}\rvert > \sqrt{n}\epsilon)
 \mathbbm{1}\{\sup_{t\in\mathcal{T},i\in \{1,\ldots,n\}}\lVert \textbf{h}_{n,i}(t,\hat{\bs{\beta}}_n)-\tilde{\textbf{h}}_{i}(t,\bs{\beta}_0) \rVert_\infty < \delta\}\\
&=
\mathbb{P}_0(\sup_{t\in\mathcal{T},i\in \{1,\ldots,n\}}\lvert \bs\lambda^\top \textbf{h}_{n,i}(t,\hat{\bs{\beta}}_n)
-\bs\lambda^\top \tilde{\textbf{h}}_{i}(t,\bs{\beta}_0)
+\bs\lambda^\top \tilde{\textbf{h}}_{i}(t,\bs{\beta}_0)\rvert \lvert G_{1,1}\rvert > \sqrt{n}\epsilon, \\
&\quad \sup_{t\in\mathcal{T},i\in \{1,\ldots,n\}}\lVert \textbf{h}_{n,i}(t,\hat{\bs{\beta}}_n)-\tilde{\textbf{h}}_{i}(t,\bs{\beta}_0) \rVert_\infty < \delta) \\
& \leq 
\mathbb{P}_0((p+b) \lVert \bs\lambda\rVert_\infty (\delta + \sup_{t\in\mathcal{T},i\in \{1,\ldots,n\}}\lVert \tilde{\textbf{h}}_{i}(t,\bs{\beta}_0)\rVert_\infty)  \lvert G_{1,1}\rvert > \sqrt{n}\epsilon)\\
&\quad \cdot \mathbbm{1}\{\sup_{t\in\mathcal{T},i\in \{1,\ldots,n\}}\lVert \textbf{h}_{n,i}(t,\hat{\bs{\beta}}_n)-\tilde{\textbf{h}}_{i}(t,\bs{\beta}_0) \rVert_\infty < \delta\} \\
& \leq
\mathbb{P}_0\big( \lvert G_{1,1}\rvert > \frac{\sqrt{n}\epsilon}{(p+b) \lVert \bs\lambda\rVert_\infty (\delta + \sup_{t\in\mathcal{T},i\in \{1,\ldots,n\}}\lVert \tilde{\textbf{h}}_{i}(t,\bs{\beta}_0)\rVert_\infty)}\big) \\
& \stackrel{\mathbb P}{\longrightarrow}  0,\;n\rightarrow\infty.
\end{align*}
Here, the convergence in probability of the conditional probability in the last step holds, because $\tilde{\textbf{h}}_{i}(t,\bs{\beta}_0)$ is bounded on $\mathcal{T}$ for all $i\in\mathbb{N}$, as stated in \cref{assump_general}~\ref{assump_general2}. We can conclude that $\mathbb{P}(\sup_{t\in \mathcal{T}, i\in \{1,\ldots,n\}}\lvert\bs\lambda^\top \textbf{h}_{n,i}(t,\hat{\bs{\beta}}_n)\rvert \lvert G_{1,1}\rvert > \epsilon \sqrt{n} |\mathcal{F}_2(0)) = o_p(1)$. Thus, the conditional Lindeberg condition in probability is fulfilled for $\bs\lambda^\top \textbf{D}^*_{n,h}(t)$ with $\bs\lambda\in S^{p+b-1}$. As $\lVert\bs\lambda\rVert_\infty \leq 1$, we can get an upper bound independent of $\bs\lambda$, and thus we in fact know that the asserted Lindeberg condition holds for all $\bs\lambda\in S^{p+b-1}$.
We would like to point out that the probability space can more conveniently be modelled as a product space
$ (\Omega, \mathcal A, \mathbb P) = (\Omega_1 \times \Omega_2, \mathcal A_1 \otimes \mathcal A_2, \mathbb P_1 \otimes \mathbb P_2) = (\Omega_1, \mathcal A_1, \mathbb P_1) \otimes (\Omega_2, \mathcal A_2, \mathbb P_2)$.
%From now on we distinguish between the probability measure $\mathbb P_1$ and $\mathbb P_2$ on $(\Omega_1,\mathcal{A}_1)$ and $(\Omega_2,\mathcal{A}_2)$, respectively, 
In the following we make use of this notation to explicitly refer to the probability space $(\Omega_1, \mathcal A_1, \mathbb P_1)$ underlying the data sets $\{\textbf N(t) ,\textbf Y(t) ,\textbf Z (t), t\in\mathcal{T}\}$, and the probability space $(\Omega_2, \mathcal A_2, \mathbb P_2)$ underlying the sets of multiplier processes $\{ G_1(t),\ldots ,G_n(t)$, $t\in\mathcal{T}\}$. Additionally, we denote by $\stackrel{\mathcal{L}_{\mathbb P_2}}{\longrightarrow}$ the convergence in law w.r.t the probability measure $\mathbb P_2$. Moreover, for some stochastic quantity $\textbf H_n$, we denote $\textbf H_n$ conditional on a particular data set as $\textbf H_n|\mathcal{F}_2(0)(\omega)$, $\omega\in\Omega_1$. From the conditional Lindeberg condition in probability it follows that there exists for all subsequences $n_1$ of $n$ a further subsequence $n_2$ such that $\mathbb{E}_{\mathbb P_2}(\sigma^\epsilon[\bs\lambda^\top \textbf{D}_{n_2,h}^*](\tau)|\mathcal{F}_2(0))(\omega) \stackrel{}{\longrightarrow}  0$, $n\rightarrow\infty$, for $\mathbb P_1$-almost all $\omega\in\Omega_1$ and for all $\bs\lambda\in S^{p+b-1}$. Here, $\mathbb{E}_{\mathbb P_2}(\cdot)$ indicates that the expectation is taken with respect to $\mathbb P_2$. Hence, the (unconditional) Lindeberg condition holds along the subsequence $n_2$ for $\mathbb P_1$-almost all data sets.

%We recall the result of \cref{lem:predCov_D*}, i.e.
%\[\langle \textbf{D}_{n,h}^*\rangle (t)  \stackrel{\mathbb P}{\longrightarrow}  \textbf V_{\tilde h}(t) , \text{ as } n\rightarrow\infty, \text{ for all } t\in\mathcal{T},\]
%where $\mathbb P = \mathbb P_1\times \mathbb P_2.$  $\langle \bs\lambda^\top\textbf{D}_{n,h}^*\rangle (t)$
Next, we consider the predictable covariation process of $\bs\lambda^\top\textbf{D}_{n,h}^*$ for some $\bs\lambda\in {S}^{p+b-1}$ and get, conditionally on $\mathcal{F}_2(0)$,
\[\langle \bs\lambda^\top\textbf{D}_{n,h}^*\rangle (t) = \bs\lambda^\top \langle \textbf{D}_{n,h}^*\rangle (t) \bs\lambda \stackrel{ \mathbb P_1\otimes \mathbb P_2}{\longrightarrow} \bs\lambda^\top \textbf V_{\tilde h}(t) \bs\lambda, \text{ as } n\rightarrow\infty, \text{ for all } t\in\mathcal{T},\]
according to \cref{lem:predCov_D*}. 
Furthermore, we have
\begin{align*}
& \bs\lambda^\top \big((\langle \textbf{D}_{n,h}^*\rangle  - \textbf V_{\tilde h})(t)\big) \bs\lambda = \sum_{j=1}^{p+b}\sum_{l=1}^{p+b} \lambda_j (\langle \textbf{D}_{n,h}^*\rangle  - \textbf V_{\tilde h})_{j,l}(t)  \lambda_l\\
&\leq (p+b)^2  \lVert \bs\lambda\rVert_\infty^2 \cdot \lVert \langle \textbf{D}_{n,h}^*\rangle (t) - \textbf V_{\tilde h}(t) \rVert_\infty,
\end{align*}
where $(\langle \textbf{D}_{n,h}^*\rangle  - \textbf V_{\tilde h})_{j,l}$ denotes the  $(j,l)$-th entry of the corresponding matrix. 
As $\lVert \bs\lambda\rVert_\infty\leq 1$ and $\lVert \langle \textbf{D}_{n,h}^*\rangle (t) - \textbf V_{\tilde h}(t) \rVert_\infty=o_p(1)$, in view of  \cref{lem:predCov_D*} we thus obtain 
\[\langle \bs\lambda^\top\textbf{D}_{n,h}^*\rangle (t) \stackrel{\mathbb P_1\otimes \mathbb P_2}{\longrightarrow} \bs\lambda^\top \textbf V_{\tilde h}(t) \bs\lambda, \text{ as } n\rightarrow\infty, \text{ for all } t\in\mathcal{T}, \text{ and all } \bs\lambda\in {S}^{p+b-1}.\]
Hence, there exists for every subsequenece $n_3$ of $n_2$ a further subsequence $n_4$ such that $\langle \bs\lambda^\top\textbf{D}^*_{n_4,h} \rangle | \mathcal{F}_2(0)(\omega) \stackrel{\mathbb P_2}{\longrightarrow} \bs\lambda^\top \textbf V_{\tilde h}(t) \bs\lambda , \text{ as } n\rightarrow\infty,$ for $\mathbb P_1$-almost all $\omega\in\Omega_1$, all $t\in\mathcal{T}$, and all $\bs\lambda\in {S}^{p+b-1}$. 
Clearly, it also holds that $\mathbb{E}_{\mathbb P_2}(\sigma^\epsilon[\bs\lambda^\top \textbf{D}_{n_4,h}^*](\tau)|\mathcal{F}_2(0))(\omega) \stackrel{}{\rightarrow}  0$, $n\rightarrow\infty$, for $\mathbb P_1$-almost all $\omega\in\Omega_1$ and all $\bs\lambda\in S^{p+b-1}$. 
Thus, with \cref{thm:rebolledo} it follows that 
\[\bs\lambda^\top\textbf{D}^*_{n_4,h} |\mathcal{F}_2(0)(\omega) \stackrel{\mathcal{L}_{\mathbb P_2}}{\longrightarrow} \bs\lambda^\top\textbf{D}_{\tilde h}, \text{ in } D(\mathcal{T}), \text{ as } n\rightarrow\infty,  \]
for $\mathbb P_1$-almost all $\omega\in\Omega_1$ and all $\bs\lambda\in {S}^{p+b-1}$. As the weak convergence of $\bs\lambda^\top\textbf{D}^*_{n_4,h} |\mathcal{F}_2(0)(\omega)$ holds for all $\bs\lambda\in {S}^{p+b-1}$, the Cram\'er-Wold device yields $\textbf{D}^*_{n_4,h} |\mathcal{F}_2(0)(\omega) \stackrel{\mathcal{L}_{\mathbb{P}_2}}{\longrightarrow} \textbf{D}_{\tilde h},$ in $D(\mathcal{T})^{p+b},$ as $n\rightarrow\infty,$ for $\mathbb P_1$-almost all $\omega\in\Omega_1$. Finally, we get, conditionally on $\mathcal{F}_2(0)$,
\[\textbf{D}^*_{n,h}  \stackrel{\mathcal{L}_{\mathbb{P}_2}}{\longrightarrow} \textbf{D}_{\tilde h}, \text{ in } D(\mathcal{T})^{p+b}, \text{ as } n\rightarrow\infty,\]
in $\mathbb P_1$-probability. This completes the proof of \cref{lem:D*->D}.\hfill\hfill\qedsymbol
%\end{proof}

%\smallskip
\bigskip

\noindent
\textbf{Proof of \cref{cor:optCov_D*}.}\\
%\begin{proof}[of \cref{cor:optCov_D*}]
We relate the optional covariation process $[\textbf{D}^*_{n,h}](t)$ and the predictable covariation process $\langle \textbf{D}^*_{n,h} \rangle (t)$ of $\textbf D_{n,h}^*(t)$ to each other by noting the obvious
\[[\textbf{D}^*_{n,h}](t) = [\textbf{D}^*_{n,h}](t) - \langle \textbf{D}^*_{n,h} \rangle (t) + \langle \textbf{D}^*_{n,h} \rangle (t).\]
Consequently, if the predictable covariation process $\langle \textbf{D}^*_{n,h}\rangle (t)$ converges in probability to $\textbf V_{\tilde h}(t)$, as $n\rightarrow\infty$, and it holds that $[\textbf{D}^*_{n,h}](t) - \langle \textbf{D}^*_{n,h} \rangle (t) = o_p(1)$, then also the optional covariation $[\textbf{D}^*_{n,h}](t)$ converges in probability to $\textbf V_{\tilde h}(t)$, as $n\rightarrow\infty$, and vice versa. Hence, for this proof we assume that \cref{lem:predCov_D*} holds and show that the difference between the optional covariation process and the predictable covariation  process of $\textbf D_{n,h}^*(t)$ vanishes asymptotically.

Let us consider the vectorized version $\textbf Q_n$ of the difference between the optional covariation process and the predictable covariation process of $\textbf D_{n,h}^*(t)$,  
$t\in\mathcal{T}$,
\begin{align*}
\textbf Q_n(t) &= \text{vec}([\textbf{D}^*_{n,h}](t) - \langle \textbf{D}^*_{n,h} \rangle (t)) \\
&= \frac{1}{n}\sum_{i=1}^n \int_0^t \text{vec}( \textbf h_{n,i}(u,\hat{\bs{\beta}}_n)^{\otimes 2} )(G_i^2(u)-1)dN_i(u).
\end{align*}
The $\text{vec}( \textbf h_{n,i}(t,\hat{\bs{\beta}}_n)^{\otimes 2} )$ in the  integrands are known and  locally bounded and predictable. Hence, according to Theorem II.3.1 of \cite{Andersen},
$\textbf Q_n$ is a vector of local square integrable martingales if $\int_0^\cdot (G_i^2(u)-1)dN_i(u)$ is a finite variation local square integrable martingale for all $i=1,\ldots,n$. 
This is what we show in the following three steps. %Thus, we first verify that $\int_0^\cdot (G_i^2(u)-1)dN_i(u)$ forms a finite variation local square integrable martingale for all $i=1,\ldots,n$:
\begin{enumerate}
\item The finite variation holds, because
%\begin{eqnarray*}
\[\int_0^\tau \lvert (G_i^2(u) -1)dN_i(u)\rvert \leq
 (\sup_{t \in \mathcal{T}} G_i^2(t) + 1)  N_i(\tau),\]
%\lvert G_i^2(u) -1 \rvert N_i(\tau) \\
%&\leq & G_i^2+1 < \infty \quad \text{a.s. \textcolor{red}{hier wuerde ich einfach die untere Zeile weglassen.}}
%\end{eqnarray*}
and the term on the right-hand side is almost surely finite as $N_i(\tau) < \infty$, and the supremum is a maximum of almost surely finitely many random variables.
\item It is square integrable, since
\begin{align*}
\sup_{t\in \mathcal{T}}\mathbb{E}_0\Big( \Big[\int_0^t(G_i^2(u)-1)dN_i(u)\Big]^2 \Big) 
& = \sup_{t\in \mathcal{T}}\mathbb{E}_0\Big( \Big[\sum_{j: T_{i,j}\leq t} (G_{i,j}^2-1)\Big]^2 \Big)\\
&\leq \sup_{t\in \mathcal{T}}\mathbb{E}_0\Big(\big \lvert \{j:T_{i,j}\leq t \} \big \rvert  \sum_{j: T_{i,j}\leq t} (G_{i,j}^2-1)^2 \Big)\\
&\leq  N_i(\tau )   \sum_{j=1}^{n_i} \mathbb{E} (G_{i,j}^4 -2G_{i,j}^2 + 1)\\
&\leq N_i(\tau )^2   \mathbb{E} (G_{1,1}^4)  < \infty ,
\end{align*}
where $\mathbb{E}_0(\cdot )$ denotes the conditional expectation $\mathbb{E}(\cdot|\mathcal{F}_2(0))$ and $\big\lvert \{j:T_{i,j}\leq t \} \big\rvert$ the cardinality of the corresponding set. Moreover, in the third step we have applied that the counting processes $N_i(t)$, $t\in\mathcal{T},$ are $\mathcal{F}_2(0)$-measurable, whereas the values of $G_{i,j}$ and the filtration $\mathcal{F}_2(0)$ are independent for all $j=1,\ldots , n_i$ and $ i=1,\ldots,n$. Additionally, in the fourth step we used that $G_{i,1},\ldots ,G_{i,n_i}$ are identically distributed with zero mean, unit variance and finite fourth moment for all $i = 1,\ldots,n$.
\item The martingale property is valid, as
%\begin{eqnarray*}
\begin{align*}
&\mathbb{E}\big( \int_0^t (G_i^2(u)-1)dN_i(u)|\mathcal{F}_2(s)\big)\\
& = \mathbb{E}\big( \int_0^s(G_i^2(u)-1)dN_i(u)+\int_s^t(G_i^2(u)-1)dN_i(u)|\mathcal{F}_2(s) \big)\\
& = \int_0^s(G_i^2(u)-1)dN_i(u) + \int_s^t\big(\mathbb{E}(G_i^2(u))-1\big)dN_i(u) \\
& = \int_0^s(G_i^2(u)-1)dN_i(u),
%\end{eqnarray*}
\end{align*}
where in the second step we have used that the counting process $N_i(t)$ is $\mathcal{F}_2(0)\subset \mathcal{F}_2(t)$-measurable for $t\in\mathcal{T}$, $i=1,\ldots ,n$. Furthermore, for a jump at $u \leq s$, the multiplier process $G_i(u)$ is $\mathcal{F}_2(s)$-measurable, and,
%for $u>s$ 
if $u$ is greater than or equal to the earliest jump time point, say $T_{i}(s^+)$, of process $N_i$ in $(s,\tau]$,
the values of $G_i(u)$ and the filtration $\mathcal{F}_2(s)$ are independent, $i=1,\ldots,n$. In the third step we used that the multiplier processes $G_i(t)$, $t \in \mathcal{T}$, have zero mean and unit variance, $i=1,\ldots,n$.
\end{enumerate}
In conclusion, $\textbf{Q}_n$ is a vector of local square integrable martingales. 

Next, we wish to show that $\textbf{Q}_n(t)$ converges to zero in probability, as $n\rightarrow \infty$. For this we apply Lenglart's inequality and consider the predictable covariation process $\langle \textbf{Q}_n \rangle (\tau)$ of the martingale $\textbf{Q}_n$ at $\tau$
\begin{align*}
\langle  \textbf{Q}_n \rangle (\tau) &= \Big\langle \frac{1}{n}\sum_{i=1}^n \int_0^\cdot \text{vec}( \textbf h_{n,i}(u,\hat{\bs{\beta}}_n)^{\otimes 2} )(G_i^2(u)-1)dN_i(u) \Big\rangle (\tau)\\
&=\frac{1}{n^2}\sum_{i=1}^n \int_0^\tau \text{vec}( \textbf h_{n,i}(u,\hat{\bs{\beta}}_n)^{\otimes 2} )^{\otimes 2}  d\Big\langle \int_0^\cdot (G_i^2(v)-1)dN_i(v)\Big\rangle (u)\\
&= \frac{1}{n^2}\sum_{i=1}^n\int_0^\tau \text{vec}( \textbf h_{n,i}(u,\hat{\bs{\beta}}_n)^{\otimes 2} )^{\otimes 2} (\mathbb{E}(G_i^4(u))-1)dN_i(u),
\end{align*}
where in the second step we have used that 
\begin{align*}
    &d \Big\langle \int_0^\cdot (G_i^2(u)-1)dN_i(u), \int_0^\cdot (G_l^2(u)-1)dN_l(u)\Big\rangle (t)\\
    &= \text{Cov}\big((G_i^2(t)-1)dN_i(t), (G_l^2(t)-1)dN_l(t) |\mathcal{F}_{t\text{-}}\big)\\
    &=  \text{Cov}\big(G_i^2(t), G_l^2(t) \big)dN_i(t)dN_l(t) \\
    &=0,
\end{align*}
because $G_1(t),\ldots,G_n(t)$, $t\in\mathcal{T}$, are pairwise independent and no two counting processes jump simultaneously. The third step holds due to
\begin{align*}
d \Big\langle \int_0^\cdot (G_i^2(u)-1)dN_i(u)\Big\rangle (t) &= \mathbb{E}\big([(G_i^2(t)-1)dN_i(t)]^2|\mathcal{F}_{t\text{-}}\big)\\
&=  \big(\mathbb{E}(G_i^4(t))-2\mathbb{E}(G_i(t)^2)+1\big)dN_i(t) \\
&= \big(\mathbb{E}(G_i^4(t))-1\big)dN_i(t).
\end{align*}
We continue by stating that
\begin{align}
\label{eq:Q0}
%\begin{split}
\langle  \textbf{Q}_n \rangle (\tau) &= \frac{1}{n^2}\sum_{i=1}^n \int_0^\tau \text{vec}( \tilde{\textbf h}_{i}(u,\bs{\beta}_0)^{\otimes 2} )^{\otimes 2} (\mathbb{E}(G_i^4(u))-1)dN_i(u)\\
&+ \frac{1}{n^2}\sum_{i=1}^n \int_0^\tau [\text{vec}( \textbf h_{n, i}(u,\hat{\bs{\beta}}_n)^{\otimes 2} )^{\otimes 2} - \text{vec}( \tilde{\textbf h}_{i}(u,\bs{\beta}_0)^{\otimes 2} )^{\otimes 2}](\mathbb{E}(G_i^4(u))-1)dN_i(u).\nonumber
%\end{split}
\end{align}
For  the first term on the right-hand side we have
\begin{align}
\label{eq:Q1}
\begin{split}
& \frac{1}{n^2}\sum_{i=1}^n \int_0^\tau \text{vec}( \tilde{\textbf h}_{i}(u,\bs{\beta}_0)^{\otimes 2} )^{\otimes 2} (\mathbb{E}(G_i^4(u))-1)dN_i(u)\\
&\leq \frac{1}{n}  \sup_{i\in\{1,\ldots,n\},t\in\mathcal{T}}\lVert  \tilde{\textbf h}_{i}(t,\bs{\beta}_0)\rVert_\infty^4  (\mathbb{E}(G_{1,1}^4)-1) \frac{1}{n}\sum_{i=1}^n N_i(\tau ) = o_p(1), \text{ as } n\rightarrow\infty,
\end{split}
\end{align}
since  $\mathbb{E}(G_{1,1}^4) < \infty$ according to \cref{assump_general}~\ref{assump_general2}, and $\frac{1}{n}\sum_{i=1}^n N_i(\tau) = O_p(1)$, as was derived at the beginning of the proof of \cref{lem:Cn-C}.
% under \cref{assump_general}~\ref{assump_general3} and due to the integrability of the cumulative intensity processes. 
%As pointed out earlier, we can transfer the results on convergence in probability from \cref{sec:Notation} to the convergence in conditional probability with the help of Fact 1 stated in the supplement of \cite{dobler19}.
%\newpage
Additionally, for the second term on the right-hand side we find
\begin{align}
\begin{split}
\label{eq:Q2}
&\frac{1}{n^2}\sum_{i=1}^n \int_0^\tau [\text{vec}( \textbf h_{n, i}(u,\hat{\bs{\beta}}_n)^{\otimes 2} )^{\otimes 2} - \text{vec}( \tilde{\textbf h}_{i}(u,\bs{\beta}_0)^{\otimes 2} )^{\otimes 2}](\mathbb{E}(G_i^4(u))-1)dN_i(u)\\
    &\leq \frac{1}{n}\sup_{i\in\{1,\ldots,n\},t\in\mathcal{T}} \lVert \text{vec}( \textbf h_{n, i}(t,\hat{\bs{\beta}}_n)^{\otimes 2} )^{\otimes 2} - \text{vec}( \tilde{\textbf h}_{i}(t,\bs{\beta}_0)^{\otimes 2} )^{\otimes 2}\rVert_\infty \\
&\quad \cdot  (\mathbb{E}(G_{1,1}^4)-1) \frac{1}{n}\sum_{i=1}^n N_i(\tau )\\
    &\leq 
%\sup_{i\in\{1,\ldots,n\},t\in\mathcal{T}} \Big(
    \ \rVert \textbf{h}_{n,i}(t,\hat{\bs{\beta}}_n) \lVert_\infty^2 
     \Big [ \lVert \textbf{h}_{n,i}(t,\hat{\bs{\beta}}_n) -
    \tilde{\textbf{h}}_{i}(t,{\bs{\beta}_0})\rVert_\infty 
    \lVert \textbf{h}_{n,i}(t,\hat{\bs{\beta}}_n) \rVert_\infty\\
&{} \quad  + 
    \lVert 
    \tilde{\textbf{h}}_{i}(t,{\bs{\beta}_0})\rVert_\infty 
    \lVert\textbf{h}_{n,i}(t,\hat{\bs{\beta}}_n) -
    \tilde{\textbf{h}}_{i}(t,{\bs{\beta}_0})\rVert_\infty
  \Big ]\\
     &{} \quad  + \lVert \tilde{\textbf{h}}_{i}(t,{\bs{\beta}_0}) \rVert_\infty^2 
   \Big [ \lVert \textbf{h}_{n,i}(t,\hat{\bs{\beta}}_n) -
    \tilde{\textbf{h}}_{i}(t,{\bs{\beta}_0})\rVert_\infty 
    \lVert \textbf{h}_{n,i}(t,\hat{\bs{\beta}}_n) \rVert_\infty \\
&{} \quad + 
    \lVert  \tilde{\textbf{h}}_{i}(t,{\bs{\beta}_0})\rVert_\infty
    \lVert \textbf{h}_{n,i}(t,\hat{\bs{\beta}}_n) -
    \tilde{\textbf{h}}_{i}(t,{\bs{\beta}_0})\rVert_\infty
   \Big ] \Big)\\
 &\quad \cdot \frac{1}{n} (\mathbb{E}(G_{1,1}^4)-1) \frac{1}{n}\sum_{i=1}^n N_i(\tau )\\
 & = o_p(1),\text{ as } n\rightarrow\infty,
\end{split}
\end{align}
where we used $\mathbb{E}(G_{1,1}^4) < \infty$, $\frac{1}{n}\sum_{i=1}^n N_i(\tau) = O_p(1)$, $\rVert \textbf{h}_{n,i}(t,\hat{\bs{\beta}}_n) \lVert_\infty <\infty$, \cref{assump_general}~\ref{assump_general1},~\ref{assump_general2}, and \eqref{eq:otimes_2_twice} in combination with the triangle inequality. In particular, the terms in brackets vanish asymptotically, as $n\rightarrow\infty$. 
Combining \eqref{eq:Q0}, \eqref{eq:Q1} and \eqref{eq:Q2}, we get $\langle  \textbf{Q}_n \rangle (\tau) = o_p(1)$, as $n\rightarrow\infty$, and with Lenglart's inequality it follows that
\[ \textbf{Q}_n (t) = \text{vec}\big([\textbf{D}_{n,h}^*](t) - \langle\textbf{D}_{n,h}^*\rangle (t)\big ) \stackrel{\mathbb P}{\longrightarrow} 0, \text{ as } n\rightarrow\infty, \text{ for all } t\in\mathcal{T}.\]
In combination with \cref{lem:predCov_D*}, we have $[\textbf{D}_{n,h}^*](t) \stackrel{\mathbb P}{\longrightarrow} \textbf V_{\tilde h}(t) $, as $n \to \infty$, for all $t\in\mathcal{T}$. This completes the proof of \cref{cor:optCov_D*}. \hfill\hfill\qedsymbol
%\end{proof}

%\smallskip
%\bigskip
%\newpage

\noindent
\textbf{Proof of \cref{lem:Cn*-->C}.}\\
%\begin{proof}[of \cref{lem:Cn*-->C} ]
Recall from \eqref{eq:C*} that $ \textbf B_n^*(t) =\frac1n \sum_{i=1}^n \int_0^t \textnormal D \textbf k_{n,i}(u, \hat{\bs{\beta}}_n) \big( G_i(u) + 1\big) dN_i(u)$. Then, we have
%\begin{equation}\label{eq:B*n_B}
%\begin{array}{r@{}l}
\begin{align}
\begin{split}
    \label{eq:B*n_B}
    &{} \sup_{t\in\mathcal{T}} \lVert \textbf B^*_n(t) - \textbf B(t) \rVert \\
 &{}\leq \sup_{t\in\mathcal{T}} \Big\lVert  \frac{1}{n}\sum_{i=1}^n \int_0^t [\textnormal D\textbf k_{n,i}(u,\bs{\beta}_0) -  \tilde{\textbf K}_{i} (u,\bs{\beta}_0)] (G_i(u) + 1) dN_i(u)\Big\rVert\\
&{} \quad + \sup_{t\in\mathcal{T}} \Big\lVert  \frac{1}{n}\sum_{i=1}^n \int_0^t  \tilde{\textbf K}_{i} (u,\bs{\beta}_0) (G_i(u) + 1) dN_i(u) - \textbf B(t) \Big\rVert \\
&{} \leq \sup_{t\in\mathcal{T}} \Big\lVert  \frac{1}{n}\sum_{i=1}^n \int_0^t [\textnormal D\textbf k_{n,i}(u,\bs{\beta}_0) -  \tilde{\textbf K}_{i} (u,\bs{\beta}_0)] (G_i(u) + 1) dN_i(u)\Big\rVert\\
&{} \quad + \sup_{t\in\mathcal{T}} \Big\lVert  \frac{1}{n}\sum_{i=1}^n \int_0^t  \tilde{\textbf K}_{i} (u,\bs{\beta}_0) G_i(u) dN_i(u) \Big\rVert\\
&{}\quad + \sup_{t\in\mathcal{T}} \Big\lVert \frac{1}{n}\sum_{i=1}^n  \int_0^t \tilde{\textbf K}_{i} (u,\bs{\beta}_0) dN_i(u) - \textbf B(t) \Big\rVert .
\end{split}
\end{align}
%\end{array}
%\end{equation}
We consider the second term on the right-hand side of the second step of \eqref{eq:B*n_B} first. According to \cref{lemma:mgale} with $ h_{n, i}(t, \hat{{\beta}}_n)\equiv 1 $, $\int_0^t G_i(u)\, dN_i(u)$ is a square integrable martingale w.r.t. $\mathcal{F}_2$. Moreover, it holds that $\int_0^\tau \lvert G_i(u) \,dN_i(u)\rvert \leq \max_{j=1,\ldots ,n_i} \lvert G_{i,j}\rvert   N_i(\tau) < \infty$ almost surely, as the maximum is taken over finitely many almost surely finite random variables. Thus, the martingale is also of finite variation. Due to \cref{ass:Bn_Cn}~\ref{item:ass_Bn_2} and with Theorem II.3.1. of \cite{Andersen}, it follows that $\frac{1}{ n}\sum_{i=1}^n \int_0^t  \tilde{\textbf K}_{i} (u,\bs{\beta}_0) G_i(u) dN_i(u) $ is a local square integrable martingale w.r.t. $\mathcal{F}_2$. Furthermore, its predictable covariation process at $\tau$ is given by
%\begin{equation}\label{eq:lem_3.8_secondTerm}
%\begin{array}{r@{}l}
\begin{align}
\begin{split}
    \label{eq:lem_3.8_secondTerm}
     &{} \Big\langle \frac{1}{ n}\sum_{i=1}^n \int_0^{\cdot}  \tilde{\textbf K}_{i} (u,\bs{\beta}_0) G_i(u) dN_i(u) \Big\rangle (\tau )\\
    &{}= \frac{1}{ n^2}\sum_{i=1}^n\sum_{j=1}^n \Big\langle   \int_0^{\cdot}  \tilde{\textbf K}_{i} (u,\bs{\beta}_0) G_i(u) dN_i(u) ,  \int_0^{\cdot}  \tilde{\textbf K}_{j} (u,\bs{\beta}_0) G_j(u) dN_j(u)\Big\rangle (\tau )\\
     &{} =  \frac{1}{ n^2}\sum_{i=1}^n\sum_{j=1}^n \int_0^\tau  \tilde{\textbf K}_{i} (u,\bs{\beta}_0) \, d \Big\langle \int_0^{\cdot} G_i(s) dN_i(s) ,  \int_0^{\cdot} G_j(s) dN_j(s)\Big\rangle  (u )\, \tilde{\textbf K}_{j} (u,\bs{\beta}_0)^\top\\
     &{} =   \frac{1}{n^2}\sum_{i=1}^n \int_0^\tau  \tilde{\textbf K}_{i} (u,\bs{\beta}_0)^{\otimes 2} \, dN_i(u),
\end{split}
\end{align}
%\end{array}
%\end{equation}
because $\langle \int_0^{\cdot} G_i(s) dN_i(s) ,  \int_0^{\cdot} G_j(s) dN_j(s)\rangle (u ) = N_i(u)$, for $i=j$, and zero otherwise, according to \cref{lemma:mgale}. Additionally, in the second step of \eqref{eq:lem_3.8_secondTerm} the aforementioned Theorem II.3.1. has been used. For the remaining part of this proof we use unconditional convergence in probability instead of conditionally on $\mathcal{F}_2(0)$, because due to Fact 1 of the supplement of \cite{dobler19} these two types of convergence are equivalent.
We wish to show that the last term on the right-hand side of \eqref{eq:lem_3.8_secondTerm} converges to zero in probability, as $n\rightarrow\infty$. For this, we bound that term from above by $ \frac{1}{n} \sup_{i\in\{1,\ldots,n\}, t\in\mathcal{T}} \lVert  \tilde{\textbf K}_{i} (t,\bs{\beta}_0) \rVert_\infty^2  \frac{1}{n} \sum_{i=1}^n N_i(\tau)$. Recall that $\sup_{i\in\{1,\ldots,n\}, t\in\mathcal{T}} \lVert  \tilde{\textbf K}_{i} (t,\bs{\beta}_0) \rVert_\infty^2 < \infty$, by \cref{ass:Bn_Cn}~\ref{item:ass_Bn_2}, and $\frac{1}{n} \sum_{i=1}^n N_i(\tau) = O_p(1)$, by the integrability of $\Lambda_i(\tau, \mathbf{{\beta}}_0)$ %\cref{assump_general}~\ref{item:ass_Bn_4} 
and \cref{assump_general}~\ref{assump_general3}, as stated at the beginning of \cref{lem:Cn-C}. Hence, the predictable covariation process of $\frac{1}{n}\sum_{i=1}^n \int_0^t  \tilde{\textbf K}_{i} (u,\bs{\beta}_0) G_i(u) dN_i(u)$ at $\tau$ converges to zero in probability, as $n\rightarrow\infty$. With Lenglart's inequality it follows that the corresponding martingale converges to zero in probability, as $n\rightarrow\infty$, for all $t\in\mathcal{T}$. In other words, the second term on the right-hand side of the second step of \eqref{eq:B*n_B} vanishes asymptotically. 

Next, we consider the first term on the right-hand side of the second step of \eqref{eq:B*n_B}. For this term we get 
%\begin{equation*}
%\begin{array}{r@{}l}
\begin{align*}
    &{} \sup_{t\in\mathcal{T}} \lVert  \frac{1}{n}\sum_{i=1}^n \int_0^t [\textnormal D\textbf k_{n,i}(u,\bs{\beta}_0) -  \tilde{\textbf K}_{i} (u,\bs{\beta}_0)] (G_i(u) + 1) dN_i(u)\rVert\\
    &{}\leq \sup_{i\in\{1,\ldots,n\}, t\in\mathcal{T}}  \lVert\textnormal D \textbf k_{n,i}(t,\hat{\bs {\beta}}) - \tilde{\textbf K}_{i}(t,\bs {\beta}_0)\rVert   \frac{1}{n}\sum_{i=1}^n \int_0^\tau \vert G_i(u) + 1 \rvert\, dN_i(u).
\end{align*}
%\end{array}
%\end{equation*}
According to \cref{ass:Bn_Cn}~\ref{item:ass_Bn_1}, the first term on the right-hand side of the inequality above converges to zero in probability, as $n\rightarrow\infty$. We now address the corresponding second term, which can be rewritten as $\frac{1}{n} \sum_{i=1}^n \sum_{j=1}^{n_i}  \lvert G_{i,j} + 1\rvert$. Furthermore, we have
\begin{align}
\begin{split}\label{eq:integrable}
    \mathbb E\Big(\sum_{j=1}^{n_i}   \rvert G_{i,j} + 1 \rvert \Big) &{} = \mathbb E \Big( \mathbb E \Big( \sum_{j=1}^{n_i}   \rvert G_{i,j} + 1 \rvert|\mathcal{F}_2(0)\Big)\Big)\\
     &{} = \mathbb E \Big( \sum_{j=1}^{n_i} \mathbb E (  \rvert G_{i,j} + 1 \rvert)\Big) \\
    &{} \leq  2 \mathbb E ( N_i(\tau )) < \infty,
    \end{split}
\end{align}
where
% in the first step of \eqref{eq:integrable} the tower property has been applied, and 
in the second step we have used that $N_i(t)$ with $N_i(\tau ) = n_i$ is $\mathcal{F}_2(0)$-measurable and $G_i(t)$, $t\in\mathcal{T}$, is independent of $\mathcal{F}_2(0)$. Additionally,  in the  last step of \eqref{eq:integrable} we employed that $\text{Var}(\lvert G_{i,j}\rvert) = \mathbb E (G_{i,j}^2) - \mathbb E (\lvert G_{i,j}\rvert)^2 \geq 0$ and $\mathbb E (G_{i,j}^2)=1$ implies $\mathbb E (\lvert G_{i,j}\rvert) \leq 1$. As the pairs $(G_i(t),N_i(t))$ are pairwise independent and identically distributed, it follows with \eqref{eq:integrable} and the law of large numbers that $\frac{1}{n} \sum_{i=1}^n \sum_{j=1}^{n_i}  \lvert G_{i,j} + 1 \rvert \stackrel{\mathbb{P}}{\longrightarrow} \mathbb{E}(\sum_{j=1}^{n_1}  \lvert G_{1,j} + 1 \rvert )$, as $n\rightarrow\infty$.
Finally, we conclude that $\frac{1}{n}\sum_{i=1}^n \int_0^\tau \vert G_i(u) + 1 \rvert\, dN_i(u) = O_p(1)$, which is why also the first term on the right-hand side of the second step of \eqref{eq:B*n_B} converges to zero in probability, as $n\rightarrow\infty$.  It is only left to consider the third term on the right-hand side of the second step of \eqref{eq:B*n_B}. In fact, we have already shown in the proof of \cref{lem:Cn-C} that this term  converges to zero in probability, as $n\rightarrow\infty$.
Thus, we have proven that all three terms of \eqref{eq:B*n_B} converge to zero in probability, as $n\rightarrow\infty$, which completes the proof of \cref{lem:Cn*-->C}.
\hfill\hfill\qedsymbol

%\bigskip
\bigskip
%\end{proof}
\noindent
\textbf{Proof of \cref{thm:asyEquivalence}.}\\
%\begin{proof}[Proof of \cref{thm:asyEquivalence}]
We aim to derive the weak limit of the term $\textbf D^*_{n,k} + \textbf B^*_n  \textbf C^*_n \textbf D^*_{n,g}(\tau)$, as $n\rightarrow\infty$, where $\textbf D^*_{n,k}$ and $\textbf D^*_{n,g}$ are vector-valued stochastic processes, $\textbf B^*_n$ is a matrix-valued stochastic process and $\textbf C^*_n$ is a random matrix. Recall the notation introduced in the proof of \cref{lem:D*->D} regarding the product probability space $ (\Omega_1, \mathcal A_1, \mathbb P_1)~\otimes~(\Omega_2, \mathcal A_2, \mathbb P_2)$, the convergence in law w.r.t $\mathbb P_2$, $\stackrel{\mathcal{L}_{\mathbb P_2}}{\longrightarrow}$, and $\cdot|\mathcal{F}_2(0)(\omega)$. According to \cref{lem:D*->D}, 
we have, conditionally on $\mathcal{F}_2(0)$, 
$ ({\textbf D^*_{n,k}}^\top, {\textbf D^*_{n,g}}^\top)^\top = \textbf{D}^*_{n, h}  \stackrel{\mathcal{L}_{\mathbb P_2}}{\longrightarrow} \textbf{D}_{\tilde{h}}$, in $(D(\mathcal{T}))^{p+b}$, as $n\rightarrow \infty,$ in $\mathbb P_1$-probability, where $\textbf D_{\tilde{h}}$ is given in \cref{thm:Xn-X_convergence}. Thus, for every subsequence $n_1$ of $n$ there exists a further subsequence $n_2$ such that 
\begin{align}\label{eq:D*_a.s.}
\textbf{D}^*_{n_2, h}|\mathcal{F}_2(0)(\omega )  \stackrel{\mathcal{L}_{\mathbb P_2}}{\longrightarrow} \textbf{D}_{\tilde{h}},\text{ in } (D(\mathcal{T}))^{p+b}, \text{ as } n\rightarrow \infty,
\end{align}
for $\mathbb P_1$-almost all $\omega \in \Omega_1$. Moreover, with \cref{lem:Cn*-->C} it follows that, conditionally on $\mathcal{F}_2(0)$, $\textbf B_{n_2}^*(t) \stackrel{\mathbb P_1\otimes\mathbb P_2}{\longrightarrow} \textbf B(t)$ uniformly in $t\in\mathcal{T}$, as $n\rightarrow\infty$. Hence, for every subsequence $n_3$ of $n_2$ there exists a further subsequence $n_4$ such that $\textbf B_{n_4}^*(t)|\mathcal{F}_2(0)(\omega ) \stackrel{\mathbb P_2}{\longrightarrow} \textbf B(t)$, as $n\rightarrow\infty$, uniformly in $t\in\mathcal{T}$, for $\mathbb P_1$-almost all $\omega \in \Omega_1$. Consequently, we have 
\begin{align}\label{eq:B_n4-inLaw}
    \textbf B_{n_4}^*|\mathcal{F}_2(0)(\omega ) \stackrel{\mathcal{L}_{\mathbb P_2}}{\longrightarrow} \textbf B, \text{ in } \mathcal{D}(\mathcal{T}))^{p q}, \text{ as } n\rightarrow\infty,
\end{align}
for $\mathbb P_1$-almost all $\omega \in \Omega_1$. 
Clearly, \eqref{eq:D*_a.s.} also holds along the subsequence $n_4$. 
Furthermore, we assume that, conditionally on $\mathcal{F}_2(0)$, $\textbf C^*_n$ converges in $\mathbb P_1\otimes \mathbb P_2$-probability to $\textbf C$, i.e., the limits of $\textbf C^*_n$ and $\textbf C_n$, given in \cref{sec:Notation}, are identical. Thus, for every subsequence $n_5$ of $n_4$ there exists a further subsequence $n_6$ such that $\textbf C_{n_6}^*|\mathcal{F}_2(0)(\omega ) \stackrel{\mathbb P_2}{\longrightarrow} \textbf C$, as $n\rightarrow\infty$, for $\mathbb P_1$-almost all $\omega \in \Omega_1$.
 Again, it follows that 
\[\textbf C^*_{n_6}|\mathcal{F}_2(0)(\omega ) \stackrel{\mathcal{L}_{\mathbb P_2}}{\longrightarrow} \textbf C, \text{ as } n\rightarrow\infty,  \]
for $\mathbb P_1$-almost all $\omega \in \Omega_1$. 
Obviously, \eqref{eq:D*_a.s.} and \eqref{eq:B_n4-inLaw} also hold along the subsequence $n_6$. 
%{\vio [explanation: I forgot to consider a further sub-subsequence $n_6$.]} 
Then,
\[(\textbf{D}^*_{n_6, h},\textbf B_{n_6}^*,\textbf C^*_{n_6})|\mathcal{F}_2(0)(\omega ) \stackrel{\mathcal{L}_{\mathbb P_2}}{\longrightarrow} (\textbf{D}_{\tilde{h}},\textbf B,\textbf C) \text{ in } \mathcal D[0,\tau]^{p+b+pq}\times \mathbb{R}^{pq}, \text{ as } n\rightarrow\infty,\]
for $\mathbb P_1$-almost all $\omega \in \Omega_1$ follows analogously to the proof of \cref{thm:Xn-X_convergence}. 
Eventually, the continuous mapping theorem with, successively,  the functions $f_1,f_2$, and $f_3$ given in the proof of \cref{thm:Xn-X_convergence} is applied  to $(\textbf{D}^*_{n_6, h},\textbf B_{n_6}^*,\textbf C^*_{n_6})|\mathcal{F}_2(0)(\omega )$. In particular, we get $\textbf D^*_{n_6,k} + \textbf B^*_{n_6}  \textbf C^*_{n_6} \textbf D^*_{n_6,g}(\tau)|\mathcal{F}_2(0)(\omega )\stackrel{\mathcal{L}_{\mathbb P_2}}{\longrightarrow} \textbf D_{\tilde k} + \textbf B  \textbf C \textbf D_{\tilde g}(\tau)$ for $\mathbb P_1$-almost all $\omega \in \Omega_1$. Finally, by invoking the help of the subsequence principle again, we can conclude that, conditionally
on $\mathcal{F}_2(0)$,
\[\textbf D^*_{n,k} + \textbf B^*_{n}  \textbf C^*_{n} \textbf D^*_{n,g}(\tau)\stackrel{\mathcal{L}_{\mathbb P_2}}{\longrightarrow} \textbf D_{\tilde k} + \textbf B  \textbf C \textbf D_{\tilde g}(\tau),  \text{ in } \mathcal{D}(\mathcal{T}))^{p}, \text{ as } n\rightarrow\infty,\]
in $\mathbb P_1$-probability. Moreover, we can summarize the results of \cref{thm:Xn-X_convergence} and \cref{thm:asyEquivalence} with the following statement
\[d[\mathcal{L}_{\mathbb P_2} (\sqrt{n}(\textbf{X}_{n}^*-\textbf{X}_{n})|\mathcal{F}_2(0)),\mathcal{L}_{\mathbb P_1} (\sqrt{n}(\textbf{X}_{n}-\textbf{X}))]\stackrel{\mathbb P_1}{\longrightarrow} 0, \text{ as } n\rightarrow \infty. \]
\hfill \hfill\qedsymbol

%% file: Chapter2.tex
%\noindent
\textbf{Part II: Application in Fine-Gray Models}

%
%%
%%%
%%%%
%%%%%
%%%%%%
%%%%%%%
\section{Introduction}
%%%%%%%
%%%%%%
%%%%%
%%%%
%%%
%%
%

In this Part~II, we apply the wild bootstrap as described in Part~I to the estimators involved in the Fine-Gray model \citep{Fine-Gray} under censoring-complete data. The Fine-Gray model, which is also called the subdistribution hazards model, has been developed for the competing risks setting. In competing risks analyses, the considered survival outcome is divided into several endpoints that preclude each other. This means that for each individual only one transition out of the initial state into one of the competing endpoints is possible. Although one often is primarily interested in only one particular endpoint, the so-called event of interest, it is important to choose a model that appropriately adjusts for the competing risks. For example, in \cite{application_compRisk} the authors compared the results of a data set analysed with and without accounting for competing risks and thereby illustrated the bias that is introduced when the competing event is ignored.

The two perhaps most popular types of regression models that take competing risks into account, are the cause-specific hazard model---based on fitting multiple Cox-models \citep{cox72}---and the subdistribution hazard model, which is also called the Fine-Gray model. As stated in \cite{Intro_compRisk}, in the cause-specific hazard model ``the effect of the covariates on the rate of occurrence of the outcome'' is modeled, whereas in the subdistribution hazard model ``the effect of covariates on the cumulative incidence function'' is described. As a consequence, in the subdistribution hazard model there is a direct and easily interpretable link between the covariates and the cumulative incidence function for one type of event. This is beneficial, especially because the cumulative incidence function is often used to summarize competing risks data. In the cause-specific hazard model the cumulative incidence function depends on the cause-specific hazard of all event types. Thus, in this model the effect of a covariate on the cause-specific hazard of the event of interest may differ from the effect of the covariate on the corresponding cumulative incidence function due to the effect of the covariate on the cause-specific hazard(s) of the competing event(s) \citep{gray88}. In the subdistribution hazard model this is avoided by directly modeling the cumulative incidence function. In \cite{Fine-Gray} a Cox proportional hazards model is proposed for this. Although the Fine-Gray model enjoys great popularity due to this direct relation, in \cite{Hein_CIF>1} it has been found that in certain situations the sum of multiple estimated cumulative incidence functions following Fine-Gray models might exceed 1. When to use which of the two models is discussed in \cite{Intro_compRisk} and illustrated by means of a simulation study in \cite{simulation_compRisk}. Further comparison of the cause-specific hazard model and the Fine-Gray model can be found in \cite{hein_tutorial} and \cite{Hein_reduction_factor}, where in the former paper the comparison is handled from a practical point of view and in the latter the so-called reduction factor has been introduced in order to relate the two models from a theoretical perspective. 

Several ways to extend the subdistribution hazards model have been introduced. For example, in \cite{Fine-Gray} complete data, censoring-complete data and right-censored data have been considered, while in \cite{Li_interval_cens} the subdistribution hazards model is extended to the case of interval censored data. Furthermore, instead of using the Cox proportional hazards model for the subdistribution it has been suggested to use an additive hazards model in \cite{additive_sun}.

All in all, the Fine-Gray model as proposed in \cite{Fine-Gray} plays an important role in the competing risks setting, which is why in the present Part~II we chose to justify the use of the wild bootstrap as an approximation procedure for the associated estimators under censoring-complete data. At the same time, this exemplifies how to apply the theory developed in Part~I. In comparison to the examples given in that Part~II, the present application is more involved: we show in detail that the proposed assumptions hold and we extend the theory to the cumulative incidence function as a functional of counting process-based estimators. In this regard, the estimators of the Fine-Gray model are either of the general counting process-based form we assumed in Part~I or they have the asymptotic martingale representation we considered in that chapter. In both cases the theory established in Part~I is applicable. Additionally, the exact distributions of these estimators around their target quantities are unknown which is why approximating the distribution is a natural solution, e.g., when the aim is an interval or band estimation. Due to the structure of the estimators and the need for an approximation procedure, this situation is exemplary for the general setting in which the wild bootstrap has been studied in Part~I.

The present chapter is organized as follows. The Fine-Gray model and the underlying notation is introduced in \cref{subsec:Fine-Gray}. In \cref{subsec:F-G-estimators} we employ the theory developed in Part~I to derive the limiting distribution of all relevant basic estimators. Furthermore, in \cref{subsec:Fine-Gray_WB}, we define the wild bootstrap estimators according to  Part~I and use the theory provided in that chapter to derive the corresponding limiting distributions. Additionally, in \cref{subsec:Fine-Gray_CIF} we extend the theory of Part~I by considering a functional of the corresponding estimators, the cumulative incidence function. In particular, we study the weak limit of the cumulative incidence function by means of the functional $\delta$-method. In \cref{sec:CBs} we derive   time-simultaneous confidence bands for the cumulative incidence function.  \cref{sec:simulationStudy} contains the results of an extensive simulation study with which various resampling details for small sample sizes are evaluated. A real data example is given in \cref{sec:data_example} to illustrate the usefulness of wild bootstrap-based confidence bands. We conclude this chapter with a short discussion in \cref{sec:disc}.
All proofs are given in the Appendix.

%
%%
%%%
%%%%
%%%%%
%%%%%%
%%%%%%%
\section{Application of the Wild Bootstrap to Fine-Gray Models}\label{sec:Fine-Gray}
%%%%%%%
%%%%%%
%%%%%
%%%%
%%%
%%
%

%
%%
%%%
%%%%
%%%%%
%%%%%%
%%%%%%%
\subsection{The Fine-Gray Model under Censoring-Complete Data: Preliminaries and Notation}
\label{subsec:Fine-Gray}
%%%%%%%
%%%%%%
%%%%%
%%%%
%%%
%%
%
\noindent
For each of $n$ individuals $i=1, \dots, n$, we let $T_i$ be the survival time in a competing risk setting with $K$ event types, and $C_i$ the right-censoring time which are both defined on a probability space $(\Omega, \mathcal{A}, \mathbb{P})$.
%Furthermore, the
The individuals may be observed within the time frame $\mathcal{T}=[0,\tau]$, where $\tau$ is the maximum follow-up time, but
$T_i$ is only observable if $T_i \leq C_i$.
On the other hand, $C_i$ is assumed to be always observable, 
e.g., %if 
there is only administrative loss to follow-up. In other words, we consider in this chapter the case of censoring-complete data only. Moreover, for each $i$ we observe bounded $q$-dimensional vectors of time-constant covariates $\textbf{Z}_{i} $, measured at baseline,  and, if $T_i \leq C_i$, the type of the occurred event $\epsilon_i \in \{1,\ldots,K\}$. In the competing risks setting, the event types are mutually exclusive.  
It is  assumed that the data  $(\min(T_i,C_i), \mathbbm{1}(T_i\leq C_i), \mathbbm{1}(T_i\leq C_i)\epsilon_i, C_i, \textbf{Z}_i)$, $i=1,\ldots, n$, are independent and identically distributed, 
and that the event times and events types are conditionally independent of the censoring times given the covariates.
In the Fine-Gray model setting we focus on events of type 1 only, and individuals who have experienced an event of type  other than 1 remain in the so-called risk set until their censoring times. Thus, the risk set at the event time of individual $i$ is 
$$R_i=\{j:(\min(C_j,T_j) \geq T_i)\text{ or } (T_j \leq T_i \leq C_j \text{ and } \epsilon_j \neq 1)\}.$$ 
Note that, as Fine and Gray discussed in their original paper \citep{Fine-Gray}, the notion ``risk set'' is actually misleading because if an individual $i$ has experienced some event of type other than 1, it is of course impossible that this individual experiences the event of type 1 in the future. However, this definition of the risk set leads to the particular form of the cumulative incidence function under the Fine-Gray model, see \eqref{eq:CIF} below.   
Finally, multivariate quantities are written in bold type and whenever there is no ambiguity or no need for specification, we will suppress the subscript $i$ that indicates the individual. 

The central role in the Fine-Gray model is played by the cumulative incidence function (CIF) of the event of type 1 which is denoted by $F_1 $ and defined as the probability that the event of type 1 has already occurred by time $t$, given a particular covariate vector $\textbf Z$, this is,
\[F_1(t| \textbf{Z}) = \mathbb{P}(T\leq t, \epsilon = 1 | \textbf Z), \quad t\in\mathcal{T}.\]
Moreover, the instantaneous risk of a type 1 event, given that one is ``at risk'' and given the covariate vector $\textbf Z$, is quantified by the so-called subdistribution hazard $\alpha_{1}$. The 
subdistribution hazard is defined as
\begin{align*}
  &\alpha_{1}(t|\textbf Z)\\ 
  &  = \lim\limits_{\Delta t \rightarrow 0} \frac{1}{\Delta t} \mathbb{P}[t\leq T\leq t+\Delta t,\epsilon=1|\{\min(C,T) \geq t \} \cup (\{T\leq t \leq C \} \cap \{\epsilon\neq 1\}),\textbf Z] \\
   & = \lim\limits_{\Delta t \rightarrow 0} \frac{1}{\Delta t} \mathbb{P}[t\leq T\leq t+\Delta t,\epsilon=1|\{T \geq t \} \cup (\{T\leq t \} \cap \{\epsilon\neq 1\}),\textbf Z], \quad t\in\mathcal{T};
\end{align*}
cf.\ \cite{gray88} and \cite{Fine-Gray}. Due to the particular definition of the risk set, there is a direct relation between $F_1$ and $\alpha_1$, which is $\alpha_{1}(t|\textbf Z) = -d\log\{1-F_1(t|\textbf Z)\}/dt$ or equivalently,
\begin{align}\label{eq:CIF}     
     F_1(t|\textbf Z) = 1-\exp\Big\{-\int_0^t\alpha_{1}(u|\textbf Z ) du\Big\}, \quad t\in\mathcal{T}.
\end{align}
As proposed by the authors of \cite{Fine-Gray}, we choose the following proportional hazards model for the subdistribution through which the covariates are included in a semiparametric manner:
\begin{align}\label{eq:semiparametric_alpha_1}
    \alpha_1(t|\textbf Z)
    =\alpha_1(t,\boldsymbol{\beta}_0|\textbf Z)
    =\alpha_{1;0}(t)\exp(\textbf Z^{\top}\boldsymbol{\beta}_0), \quad t\in\mathcal{T}, 
\end{align}
where $\alpha_{1;0}(t)$ denotes the unknown non-negative baseline subdistributional hazard of event type 1 at time $t$, and $\boldsymbol{\beta}_0$ denotes the unknown vector of regression coefficients. Combining \eqref{eq:CIF} and \eqref{eq:semiparametric_alpha_1}, we specify the cumulative incidence function of event type 1 as follows
\begin{align}\label{eq:F_1_alpha_1}
    F_1(t|\textbf Z)= 1-\exp\{-\exp(\textbf Z^{\top}\boldsymbol{\beta}_0)\cdot A_{1;0}(t)\}, \quad t\in\mathcal{T},
\end{align}
where $A_{1;0}(t)=\int_0^t \alpha_{1;0}(u) du$ is the cumulative baseline subdistribution hazard. We assume that $A_{1;0}(\tau) < \infty$. Note that $F_1$ is a functional, say $\Gamma$, of $\bs{\theta}_0(t)=(\boldsymbol{\beta}_0^\top,A_{1;0}(t))^\top$, $t\in\mathcal{T}$, i.e.,  $$F_1(t|\textbf Z) = \Gamma (\bs{\theta}_0(t)|\textbf Z),\quad t\in\mathcal{T}. $$
 As a consequence, we may obtain an estimator $\hat{F}_{1,n}$ for $F_1$ via estimators $\hat{\boldsymbol{\beta}}_n$ and $\hat A_{1;0,n}$  for $\bs{{\beta}}_0$ and $A_{1;0}$, respectively. For $\hat{\boldsymbol{\beta}}_n$ we will take the well-known maximum partial likelihood estimator (MPLE), and for $\hat A_{1;0,n}$ the Breslow estimator (see  \cref{subsec:F-G-estimators}).
Thus, $\hat{F}_{1,n}$ is given as the functional $\Gamma$ of $\hat {\bs\theta}_n(t) = (\hat{\boldsymbol{\beta}}_n^\top, \hat A_{1;0,n}(t,\hat {\bs{\beta}}_n))^\top$, $t\in\mathcal{T}$, so that 
\[\hat{F}_{1,n}(t|\textbf Z) =\Gamma (\hat {\bs\theta}_n(t)|\textbf Z) = 1-\exp\{-\exp(\textbf Z^{\top}\hat{\boldsymbol{\beta}}_n)\cdot \hat{A}_{1;0,n}(t,\hat{\boldsymbol{\beta}}_n)\}, \quad t\in\mathcal{T}.\]
 Considering $F_1$ and $\hat{F}_{1,n}$ as functionals of ${\bs\theta}_0$ and $\hat {\bs\theta}_n$, respectively, will be of use when studying the (limiting) distribution of the stochastic process $\sqrt{n}(\hat{F}_{1,n} - F_1)$.

From a practical point of view, one is typically  interested in an interval or band estimate 
of $F_1$. For this, one needs the distribution of $\hat{F}_{1,n} - F_1$. As the exact distribution of the corresponding stochastic process is unknown, we suggest to approximate it via the wild bootstrap. Therefore, we will introduce a wild bootstrap estimator $\hat {\bs\theta}^*_n(t)= (\hat{\boldsymbol{\beta}}^{*\top}_n, \hat{A}^*_{1;0,n}(t,\hat {\bs{\beta}}^*_n))^\top$, $t\in\mathcal{T}$, for ${\bs\theta}_0$ in \cref{subsec:Fine-Gray_WB}. Based on $\hat {\bs\theta}^*_n$, we define the resampled cumulative incidence function $\hat{F}_{1,n}^*$ by
\[\hat{F}^*_{1,n}(t|\textbf Z) =\Gamma (\hat {\bs\theta}^*_n(t)|\textbf Z) = 1-\exp\{-\exp(\textbf Z^{\top}\hat{\boldsymbol{\beta}}^*_n)\cdot \hat{A}^*_{1;0,n}(t,\hat{\boldsymbol{\beta}}^*_n)\}, \quad t\in\mathcal{T}.\]
Furthermore, we approximate the distribution of $\sqrt{n}(\Gamma (\hat {\bs\theta}_n|\textbf Z) - \Gamma ({\bs\theta}_0|\textbf Z))$ by the conditional distribution, given the data, of $\sqrt{n}(\Gamma (\hat {\bs\theta}^*_n|\textbf Z) - \Gamma (\hat {\bs\theta}_n|\textbf Z))$. In fact, we will show with \cref{thm:asyEquiv_CIF} in \cref{subsec:Fine-Gray_CIF} that the (conditional) distributions of these two stochastic processes are asymptotically equivalent. The derivation of this result relies on results on the level of the estimators and on the functional $\delta$-method. For this reason, we will first study the limiting distribution of $\sqrt{n}(\hat{\bs{\beta}}_n - \bs{{\beta}}_0)$ and $\sqrt{n}(\hat A_{1;0,n}(\cdot,\hat{\boldsymbol{\beta}}_n) - A_{1;0}(\cdot)$ in \cref{subsec:F-G-estimators} and the limit distribution of their wild bootstrap counterparts $\sqrt{n}(\hat{\bs{\beta}}^*_n - \hat{\bs{\beta}}_n)$ and $\sqrt{n}(\hat A_{1;0,n}^*(\cdot,\hat{\boldsymbol{\beta}}^*_n) - \hat{A}_{1;0,n}(\cdot, \hat{\bs{\beta}}_n))$ in \cref{subsec:Fine-Gray_WB}. Then, with \cref{thm:F-G_asyEquivalence} of \cref{subsec:Fine-Gray_WB} we will prove  that the (conditional) distributions of $\sqrt{n}(\hat {\bs\theta}_n - {\bs\theta}_0)$ and $\sqrt{n}(\hat {\bs\theta}^*_n - \hat {\bs\theta}_n)$ are asymptotically equivalent.

\begin{rem}\label{rem:F-G_Cox}
%The filtration $\mathcal{F}_1(t)$ does not contain information on event times related to event types other than 1.
%The filtration only contains information on whether or not the event of interest, that of type 1, has already been experienced. In that sense the definition of the filtration is in alignment with the definition of the risk set $R_i$, which is crucial for the preservation of the martingale property. The particular definition of the risk set in turn relates the Fine-Gray model in the competing risks setting (with censoring complete data) to the Cox model without competing events. Without competing events, an individual is at risk of experiencing the event whenever (s)he did not yet experience the event and is still under follow-up. 
In this remark, we wish to distinguish the Fine-Gray model in the competing risks setting under censoring-complete data and the ordinary Cox model without competing events. In both models one describes the transition of an individual from the state ``event (of interest) has not yet happened and individual has not yet been censored'' to the state ``event (of interest) has already occurred''. In that sense, the Fine-Gray model can be understood as a reduction of a competing risks model in which the transitions to all competing events are considered separately and simultaneously to a model in which, like the ordinary Cox survival model, only one type of state transition is modelled. %This provides an intuitive explanation for the direct link between the cumulative incidence function of event type 1 and the subdistribution hazard of event type 1, without the necessity to incorporate the other competing risk(s).
Additionally, in both models the (subdistribution) hazard is based on the same proportional model. %semiparametrically and multiplicatively with the exponential relative risk function. 
The differences between the two models are in the definition of the counting process, the at-risk set, the at-risk indicator, and the filtration, while the remaining structures stay the same. In fact, for $K=1$ the Fine-Gray model reduces to the ordinary Cox model. As a consequence, the structure of the theoretical results for the (wild bootstrap) estimators in the context of the Fine-Gray model coincides with the structure of the results for the (wild bootstrap) estimators in Cox models. Hence, one may compare the results presented in this chapter for the Fine-Gray model with those stated in Chapter VII of \cite{Andersen} for the standard estimators in Cox models and with those  in \cite{dobler19} for their wild bootstrap counterparts.
\end{rem}
%
%%
%%%
%%%%
%%%%%
%%%%%%
%%%%%%%
\subsection{The Estimators involved in the Fine-Gray Model and Weak Convergence Results}\label{subsec:F-G-estimators}
%%%%%%%
%%%%%%
%%%%%
%%%%
%%%
%%
%

We will now introduce the counting process notation by means of which the estimators are formulated. The counting process $N_i(t)=\mathbbm{1}\{\min(T_i, C_i) \leq t,T_i \leq C_i, \epsilon_i=1\}$ records for individual $i$ the observable type 1 event time  and
$Y_i(t) = \mathbbm{1}\{C_i \geq t\}(1-N_i(t-))$ is the at-risk indicator of individual $i$, $i = 1,\ldots,n$, $t\in\mathcal{T}$. 
Note that each counting process jumps at most once in the present competing risks setting. 
Moreover, given $\textbf Z$, the cumulative intensity process for individual $i$ is given by $\Lambda_i(t,\boldsymbol{\beta}_0)=\Lambda_i(t,\boldsymbol{\beta}_0| \textbf Z) = \int_0^t Y_i(u)\alpha_{1}(u,\boldsymbol{\beta}_0|\textbf Z_i)du$, which can be shown to be the compensator of the counting process $N_i(t)$. 
In other terms, conditionally on $\textbf Z$ the process
\[M_i(t) = N_i(t) - \Lambda_i(t,\boldsymbol{\beta}_0)\]
is a square integrable martingale with respect to the filtration 
\begin{align}\label{eq:filtration_1}
    \mathcal{F}_1(t) = \sigma\{\mathbbm{1}\{C_i \geq u\}, N_i(u), Y_i(u), \textbf{Z}_i, 0< u\leq t, i=1,\ldots ,n \}, t\in\mathcal{T};
\end{align}
cf.\ \cite{Fine-Gray}.

\noindent
Furthermore, denoting $\textbf Z_{i}^{\otimes 0} = 1$, $\textbf Z_{i}^{\otimes 1} = \textbf Z_{i}$, and $\textbf Z_{i}^{\otimes 2} = \textbf Z_{i}\cdot \textbf Z_{i}^\top$, we define for $m \in \{0,1,2\}$ (in non-bold-type for $m=0$),
\begin{align}
\begin{split}
\label{eq:S-E-R}
\textbf S^{(m)}_n(t,\boldsymbol{\beta}) &= \frac{1}{n}\sum_{i=1}^n \textbf Z_{i}^{\otimes m} Y_i(t)\exp\{\textbf Z_{i}^{\top}\boldsymbol {\beta}\},\\ 
\textbf E_n(t,\boldsymbol{\beta}) &= \textbf S^{(1)}_n(t,\boldsymbol{\beta})\cdot S^{(0)}_n(t,\boldsymbol{\beta})^{-1},\\
\textbf R_n(t,\boldsymbol{\beta}) &= \textbf S^{(2)}_n(t,\boldsymbol{\beta})\cdot S^{(0)}_n(t,\boldsymbol{\beta})^{-1} - \textbf E_n(t,\boldsymbol{\beta})^{\otimes 2}.
\end{split}
\end{align}

In preparation for the upcoming results we state the following regularity assumptions. 
\begin{ass}\label{assump2}
There exists a bounded neighborhood $\mathcal{B} \subset \mathbb{R}^q$ of $\boldsymbol {\beta}_0$ and deterministic functions $s^{(0)}$, $\textbf s^{(1)}$, and $\textbf s^{(2)}$ defined on $\mathcal{T}\times\mathcal{B}$ such that for $m=0,1,2$,
\begin{thmlist}
\item \[\sup_{t\in \mathcal{T},\bs{\beta}\in \mathcal{B}} \left\lVert \textbf S^{(m)}(t,\boldsymbol{\beta})-\textbf s^{(m)}(t,\boldsymbol{\beta})\right\rVert \underset{n\rightarrow\infty}{\overset{\text{P}}{\longrightarrow}}0;\] \label{assump2_1}
\item $\textbf s^{(m)}$ is a continuous function of $\bs{\beta} \in \mathcal{B}$ uniformly in $t\in\mathcal{T}$ and bounded on $\mathcal{T}\times\mathcal{B}$;\label{assump2_2}
\item $s^{(0)}(\cdot,\boldsymbol{\beta})$ is bounded away from zero on $\mathcal{T}$;\label{assump2_3}
\item $(Y_i,N_i, \textbf Z_i)$, $i=1,\ldots ,n$, are pairwise independent and identically distributed;\label{assump2_4_new}
\item $\textbf V_{\tilde g}(\tau) =\int_0^\tau \textbf r(u,\boldsymbol{\beta}_0) s^{(0)}(u,\boldsymbol{\beta}_0)dA_{1;0}(u)$ is positive definite, where $\textbf r(t,\boldsymbol{\beta}) = \textbf s^{(2)}(t,\boldsymbol{\beta})\cdot s^{(0)}(t,\boldsymbol{\beta})^{-1} - \textbf e(t,\boldsymbol{\beta})^{\otimes 2}$ and $\textbf e(t,\boldsymbol{\beta}) = \textbf s^{(1)}(t,\boldsymbol{\beta})\cdot s^{(0)}(t,\boldsymbol{\beta})^{-1}$. \label{assump2_4}
\end{thmlist}
\end{ass}

\noindent
Note that, due to the continuous mapping theorem, $\textbf e(t,\boldsymbol{\beta}) = \textbf s^{(1)}(t,\boldsymbol{\beta})\cdot s^{(0)}(t,\boldsymbol{\beta})^{-1}$ and $\textbf r(t,\boldsymbol{\beta}) = \textbf s^{(2)}(t,\boldsymbol{\beta})\cdot s^{(0)}(t,\boldsymbol{\beta})^{-1} - \textbf e(t,\boldsymbol{\beta})^{\otimes 2}$ are the respective limits in probability of $\textbf E_n(t,\boldsymbol{\beta}) $ and $\textbf R_n(t,\boldsymbol{\beta})$ as $n\rightarrow\infty$. In fact, with \cref{assump2}~\ref{assump2_4_new}, the boundedness of the covariates, and the law of large numbers, we have 
\begin{align}\label{eq:form_of_s}
    \textbf s^{(m)}(t,\boldsymbol{\beta}) = \mathbb{E}(Y_1(t)\textbf Z_1^{\otimes m}\exp(\textbf Z_1^\top\bs{\beta})),
\end{align}
for all fixed $t\in\mathcal{T}$, $m \in \{0,1,2\}$ (in non-bold-type for $m=0$), and $\bs{\beta}\in\mathcal{B}$. Furthermore, with the following \cref{ass:A1} we connect \cref{assump2} above with \cref{assump_general} and \cref{ass:Bn_Cn} of Part~I, and we connect \cref{assump2} with the assumptions stated in Condition~VII.2.1 of \cite{Andersen}. The relation with the assumptions made in Part~I is needed when employing the corresponding results and the relation made with the Condition of \cite{Andersen} is needed for the asymptotic representation of the MPLE.

\begin{lem}\label{ass:A1}\quad 
\begin{thmlist}
    \item If \cref{assump2}~\ref{assump2_1}~-~\ref{assump2_4_new} hold, then \cref{assump_general} and \cref{ass:Bn_Cn} of Part~I hold.\label{ass:A1_1}
    \item If \cref{assump2} holds, then \cref{ass:Cn-C} and \cref{ass:Cn_star-Cn} of Part~I hold. \label{ass:A1_3}
    \item If \cref{assump2} holds, then Condition VII.2.1 of \cite{Andersen} holds.\label{ass:A1_2}
\end{thmlist}
\begin{proof}
See Appendix. %\hfill\qedsymbol
\end{proof}
\end{lem}

%{\green argue that Cond. VII.2.1. d) holds, give it a number and refer to it below the asymptotic representation of beta. Check out Thm Differenzieren nach Parametern.}

As we aim at translating the results of the general setting into results for (the estimators involved in) the Fine-Gray model, we recall the essential notation of Part~I:
\begin{align} \label{eq:Xn_copyII}
    \textbf X_n(t) =\frac1n \sum_{i=1}^n \int_0^t \textbf k_{n, i}(u,\tilde {\bs{\beta}}_n) d N_i(u), \quad t\in\mathcal{T},
\end{align}
that is, the statistic $\textbf X_n$ is a counting process integral with respect to a locally bounded stochastic process $\textbf k_{n, i}(\cdot , \bs{\beta})$ evaluated at a consistent estimator $\bs{\beta} = \tilde{\bs{\beta}}_n$ of the true model parameter $\bs{{\beta}}_0$, cf.\ \eqref{eq:Xn}. Under mild regularity assumptions, the asymptotic representation of $\sqrt{n}  (\textbf X_n - \textbf X) $ is given by
\begin{equation}  \label{eq:Xn-X_inD_copyII}
\sqrt{n}  (\textbf X_n - \textbf X)  
= \textbf{D}_{n, k} +\textbf B_n\cdot \textbf C_n\cdot \textbf{D}_{n, g}(\tau) + o_p(1) ,
\end{equation}
where $\textbf{D}_{n, k}$ and $\textbf{D}_{n, g}$ are local square integrable martingales with respect to $\mathcal{F}_1$,  cf.\ \eqref{eq:Dn} and \eqref{eq:Xn-X_inD} of Part~I. In particular, $\textbf{D}_{n, k}$ and $\textbf{D}_{n, g}$ are
martingale integrals with respect to locally bounded stochastic processes $\textbf k_{n,i}(\cdot,\bs{\beta})$ and $\textbf g_{n,i}(\cdot , \bs{\beta})$ evaluated at $\bs{\beta} = \bs{{\beta}}_0$, respectively, that are predictable for $\bs{\beta} = \bs{{\beta}}_0$. Moreover, $\textbf B_n$ is a matrix-valued counting-process integral and $\textbf C_n$ is a random matrix, cf.\ \eqref{notation:Bn} of Part~I. Lemmas~\ref{lem:Dn} and \ref{lem:Cn-C}, and Assumption~\ref{ass:Cn-C} of Part~I give the conditions for $\textbf{D}_{n, k}$, $\textbf{D}_{n, g}$, $\textbf{B}_n$, and $\textbf{C}_n$ to converge to a continuous zero-mean Gaussian vector martingale $\textbf{D}_{\tilde k}$, a continuous zero-mean Gaussian vector martingale $\textbf{D}_{\tilde g}$, a continuous matrix-valued deterministic function $\textbf{B}(t)$, and a deterministic matrix $\textbf C$, respectively.

Since we will use the general notation of \eqref{eq:Xn_copyII} and \eqref{eq:Xn-X_inD_copyII} for both the MPLE $\hat{\bs{\beta}}_n$ and the Breslow estimator $\hat A_{1;0,n}(\cdot , \hat{\bs{\beta}}_n)$, we will add superscripts to the corresponding components to specify whether they refer to the MPLE (superscript $(1)$) or to the Breslow estimator (superscript $(2)$). The notation of the asymptotic results is not ambiguous, which is why we omit the superscripts there.
Finally, we write $D(\mathcal{T})^{p}$ for the space of cadlag functions mapping from $\mathcal{T}$ to $\mathbb R^{p}$ equipped with the product Skorohod topology, $p\in\mathbb N$.
%\begin{align}
%\label{eq:Dn_copyII}
%    \textbf D_{n, h}(t) =   \frac{1}{\sqrt{n}} \sum_{i=1}^n \int_0^t \textbf h_{n, i}(u, \bs {\beta}_0) d M_i(u),\quad t\in\mathcal{T}, 
%\end{align}

We now investigate the MPLE $\hat{\bs{\beta}}_n$ and the Breslow estimator $\hat A_{1;0,n}(\cdot , \hat{\bs{\beta}}_n)$.
As the name suggests, the MPLE $\hat{\bs{\beta}}_n$ maximizes a partial likelihood, which has a counting process-based expression. In other words, the estimator $\hat{\bs{\beta}}_n$ of $\bs{{\beta}}_0$ is defined as the root of the score statistic 
\[\bs U_{n} (t,\bs{\beta}) = \sum_{i=1}^n \int_0^t ( \bs Z_i - \bs E_n(u,\bs{\beta}) )dN_i(u)\] 
at $t=\tau$, see (7.2.16) on p.~486 of \cite{Andersen}. With a Taylor expansion of $\boldsymbol 0 = \boldsymbol U_n(\tau, \hat{\boldsymbol {\beta}}_n)$ around $\boldsymbol {\beta}_0$ and due to the consistency of  $\hat{\boldsymbol {\beta}}_n$ according to \cref{ass:A1}~\ref{ass:A1_2} in combination with Theorem VII.2.1 of \cite{Andersen} (see \cref{rem:towards_asyRep_beta}), we obtain under \cref{assump2} that
\begin{align}\label{eq:asy_rep_2}
    \sqrt{n}(\hat{\boldsymbol{\beta}}_n-\boldsymbol{\beta}_0) = \big(\frac{1}{n}\textbf I_n(\tau ,\boldsymbol{\beta}_0)\big)^{-1}\frac{1}{\sqrt{n}}\textbf U_n(\tau,\boldsymbol{\beta}_0) + o_p(1),
\end{align}
where $\textbf I_n(t ,\boldsymbol{\beta}) = \sum_{i=1}^n\int_0^t \textbf R_n(u,\bs{\beta})dN_i(u) $ is the negative Jacobian of the score statistic at $\bs{\beta} = \bs{{\beta}}_0$. Note that, although the MPLE $\hat{\bs{\beta}}_n$ is related to a counting process-based statistic via the score statistic, it does not have the general counting process-based form \eqref{eq:Xn_copyII} itself. However, the general results established in Part~I hold as long as the asymptotic representation \eqref{eq:Xn-X_inD_copyII} is retrieved. Thus, we wish to relate the asymptotic representation of $\sqrt{n}(\hat{\boldsymbol{\beta}}_n-\boldsymbol{\beta}_0)$ on the right-hand side of \eqref{eq:asy_rep_2} with the right-hand side  of \eqref{eq:Xn-X_inD_copyII}, i.e., with $\textbf{D}_{n, k}^{(1)} +\textbf B_n^{(1)} \textbf C_n^{(1)}\textbf{D}^{(1)}_{n, g}(\tau)$. In particular, we identify the corresponding components as follows:
\begin{align}\label{eq:C_n-part2}
    \textbf C_n^{(1)} = \big(\frac{1}{n}\textbf I_n(\tau ,\boldsymbol{\beta}_0)\big)^{-1} ,
\end{align}
which is to be understood as a generalized inverse of $\textbf I_n(\tau ,\boldsymbol{\beta}_0)$, e.g., the corresponding Moore-Penrose inverse, if the inverse does not exist, and
\[\textbf D_{n,g}^{(1)}(t) = \frac{1}{\sqrt{n}}\textbf U_n(t ,\boldsymbol{\beta}_0),\quad t\in\mathcal{T},\]
where $\bs U_{n} (\cdot,\bs{\beta})$ evaluated at $\bs{\beta} = \bs{{\beta}}_0$ is a local square integrable martingale with respect to $\mathcal{F}_1$ according to \cref{rem:U_a_martingale}. Additionally, the integrands $\textbf g_{n,i}^{(1)}$ of $\textbf D_{n,g}^{(1)}$ evaluated at $\bs{\beta} = \bs{{\beta}}_0$ are given via 
$$\textbf g_{n,i}^{(1)}(t,\bs{\beta})=\textbf Z_i - \textbf E_n(t,\bs{\beta}), \quad t\in\mathcal{T}, $$
for $i=1,\ldots ,n$.
The remaining components on the right-hand side of \eqref{eq:Xn-X_inD_copyII} are superfluous and we define $\textbf D_{n,k}^{(1)}$ as the $q$-dimensional zero process and we set $\textbf B_n^{(1)}$ equal to the $(q\times q)$-dimensional identity matrix, cf.\ \eqref{eq:theta_hat-theta}. Finally, with the notation introduced above, we rewrite \eqref{eq:asy_rep_2} as
\begin{align}\label{eq:asy_rep_beta_D_2}
    \sqrt{n}(\hat{\boldsymbol{\beta}}_n-\boldsymbol{\beta}_0)   =  \textbf C_n^{(1)}\cdot \textbf D_{n,g}^{(1)}(\tau ) +o_p(1).
\end{align}
With \eqref{eq:asy_rep_beta_D_2} we retrieved the desired asymptotic martingale representation  \eqref{eq:Xn-X_inD_copyII}, for which we have derived asymptotic results in Part~I. In the following lemma the corresponding asymptotic distribution is given.

\begin{lem}\label{lem:results_beta}
If \cref{assump2} holds, then 
$$\sqrt{n}(\hat{\boldsymbol{\beta}}_n-\boldsymbol{\beta}_0) \stackrel{\mathcal{L}}{\longrightarrow} \textbf C\cdot \textbf D_{\tilde g}(\tau), \text{ as } n\rightarrow\infty,$$
where $\textbf C = \textbf V_{\tilde g}(\tau)^{-1} $ and $\textbf D_{\tilde g}(\tau) \sim \mathcal{N}(0,\textbf V_{\tilde g}(\tau))$ with 
\begin{align}\label{eq:covar_g_tilde}
    \textbf V_{\tilde g}(\tau) = \int_0^\tau\mathbb{E}\big(  (\textbf Z_1 - \textbf e(u,\bs{{\beta}}_0) )^{\otimes 2} \lambda_1(u,\bs{{\beta}}_0)  \big) du = \int_0^\tau \textbf r(u,\boldsymbol{\beta}_0) s^{(0)}(u,\boldsymbol{\beta}_0)dA_{1;0}(u).
\end{align} 
Thus, $\textbf C\cdot \textbf D_{\tilde g}(\tau) \sim \mathcal{N}(0,\textbf V_{\tilde g}(\tau)^{-1})$.
\begin{proof}
The statement follows from \eqref{eq:asy_rep_beta_D_2} by means of \cref{ass:A1}~\ref{ass:A1_1}~\&~\ref{ass:A1_3} in combination with \cref{thm:Xn-X_convergence} of Part~I. Moreover, the limit in probability of $\textbf C_n^{(1)}$ as $n\rightarrow \infty$ is derived in the proof of \cref{ass:A1}\ref{ass:A1_3}. %\hfill\qedsymbol
\end{proof}
\end{lem}

Next, we consider the Breslow estimator $\hat A_{1;0,n}(\cdot,\hat{\bs{\beta}}_n) $ of $A_{1;0}(\cdot)$ which is given by
\begin{align} \label{eq:breslow}
\hat A_{1;0,n}(t,\hat{\bs{\beta}}_n) = \frac{1}{n} \sum_{i=1}^n \int_0^t \frac{J_n(u)}{S^{(0)}_n(u,\hat{\bs{\beta}}_n)} dN_i(u) ,\quad t\in\mathcal{T},
\end{align}
where $J_n(t)=\mathbbm{1}\{\sum_{i=1}^n Y_i(t) > 0\}$ equals zero if and only if no individual is at-risk anymore. As this estimator has the general counting process-based form considered in \eqref{eq:Xn_copyII}, we identify $\hat A_{1;0,n}(\cdot,\hat{\bs{\beta}}_n) = X_n^{(2)}(\cdot)$ and $A_{1;0}(\cdot)=X^{(2)}(\cdot)$. In particular, the integrand $k^{(2)}_{n}(\cdot,\hat{\bs{\beta}}_n)$ of $ X_n^{(2)}$ is given by
$$k_{n}^{(2)}(t,{\bs{\beta}}) = J_n(t)\cdot S^{(0)}_n(t,{\bs{\beta}})^{-1},\quad t\in\mathcal{T}, \bs{\beta} \in\mathbb{R}^q.$$ 
According to \cref{rem:towards_asyRep_A} in the appendix, $\sqrt{n}(\hat A_{1;0,n}(\cdot,\hat{\bs{\beta}}_n) - A_{1;0}(\cdot)) = \sqrt{n}(X_n^{(2)}(\cdot) - X^{(2)}(\cdot))$ exhibits the desired asymptotic representation given in \eqref{eq:Xn-X_inD_copyII} as we have
\begin{align}\label{eq:asy_rep_D_2}
    \sqrt{n}(\hat A_{1;0,n}(\cdot,\hat{\bs{\beta}}_n) - A_{1;0}(\cdot)) =
    D_{n,k}^{(2)}(\cdot) + \textbf B_n^{(2)}(\cdot)\cdot \textbf C_n^{(2)}\cdot \textbf D_{n,g}^{(2)}(\tau ) +o_p(1), 
\end{align}
with 
\[ D_{n,k}^{(2)}(t) = \frac{1}{\sqrt{n}} \sum_{i=1}^n \int_0^t \frac{J_n(u)}{S^{(0)}_n(u,{\bs{{\beta}}_0})} d M_i(u),\quad t\in\mathcal{T},\]
and
\begin{align}\label{eq:B_n-part2}
    \textbf B_n^{(2)}(t) = - \frac{1}{{n}} \sum_{i=1}^n \int_0^t J_n(u) \textbf E_n(u,\bs{{\beta}}_0)^\top \cdot S^{(0)}_n(u,{\bs{{\beta}}_0})^{-1} dN_i(u),\quad t\in\mathcal{T}.
\end{align}
Here, $-J_n(t)\cdot \textbf E_n(t,\bs{{\beta}}_0)^\top \cdot S^{(0)}_n(t,{\bs{{\beta}}_0})^{-1} $ is the Jacobian of $k_n(t,\bs{\beta})$ with respect to $\bs{\beta}$ at $\bs{\beta}= \bs{{\beta}}_0$. Note that $ D_{n,k}^{(2)}$ is a local square integrable martingale with respect to $\mathcal{F}_1$ according to Proposition II.4.1 of \cite{Andersen}, as $k_{n}(\cdot,{\bs{\beta}})$ at $\bs{\beta} = \bs{{\beta}}_0$ is predictable and locally bounded. Additionally, $\textbf C_n^{(2)}\cdot \textbf D_{n,g}^{(2)}(\tau ) = \textbf C_n^{(1)}\cdot \textbf D_{n,g}^{(1)}(\tau )$, because the MPLE $\hat{\bs{\beta}}_n$ has been used as the consistent estimator $\tilde{\bs{\beta}}_n$ of $\bs{{\beta}}_0$ in the context of the Breslow estimator, cf.\ \eqref{eq:Xn_copyII}.
We are now ready to state the limiting distribution of $\sqrt{n}(\hat A_{1;0,n}(\cdot,\hat{\bs{\beta}}_n) - A_{1;0}(\cdot))$.

\begin{lem}\label{lem:results_A}
If \cref{assump2} holds, then 
$$\sqrt{n}(\hat A_{1;0,n}(\cdot,\hat{\bs{\beta}}_n) - A_{1;0}(\cdot))  \stackrel{\mathcal{L}}{\longrightarrow} D_{\tilde k}(\cdot) + \textbf B(\cdot)\cdot \textbf C \cdot \textbf D_{\tilde g}(\tau), \text{ in } D(\mathcal{T}) , \text{ as } n\rightarrow\infty,  $$
where the zero-mean Gaussian martingale $ D_{\tilde k}$ is the weak limit of $ D_{n,k}^{(2)}$ and $ D_{\tilde k}$ has the variance function 
\begin{align}\label{eq:covar_k_tilde}
     V_{\tilde k}(t) = \int_0^t \mathbb{E}(  s^{(0)}(u,\bs{{\beta}}_0)^{ -2} \lambda_1(u,\bs{{\beta}}_0) )du = \int_0^t  s^{(0)}(u,\bs{{\beta}}_0)^{ -1}dA_{1;0}(u), \quad t\in\mathcal{T}.
\end{align}
Additionally, $\textbf B$ is the uniform limit in probability of $\textbf B_n^{(2)}$ with 
$$\textbf B(t) = \int_0^t \mathbb{E}(-e(u,\bs{{\beta}}_0)^\top \cdot s^{(0)}(u,{\bs{{\beta}}_0})^{-1} \lambda_i(u,\bs{{\beta}}_0))du = \int_0^t - e(u,\bs{{\beta}}_0)^\top dA_{1;0}(u), \quad t\in\mathcal{T},$$ 
and $\textbf C \cdot \textbf D_{\tilde g}(\tau)$ is as in \cref{lem:results_beta}.
Moreover, the covariance function of $  D_{\tilde k} + \textbf B  \cdot \textbf C\cdot \textbf D_{\tilde g}(\tau)$ is given by
\begin{align*}
    t &\mapsto V_{\tilde k}(t) + \textbf B(t) \cdot \textbf C \cdot \textbf B(t)^\top.
\end{align*}
\begin{proof}
This statement follows from \eqref{eq:asy_rep_D_2} by means of \cref{ass:A1}~\ref{ass:A1_1}~\&~\ref{ass:A1_3} in combination with \cref{thm:Xn-X_convergence} of Part~I. For the covariance function of $  D_{\tilde k} + \textbf B  \cdot \textbf C\cdot \textbf D_{\tilde g}(\tau)$ we have
\begin{align*}
    t &\mapsto  V_{\tilde k}(t) + \textbf B(t) \cdot \textbf C\cdot\textbf V_{\tilde g}(\tau)\cdot \textbf C^\top \cdot \textbf B(t)^\top +\textbf V_{\tilde k,\tilde g}(t)\cdot \textbf C^\top\cdot\textbf B( t)^\top  + \textbf B(t) \cdot \textbf C\cdot \textbf V_{\tilde g, \tilde k}(t)\\
     &= V_{\tilde k}(t) + \textbf B(t) \cdot \textbf C \cdot \textbf B(t)^\top. 
\end{align*}
The last equality follows from $\textbf C = \textbf V_{\tilde g}(\tau)^{-1} $ and due to 
\begin{align}\label{eq:covar_k_g_tilde}
\begin{split}
\textbf V_{\tilde k,\tilde g}(t)^\top &= \textbf V_{\tilde g,\tilde k}(t) = \langle {\textbf D}_{\tilde{{ g}}}, {\textbf D}_{\tilde{{ k}}} \rangle \\
    &\int_0^t \mathbb{E}((\textbf Z_1 - \textbf e(u,\bs{{\beta}}_0))s^{(0)}(u,\bs{{\beta}}_0)^{-1}\lambda_1(u,\bs{{\beta}}_0))du \\
     &= \int_0^t \mathbb{E}(\textbf Z_1 Y_1(u)\exp(\textbf Z_1^\top\bs{{\beta}}_0))s^{(0)}(u,\bs{{\beta}}_0)^{-1} dA_{1;0}(u) -  \int_0^t  \textbf e(u,\bs{{\beta}}_0) dA_{1;0}(u)\\
     &=\bs 0_{q\times 1},
\end{split}
\end{align}
where $ \bs 0_{q\times 1}$ denotes the $q$-dimensional vector of zeros. In other words, $ \textbf D_{n, g} $ and $D_{n, k}$ are asymptotically orthogonal.
%\hfill \qed
\end{proof}
\end{lem}

With \cref{lem:results_beta} and \cref{lem:results_A} we retrieved the well-known results on the limiting distribution of $\sqrt{n}(\hat{\boldsymbol{\beta}}_n-\boldsymbol{\beta}_0)$ and $\sqrt{n}(\hat A_{1;0,n}(\cdot,\hat{\bs{\beta}}_n) - A_{1;0}(\cdot))$, respectively, by means of the theory established in Part~I. Thereby we illustrated  how to translate the general results into results for the basic estimators of a particular model.

%
%%
%%%
%%%%
%%%%%
%%%%%%
%%%%%%%
\subsection{The Wild Bootstrap Estimators and Weak Convergence Results}\label{subsec:Fine-Gray_WB}
%%%%%%%
%%%%%%
%%%%%
%%%%
%%%
%%
%
We will now apply the wild bootstrap to the MPLE $\hat{\bs{\beta}}_n$ and to the Breslow estimator $\hat A_{1;0,n}(\cdot,\hat{\bs{\beta}})$. Detailed information on  this resampling scheme can be found in \cref{sec:generalBootstrap}. At this point, we merely want to draw attention to the most important ingredient of the wild bootstrap: the multiplier processes $G_1(t),\ldots , G_n(t)$, $t\in\mathcal{T}$. In the present context, in which the counting processes jump only once, the multiplier processes reduce to random variables $G_1,\ldots,G_n$ that are i.i.d.\ with mean zero, unit variance and finite fourth moment. Moreover, the filtration corresponding to the wild bootstrap is  constructed such that at time zero it contains the data collected during follow-up, like $\mathcal{F}_1(\tau)$, and that at the event times of type 1, the wild bootstrap multipliers $G_i$ that belong to the individuals who experienced the event of type 1 are included, $i=1,\ldots,n$. This results in the filtration
\begin{align}\label{eq:filtration_2}
    \mathcal{F}_2(t) =\sigma\{G_i\cdot N_i(s), \mathbbm{1}\{C_i \geq u\}, N_i(u), Y_i(u), \textbf{Z}_i, 0<s\leq t, u\in\mathcal{T},  i=1,\ldots,n\}, \quad t\in \mathcal{T}, 
\end{align}
from the resampling-point of view. 

Let us turn to the wild bootstrap counterparts $\hat{\bs{\beta}}_n^*$ of $\hat{\bs{\beta}}_n$ and $\hat{A}^*_{1;0,n}(\cdot,\hat{\bs{\beta}}_n^*) $ of $\hat{A}_{1;0,n}(\cdot,\hat{\bs{\beta}}_n)$. For this, we recall the wild bootstrap counterpart $\textbf{X}^*_n$ of $\textbf{X}_n$ introduced in Part~I:
\begin{align} \label{eq:Xn*_copyII}
  \textbf{X}^*_n (t) = \frac{1}{n} \sum_{i=1}^n \int_0^t \textbf{k}_{n,i}(u,\tilde{\bs{\beta}}^*_n)\big(G_i(u)+1 \big) dN_i(u),\quad t\in \mathcal{T},
\end{align} 
where $\textbf{X}^*_n$ is obtained by applying \cref{WB_replacement} of Part~I to $\textbf{X}_n$, cf.\ \eqref{eq:Xn*} of Part~I. Note, $\tilde{\bs{\beta}}^*_n$ is the wild bootstrap counterpart of $\tilde{\bs{\beta}}_n$. Under mild regularity assumptions, the asymptotic representation of $\sqrt{n}  (\textbf X_n^* - \textbf X_n) $ is given by
\begin{align}\label{eq:X*-Xn_combined_II}
   \sqrt{n}  (\textbf X_n^* - \textbf X_n)= \textbf{D}^*_{n, k} +\textbf B^*_n\cdot \textbf C^*_n\cdot \textbf{D}^*_{n, g}(\tau) + o_p(1),
\end{align}
where $\textbf{D}^*_{n, k}$ and $\textbf{D}^*_{n, g}$ are square integrable martingales with respect to $\mathcal{F}_2$ according to \cref{lemma:mgale} of Part~I, cf.\ \eqref{eq:X*-Xn_4} and \eqref{eq:Dn*} of Part~I combined. Additionally, $\textbf B^*_n$ and $ \textbf C^*_n$ are the wild bootstrap counterparts of $\textbf B_n$ and $ \textbf C_n$, respectively.

As mentioned in \cref{subsec:F-G-estimators}, the estimator $\hat{\bs{\beta}}_n$ does not have the general counting process-based form of the right-hand side of \eqref{eq:Xn_copyII}, but the corresponding asymptotic representation $\textbf{D}_{n, k} +\textbf B_n \textbf C_n\textbf{D}_{n, g}(\tau) + o_p(1)$ of  \eqref{eq:Xn-X_inD_copyII} is retrieved by  \eqref{eq:asy_rep_beta_D_2}. Thus, we apply the wild bootstrap to the asymptotic representation of $\sqrt{n}(\hat{\bs{\beta}}_n - \bs{{\beta}}_0)$  on the right-hand side of \eqref{eq:asy_rep_beta_D_2} in order to obtain its wild bootstrap counterpart $\sqrt{n}(\hat{\boldsymbol{\beta}}^*_n-\hat{\boldsymbol{\beta}}_n)$. In particular, we will apply \cref{WB_replacement} of Part~I to $\textbf D_{n,g}^{(1)}$ to obtain the wild bootstrap version $\textbf D^{*(1)}_{n,g}$, we replace $\textbf C_n^{(1)}$ by a wild bootstrap counterpart $\textbf C_n^{*(1)}$ such that \cref{ass:Cn_star-Cn} of Part~I holds, and we set $o_p(1)$ to zero. These steps yield
\begin{align}\label{eq:asyRep_beta_star}
    \sqrt{n}(\hat{\boldsymbol{\beta}}^*_n-\hat{\boldsymbol{\beta}}) = \textbf C^{*(1)}_n\cdot \textbf D^{*(1)}_{n,g}(\tau ) + 0,
\end{align}
where the wild bootstrap counterpart $\textbf D^{*(1)}_{n,g}$ of $\textbf D^{(1)}_{n,g}$ is given by
\[\textbf D^{*(1)}_{n,g}(t) = \frac{1}{\sqrt{n}} \sum_{i=1}^n \int_0^t (\textbf Z_i - \textbf E_n(u,\hat{\boldsymbol{\beta}}_n)) G_i dN_i(u), \quad t\in\mathcal{T},\]
and the wild bootstrap counterpart $\textbf C^{*(1)}_n$ of $\textbf C^{(1)}_n =  \big(\frac{1}{n}\textbf I_n(\tau ,\boldsymbol{\beta}_0)\big)^{-1}$ is defined through the optional covariation process $[ \textbf D^{*(1)}_{n,g} ] (\tau) $ of $\textbf D^{*(1)}_{n,g}$ at $\tau$, i.e., 
$$\textbf C^{*(1)}_n =\big( [\textbf D^{*(1)}_{n,g}](\tau)\big)^{-1} = \big(\frac{1}{n}\sum_{i=1}^n \int_0^\tau (\bs Z_{i}-\bs E_n(u,\hat{\boldsymbol{\beta}}_n))^{\otimes 2}G_i^2dN_i(u) \big)^{-1};$$
cf.\ \cref{lemma:mgale} of Part~I. According to \cref{ass:A1}~\ref{ass:A1_3}, \cref{ass:Cn_star-Cn} of Part~I is fulfilled for this choice for $\textbf C^{*(1)}_n$ under \cref{assump2}. Moreover, we note that $\textbf D^{*(1)}_{n,g}$ is a local square integrable martingale with respect to $\mathcal{F}_2$ according
to \cref{lemma:mgale} of Part~I, since the integrands $\textbf g_{n,i}^{(1)}(t,\hat{\boldsymbol{\beta}}_n) = (\textbf Z_i - \textbf E_n(u,\hat{\boldsymbol{\beta}}_n))$ of $\textbf D^{*(1)}_{n,g}$ are known, $\mathcal{F}_1(\tau)$-measurable functions, $i=1,\ldots n$.
In this way, we retrieved the asymptotic martingale representation \eqref{eq:X*-Xn_combined_II} for $\sqrt{n}(\hat{\boldsymbol{\beta}}^*_n-\hat{\boldsymbol{\beta}}_n)$ with $o_p(1)$ set to zero, namely $ \textbf{D}_{n, k}^{*(1)} +\textbf B_n^{*(1)} \textbf C_n^{*(1)}\textbf{D}_{n, g}^{*(1)}(\tau)$ with $\textbf D_{n,k}^{*(1)}$ defined as the $q$-dimensional zero process and $\textbf B_n^{*(1)}$ set equal to the $(q\times q)$-dimensional identity matrix.
%, cf.\ \eqref{eq:theta*-theta_hat}. 
Finally, we obtain the wild bootstrap counterpart $\hat{\boldsymbol{\beta}}^*_n$ of $\hat{\boldsymbol{\beta}}_n$. By solving  \eqref{eq:asyRep_beta_star} for $\hat{\boldsymbol{\beta}}^*_n$, we find
\begin{align}\label{eq:beta_star}
    \hat{\boldsymbol{\beta}}^*_n = \frac{1}{\sqrt{n}} \textbf C^{*(1)}_n\cdot \textbf D^{*(1)}_{n,g}(\tau ) + \hat{\boldsymbol{\beta}}_n.
\end{align}
We are now ready to present the asymptotic distribution of $\sqrt{n}(\hat{\boldsymbol{\beta}}^*_n-\hat{\boldsymbol{\beta}})$.

\begin{lem}\label{lem:results_beta_star}
If \cref{assump2} holds, then, conditionally on $\mathcal{F}_2(0)$,
$$\sqrt{n}(\hat{\boldsymbol{\beta}}^*_n-\hat{\boldsymbol{\beta}}_n) \stackrel{\mathcal{L}}{\longrightarrow} \textbf C\cdot \textbf D_{\tilde g}(\tau), \text{ in probability, as } n\rightarrow\infty,$$
with $\textbf C\cdot \textbf D_{\tilde g}(\tau) $ as  in \cref{lem:results_beta}. 
\begin{proof}
This statement follows from \eqref{eq:asyRep_beta_star} by means of \cref{ass:A1}~\ref{ass:A1_1}~\&~\ref{ass:A1_3} in combination with \cref{thm:asyEquivalence} of Part~I.
\end{proof}
\end{lem}
We see from \eqref{eq:breslow} that the Breslow estimator $\hat{A}_{1;0,n}(\cdot,\hat{\bs{\beta}}_n)$ has the general counting process-based form on the right-hand side of \eqref{eq:Xn_copyII}. By applying \cref{WB_replacement} of Part~I directly to $\hat{A}_{1;0,n}(\cdot,\hat{\bs{\beta}}_n)$, we find that its wild bootstrap counterpart $\hat{A}_{1;0,n}^*(\cdot,\hat{\boldsymbol{\beta}}^*_n)$ is given by
\[\hat{A}_{1;0,n}^*(t,\hat{\boldsymbol{\beta}}^*_n)=\frac{1}{n}\sum_{i=1}^n\int_0^t\frac{J_n(u)}{S_n^{(0)}(u,\hat{\boldsymbol{\beta}}_n^*)}(G_i +1) dN_i(u),\quad t\in\mathcal{T},\]
and we identify $\hat{A}_{1;0,n}^*(\cdot,\hat{\boldsymbol{\beta}}^*_n) = X^{*(2)}_n$.
%\sout{\add{[DD: is that true?] {\vio [MD: yes, because $\frac{1}{n}\sum_{i=1}^n\hat{\Lambda}_i(t,{\hat{\beta}}) = \frac{1}{n}\sum_{i=1}^n Y_i(t) \exp(Z_i\hat{\beta})d\hat{A}_{1;0,n}(t,{\hat{\beta}})= S_n^{(0)}(u,\hat{\boldsymbol{\beta}}_n)/ S_n^{(0)}(u,\hat{\boldsymbol{\beta}}_n) \frac{1}{n}\sum_{i=1}^n dN_i(t)$]}}}, as proposed by \cite{spiekerman98} {\green check ref!}. 
According to \cref{rem:assump_4_I_18} in the appendix, $\sqrt{n}(\hat{A}_{1;0,n}^*(\cdot,\hat{\bs{\beta}}^*_n)-\hat{A}_{1;0,n}(\cdot,\hat{\bs{\beta}}_n))=\sqrt{n}(X^{*(2)}_n - X^{(2)}_n)$ has the desired asymptotic representation \eqref{eq:X*-Xn_combined_II} with $o_p(1)$ set to zero. 
Indeed, we have
\begin{align}\label{eq:X*-Xn_3_part2}
 \sqrt{n}(\hat{A}_{1;0,n}^*(\cdot,\hat{\bs{\beta}}^*_n)-\hat{A}_{1;0,n}(\cdot,\hat{\bs{\beta}}_n))
 =D^{*(2)}_{n,k}(\cdot) + \textbf B_n^{*(2)}(\cdot)\textbf C_n^{*(2)}\textbf D^{*(2)}_{n,g}(\tau),
\end{align}
where  the wild bootstrap counterpart $ D^{*(2)}_{n,k}$ of $ D^{(2)}_{n,k}$ is given by
\[ D^{*(2)}_{n,k} (t)= \frac{1}{\sqrt{n}} \sum_{i=1}^n  \int_0^t \frac{J_n(u)}{S^{(0)}_n(u,\hat{\bs{\beta}}_n)}  G_i(u)dN_i(u),\quad t\in\mathcal{T},\]
and the wild bootstrap counterpart $\textbf B^{*(2)}_n$ of $\textbf B^{(2)}_n $ equals
\begin{align}\label{eq:B_n^*-part2}
    \textbf B_n^{*(2)}(t) = -  \frac{1}{n} \sum_{i=1}^n  \int_0^t J_n(u)\cdot\textbf E_n(u,\hat{\bs{\beta}}_n)^\top \cdot S^{(0)}_n(u,\hat{\bs{\beta}}_n)^{-1} (G_i(u) + 1) dN_i(u),
\end{align}
$t\in\mathcal{T}$. Note that $ D^{*(2)}_{n,k}$ is a local square integrable martingale with respect to $\mathcal{F}_2$ according to \cref{lemma:mgale} of Part~I, because the integrand $k^{(2)}_{n}(t,\hat{\boldsymbol{\beta}}_n) = \frac{J_n(u)}{S^{(0)}_n(u,\hat{\bs{\beta}}_n)}$ of $ D^{*(2)}_{n,k}$ is a known, $\mathcal{F}_1(\tau)$-measurable function. Additionally, $\textbf C_n^{*(2)}\cdot \textbf D_{n,g}^{*(2)}(\tau ) = \textbf C_n^{*(1)}\cdot \textbf D_{n,g}^{*(1)}(\tau )$, because the wild bootstrap counterpart $\hat{\bs{\beta}}_n^*$ of the MPLE $\hat{\bs{\beta}}_n$ has been used as wild bootstrap estimator $\tilde{\bs{\beta}}_n^*$ of $\tilde{\bs{\beta}}_n$ in the context of the Breslow estimator, cf.\ \eqref{eq:Xn*_copyII}. Finally, we  present the asymptotic distribution of $\sqrt{n}(\hat{A}_{1;0,n}^*(\cdot,\hat{\bs{\beta}}_n^*)-\hat{A}_{1;0,n}(\cdot,\hat{\bs{\beta}}_n))$.
\begin{lem}\label{lem:results_A_star}
If \cref{assump2} holds, then, conditionally on $\mathcal{F}_2(0)$,
$$ \sqrt{n}(\hat{A}^*_{1;0,n}(\cdot,\hat{\bs{\beta}}^*_n) - \hat{A}_{1;0,n}(\cdot, \hat{\bs{\beta}}_n)) \stackrel{\mathcal{L}}{\longrightarrow} D_{\tilde k}(\cdot) + \textbf B(\cdot)\cdot \textbf C \cdot \textbf D_{\tilde g}(\tau), \text{ in } D(\mathcal{T}) ,$$
in probability, as $n\rightarrow\infty$, where all limit components of the statement above coincide with those given in \cref{lem:results_beta} and \cref{lem:results_A}.
\begin{proof}
The lemma follows from \eqref{eq:X*-Xn_3_part2} by means of \cref{ass:A1}~\ref{ass:A1_1}~\&~\ref{ass:A1_3} in combination with \cref{thm:asyEquivalence} of Part~I.
\end{proof}
\end{lem}

As the final step of this section, we consider the joint (conditional) asymptotic distribution of the (wild bootstrap) estimators of $\bs{{\beta}}_0$ and $A_{1;0}$. This will be of use in \cref{subsec:Fine-Gray_CIF}, in which we study the (conditional) asymptotic distribution of the (wild bootstrap) estimator for $F_1$. Recall from \cref{subsec:Fine-Gray} that
\begin{align*}
    \sqrt{n}(\hat{\bs\theta}_n - {\bs{\theta}_0})(\cdot) &= (\hat{\bs{\beta}}_n^{\top} - {\bs{{\beta}}_0}^\top,\hat{A}_{1;0,n}(\cdot,\hat{\bs{\beta}}_n) - {A}_{1;0}(\cdot))^\top,\\
    \sqrt{n}(\hat{\bs\theta}_n^* - \hat{\bs\theta}_n)(\cdot) &= (\hat{\bs{\beta}}_n^{*\top} - \hat{\bs{\beta}}_n^\top,\hat{A}^*_{1;0,n}(\cdot,\hat{\bs{\beta}}_n^*) - \hat{A}_{1;0,n}(\cdot,\hat{\bs{\beta}}_n))^\top,
\end{align*}
where ${\bs{\theta}_0}$, $\hat{\bs\theta}_n  $, and $\hat{\bs\theta}_n^*$ are defined on $D(\mathcal{T})^{q+1},$ respectively. Here and below,
$d[\cdot, \cdot]$ is an appropriate distance measure between probability distributions, for example the Prohorov distance.
With this notation in mind, we can formulate the following theorem.

\begin{thm}\label{thm:F-G_asyEquivalence} 
If \cref{assump2} holds, then 
\[d[\mathcal{L}(\sqrt{n}(\hat{\bs\theta}_n^* - \hat{\bs\theta}_n)|\mathcal{F}_2(0)),\mathcal{L}(\sqrt{n}(\hat{\bs\theta}_n - {\bs{\theta}_0}))]\stackrel{\mathbb P}{\longrightarrow} 0, \text{ as } n\rightarrow \infty.\]
\begin{proof}
See Appendix.%\hfill\qedsymbol
\end{proof}
\end{thm}

Hereby, we established the asymptotic validity of the wild bootstrap as an approximation procedure for the estimators of the Fine-Gray model under censoring-complete data.  %The details and results of the simulation study can be found in the following \cref{sec:simulationStudy}.

%
%%
%%%
%%%%
%%%%%
%%%%%%
%%%%%%%
\subsection{A Weak Convergence Result for CIFs}\label{subsec:Fine-Gray_CIF}
%%%%%%%
%%%%%%
%%%%%
%%%%
%%%
%%
%

We will now infer the (conditional) limiting distributions of $\sqrt{n}(\hat F_{1,n} - F_1) = \sqrt{n}(\Gamma (\hat{\bs\theta}_n) - \Gamma (\bs{\theta}_0))$ and $\sqrt{n}(\hat F_{1,n}^* - \hat F_{1,n}) = \sqrt{n}(\Gamma (\hat{\bs\theta}_n^*) - \Gamma (\hat{\bs\theta}_n))$ from the (conditional) limiting distributions of $\sqrt{n}(\hat{\bs\theta}_n - \bs{\theta}_0)$ and $\sqrt{n}(\hat{\bs\theta}_n^* - \hat{\bs\theta}_n)$, respectively, with the functional $\delta$-method. In particular, we have for $j = 1,2$,
\begin{align}\label{eq:Gamma-hadamard}
    \sqrt{n}(\Gamma (\tilde{\bs\theta}^{(j)} ) - \Gamma (\tilde{\bs\theta}^{(j-1)} )) =\text{d}\Gamma (\tilde{\bs\theta}^{(j-1)} )\cdot \sqrt{n}(\tilde{\bs\theta}^{(j)} - \tilde{\bs\theta}^{(j-1)}) + o_p(1),
\end{align}
where $\text{d}\Gamma (\tilde{\bs\theta}^{(j-1)} )$ is the Hadamard derivative of $\Gamma$ at $\tilde{\bs\theta}^{(j-1)}$, and  $\tilde{\bs\theta}^{(j)} = (\tilde{\bs{\beta}}^{(j)},\tilde{A}^{(j)}_{1;0})$ with  $\tilde{\bs\theta}^{(0)} = \bs{\theta}_0= (\bs{{\beta}}_0^\top,A_{1;0})^\top$, $\tilde{\bs\theta}^{(1)} = \hat{\bs\theta}_n=(\hat{\bs{\beta}}_n^\top,\hat{A}_{1;0,n}(\cdot,\hat{\bs{\beta}}_n))^\top$ and $\tilde{\bs\theta}^{(2)} = \hat{\bs\theta}_n^*= (\hat{\bs{\beta}}_n^{*\top},\hat{A}^*_{1;0,n}(\cdot,\hat{\bs{\beta}}_n^*))^\top$. The corresponding Hadamard derivative is given in the following lemma.
\begin{lem}\label{lem:Hadamard}
For $j=1,2$, 
%$\text{d} \Gamma (\tilde{\bs\theta}^{(j-1)} )\cdot \sqrt{n}(\tilde{\bs\theta}^{(j)} - \tilde{\bs\theta}^{(j-1)})$ equals
\begin{align*}
    & \textnormal{d} \Gamma (\tilde{\bs\theta}^{(j-1)} )\cdot \sqrt{n}(\tilde{\bs\theta}^{(j)} - \tilde{\bs\theta}^{(j-1)})\\
    &=\exp\{-\exp(\textbf Z^\top \tilde{\bs{\beta}}^{(j-1)})\cdot \tilde{A}_{1;0}^{(j-1)}\}\exp(\textbf Z^\top \tilde{\bs{\beta}}^{(j-1)})\\
    &\qquad\cdot \big [\tilde{A}_{1;0}^{(j-1)}\cdot\textbf Z^\top\sqrt{n}(\tilde{\bs{\beta}}^{(j)} - \tilde{\bs{\beta}}^{(j-1)}) + \sqrt{n}(\tilde{A}_{1;0}^{(j)} - \tilde{A}_{1;0}^{(j-1)})  \big],\quad \text{ on } \mathcal{T}. 
\end{align*}
\begin{proof}
See Appendix.%\hfill\qedsymbol
\end{proof}
\end{lem}
 \cref{thm:F-G_asyEquivalence} and \eqref{eq:Gamma-hadamard} suggest that the conditional distribution of $\sqrt{n}(\Gamma (\hat{\bs\theta}_n^*) - \Gamma (\hat{\bs\theta}_n))$ is asymptotically equivalent to the distribution of $\sqrt{n}(\Gamma (\hat{\bs\theta}_n) - \Gamma (\bs{\theta}_0))$. This is in fact what we prove with the following theorem. 
\begin{thm}\label{thm:asyEquiv_CIF}
If \cref{assump2} holds, then
\[d[\mathcal{L}(\sqrt{n}(\Gamma (\hat{\bs\theta}_n^*) - \Gamma( \hat{\bs\theta}_n))|\mathcal{F}_2(0)),\mathcal{L}(\sqrt{n}(\Gamma (\hat{\bs\theta}_n) - \Gamma ({\bs{\theta}_0})))]\stackrel{\mathbb P}{\longrightarrow} 0, \text{ as } n\rightarrow \infty.\]
\begin{proof}
See Appendix.%\hfill\qedsymbol
\end{proof}
\end{thm}

\noindent
Due to the asymptotic result of \cref{thm:asyEquiv_CIF} we validated the wild bootstrap as an appropriate procedure to approximate the distribution of $\sqrt{n}(\Gamma (\hat{\bs\theta}_n) - \Gamma ({\bs{\theta}_0}))=\sqrt{n}(\hat{F}_{1,n} - F_1)$ under censoring-complete data. 
%We will use this information in \cref{sec:CBs} for the construction of asymptotically valid time-simultaneous confidence bands for ${ F}_1$. The small sample performance of the wild bootstrap is studied by means of an extensive simulation study presented in \cref{sec:simulationStudy}. 

%Here, the basic idea is to resample the CIF and to use the empirical knowledge of the behaviour of the resampled CIF around the estimated CIF as a replacement for the unknown distribution of the estimated CIF around the true CIF. 

%
%%
%%%
%%%%
%%%%%
%%%%%%
%%%%%%%
\section{Time-Simultaneous Confidence Bands for CIFs}\label{sec:CBs}
%%%%%%%
%%%%%%
%%%%%
%%%%
%%%
%%
%

Our aim is the prediction of $F_1(\cdot | \textbf Z) = \Gamma ({\bs{\theta}_0}(\cdot))$ for an individual with covariate vector $\textbf Z$, including an asymptotically valid time-simultaneous $(1-\alpha)$-confidence band, on a time interval $[t_1,t_2]\subset[0,\tau]$. The band will be based on the estimator $\hat F_{1,n}(\cdot | \textbf Z) = \Gamma(\hat{\boldsymbol{\theta}}_n(\cdot))$ of $ F_1(\cdot | \textbf Z)$ and a wild bootstrap-based quantile. Such a quantile replaces the unknown quantile related to the stochastic process 
\[W_n(t) = \sqrt{n}(\hat F_{1,n}(t | \textbf Z) - F_1(t | \textbf Z)), \quad t \in [t_1,t_2].\] We will investigate the use of several types of quantiles, related to six different approximations of the distribution of $W_n$.
First, we approximate the distribution of $W_n$ with that of the following three wild bootstrap counterparts:
\begin{align*}
    W^{*,0}_n(\cdot) &= \sqrt{n}(\hat{F}^*_{1,n}(\cdot | \textbf{Z}) - \hat{F}_{1,n}(\cdot | \textbf{Z}))\\
    & = \sqrt{n}(\Gamma (\hat{\bs\theta}_n^*(\cdot)) - \Gamma( \hat{\bs\theta}_n(\cdot))),\\
    W_n^{*,1}(\cdot) &= \textnormal{d}\Gamma(\hat{\bs\theta}_n)(\cdot)\cdot \sqrt{n}(\hat{\bs\theta}^*_n(\cdot)-\hat{\bs\theta}_n(\cdot)),\\
   %\text{or } 
   W_n^{*,2}(\cdot) &= \textnormal{d}\Gamma(\hat{\bs\theta}^*_n)(\cdot)\cdot \sqrt{n}(\hat{\bs\theta}^*_n(\cdot)-\hat{\bs\theta}_n(\cdot)),
\end{align*}
where $\hat{F}^*_{1,n}(\cdot | \textbf{Z}) = \Gamma(\hat{\bs\theta}^*_n(\cdot))$. The two wild bootstrap counterparts $W_n^{*,1}$ and $W_n^{*,2}$ of $W_n$ are motivated by 
%the functional $\delta$-method and the continuous mapping theorem, cf.\ 
\eqref{eq:Gamma-hadamard}. For $j=0,1,2$, we define the wild bootstrap-based $(1-\alpha)$-quantile $q_{1-\alpha,n}^{*,j}$ related to $W_n^{*,j}$, as the conditional $(1-\alpha)$-quantile of $\sup_{t\in[t_1,t_2]}\lvert W_n^{*,j}(t) \rvert $, given the data. Due to \cref{thm:asyEquiv_CIF} in combination with \eqref{eq:Gamma-hadamard},  the corresponding unweighted and untransformed time-simultaneous $(1-\alpha)$-confidence bands for $F_1(\cdot | \textbf{Z})$, denoted by $CB^*_{1,n,j}$, are asymptotically valid and they are given by
\begin{align}\label{eq:CB1}
CB^*_{1,n,j}(t|\textbf Z) = \hat{F}_{1,n}(t | \textbf Z) \mp q_{1-\alpha,n}^{*,j}/\sqrt{n}, \quad t\in [t_1,t_2],\quad j=0,1,2.
\end{align}
Next, in order to improve the performance of the confidence bands, especially for small sample sizes, it is advocated in \cite{lin97} to use a transformed process $W_{n,\phi,1} =\sqrt{n}(\phi(\hat{F}_{1,n}(\cdot | \textbf{Z})) - \phi(F_1(\cdot| \textbf{Z})))$,
instead of $W_n$.
Here, $\phi: [0,1] \to \mathbb{R}$ is a continuously differentiable one-to-one mapping. So we will use three approximations based on this idea as well.
For the case at hand, we chose for $\phi$ the complementary log-log transformation $\phi(t) = \log(-\log(1-t))$, cf.\ \cite{lin97} and \cite{beyersmann13}. Additionally to this transformation, we incorporate the weight function $g_n(t) = 1/\hat{\sigma}_n(t),$ 
where $\hat{\sigma}_n^2(t)$ is a consistent estimator of the variance of $W_{n,\phi,1}(t)$. More concretely, we consider the weighted and transformed process 
$$W_{n,\phi,g_n}(t) =  \sqrt{n}g_n(t)\big(\phi(\hat {F}_{1,n}(t | \textbf{Z})) - \phi({F}_1(t | \textbf{Z}))\big),\quad t\in[t_1,t_2],$$ 
based on which we construct the so-called equal-precision wild bootstrap confidence bands. For this, we approximate the distribution of $W_{n,\phi,g_n}$ by the distribution of either one of the following three wild bootstrap counterparts: 
\begin{align*}
 W^{*,0}_{n,\phi,g^*_n}(\cdot) & = \sqrt{n}g^*_n(\cdot)(\phi(\hat{F}_{1,n}^*(\cdot| \textbf{Z})) - \phi(\hat F_{1,n}(\cdot| \textbf{Z}))) \\
 & = \sqrt{n}g^*_n(\cdot)(\textbf Z^\top(\hat{\bs{\beta}}^*_n-\hat{\bs{\beta}}_n) 
 + \log(\hat{A}^*_{1;0,n}(\cdot,\hat{\bs{\beta}}^*_n))-\log(\hat{A}_{1;0,n}(\cdot,\hat{\bs{\beta}}_n))), \\
W^{*,1}_{n,\phi,g^*_n}(\cdot) & = \sqrt{n}g^*_n(\cdot)(\textbf Z^\top(\hat{\bs{\beta}}^*_n-\hat{\bs{\beta}}_n) + \hat{A}_{1;0,n}(\cdot,\hat{\bs{\beta}}_n)^{-1}(\hat{A}^*_{1;0,n}(\cdot,\hat{\bs{\beta}}^*_n)-\hat{A}_{1;0,n}(\cdot,\hat{\bs{\beta}}_n))),\\
%\text{or } 
W^{*,2}_{n,\phi,g^*_n}(\cdot) & = \sqrt{n}g^*_n(\cdot)(\textbf Z^\top(\hat{\bs{\beta}}^*_n-\hat{\bs{\beta}}_n) + \hat{A}^*_{1;0,n}(\cdot,\hat{\bs{\beta}}^*_n)^{-1}(\hat{A}^*_{1;0,n}(\cdot,\hat{\bs{\beta}}^*_n)-\hat{A}_{1;0,n}(\cdot,\hat{\bs{\beta}}_n))),
\end{align*}
where all three versions are asymptotically equivalent according to the functional $\delta$-method and the continuous mapping theorem. Additionally, the bootstrapped weight function $g^*_n(t) = 1/\hat{\sigma}^*_n(t)$ involves a bootstrap version $\hat{\sigma}^{*2}_n(t)$ of  $\hat{\sigma}^{2}_n(t)$. Both estimators, $\hat{\sigma}^{2}_n(t)$ and $\hat{\sigma}^{*2}_n(t)$, are given in the lemma below.

\begin{lem}\label{lem:sigma_hat_sigma_star}
If \cref{assump2} holds, then, for a given covariate vector $\textbf Z$,
\begin{align}\label{eq:cov_sigma}
\begin{split}
\hat{\sigma}^2_n(t) = \hat{A}_{1;0,n}(t,\hat{\bs{\beta}}_n)^{-2} &\Big[ \int_0^{t} S^{(0)}_n(u,\hat{\bs{\beta}}_n)^{-1}d\hat{A}_{1;0,n}(u,\hat{\bs{\beta}}_n)\\
    & +\int_0^t (\textbf Z - \textbf E_n(u,\hat{\bs{\beta}}_n))^\top d\hat{A}_{1;0,n}(u,\hat{\bs{\beta}}_n) \Big(\frac{1}{n}\bs I_n(\tau,\hat{\bs{\beta}}_n)\Big)^{-1}\\
    & \cdot \int_0^t (\textbf Z - \textbf E_n(u,\hat{\bs{\beta}}_n)) d\hat{A}_{1;0,n}(u,\hat{\bs{\beta}}_n)  \Big],
\end{split}
\end{align} 
and
\begin{align}\label{eq:WB_variance_estimator}
\begin{split}
 \hat{\sigma}^{*2}_n(t) =  \hat{A}^*_{1;0,n}(t,\hat{\bs{\beta}}^*_n))^{-2} &\big[ \frac{1}{n}\sum_{i=1}^n \int_0^{t} S^{(0)}_n(u,\hat{\bs{\beta}}^*_n)^{-2}G_i^2 dN_i(u)\\
& +\int_0^t (\textbf Z - \textbf E_n(u,\hat{\bs{\beta}}^*_n))^\top d\hat{A}^*_{1;0,n}(u,\hat{\bs{\beta}}^*_n) (\frac{1}{n}\bs I^*_n(\tau,\hat{\bs{\beta}}^*_n))^{-1}\\
& \cdot\int_0^t (\textbf Z - \textbf E_n(u,\hat{\bs{\beta}}^*_n)) d\hat{A}^*_{1;0,n}(u,\hat{\bs{\beta}}^*_n)  \big]
\end{split}
\end{align} 
are consistent (wild bootstrap) estimators for the variance of $W_{n,\phi , 1}(t)$.
\begin{proof}
See Appendix.%\hfill\qedsymbol
\end{proof}
\end{lem}

\noindent
Like before, we replace the unknown $(1-\alpha )$-quantile corresponding to $W_{n,\phi,g_n}$ by either one of the wild bootstrap-based quantiles $\tilde{q}_{1-\alpha,n}^{*,j}$ corresponding to $W_{n,\phi,g^*_n}^{*,j}$, where $\tilde{q}_{1-\alpha,n}^{*,j}$ is the conditional $(1-\alpha )$-quantile of $\sup_{t\in[t_1,t_2]}\lvert W_{n,\phi,g^*_n}^{*,j}(t)\rvert$ given the data, $j=0,1,2$. From \cref{thm:F-G_asyEquivalence}, \cref{lem:sigma_hat_sigma_star} and the continuous mapping theorem, it follows analogously to the proof of \cref{thm:asyEquiv_CIF} that these wild bootstrap-based quantiles are asymptotically valid. The corresponding $\log$-$\log$-transformed time-simultaneous equal-precision $(1-\alpha)$ confidence bands for ${F}_1(\cdot | \textbf{Z})$, denoted by $CB^{*,EP}_{1,n,j}$, are given by
\begin{align}\label{eq:CB2}
\begin{split}
    CB^{*,EP}_{1,n,j}(t|\textbf Z) &= \phi^{-1}\big( \phi (\hat{F}_{1,n}(t | \textbf{Z})) \mp \tilde{q}_{1-\alpha,n}^{*,j}/\{\sqrt{n}g_n(t)\} \big )\\
    &= 1 - (1-\hat{F}_{1,n}(t | \textbf{Z}))^{\exp(\mp \tilde{q}_{1-\alpha,n}^{*,j} \hat \sigma_n/\sqrt{n})} ,\quad t\in [t_1,t_2], \quad j=0,1,2,
    \end{split}
\end{align}
where $\phi^{-1}(y) = 1-\exp(-e^y)$.

%
%%
%%%
%%%%
%%%%%
%%%%%%
%%%%%%%
\section{Simulation Study on Wild Bootstrap-Based Confidence Bands} 
\label{sec:simulationStudy}
%%%%%%%
%%%%%%
%%%%%
%%%%
%%%
%%
%
\subsection{Simulation Set-Up}\label{sec:set-up}
Our simulation study is inspired by the \textit{sir.adm} data set of the \texttt{mvna} R-package and is conducted using R-3.5.1, cf.\ \cite{R}. The aim is to assess the reliability of the six types of wild bootstrap $95\%$ confidence bands for $F_1(\cdot | \textbf Z)$, as given in \eqref{eq:CB1} and \eqref{eq:CB2}, in a non-asymptotic, real life setting. For this we evaluated 144 simulation settings and we simulated 5,000 studies  for each simulation setting based on which the empirical coverage probability was calculated. Moreover, the wild bootstrap-based quantiles $q^{*,j}_{0.95,n}$ and $\tilde{q}^{*,j}_{0.95,n}$, $j=0,1,2$ are based on 2,000 wild bootstrap iterations. The simulation settings were chosen as follows:
\begin{itemize}
    \item sample sizes: $n = 100, 200, 300$;
    \item multiplier distributions: $\mathcal{N}(0,1), \text{Exp}(1)-1$, or  $\text{Pois}(1)-1$;
    \item censoring distributions: $\mathcal{U}(0,c)$  with varying maximum parameters $c$ resulting in censoring rates of about $20\%$ to $25\%$ (light censoring) or about $37\%$ to $43\%$ (strong censoring);
    \item covariates: univariate $Z \sim \textnormal{Bernoulli}(0.2)$ or trivariate $(Z_{ij})_{j=1}^3$ with independent $Z_{i1} \sim \mathcal{N}(0,1)$, $Z_{i2} \sim \textnormal{Bernoulli}(0.15)$, $Z_{i3} \sim \textnormal{Bernoulli}(0.4)$, which stand for the standardized age ($j=1$), the pneumonia status ($j=2$), and the gender ($j=3$) of a patient $i$, $i=1,\ldots ,n$;
    \item time-constant \textit{cause-specific} baseline hazard rates of event type 1 and of event type 2: in the univariate covariate case, $\alpha_{01;0} =0.5$ and $\alpha_{02;0} \in \{0.05, 0.5\}$;
    in the trivariate covariate case, $(\alpha_{01;0}, \alpha_{02;0}) \in \{(0.05, 0.05),$ $ (0.08, 0.008)\}$ the latter of which is motivated from the \textit{sir.adm} data set that will be introduced in \cref{sec:data_example} below;
    \item parameter (vector): in the univariate covariate case, ${\beta}_0 \in \{-0.5, -0.25, 0.25\}$;
    $\bs{\beta}_0 \in \{ ({-}0.05,{-}0.5,{-}0.05),({-}0.05,{-}0.25,{-}0.05),({-}0.05,0.25,{-}0.05)\}$ in the trivariate covariate case;
    \item covariate choices for the confidence bands: in the univarate covariate case, $Z \in \{0,1\}$; in the trivariate covariate case, 
     $\textbf Z \in \{ (-2/3, 0, 1), (2/3, 1, 0)\}$, i.e., a 45 years old female without pneumonia and a 70 years old male with pneumonia on hospital admission, respectively.
\end{itemize}
Based on the above parameter choices, we simulated survival times and event types according to the Fine-Gray model. For this we used the algorithms described in \cite{Beyersmann}, in which it is suggested to simulate the corresponding survival data by exploiting the cause-specific hazards in the following way.
\begin{itemize}
    \item Given time-constant cause-specific hazards $\alpha_{01;0}$, $\alpha_{02;0}$, the baseline subdistribution hazard of event type 1 is $\displaystyle{\alpha_{1;0}(t) = \frac{\alpha_{01;0} + \alpha_{02;0}}{1+\alpha_{02;0}/\alpha_{01;0}\cdot \exp\{(\alpha_{01;0} + \alpha_{02;0})t\}} }$.
    \item For the cause-specific hazard of event type 1 we chose a time-constant Cox proportional hazards model, i.e., $\alpha_{01| \textbf{Z}} = \alpha_{01;0}\cdot \exp\{\textbf{Z}^\top \bs{\beta}_0\}$. Recall from \eqref{eq:semiparametric_alpha_1} that the subdistributional hazard of event type 1 is given by $\alpha_{1}(t| \textbf{Z}) = \alpha_{1;0}(t)\cdot \exp\{\textbf{Z}^\top \bs{\beta}_0\}$. 
    \item Given $\alpha_{01| \textbf{Z}}$, $\alpha_{1}(t| \textbf{Z})$, the cause-specific hazard rate of event type 2 is 
    $$\alpha_{02}(t|\textbf Z) = \alpha_1(t|\textbf Z) - \alpha_{01| \textbf{Z}} - \displaystyle{\frac{d}{dt} }\log(\alpha_{1}(t|\textbf Z))$$ 
    with $\displaystyle{\frac{d}{dt} \log(\alpha_{1}(t|\textbf Z)) = -\frac{\alpha_{01;0} + \alpha_{02;0}}{1+\alpha_{01;0}/\alpha_{02;0}\cdot \exp\{-(\alpha_{01;0} + \alpha_{02;0})t\}}}$.
\end{itemize}
The time intervals $[t_1,t_2]$ with respect to which the confidence bands were determined, correspond to the first and the last decile of the observed survival times of event type 1 across all simulated studies of a kind, where for each realized data set, $t_1$ was also taken to be at least the first observed survival time of type 1. This has been done to avoid poor approximation due to proximity of the band's boundary time points to the extremes of the event times, cf.\ \cite{lin97}.

\subsection{Results of the Simulation Study}\label{subsec:results_simStudy}

In our simulation study, we assessed the actual coverage probability of several wild bootstrap 95\% confidence bands for $F_1(\cdot | \textbf Z)$ based on the wild bootstrap 95\%-quantiles $q_{1-\alpha , n}^{*,j}$, $\tilde{q}_{1-\alpha , n}^{*,j}$, $j=0,1,2$, and three different distributions for the multipliers. The corresponding results are summarized in \cref{tab:results_combined}.
%as described in \cref{sec:CBs}~and~\ref{sec:set-up}. 
As described in \cref{sec:set-up}, we have simulated 144 settings with varying sample sizes, varying censoring rates and varying covariate effects, among others. The simulated coverage probabilities for each setting can be found in the appendix, see Tables~\ref{tab:Norm_uni_0}--\ref{tab:Poi_tri_1}. In order to illustrate the results of all simulated settings at a glance, we calculated for every combination of multipliers and quantiles the percentages of settings with a coverage probability in between 93.0\% and 97.0\% (\cref{tab:resluts_93-97}),  at most 92.0\% (\cref{tab:resluts_0-92}), and at least 98.0\% (\cref{tab:resluts_98-100}). Overall, the combination of multiplier distribution and type of quantile seems to have a major impact on the reliability of the confidence bands. In particular, none of the tested distributions work well in combination with all type of quantiles and vice versa. There are several combinations that turned out too liberal or too conservative. In this respect, we only mention those combinations for which the bands of at least 15\% of the 144 settings are either too liberal or too conservative. The combination of centered exponential multipliers and quantile $\tilde{q}_{1-\alpha , n}^{*,0}$ resulted in too low coverage probabilities, as 25.7\% of the 144 settings have a coverage probability between 0\% and 92\%. Too high coverage probabilities were found for the combinations of standard normal multipliers with quantile $\tilde{q}_{1-\alpha , n}^{*,1}$, centered exponential multipliers with quantile $q_{1-\alpha , n}^{*,1}$, centered Poisson multipliers with quantile $\tilde{q}_{1-\alpha , n}^{*,1}$, and centered exponential multipliers with quantile $\tilde{q}_{1-\alpha , n}^{*,2}$, as 38\%, 30.6\%, 22.9\%, and 18.1\%, respectively,  of their 144 settings have a coverage probability between 98\% and 100\%. 

We consider nominal 95\% confidence bands with actual coverage probability between 93\% and 97\% as acceptable. There are 4 combinations of multipliers and quantiles such that in at least 90\% of the 144 simulated settings coverage probabilities between 93\% and 97\% were achieved.
%see the following percentages in parentheses for the corresponding proportions. 
The results of the following combinations are in this sense  satisfactory: standard normal multipliers in combination with quantile $\tilde{q}_{1-\alpha , n}^{*,0}$ (97.9\%), standard normal multipliers with quantile $q_{1-\alpha , n}^{*,1}$ (95.1\%), centered Poisson multipliers with quantile $q_{1-\alpha , n}^{*,1}$ (93.8\%), and standard normal multipliers with quantile $q_{1-\alpha , n}^{*,0}$ (91\%). Note that for those combinations none of the 144 settings showed a too low coverage probability below 92\%. Additionally, for the combination of standard normal multipliers with quantile $\tilde{q}_{1-\alpha , n}^{*,0}$ none of the settings led to a too high coverage probability, i.e., above 98\%. In conclusion, we recommend to use the 95\% equal-precision confidence band based on $\tilde{q}_{1-\alpha , n}^{*,0}$ with standard normal multipliers, as in 97.9\% of the simulated settings the coverage probability was between 93\% and 97\%, and 100\% of the settings resulted in coverage probabilities between 92.1\% and 97.9\%. 

\begin{table}
\begin{subtable}[t]{\textwidth}
\centering
\begin{tabular}[t]{lllllll}
  \hline
  \hline
& $q_{1-\alpha , n}^{*,0}$ & $q_{1-\alpha , n}^{*,1}$ & $q_{1-\alpha , n}^{*,2}$ & $\tilde{q}_{1-\alpha , n}^{*,0}$ & $\tilde{q}_{1-\alpha , n}^{*,1}$ & $\tilde{q}_{1-\alpha , n}^{*,2}$  \\ 
  N(0,1) & 91 & 95.1 & 75.7 & 97.9 & 47.9 & 77.8 \\ 
  Exp(1)-1 & 84 & 51.4 & 69.4 & 45.8 & 77.1 & 53.5 \\ 
  Poi(1)-1 & 89.6 & 93.8 & 77.8 & 68.8 & 56.9 & 73.6 \\ 
   \hline
\end{tabular}
\caption{\textit{93\% - 97\%}}
\label{tab:resluts_93-97}
\end{subtable}

\bigskip 

\begin{subtable}[t]{\textwidth}
\centering
\begin{tabular}[t]{lllllll}
  \hline
  \hline
& $q_{1-\alpha , n}^{*,0}$ & $q_{1-\alpha , n}^{*,1}$ & $q_{1-\alpha , n}^{*,2}$ & $\tilde{q}_{1-\alpha , n}^{*,0}$ & $\tilde{q}_{1-\alpha , n}^{*,1}$ & $\tilde{q}_{1-\alpha , n}^{*,2}$  \\ 
  N(0,1) & 0 & 0 & 0 & 0 & 0 & 0.7 \\ 
  Exp(1)-1 & 0.7 & 0 & 12.5 & 25.7 & 0 & 0 \\ 
  Poi(1)-1 & 0.7 & 0 & 8.3 & 4.2 & 0 & 0 \\ 
   \hline
\end{tabular}
\caption{\textit{0\% - 92\% }}
\label{tab:resluts_0-92}
\end{subtable}

\bigskip 

\begin{subtable}[t]{\textwidth}
\centering
\begin{tabular}[t]{lllllll}
  \hline
  \hline
& $q_{1-\alpha , n}^{*,0}$ & $q_{1-\alpha , n}^{*,1}$ & $q_{1-\alpha , n}^{*,2}$ & $\tilde{q}_{1-\alpha , n}^{*,0}$ & $\tilde{q}_{1-\alpha , n}^{*,1}$ & $\tilde{q}_{1-\alpha , n}^{*,2}$  \\ 
  N(0,1) & 2.8 & 2.8 & 10.4 & 0 & 38.2 & 6.9 \\ 
  Exp(1)-1 & 3.5 & 30.6 & 3.5 & 0 & 8.3 & 18.1 \\ 
  Poi(1)-1 & 2.8 & 4.2 & 2.8 & 0 & 22.9 & 12.5 \\ 
   \hline
\end{tabular}
\caption{\textit{98\% - 100\%}}
\label{tab:resluts_98-100}
\end{subtable}
\caption{\textit{Percentage of the 144 simulated settings with simulated coverage probability between 93\% - 97\% (\subref{tab:resluts_93-97}), between 0\% - 92\% (\subref{tab:resluts_0-92}), between 98\% - 100\% (\subref{tab:resluts_98-100}). The simulated coverage probabilities refer to confidence bands for the cumulative incidence function calculated under the indicated distribution of the multipliers and the specified quantile.}}
% \subref{tab:table1_a}, \subref{tab:table1_b}, \subref{tab:table1_c} and \subref{tab:table1_d}.
\label{tab:results_combined}
\end{table}

%As individuals, who experience an event other than of type one, remain in the at-risk sets until they are censored, it suffices to 

%
%%
%%%
%%%%
%%%%%
%%%%%%
%%%%%%%
\section{Real Data Example: Impact of Pneumonia on the CIF}
\label{sec:data_example}

In this section, we illustrate the wild bootstrap-based 95\% confidence bands for a real data set. The data set was obtained by merging the \textit{sir.adm} data set from the R-package \texttt{mvna} with the \textit{icu.pneu} data set from the R-package \texttt{kmi} by matching the patient ID. These data sets are random subsamples of the data that originate from the SIR 3 cohort study conducted at the Charité university hospital in Berlin, Germany, during a period of 18 month from January 2000 until July 2001. The goal of that study was to determine the incidence of hospital-acquired infection in intensive care units (ICU). See \cite{SIR_results} and \cite{SIR_how_many} for a detailed description of the study and the corresponding results. One may find further statistical analyses of the data in, e.g., \cite{SIR_beyersmann} and \cite{SIR_wolkewitz}. 
As described in \cite{cmprsk}, the \textit{sir.adm} data set contains 747 patients for whom their pneumonia status on admission to the ICU, age, and sex are given as baseline covariates. 
The data set \textit{icu.pneu} contains 1,313 patients for whom a nosocomial pneumonia indicator, their age, and sex are available as covariates. The nosocomial pneumonia indicator switches from zero to one at the time of infection. However, we have established the wild bootstrap only for the case of time-constant (i.e.\ baseline) covariates in this chapter. Thus, we exclude the nosocomial pneumonia indicator from our analysis. Practical guidelines for the inclusion of time-dependent covariates in Fine-Gray models are given by \cite{beyersmann_timeDep_F-G}. By merging the two data sets, we obtained a data set of 524 patients for whom the covariates are comprised of their pneumonia status on admission to the ICU, age, and sex. For example, the merged data set contains 63 patients with pneumonia on admission, 221 female patients and the average age of a patient was 57.62 years (with quartiles 46.55, 61.35, 70.95 years). Moreover, the outcome of the ICU-stay of each patient---alive discharge from hospital, death, or censoring---was recorded. Thus, we have discharge from hospital and death as the competing risks. In our study, we took the status \emph{death} as the event of interest, i.e., as event of type 1. Note that censoring occurred only due to administrative loss to follow-up. In the data set at hand, 459 patients were discharged from hospital, 54 patients died and 11 were censored. Additionally, the data set contains for each patient the time in ICU till either occurrence of an event or censoring. Furthermore, the data set includes the administrative censoring times for all patients that have been discharged alive from the hospital, but not for the deceased individuals. 
That is, the data set holds the censoring times for all individuals except for those who experienced the event of interest. We will call such data sets \textit{partially-censoring-complete}. In contrast, a data set with censoring times for all individuals is called \textit{censoring-complete}. 

Nevertheless, from a practical point of view, partially-censoring-complete data are sufficient, because individuals are considered to be at-risk until either they experience the event of interest or until they are censored. 
Thus, the at-risk indicator is computable for all individuals based on partially-censoring-complete data. 
From a theoretical point of view, the underlying $\sigma$-algebras for our martingale arguments have to be modified in order to be in line with partially-censoring-complete data. In particular, in \eqref{eq:filtration_1} and \eqref{eq:filtration_2} we replace $\mathbbm 1 \{C_i\geq u\}$ by $\mathbbm 1 \{C_i\geq u\}(1-N_i(u))$, where $N_i$ counts the observed events of interest of individual i and $C_i$ is the censoring time of individual i. In this way, the censoring information is available unless the individual has experienced the event of interest.

%In our study, we took the status \emph{alive discharge from hospital} as the event of interest, i.e., as event of type 1. 
In our present data example, we computed the wild bootstrap confidence band for the cumulative incidence function of event type 1, $F_1(\cdot | \textbf Z)$, %according to the Fine-Gray model 
for two covariate vectors ($\textbf Z=\textbf{z}_1$ and $\textbf Z=\textbf{z}_2$). First, for a female individual of average age \textit{without} pneumonia on admission (encoded by the covariate vector $\textbf{z}_1$). Second, for a female individual of average age \textit{with} pneumonia on admission (encoded by the covariate vector $\textbf{z}_2$). In particular, we computed the log-log-transformed 95\% equal-precision wild bootstrap confidence bands $CB^{*,EP}_{1,524,0}(t| \textbf{z}_1)$ and $CB^{*,EP}_{1,524,0}(t| \textbf{z}_2)$ on the interval $t\in [t_1,t_2] = [6,48]$ (time in days) with standard normal multipliers and quantile $\tilde q^{*,0}_{0.95,524}$. As in \cref{sec:set-up}, the boundary values $t_1$ and $t_2$ correspond to the first and the last decile of the observed survival times of the event of interest.
%such that poor approximations due to proximity to the extremes of the event times are avoided. 
Note that no event of interest occurs during the time interval $(44,48]$ and therefore, the figures will be plotted with respect to the time interval $[6,44]$. Because we only consider this particular type of band  for the present data example, we simplify the corresponding notation to $CB^{*}_{1,524}(\cdot| \textbf{z}_j)$, $j=1,2$. The choice of standard normal multipliers in combination with the quantile $\tilde q^{*,0}_{0.95,524}$ has been made in accordance with the results of the simulation study of \cref{sec:simulationStudy}. The wild bootstrap-based quantile has been calculated using 2,000 wild bootstrap iterations. 
{
}

 \begin{figure}
    \centering
    \begin{subfigure}[H]{0.45\textwidth}    
        \includegraphics[width=\textwidth]{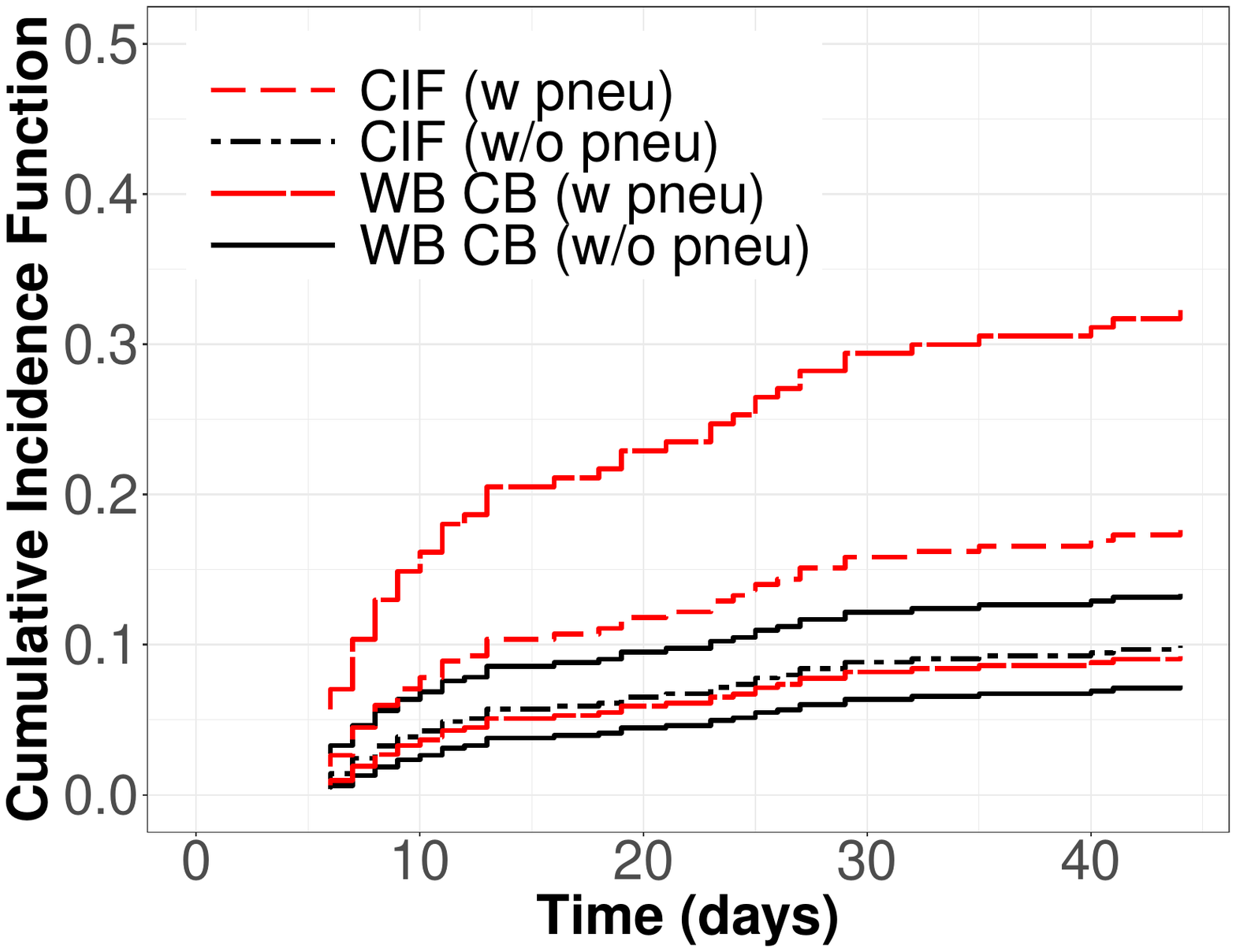}
    \end{subfigure}
        %\caption{}
        %$\overline{F}_{1,n}(\cdot|\textbf z_1)$, %$\overline{F}_{1,n}(\cdot|\textbf z_2)$, $CB^{*,l}_{1,n}$, and %$\overline{CB}^{*}_{1,n}$, $l=1,\ldots,100$.}
        \caption{\textit{The estimated cumulative incidence function $\hat{F}_{1,524}(\cdot | \textbf{z}_j)$ for a female individual of average age without pneumonia (CIF w/o pneu) and a female individual of average age with pneumonia (CIF w pneu) with lower and upper bounds of the corresponding wild bootstrap confidence bands ${CB}^{*}_{1,524}(\cdot | \textbf{z}_j)$ (WB CB), $j=1,2$.}}
    \label{fig:dataEx_Chp2}
\end{figure}

In Figure~\ref{fig:dataEx_Chp2} the estimated cumulative incidence function $\hat{F}_1(\cdot | \textbf{z}_j)$ is plotted on the time interval $[6,44]$ for the individual without pneumonia on admission ($\textbf{z}_1$) and for the individual with pneumonia on admission ($\textbf{z}_2$), together with the lower bounds and upper bounds of the corresponding wild bootstrap confidence bands ${CB}^{*}_{1,524}(\cdot | \textbf{z}_j)$, $j=1,2$. 
The lower and upper bounds of ${CB}^{*}_{1,524}(44 | \textbf{z}_1)$ and ${CB}^{*}_{1,524}(44 | \textbf{z}_2)$ equal $(0.073,0.134)$ and $(0.092,0.323)$, respectively. Thus, the wild bootstrap confidence band after 44 days for the individual with pneumonia is considerably wider than the wild bootstrap confidence band after 44 days for the individual without pneumonia. This is most likely caused by a larger variance estimate due to the relatively few patients with pneumonia on admission to the hospital (63 out of 524 in the whole data set). In other words, for a female individual of average age without pneumonia on admission, the predicted chances of dying in the ICU is not only lower but also more precise than the predicted chances of experiencing the event of interest for a female individual of average age with pneumonia on admission.
Moreover, one can see from the figure that the two confidence bands are overlapping on the entire time interval.

{
} %old plot

\begin{figure}%\label{fig:WB_variation}
    \centering
    \begin{subfigure}[H]{0.45\textwidth}
        \includegraphics[width=\textwidth]{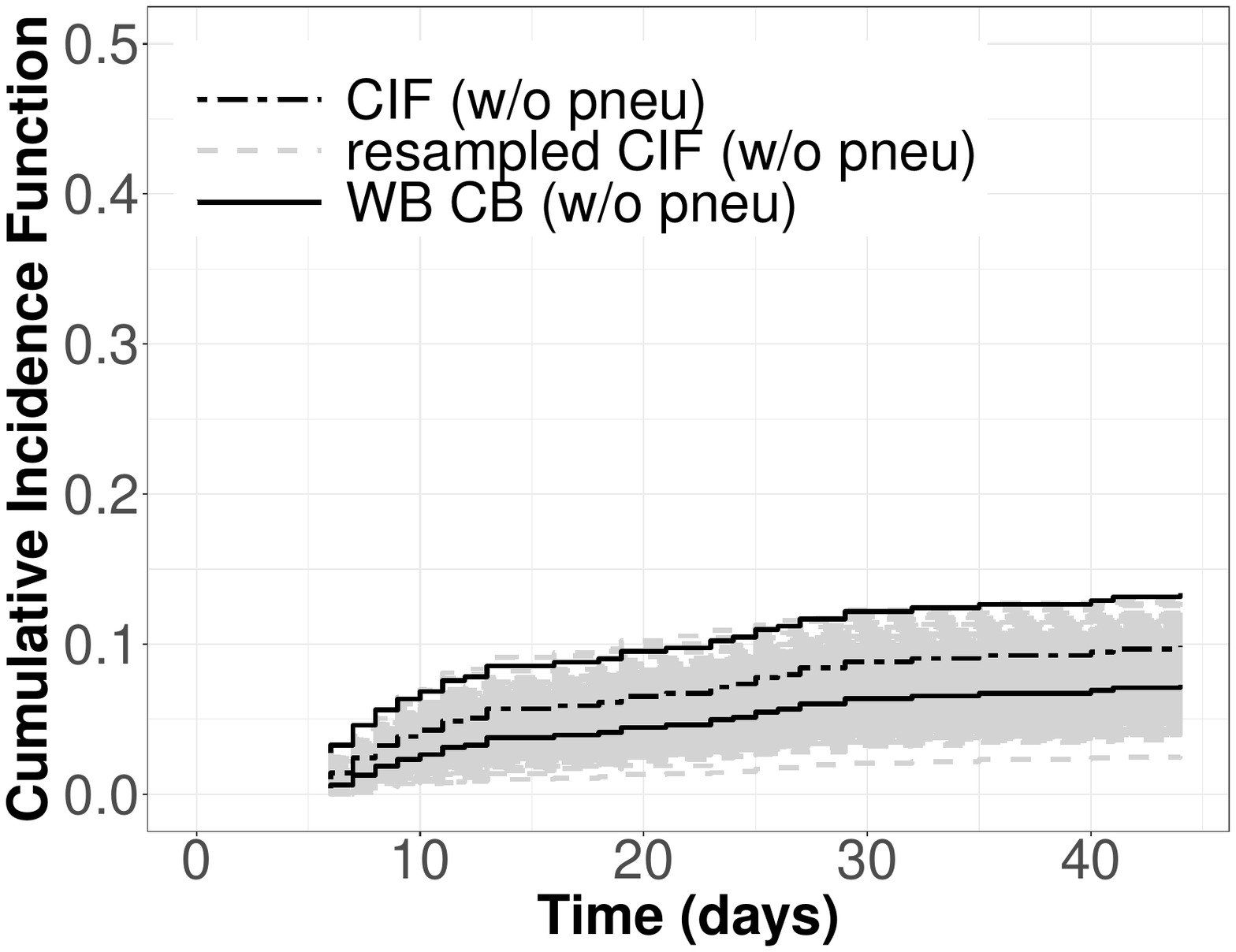}
        \caption{\textit{A female individual of average age without pneumonia on hospital admission $(\textbf{z}_1)$.}}
        \label{fig:WB-100}
    \end{subfigure}
    ~ %add desired spacing between images, e. g. ~, \quad, \qquad, \hfill etc. 
    \begin{subfigure}[H]{0.45\textwidth}
        \includegraphics[width=\textwidth]{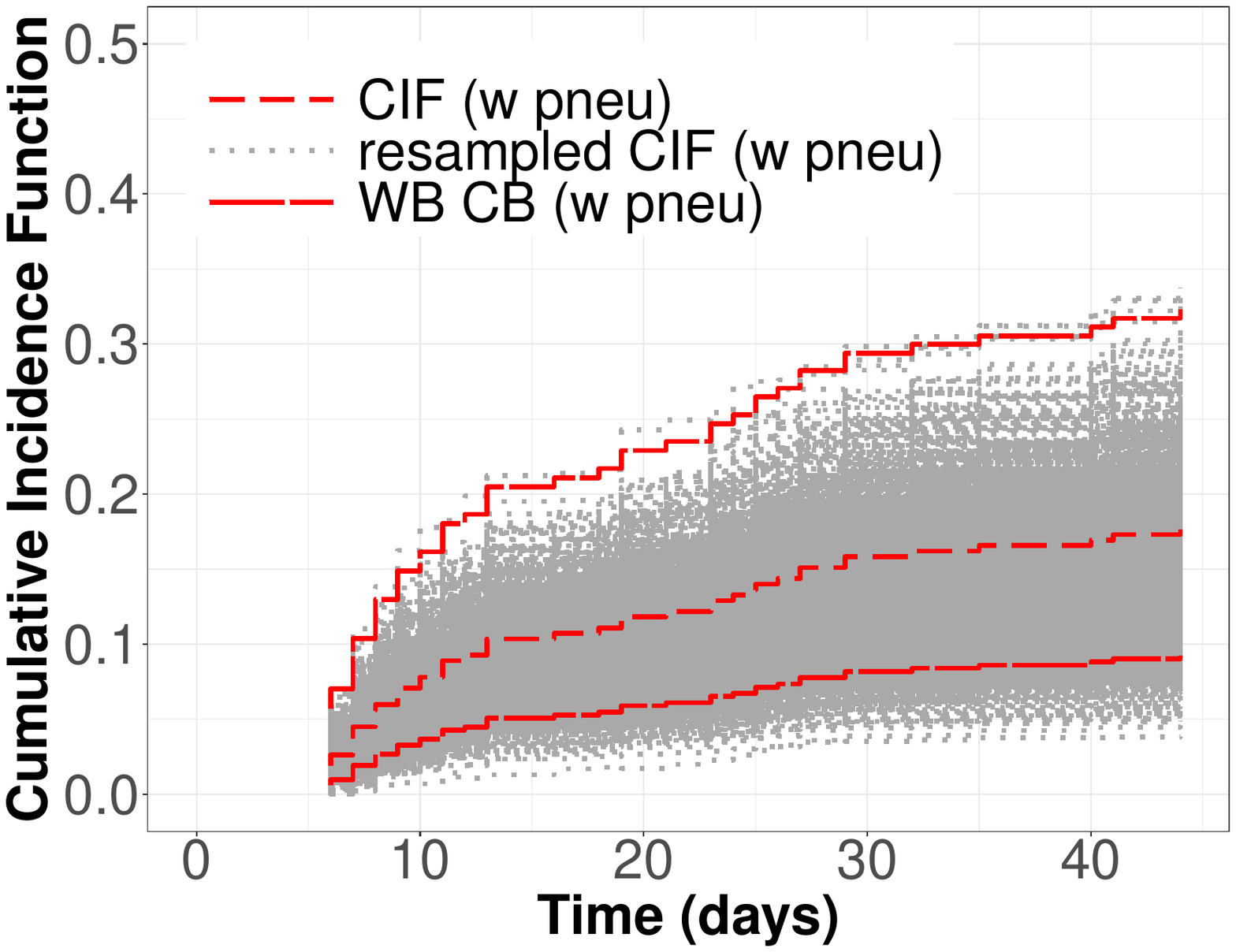}
        \caption{\textit{A female individual of average age with pneumonia on hospital admission $(\textbf{z}_2)$.}}
        \label{fig:WB-200}
    \end{subfigure}
    \caption{\textit{Each plot contains the estimated cumulative incidence function (CIF) $\hat{F}_{1,524}$, 2,000 realizations of the resampled cumulative incidence function (resampled CIF) $\hat{F}^*_{1,524}$, and the corresponding $95\%$ equal-precision wild bootstrap confidence band (WB CB) $CB^{*}_{1,524}$ for a female individual of average age without pneumonia on admission (\subref{fig:WB-100}) and a female individual of average age with pneumonia on admission (\subref{fig:WB-200}).}}\label{fig:WB_variation}
\end{figure}

In Figure~\ref{fig:WB_variation} we present the relationship between the estimated cumulative incidence function $\hat{F}_{1,524}(\cdot | \textbf{z}_j)$, the resampled cumulative incidence functions $\hat{F}^*_{1,524}(\cdot | \textbf{z}_j)$, and the equal-precision $95\%$ wild bootstrap confidence band $CB^{*}_{1,524}(\cdot | \textbf{z}_j)$ for an individual without pneumonia on admission ($\textbf{z}_1$) and for an individual with pneumonia on admission ($\textbf{z}_2$), $j=1,2$.  It can be seen that the resampled cumulative incidence functions fluctuate vertically around the estimated cumulative incidence function. This illustrates the randomness induced by the multipliers which is supposed to mimic the randomness that one would observe if several data sets would have been used for the estimation of the cumulative incidence function. Furthermore, the resampled cumulative incidence functions are asymmetrically distributed around the estimated cumulative incidence function. This is likely due to the complementary $\log-\log$-transformation of the equal-precision wild bootstrap confidence band.

%{Comparison_pneu_CIF_CB_WB-100_2.png}
%{Comparison_pneu_CIF_CB_WB-200_2.png}
%{Comparison_pneu_CIF_CB_WB-300_2.png}
%{Comparison_pneu_CIF_CB_WB-500_2.png}

%
%%
%%%
%%%%
%%%%%
%%%%%%
%%%%%%%
\section{Discussion}
\label{sec:disc}
%%%%%%%
%%%%%%
%%%%%
%%%%
%%%
%%
%

In the above, we have demonstrated in detail how the martingale-based theory of Part~I can be applied to justify the wild bootstrap for the estimators involved in the Fine-Gray model under censoring-complete data. The key role in this is played by the asymptotic (wild bootstrap) martingale representation considered in Part~I and the asymptotic results on the corresponding distribution derived in that chapter. In the present chapter we retrieved the  representation for the MPLE, the Breslow estimator, and their wild bootstrap counterparts. We then used the results on the asymptotic distribution from Part~I to infer the asymptotic distribution of the (wild bootstrap) estimators involved in the Fine-Gray model. Moreover, we extended the results to a functional of those estimators in order to justify the wild bootstrap for the cumulative incidence function, which is typically the function of interest in the context of this model. Based on these results, we presented two types of asymptotically valid time-simultaneous confidence bands that can be used to predict the cumulative incidence function for given covariate combinations.

We also conducted an extensive simulation study to evaluate the reliability of different resampling details for small sample size. We discovered that the coverage probability depends on both the chosen distribution of the multipliers and the type of wild bootstrap-based quantile. In summary, the choice of standard normal in combination with either of the quantiles $q_{1-\alpha , n}^{*,0}$ or $q_{1-\alpha , n}^{*,1}$, and centered Poisson multipliers in combination with the quantiles $q_{1-\alpha , n}^{*,1}$ resulted in the most reliable bands based on the untransformed cumulative incidences.
Additionally, for bands based on the complementary $\log-\log$-transformation, which additionally have the advantage of including only values between 0 and 1, normal multipliers in combination with $\tilde q_{1-\alpha , n}^{*,0}$ resulted in the most reliable confidence bands of all.

Furthermore, we illustrated the wild bootstrap confidence band corresponding to the best choice of multiplier distribution and type of quantile found via the simulation study for a real data set. In particular, we predicted the band estimate of the cumulative incidence function for death as the event of interest for female individuals of average age with and without pneumonia on admission. Thereby, the chances of dying could be compared for those two covariate combinations.

We have introduced the Fine-Gray model for time-constant covariates only. A practical solution to the question of how to extend the Fine-Gray model to time-dependent covariates can be found in \cite{beyersmann_timeDep_F-G}. In that paper the authors suggested the usage of multistate models in combination with discrete covariates in order to treat time-dependent covariates in Fine-Gray models. Moreover, the general case of independently right-censored data is not covered by our theory developed in Part~I. This is due to the fact that for the general case, the score function does not exhibit a martingale property anymore (see Appendix A of \cite{Fine-Gray}). In a forthcoming paper, we will develop a wild bootstrap-based confidence band for the cumulative incidence function which is adjusted to independently right-censored data via multiple imputation.
\begin{table}[ht]
\centering
\begin{tabular}{llllllllll}
  \hline
  \hline
  &  &  &  & $q^*_0$ & $q^*_1$ & $q^*_2$ & $\tilde{q}^*_0$ & $\tilde{q}^*_1$ & $\tilde{q}^*_2$ \\ 
    &  &  &  &  &  &  &  &  &  \\ 
    n & cens. & ${\beta}_0$ & $(\alpha_{010},\alpha_{020})$ &  &  & &  &  &  \\ 
  100 & low & -0.5 & (0.5,0.05) & 96.2 & 94.5 & 98.4 & 94.7 & 99 & 92.3 \\ 
     &  &  & (0.5,0.5) & 95.3 & 94.7 & 96.5 & 95 & 99.1 & 93.8 \\ 
     &  & -0.25 & (0.5,0.05) & 95.5 & 93.8 & 98.4 & 93.7 & 98.8 & 91.7 \\ 
     &  &  & (0.5,0.5) & 95.1 & 94.3 & 96.2 & 95.3 & 99.2 & 93.7 \\ 
     &  & 0.25 & (0.5,0.05) & 95.2 & 93.3 & 98.2 & 94.2 & 99.2 & 92.6 \\ 
     &  &  & (0.5,0.5) & 94.5 & 94 & 95.5 & 94.9 & 99.3 & 93.9 \\ 
     & high & -0.5 & (0.5,0.05) & 95.9 & 94.2 & 98.5 & 94.9 & 99.2 & 92.8 \\ 
     &  &  & (0.5,0.5) & 95.9 & 95.3 & 96.8 & 95.7 & 98.9 & 94.8 \\ 
     &  & -0.25 & (0.5,0.05) & 95.6 & 94.1 & 98.2 & 94.9 & 99.3 & 92.8 \\ 
     &  &  & (0.5,0.5) & 95.3 & 94.6 & 96.1 & 94.9 & 99.1 & 94.2 \\ 
     &  & 0.25 & (0.5,0.05) & 95.1 & 93.8 & 97.5 & 94.3 & 99.1 & 92.6 \\ 
     &  &  & (0.5,0.5) & 94.7 & 94.4 & 95.4 & 94.8 & 99 & 94.2 \\ 
    200 & low & -0.5 & (0.5,0.05) & 95.7 & 94.8 & 97.5 & 94.7 & 98.5 & 93.4 \\ 
     &  &  & (0.5,0.5) & 95.2 & 95 & 95.9 & 95.7 & 99 & 93.6 \\ 
     &  & -0.25 & (0.5,0.05) & 95.5 & 94.6 & 97.4 & 94.8 & 98.6 & 93.5 \\ 
     &  &  & (0.5,0.5) & 94.2 & 93.9 & 94.9 & 94.7 & 98.8 & 92.9 \\ 
     &  & 0.25 & (0.5,0.05) & 95.5 & 94.9 & 97 & 94.7 & 98.7 & 93.7 \\ 
     &  &  & (0.5,0.5) & 94.8 & 94.6 & 95.3 & 94.3 & 98.8 & 93.2 \\ 
     & high & -0.5 & (0.5,0.05) & 95.3 & 94.3 & 97.3 & 94.6 & 98.7 & 93.1 \\ 
     &  &  & (0.5,0.5) & 95.2 & 95 & 95.9 & 94.4 & 98.8 & 92.7 \\ 
     &  & -0.25 & (0.5,0.05) & 95.3 & 94.5 & 97.1 & 94.8 & 98.9 & 93.4 \\ 
     &  &  & (0.5,0.5) & 94.3 & 94.1 & 94.8 & 94.5 & 98.8 & 93.1 \\ 
     &  & 0.25 & (0.5,0.05) & 95.3 & 94.5 & 97 & 94.3 & 98.6 & 93.3 \\ 
     &  &  & (0.5,0.5) & 94.5 & 94.4 & 94.7 & 94.9 & 99.1 & 93.5 \\ 
    300 & low & -0.5 & (0.5,0.05) & 95.2 & 94.7 & 97 & 94.8 & 98.3 & 93.8 \\ 
     &  &  & (0.5,0.5) & 95 & 94.8 & 95.5 & 94.7 & 98.6 & 93.1 \\ 
     &  & -0.25 & (0.5,0.05) & 95.7 & 94.9 & 96.8 & 95.2 & 98.5 & 94.2 \\ 
     &  &  & (0.5,0.5) & 94.8 & 94.5 & 95 & 94.9 & 98.7 & 93.6 \\ 
     &  & 0.25 & (0.5,0.05) & 95.6 & 95.2 & 96.6 & 94.5 & 98.3 & 93.8 \\ 
     &  &  & (0.5,0.5) & 94.2 & 94.2 & 94.5 & 94.5 & 98.5 & 93.2 \\ 
     & high & -0.5 & (0.5,0.05) & 95.2 & 94.5 & 96.8 & 95.1 & 98.5 & 93.6 \\ 
     &  &  & (0.5,0.5) & 94.2 & 93.9 & 94.7 & 94.6 & 98.8 & 92.5 \\ 
     &  & -0.25 & (0.5,0.05) & 94.7 & 94.2 & 96.2 & 94.3 & 98.4 & 93.2 \\ 
     &  &  & (0.5,0.5) & 94.8 & 94.7 & 95.2 & 94.8 & 98.8 & 93.5 \\ 
     &  & 0.25 & (0.5,0.05) & 94.6 & 94.2 & 96 & 93.7 & 98 & 92.8 \\ 
     &  &  & (0.5,0.5) & 94 & 93.8 & 94.2 & 94.3 & 98.8 & 93.2 \\ 
   \hline
\end{tabular}
\caption{\textit{Simulated coverage probabilities (in \%) of various 95\% confidence bands for the cumulative incidence function given an individual without pneumonia at time of hospital admission (univariate) for $\mathcal{N}(0,1)$ multiplier distribution.}} \label{tab:Norm_uni_0}
\end{table}

\begin{table}[ht]
\centering
\begin{tabular}{llllllllll}
  \hline
  \hline
  &  &  &  & $q^*_0$ & $q^*_1$ & $q^*_2$ & $\tilde{q}^*_0$ & $\tilde{q}^*_1$ & $\tilde{q}^*_2$ \\ 
     &  &  &  &  &  &  &  &  &  \\ 
     n & cens. & ${\beta}_0$ & $(\alpha_{010},\alpha_{020})$ &  &  & &  &  &  \\ 
   100 & low & -0.5 & (0.5,0.05) & 94.1 & 95.9 & 95.5 & 91.2 & 97.6 & 97.2 \\ 
     &  &  & (0.5,0.5) & 94.6 & 95.4 & 94.4 & 90.8 & 97.8 & 96.9 \\ 
     &  & -0.25 & (0.5,0.05) & 93.5 & 95.2 & 94.9 & 90.3 & 97 & 96.7 \\ 
     &  &  & (0.5,0.5) & 94.5 & 95.6 & 94 & 90.8 & 98.1 & 97 \\ 
     &  & 0.25 & (0.5,0.05) & 93.1 & 95.2 & 94.3 & 91 & 97.5 & 97 \\ 
     &  &  & (0.5,0.5) & 94.1 & 95 & 93.6 & 91.1 & 98.1 & 96.9 \\ 
     & high & -0.5 & (0.5,0.05) & 93.8 & 95.7 & 94.3 & 90.8 & 97.8 & 97.1 \\ 
     &  &  & (0.5,0.5) & 95.5 & 96.1 & 95.2 & 91.6 & 97.9 & 97.2 \\ 
     &  & -0.25 & (0.5,0.05) & 93.8 & 95.6 & 94.6 & 91 & 98.1 & 97.3 \\ 
     &  &  & (0.5,0.5) & 94.8 & 95.7 & 94.3 & 90.7 & 98.2 & 96.7 \\ 
     &  & 0.25 & (0.5,0.05) & 93.2 & 95.1 & 94.3 & 90.3 & 97.8 & 96.8 \\ 
     &  &  & (0.5,0.5) & 94.3 & 95.3 & 93.8 & 90.7 & 98.1 & 97.1 \\ 
    200 & low & -0.5 & (0.5,0.05) & 94.7 & 95.8 & 95.2 & 93.3 & 96.4 & 97.6 \\ 
     &  &  & (0.5,0.5) & 94.8 & 95.5 & 94.3 & 92.7 & 97.8 & 97.5 \\ 
     &  & -0.25 & (0.5,0.05) & 94.6 & 95.8 & 95 & 93.6 & 96.5 & 97.6 \\ 
     &  &  & (0.5,0.5) & 93.9 & 94.6 & 93.4 & 91.9 & 97.3 & 97.1 \\ 
     &  & 0.25 & (0.5,0.05) & 95 & 95.6 & 95.3 & 93.7 & 96.6 & 97.6 \\ 
     &  &  & (0.5,0.5) & 94.9 & 95.2 & 94.5 & 92.2 & 97.2 & 97.2 \\ 
     & high & -0.5 & (0.5,0.05) & 94.1 & 95.5 & 94.3 & 92.9 & 96.8 & 97.4 \\ 
     &  &  & (0.5,0.5) & 94.9 & 95.7 & 94.4 & 91.4 & 97.2 & 96.8 \\ 
     &  & -0.25 & (0.5,0.05) & 94.1 & 95.5 & 94.3 & 93.2 & 97 & 97.6 \\ 
     &  &  & (0.5,0.5) & 94.2 & 94.7 & 93.7 & 91.7 & 97.5 & 97.1 \\ 
     &  & 0.25 & (0.5,0.05) & 94.3 & 95.3 & 94.4 & 92.9 & 96.9 & 97.3 \\ 
     &  &  & (0.5,0.5) & 94.4 & 94.8 & 93.9 & 91.7 & 98 & 97.5 \\ 
    300 & low & -0.5 & (0.5,0.05) & 94.5 & 95.5 & 95 & 94 & 95.9 & 97.5 \\ 
     &  &  & (0.5,0.5) & 94.9 & 95.6 & 94.4 & 92.9 & 96.5 & 97.3 \\ 
     &  & -0.25 & (0.5,0.05) & 95 & 95.5 & 95.3 & 94.2 & 96.2 & 97.6 \\ 
     &  &  & (0.5,0.5) & 94.4 & 95.1 & 94.2 & 93.5 & 96.9 & 97.4 \\ 
     &  & 0.25 & (0.5,0.05) & 95 & 95.5 & 95.2 & 94 & 96.1 & 97.2 \\ 
     &  &  & (0.5,0.5) & 94.1 & 94.6 & 93.9 & 93 & 96.8 & 97.5 \\ 
     & high & -0.5 & (0.5,0.05) & 94.4 & 95.5 & 94.4 & 93.7 & 96.4 & 97.6 \\ 
     &  &  & (0.5,0.5) & 93.8 & 94.7 & 93.4 & 92.2 & 96.7 & 97.1 \\ 
     &  & -0.25 & (0.5,0.05) & 94.2 & 95 & 94.2 & 93.3 & 95.8 & 97.5 \\ 
     &  &  & (0.5,0.5) & 94.8 & 95.3 & 94.6 & 93 & 97.2 & 97.5 \\ 
     &  & 0.25 & (0.5,0.05) & 94.2 & 94.8 & 94 & 92.7 & 95.6 & 97 \\ 
     &  &  & (0.5,0.5) & 94.1 & 94.6 & 93.7 & 92.2 & 97 & 97.5 \\ 
   \hline
\end{tabular}
\caption{\textit{Simulated coverage probabilities (in \%) of various 95\% confidence bands for the cumulative incidence function given an individual without pneumonia at time of hospital admission (univariate) for centered $\text{Exp}(1)$ multiplier distribution.}} \label{tab:Exp_uni_0}
\end{table}

\begin{table}[ht]
\centering
\begin{tabular}{llllllllll}
  \hline
  \hline
  &  &  &  & $q^*_0$ & $q^*_1$ & $q^*_2$ & $\tilde{q}^*_0$ & $\tilde{q}^*_1$ & $\tilde{q}^*_2$ \\ 
     &  &  &  &  &  &  &  &  &  \\ 
     n & cens. & ${\beta}_0$ & $(\alpha_{010},\alpha_{020})$ &  &  & &  &  &  \\ 
  100 & low & -0.5 & (0.5,0.05) & 94.8 & 94.9 & 97 & 91.8 & 98.3 & 95 \\ 
     &  &  & (0.5,0.5) & 94.6 & 94.9 & 95.1 & 93.1 & 98.5 & 95.3 \\ 
     &  & -0.25 & (0.5,0.05) & 94 & 94.1 & 96.6 & 91 & 98 & 94.1 \\ 
     &  &  & (0.5,0.5) & 94.3 & 94.7 & 94.6 & 93.2 & 98.7 & 95.5 \\ 
     &  & 0.25 & (0.5,0.05) & 93.6 & 93.9 & 96.3 & 91.8 & 98.5 & 94.8 \\ 
     &  &  & (0.5,0.5) & 93.8 & 94.2 & 94.2 & 92.8 & 98.8 & 95.4 \\ 
     & high & -0.5 & (0.5,0.05) & 94.3 & 94.5 & 96.9 & 92.3 & 98.6 & 94.9 \\ 
     &  &  & (0.5,0.5) & 95.6 & 95.7 & 96 & 93.9 & 98.4 & 96.4 \\ 
     &  & -0.25 & (0.5,0.05) & 94.2 & 94.4 & 96.5 & 92.2 & 98.7 & 95 \\ 
     &  &  & (0.5,0.5) & 94.5 & 94.8 & 94.8 & 93.4 & 98.7 & 95.7 \\ 
     &  & 0.25 & (0.5,0.05) & 93.6 & 94 & 95.8 & 91.8 & 98.6 & 94.8 \\ 
     &  &  & (0.5,0.5) & 94.1 & 94.5 & 94.2 & 92.7 & 98.6 & 95.7 \\ 
    200 & low & -0.5 & (0.5,0.05) & 94.7 & 94.9 & 96.2 & 92.8 & 97.5 & 95.2 \\ 
     &  &  & (0.5,0.5) & 94.6 & 94.8 & 94.7 & 93.1 & 98.5 & 95.9 \\ 
     &  & -0.25 & (0.5,0.05) & 94.8 & 94.9 & 96.2 & 93.2 & 97.6 & 95.5 \\ 
     &  &  & (0.5,0.5) & 93.4 & 93.8 & 93.6 & 92.4 & 98 & 95 \\ 
     &  & 0.25 & (0.5,0.05) & 95 & 95.1 & 96 & 93.5 & 97.8 & 95.7 \\ 
     &  &  & (0.5,0.5) & 94.5 & 94.8 & 94.5 & 92.5 & 98 & 95.1 \\ 
     & high & -0.5 & (0.5,0.05) & 94.4 & 94.5 & 95.5 & 92.6 & 97.7 & 94.9 \\ 
     &  &  & (0.5,0.5) & 94.7 & 94.8 & 94.7 & 92.3 & 98.2 & 94.8 \\ 
     &  & -0.25 & (0.5,0.05) & 94.4 & 94.7 & 95.8 & 93.2 & 98.1 & 95.5 \\ 
     &  &  & (0.5,0.5) & 93.7 & 93.9 & 93.7 & 92.5 & 98.4 & 94.8 \\ 
     &  & 0.25 & (0.5,0.05) & 94.5 & 94.7 & 95.6 & 92.8 & 97.8 & 95.1 \\ 
     &  &  & (0.5,0.5) & 94.1 & 94.3 & 93.9 & 92.9 & 98.6 & 95.4 \\ 
    300 & low & -0.5 & (0.5,0.05) & 94.7 & 94.8 & 95.8 & 93.4 & 97 & 95.3 \\ 
     &  &  & (0.5,0.5) & 94.5 & 94.7 & 94.5 & 92.8 & 97.8 & 94.9 \\ 
     &  & -0.25 & (0.5,0.05) & 95 & 95.2 & 95.8 & 93.7 & 97.3 & 95.9 \\ 
     &  &  & (0.5,0.5) & 94.2 & 94.5 & 94.1 & 93.3 & 97.8 & 95.5 \\ 
     &  & 0.25 & (0.5,0.05) & 95.2 & 95.3 & 95.8 & 93.7 & 97.2 & 95.6 \\ 
     &  &  & (0.5,0.5) & 94 & 94.1 & 94 & 92.8 & 97.8 & 95.3 \\ 
     & high & -0.5 & (0.5,0.05) & 94.4 & 94.7 & 95.3 & 93.1 & 97.7 & 95.5 \\ 
     &  &  & (0.5,0.5) & 93.6 & 93.8 & 93.5 & 92.3 & 97.8 & 94.7 \\ 
     &  & -0.25 & (0.5,0.05) & 94.2 & 94.3 & 95.1 & 93 & 97.2 & 95 \\ 
     &  &  & (0.5,0.5) & 94.6 & 94.8 & 94.5 & 93.1 & 98 & 95.1 \\ 
     &  & 0.25 & (0.5,0.05) & 94.2 & 94.4 & 94.9 & 92.2 & 96.8 & 94.7 \\ 
     &  &  & (0.5,0.5) & 93.7 & 93.9 & 93.8 & 92.5 & 98.2 & 95.1 \\ 
   \hline
\end{tabular}
\caption{\textit{Simulated coverage probabilities (in \%) of various 95\% confidence bands for the cumulative incidence function given an individual without pneumonia at time of hospital admission (univariate) for centered $\text{Pois}(1)$ multiplier distribution.}} \label{tab:Poi_uni_0}
\end{table}

\begin{table}[ht]
\centering
\begin{tabular}{llllllllll}
  \hline
  \hline
  &  &  &  & $q^*_0$ & $q^*_1$ & $q^*_2$ & $\tilde{q}^*_0$ & $\tilde{q}^*_1$ & $\tilde{q}^*_2$ \\ 
     &  &  &  &  &  &  &  &  &  \\ 
     n & cens. & ${\beta}_0$ & $(\alpha_{010},\alpha_{020})$ &  &  & &  &  &  \\ 
    100 & low & -0.5 & (0.5,0.05) & 92.9 & 93.2 & 94.9 & 94.1 & 95.8 & 95.1 \\ 
     &  &  & (0.5,0.5) & 97.4 & 97.6 & 97.3 & 95.8 & 96.9 & 96.1 \\ 
     &  & -0.25 & (0.5,0.05) & 94 & 94.5 & 94.6 & 93.4 & 95.1 & 94.1 \\ 
     &  &  & (0.5,0.5) & 96.4 & 95.4 & 96.6 & 95.8 & 97.1 & 95.9 \\ 
     &  & 0.25 & (0.5,0.05) & 93.7 & 94 & 94.6 & 93.6 & 95.8 & 94.4 \\ 
     &  &  & (0.5,0.5) & 93.4 & 93.6 & 95.7 & 94.7 & 96.7 & 95 \\ 
     & high & -0.5 & (0.5,0.05) & 95.3 & 95 & 96.3 & 94.8 & 96.4 & 95.3 \\ 
     &  &  & (0.5,0.5) & 97.3 & 98.1 & 97.7 & 95.2 & 96.7 & 95.4 \\ 
     &  & -0.25 & (0.5,0.05) & 93.5 & 93.8 & 95.6 & 94.7 & 96.4 & 95.3 \\ 
     &  &  & (0.5,0.5) & 97.3 & 97.1 & 97.4 & 96.3 & 97.6 & 96.4 \\ 
     &  & 0.25 & (0.5,0.05) & 93.4 & 93.8 & 94.6 & 93.3 & 95.5 & 93.9 \\ 
     &  &  & (0.5,0.5) & 94.1 & 93.8 & 96.4 & 95.1 & 96.9 & 95.4 \\ 
    200 & low & -0.5 & (0.5,0.05) & 94.2 & 94.9 & 94.3 & 94.5 & 95.5 & 95.3 \\ 
     &  &  & (0.5,0.5) & 96.2 & 94.5 & 97.5 & 94.7 & 96.1 & 95.3 \\ 
     &  & -0.25 & (0.5,0.05) & 93.6 & 93.9 & 93.4 & 93.8 & 95.3 & 94.7 \\ 
     &  &  & (0.5,0.5) & 94.2 & 93.9 & 96.2 & 94.2 & 95.9 & 95 \\ 
     &  & 0.25 & (0.5,0.05) & 94.8 & 95 & 95.3 & 93.6 & 95 & 94.1 \\ 
     &  &  & (0.5,0.5) & 93.4 & 93.8 & 94.1 & 94 & 96 & 94.6 \\ 
     & high & -0.5 & (0.5,0.05) & 93.1 & 93.5 & 93.9 & 93.6 & 95.2 & 94.7 \\ 
     &  &  & (0.5,0.5) & 97.3 & 94.4 & 98.1 & 95 & 96.5 & 95.6 \\ 
     &  & -0.25 & (0.5,0.05) & 93.1 & 93.6 & 93.2 & 93.6 & 94.9 & 94.4 \\ 
     &  &  & (0.5,0.5) & 95.4 & 93.8 & 97.6 & 94.9 & 96.4 & 95.5 \\ 
     &  & 0.25 & (0.5,0.05) & 94.1 & 94.6 & 94.3 & 94 & 96 & 94.9 \\ 
     &  &  & (0.5,0.5) & 93.4 & 93.7 & 95 & 93.8 & 96.3 & 94.3 \\ 
    300 & low & -0.5 & (0.5,0.05) & 94.8 & 95 & 94.7 & 94.3 & 95.1 & 94.8 \\ 
     &  &  & (0.5,0.5) & 95 & 94.3 & 97 & 93.8 & 95.1 & 94.8 \\ 
     &  & -0.25 & (0.5,0.05) & 94.3 & 94.5 & 94.2 & 93.9 & 94.9 & 94.7 \\ 
     &  &  & (0.5,0.5) & 94.4 & 94.4 & 96 & 94 & 95.5 & 94.8 \\ 
     &  & 0.25 & (0.5,0.05) & 94.9 & 95 & 95.2 & 94 & 95.4 & 94.7 \\ 
     &  &  & (0.5,0.5) & 93.1 & 93.7 & 93.1 & 93.5 & 95.5 & 94.1 \\ 
     & high & -0.5 & (0.5,0.05) & 94.3 & 94.6 & 94.2 & 93.8 & 94.9 & 94.8 \\ 
     &  &  & (0.5,0.5) & 96.1 & 94.6 & 98.4 & 95.2 & 96.3 & 95.8 \\ 
     &  & -0.25 & (0.5,0.05) & 94 & 94.3 & 93.8 & 93.6 & 94.8 & 94.2 \\ 
     &  &  & (0.5,0.5) & 94.2 & 93.6 & 96.6 & 93.6 & 95.2 & 94.3 \\ 
     &  & 0.25 & (0.5,0.05) & 94.3 & 94.5 & 94.4 & 93.8 & 95.2 & 94.3 \\ 
     &  &  & (0.5,0.5) & 93.2 & 93.5 & 93.7 & 93.8 & 95.6 & 94.3 \\ 
   \hline
\end{tabular}
\caption{\textit{Simulated coverage probabilities (in \%) of various 95\% confidence bands for the cumulative incidence function given an individual with pneumonia at time of hospital admission (univariate) for $\mathcal{N}(0,1)$ multiplier distribution.}} \label{tab:Norm_uni_1}
\end{table}

\begin{table}[ht]
\centering
\begin{tabular}{llllllllll}
  \hline
  \hline
  &  &  &  & $q^*_0$ & $q^*_1$ & $q^*_2$ & $\tilde{q}^*_0$ & $\tilde{q}^*_1$ & $\tilde{q}^*_2$ \\ 
     &  &  &  &  &  &  &  &  &  \\ 
     n & cens. & ${\beta}_0$ & $(\alpha_{010},\alpha_{020})$ &  &  & &  &  &  \\ 
    100 & low & -0.5 & (0.5,0.05) & 95.3 & 98.3 & 89.8 & 92.4 & 94.2 & 94.4 \\ 
     &  &  & (0.5,0.5) & 97 & 99.1 & 94 & 93 & 95.6 & 94.8 \\ 
     &  & -0.25 & (0.5,0.05) & 96.5 & 98.2 & 92.7 & 91.3 & 93.5 & 93.7 \\ 
     &  &  & (0.5,0.5) & 93.9 & 97.6 & 93.4 & 92.8 & 95.9 & 94.7 \\ 
     &  & 0.25 & (0.5,0.05) & 97.3 & 98.3 & 95.8 & 91.2 & 93.9 & 94.3 \\ 
     &  &  & (0.5,0.5) & 92.9 & 96.9 & 90.1 & 91 & 94.7 & 93.7 \\ 
     & high & -0.5 & (0.5,0.05) & 94.1 & 98.3 & 92 & 92.3 & 94.2 & 94.2 \\ 
     &  &  & (0.5,0.5) & 97.9 & 99.5 & 94.3 & 92.8 & 95.4 & 94.1 \\ 
     &  & -0.25 & (0.5,0.05) & 94.5 & 98.2 & 89.6 & 91.9 & 94.2 & 94.2 \\ 
     &  &  & (0.5,0.5) & 96.9 & 98.4 & 93.9 & 92.9 & 96.1 & 94.7 \\ 
     &  & 0.25 & (0.5,0.05) & 96.3 & 98.2 & 93.8 & 90.3 & 93.6 & 93.2 \\ 
     &  &  & (0.5,0.5) & 91.6 & 96.7 & 91.8 & 91.3 & 95.3 & 93.9 \\ 
    200 & low & -0.5 & (0.5,0.05) & 97.6 & 98.8 & 94.6 & 94.3 & 95.3 & 96 \\ 
     &  &  & (0.5,0.5) & 93.3 & 98.4 & 90.8 & 92.9 & 94.4 & 94.9 \\ 
     &  & -0.25 & (0.5,0.05) & 97.1 & 98.2 & 95.6 & 93.4 & 94.6 & 95.5 \\ 
     &  &  & (0.5,0.5) & 94.3 & 98.1 & 89 & 92.5 & 94.3 & 94.9 \\ 
     &  & 0.25 & (0.5,0.05) & 97.5 & 97.9 & 97.3 & 92.7 & 93.9 & 94.8 \\ 
     &  &  & (0.5,0.5) & 95.4 & 97.9 & 91 & 92.2 & 94.3 & 94.9 \\ 
     & high & -0.5 & (0.5,0.05) & 96.1 & 98.3 & 90.2 & 93.1 & 94.5 & 95.2 \\ 
     &  &  & (0.5,0.5) & 92.1 & 97.4 & 92.6 & 92.7 & 94.6 & 94.6 \\ 
     &  & -0.25 & (0.5,0.05) & 96.2 & 98.1 & 93 & 92.9 & 94.1 & 94.6 \\ 
     &  &  & (0.5,0.5) & 92.1 & 97.6 & 90.3 & 92.1 & 94.2 & 94.5 \\ 
     &  & 0.25 & (0.5,0.05) & 97.5 & 98.1 & 96.6 & 92.6 & 94.8 & 95.6 \\ 
     &  &  & (0.5,0.5) & 94.2 & 98 & 89 & 91.7 & 94.3 & 94.3 \\ 
    300 & low & -0.5 & (0.5,0.05) & 97.3 & 98.3 & 95.7 & 94.2 & 94.9 & 95.7 \\ 
     &  &  & (0.5,0.5) & 95.2 & 98.8 & 89.5 & 93.5 & 94.3 & 95 \\ 
     &  & -0.25 & (0.5,0.05) & 97.2 & 97.8 & 96.2 & 94.1 & 94.5 & 95.3 \\ 
     &  &  & (0.5,0.5) & 95.9 & 98.8 & 90.4 & 93.6 & 94.8 & 95.3 \\ 
     &  & 0.25 & (0.5,0.05) & 97.1 & 97.3 & 97 & 93.8 & 94.8 & 95.7 \\ 
     &  &  & (0.5,0.5) & 95.4 & 97.7 & 92.5 & 92.7 & 94.2 & 95 \\ 
     & high & -0.5 & (0.5,0.05) & 96.8 & 98.6 & 93.5 & 93.8 & 94.7 & 95.6 \\ 
     &  &  & (0.5,0.5) & 93.9 & 98.4 & 90 & 93.9 & 94.8 & 95.6 \\ 
     &  & -0.25 & (0.5,0.05) & 97 & 98.1 & 95 & 93.4 & 94.2 & 95.1 \\ 
     &  &  & (0.5,0.5) & 94.1 & 98.3 & 88.8 & 92.5 & 93.9 & 94.5 \\ 
     &  & 0.25 & (0.5,0.05) & 96.6 & 97.1 & 96.2 & 93.4 & 94.3 & 95.2 \\ 
     &  &  & (0.5,0.5) & 95 & 97.8 & 90.8 & 92 & 94.3 & 95.1 \\ 
   \hline
\end{tabular}
\caption{\textit{Simulated coverage probabilities (in \%) of various 95\% confidence bands for the cumulative incidence function given an individual with pneumonia at time of hospital admission (univariate) for centered $\text{Exp}(1)$ multiplier distribution.}} \label{tab:Exp_uni_1}
\end{table}

\begin{table}[ht]
\centering
\begin{tabular}{llllllllll}
  \hline
  \hline
  &  &  &  & $q^*_0$ & $q^*_1$ & $q^*_2$ & $\tilde{q}^*_0$ & $\tilde{q}^*_1$ & $\tilde{q}^*_2$ \\ 
     &  &  &  &  &  &  &  &  &  \\ 
     n & cens. & ${\beta}_0$ & $(\alpha_{010},\alpha_{020})$ &  &  & &  &  &  \\ 
   100 & low & -0.5 & (0.5,0.05) & 92.9 & 95.6 & 91.3 & 93.2 & 94.8 & 95.1 \\ 
     &  &  & (0.5,0.5) & 96.8 & 98.1 & 95.5 & 94.7 & 96.2 & 95.8 \\ 
     &  & -0.25 & (0.5,0.05) & 94.7 & 96.4 & 92.2 & 92.4 & 94.2 & 93.9 \\ 
     &  &  & (0.5,0.5) & 95.1 & 96.1 & 95.3 & 94.4 & 96.4 & 95.6 \\ 
     &  & 0.25 & (0.5,0.05) & 95.1 & 96.5 & 94.6 & 92.3 & 94.9 & 94.5 \\ 
     &  &  & (0.5,0.5) & 91.9 & 94.9 & 93.1 & 92.7 & 95.9 & 94.7 \\ 
     & high & -0.5 & (0.5,0.05) & 94 & 95.9 & 95.2 & 93.6 & 95.5 & 95 \\ 
     &  &  & (0.5,0.5) & 97 & 98.8 & 96 & 94.2 & 96.1 & 95 \\ 
     &  & -0.25 & (0.5,0.05) & 92.9 & 95.7 & 92.2 & 93.3 & 95.3 & 95.1 \\ 
     &  &  & (0.5,0.5) & 96.5 & 97.9 & 95.6 & 94.7 & 96.9 & 95.9 \\ 
     &  & 0.25 & (0.5,0.05) & 94.4 & 96 & 93.1 & 91.5 & 94.5 & 93.8 \\ 
     &  &  & (0.5,0.5) & 92.4 & 94.6 & 94.5 & 93.3 & 96.3 & 95.2 \\ 
    200 & low & -0.5 & (0.5,0.05) & 95.1 & 96.7 & 93.2 & 94 & 95.3 & 95.7 \\ 
     &  &  & (0.5,0.5) & 93 & 95.9 & 95.2 & 93.7 & 95.1 & 95.2 \\ 
     &  & -0.25 & (0.5,0.05) & 94.8 & 95.9 & 93.8 & 93.4 & 94.7 & 95.2 \\ 
     &  &  & (0.5,0.5) & 92.5 & 96.1 & 92.4 & 93.2 & 95 & 95.1 \\ 
     &  & 0.25 & (0.5,0.05) & 95.9 & 96.2 & 95.8 & 92.9 & 94.4 & 94.8 \\ 
     &  &  & (0.5,0.5) & 93.8 & 95.9 & 91.3 & 92.8 & 95 & 94.9 \\ 
     & high & -0.5 & (0.5,0.05) & 93.2 & 96 & 90.2 & 93.1 & 94.6 & 95.1 \\ 
     &  &  & (0.5,0.5) & 94.6 & 95.1 & 96.3 & 94 & 95.6 & 95.6 \\ 
     &  & -0.25 & (0.5,0.05) & 93.8 & 95.5 & 91.8 & 93.1 & 94.4 & 94.7 \\ 
     &  &  & (0.5,0.5) & 92.4 & 94.9 & 94.9 & 93.3 & 95.2 & 95.1 \\ 
     &  & 0.25 & (0.5,0.05) & 95.5 & 96.1 & 94.8 & 93.2 & 95.4 & 95.4 \\ 
     &  &  & (0.5,0.5) & 92.8 & 95.9 & 91.3 & 92.7 & 95.3 & 94.6 \\ 
    300 & low & -0.5 & (0.5,0.05) & 95.3 & 96.3 & 94.5 & 94.2 & 94.9 & 95.4 \\ 
     &  &  & (0.5,0.5) & 93.6 & 96.9 & 92.5 & 93.2 & 94.7 & 95 \\ 
     &  & -0.25 & (0.5,0.05) & 95.2 & 95.9 & 94.5 & 93.8 & 94.6 & 95.1 \\ 
     &  &  & (0.5,0.5) & 93.9 & 96.7 & 91.6 & 93.7 & 95 & 95.3 \\ 
     &  & 0.25 & (0.5,0.05) & 95.7 & 95.8 & 95.8 & 93.6 & 95 & 95.2 \\ 
     &  &  & (0.5,0.5) & 93.5 & 95.5 & 91.5 & 92.8 & 94.9 & 94.6 \\ 
     & high & -0.5 & (0.5,0.05) & 94.7 & 96.3 & 92.3 & 93.6 & 94.7 & 95.2 \\ 
     &  &  & (0.5,0.5) & 92.3 & 96.4 & 94.3 & 94.3 & 95.5 & 96 \\ 
     &  & -0.25 & (0.5,0.05) & 94.8 & 96 & 93.3 & 93.5 & 94.5 & 95 \\ 
     &  &  & (0.5,0.5) & 92.4 & 96.5 & 91.9 & 93 & 94.5 & 94.6 \\ 
     &  & 0.25 & (0.5,0.05) & 95.2 & 95.6 & 94.9 & 93.2 & 94.5 & 94.8 \\ 
     &  &  & (0.5,0.5) & 93.4 & 95.4 & 91.1 & 92.7 & 95 & 94.8 \\ 
   \hline
\end{tabular}
\caption{\textit{Simulated coverage probabilities (in \%) of various 95\% confidence bands for the cumulative incidence function given an individual with pneumonia at time of hospital admission (univariate) for centered $\text{Pois}(1)$ multiplier distribution.}} \label{tab:Poi_uni_1}
\end{table}

\begin{table}[ht]
\centering
\begin{tabular}{llllllllll}
  \hline
  \hline
  &  &  &  & $q^*_0$ & $q^*_1$ & $q^*_2$ & $\tilde{q}^*_0$ & $\tilde{q}^*_1$ & $\tilde{q}^*_2$ \\ 
     &  &  &  &  &  &  &  &  &  \\ 
     n & cens. & ${\beta}_0$ & $(\alpha_{010},\alpha_{020})$ &  &  & &  &  &  \\ 
   100 & low & (-0.05,-0.5,-0.05) & (0.08,0.008) & 95.6 & 95.2 & 98 & 93.4 & 98.6 & 97.7 \\ 
     &  &  & (0.05,0.05) & 94.2 & 94.6 & 95.1 & 95.9 & 98.9 & 98.9 \\ 
     &  & (-0.05,-0.25,-0.05) & (0.08,0.008) & 95.4 & 95.1 & 97.7 & 92.9 & 98.5 & 97.4 \\ 
     &  &  & (0.05,0.05) & 94.6 & 95.1 & 95.5 & 95 & 98.5 & 98.4 \\ 
     &  & (-0.05,0.25,-0.05) & (0.08,0.008) & 95.4 & 95.1 & 97.5 & 93.4 & 98.6 & 97.5 \\ 
     &  &  & (0.05,0.05) & 93.4 & 94.2 & 94.5 & 94.1 & 98.7 & 98.5 \\ 
     & high & (-0.05,-0.5,-0.05) & (0.08,0.008) & 94.8 & 94.4 & 96.7 & 93.3 & 98.5 & 98.1 \\ 
     &  &  & (0.05,0.05) & 94.1 & 95 & 95.1 & 96.1 & 98.6 & 99.4 \\ 
     &  & (-0.05,-0.25,-0.05) & (0.08,0.008) & 94.4 & 94.4 & 96.1 & 93.5 & 98.6 & 98.3 \\ 
     &  &  & (0.05,0.05) & 94.1 & 94.6 & 95.4 & 95.8 & 98.5 & 98.9 \\ 
     &  & (-0.05,0.25,-0.05) & (0.08,0.008) & 94.3 & 94.5 & 96 & 93.9 & 98.6 & 98.2 \\ 
     &  &  & (0.05,0.05) & 93.6 & 94.4 & 94.6 & 95.6 & 98.9 & 99.1 \\ 
    200 & low & (-0.05,-0.5,-0.05) & (0.08,0.008) & 95.3 & 95.1 & 97.1 & 93.2 & 97.7 & 95.7 \\ 
     &  &  & (0.05,0.05) & 94.5 & 94.8 & 95 & 94.2 & 98.1 & 97.3 \\ 
     &  & (-0.05,-0.25,-0.05) & (0.08,0.008) & 94.9 & 94.4 & 97 & 92.9 & 97.8 & 95.7 \\ 
     &  &  & (0.05,0.05) & 93.8 & 94.2 & 94.1 & 93.5 & 97.9 & 96.7 \\ 
     &  & (-0.05,0.25,-0.05) & (0.08,0.008) & 94.8 & 94.6 & 96.7 & 93.6 & 98 & 96.2 \\ 
     &  &  & (0.05,0.05) & 93.7 & 94.3 & 94.2 & 93.3 & 98.1 & 97.1 \\ 
     & high & (-0.05,-0.5,-0.05) & (0.08,0.008) & 94.3 & 94.2 & 95.9 & 93.3 & 97.8 & 96.6 \\ 
     &  &  & (0.05,0.05) & 94.3 & 94.6 & 94.8 & 93.4 & 98.2 & 97.8 \\ 
     &  & (-0.05,-0.25,-0.05) & (0.08,0.008) & 93.4 & 93.6 & 95.4 & 92.9 & 97.6 & 96.2 \\ 
     &  &  & (0.05,0.05) & 93.5 & 94 & 93.9 & 93.2 & 98.1 & 97.6 \\ 
     &  & (-0.05,0.25,-0.05) & (0.08,0.008) & 94.2 & 94.4 & 95.4 & 93.1 & 97.6 & 96.3 \\ 
     &  &  & (0.05,0.05) & 93.9 & 94.2 & 94.1 & 93.6 & 98.4 & 97.9 \\ 
    300 & low & (-0.05,-0.5,-0.05) & (0.08,0.008) & 94.8 & 94.8 & 96.5 & 93.5 & 97.1 & 95.2 \\ 
     &  &  & (0.05,0.05) & 93.7 & 94 & 93.9 & 93.5 & 97.8 & 96.2 \\ 
     &  & (-0.05,-0.25,-0.05) & (0.08,0.008) & 95.3 & 95 & 96.6 & 94.1 & 97.6 & 95.6 \\ 
     &  &  & (0.05,0.05) & 94 & 94.2 & 94.3 & 93.8 & 97.8 & 96.2 \\ 
     &  & (-0.05,0.25,-0.05) & (0.08,0.008) & 95 & 94.7 & 96.3 & 93.7 & 97 & 95.5 \\ 
     &  &  & (0.05,0.05) & 94.3 & 94.4 & 94.5 & 93.7 & 97.8 & 96.3 \\ 
     & high & (-0.05,-0.5,-0.05) & (0.08,0.008) & 94.7 & 94.5 & 95.5 & 93.3 & 97.1 & 95.7 \\ 
     &  &  & (0.05,0.05) & 94.6 & 94.8 & 94.9 & 93.8 & 97.9 & 97 \\ 
     &  & (-0.05,-0.25,-0.05) & (0.08,0.008) & 94.6 & 94.6 & 95.7 & 93.8 & 97.3 & 96 \\ 
     &  &  & (0.05,0.05) & 94.5 & 94.7 & 94.5 & 93.8 & 97.7 & 97 \\ 
     &  & (-0.05,0.25,-0.05) & (0.08,0.008) & 94.2 & 94.4 & 95.1 & 93.6 & 97.2 & 95.9 \\ 
     &  &  & (0.05,0.05) & 94.9 & 95 & 95 & 94 & 98 & 96.9 \\ 
   \hline
\end{tabular}
\caption{\textit{Simulated coverage probabilities (in \%) of various 95\% confidence bands for the cumulative incidence function given a 45 years old female individual without pneumonia at time of hospital admission (trivariate) for $\mathcal{N}(0,1)$ multiplier distribution.}} \label{tab:Norm_tri_0}
\end{table}

\begin{table}[ht]
\centering
\begin{tabular}{llllllllll}
  \hline
  \hline
  &  &  &  & $q^*_0$ & $q^*_1$ & $q^*_2$ & $\tilde{q}^*_0$ & $\tilde{q}^*_1$ & $\tilde{q}^*_2$ \\ 
     &  &  &  &  &  &  &  &  &  \\ 
      n & cens. & ${\beta}_0$ & $(\alpha_{010},\alpha_{020})$ &  &  & &  &  &  \\ 
 100 & low & (-0.05,-0.5,-0.05) & (0.08,0.008) & 96.2 & 95.9 & 98.7 & 91.3 & 97.8 & 98.7 \\ 
     &  &  & (0.05,0.05) & 94.7 & 94.8 & 95.8 & 94.6 & 98.2 & 99.5 \\ 
     &  & (-0.05,-0.25,-0.05) & (0.08,0.008) & 95.8 & 95.6 & 98.2 & 90.8 & 97.2 & 98.6 \\ 
     &  &  & (0.05,0.05) & 95.2 & 95.5 & 96.2 & 93.3 & 97.9 & 99 \\ 
     &  & (-0.05,0.25,-0.05) & (0.08,0.008) & 96 & 96 & 98.3 & 91.3 & 97.6 & 98.8 \\ 
     &  &  & (0.05,0.05) & 94.2 & 94.8 & 95.5 & 92.2 & 98 & 99.1 \\ 
     & high & (-0.05,-0.5,-0.05) & (0.08,0.008) & 95.2 & 95 & 97.4 & 91.3 & 97.5 & 99 \\ 
     &  &  & (0.05,0.05) & 94.2 & 95 & 95.6 & 95.1 & 98.2 & 99.8 \\ 
     &  & (-0.05,-0.25,-0.05) & (0.08,0.008) & 95.3 & 95.1 & 97.2 & 91.4 & 98 & 98.9 \\ 
     &  &  & (0.05,0.05) & 94.3 & 95 & 95.8 & 94.9 & 98 & 99.3 \\ 
     &  & (-0.05,0.25,-0.05) & (0.08,0.008) & 95.2 & 95.2 & 97 & 92.1 & 97.9 & 98.8 \\ 
     &  &  & (0.05,0.05) & 93.9 & 94.7 & 95.2 & 93.9 & 98.5 & 99.4 \\ 
    200 & low & (-0.05,-0.5,-0.05) & (0.08,0.008) & 95.7 & 95.6 & 97.5 & 92.2 & 96.1 & 97.6 \\ 
     &  &  & (0.05,0.05) & 95.1 & 95.2 & 95.7 & 92.4 & 97.2 & 98.6 \\ 
     &  & (-0.05,-0.25,-0.05) & (0.08,0.008) & 95.4 & 95.4 & 97.4 & 92 & 96.1 & 97.8 \\ 
     &  &  & (0.05,0.05) & 94.7 & 94.8 & 95.1 & 92 & 96.5 & 98 \\ 
     &  & (-0.05,0.25,-0.05) & (0.08,0.008) & 95.4 & 95.3 & 97.2 & 92.5 & 96.6 & 97.9 \\ 
     &  &  & (0.05,0.05) & 94.9 & 95.1 & 95.4 & 91.9 & 96.9 & 98.1 \\ 
     & high & (-0.05,-0.5,-0.05) & (0.08,0.008) & 95 & 95.1 & 96.7 & 92.2 & 96.3 & 98.1 \\ 
     &  &  & (0.05,0.05) & 95.1 & 95.2 & 95.4 & 92 & 97 & 98.6 \\ 
     &  & (-0.05,-0.25,-0.05) & (0.08,0.008) & 94.2 & 94.4 & 96.1 & 91.6 & 96.1 & 97.7 \\ 
     &  &  & (0.05,0.05) & 94.3 & 94.5 & 94.6 & 91.9 & 96.9 & 98.5 \\ 
     &  & (-0.05,0.25,-0.05) & (0.08,0.008) & 94.9 & 95 & 96.2 & 92.5 & 96 & 97.8 \\ 
     &  &  & (0.05,0.05) & 94.6 & 94.9 & 94.9 & 92.2 & 97.4 & 98.7 \\ 
    300 & low & (-0.05,-0.5,-0.05) & (0.08,0.008) & 95.3 & 95.4 & 96.8 & 92.9 & 95.3 & 97.1 \\ 
     &  &  & (0.05,0.05) & 94.4 & 94.6 & 94.8 & 92.5 & 96.3 & 98.1 \\ 
     &  & (-0.05,-0.25,-0.05) & (0.08,0.008) & 95.6 & 95.6 & 96.9 & 93.6 & 95.6 & 97.8 \\ 
     &  &  & (0.05,0.05) & 95 & 95.1 & 95.3 & 92.8 & 96.3 & 98 \\ 
     &  & (-0.05,0.25,-0.05) & (0.08,0.008) & 95.4 & 95.4 & 96.5 & 93.5 & 95.6 & 97 \\ 
     &  &  & (0.05,0.05) & 95.1 & 95.2 & 95.3 & 92.9 & 96.5 & 98 \\ 
     & high & (-0.05,-0.5,-0.05) & (0.08,0.008) & 95.1 & 95.2 & 96 & 92.9 & 95.5 & 97.3 \\ 
     &  &  & (0.05,0.05) & 95.1 & 95.3 & 95.5 & 92.8 & 96.7 & 98.2 \\ 
     &  & (-0.05,-0.25,-0.05) & (0.08,0.008) & 95.1 & 95 & 96.1 & 93.2 & 95.6 & 97.5 \\ 
     &  &  & (0.05,0.05) & 95.2 & 95.2 & 95.3 & 92.4 & 96.6 & 98 \\ 
     &  & (-0.05,0.25,-0.05) & (0.08,0.008) & 94.9 & 95 & 95.8 & 93.2 & 95.8 & 97.2 \\ 
     &  &  & (0.05,0.05) & 95.6 & 95.7 & 95.8 & 93.3 & 96.8 & 98.1 \\ 
   \hline
\end{tabular}
\caption{\textit{Simulated coverage probabilities (in \%) of various 95\% confidence bands for the cumulative incidence function given a 45 years old female individual without pneumonia at time of hospital admission (trivariate) for centered $\text{Exp}(1)$ multiplier distribution.}} \label{tab:Exp_tri_0}
\end{table}

\begin{table}[ht]
\centering
\begin{tabular}{llllllllll}
  \hline
  \hline
  &  &  &  & $q^*_0$ & $q^*_1$ & $q^*_2$ & $\tilde{q}^*_0$ & $\tilde{q}^*_1$ & $\tilde{q}^*_2$ \\ 
     &  &  &  &  &  &  &  &  &  \\ 
     n & cens. & ${\beta}_0$ & $(\alpha_{010},\alpha_{020})$ &  &  & &  &  &  \\ 
 100 & low & (-0.05,-0.5,-0.05) & (0.08,0.008) & 95.8 & 95.3 & 98.2 & 92.6 & 98.2 & 98.4 \\ 
     &  &  & (0.05,0.05) & 94.5 & 94.8 & 95.5 & 96 & 98.6 & 99.4 \\ 
     &  & (-0.05,-0.25,-0.05) & (0.08,0.008) & 95.4 & 95 & 97.7 & 92 & 98 & 98.3 \\ 
     &  &  & (0.05,0.05) & 94.8 & 95.1 & 95.6 & 94.8 & 98.2 & 98.9 \\ 
     &  & (-0.05,0.25,-0.05) & (0.08,0.008) & 95.2 & 95.1 & 97.6 & 92.3 & 98.2 & 98.3 \\ 
     &  &  & (0.05,0.05) & 93.1 & 93.9 & 94.2 & 93.8 & 98.4 & 99 \\ 
     & high & (-0.05,-0.5,-0.05) & (0.08,0.008) & 94.6 & 94.4 & 97 & 92.8 & 98 & 98.8 \\ 
     &  &  & (0.05,0.05) & 94.1 & 95 & 95.3 & 96.7 & 98.5 & 99.8 \\ 
     &  & (-0.05,-0.25,-0.05) & (0.08,0.008) & 94.3 & 94.3 & 96.4 & 92.8 & 98.4 & 98.9 \\ 
     &  &  & (0.05,0.05) & 94.1 & 94.8 & 95.4 & 95.7 & 98.2 & 99.4 \\ 
     &  & (-0.05,0.25,-0.05) & (0.08,0.008) & 94.1 & 94.3 & 96 & 93.3 & 98.2 & 98.8 \\ 
     &  &  & (0.05,0.05) & 93.2 & 94.4 & 94.6 & 95.2 & 98.6 & 99.4 \\ 
    200 & low & (-0.05,-0.5,-0.05) & (0.08,0.008) & 95.1 & 95 & 97.1 & 92.5 & 97 & 97.1 \\ 
     &  &  & (0.05,0.05) & 94.6 & 94.7 & 94.9 & 93.3 & 97.9 & 98.2 \\ 
     &  & (-0.05,-0.25,-0.05) & (0.08,0.008) & 94.6 & 94.7 & 97 & 92.5 & 97.2 & 97.1 \\ 
     &  &  & (0.05,0.05) & 93.8 & 94.2 & 94.2 & 92.9 & 97.2 & 97.6 \\ 
     &  & (-0.05,0.25,-0.05) & (0.08,0.008) & 94.9 & 94.8 & 96.6 & 92.9 & 97.4 & 97.3 \\ 
     &  &  & (0.05,0.05) & 93.8 & 94.2 & 94.3 & 92.6 & 97.5 & 97.8 \\ 
     & high & (-0.05,-0.5,-0.05) & (0.08,0.008) & 94.4 & 94.3 & 96.1 & 92.9 & 97.3 & 97.6 \\ 
     &  &  & (0.05,0.05) & 94.6 & 94.8 & 95 & 93 & 97.8 & 98.5 \\ 
     &  & (-0.05,-0.25,-0.05) & (0.08,0.008) & 93.4 & 93.5 & 95.3 & 92.1 & 96.8 & 97.4 \\ 
     &  &  & (0.05,0.05) & 93.6 & 94 & 94.1 & 92.9 & 97.6 & 98.3 \\ 
     &  & (-0.05,0.25,-0.05) & (0.08,0.008) & 94.2 & 94.4 & 95.3 & 92.8 & 96.9 & 97.3 \\ 
     &  &  & (0.05,0.05) & 93.8 & 94.2 & 94.2 & 93.1 & 97.9 & 98.4 \\ 
    300 & low & (-0.05,-0.5,-0.05) & (0.08,0.008) & 95 & 94.9 & 96.5 & 93.2 & 96.3 & 96.4 \\ 
     &  &  & (0.05,0.05) & 93.7 & 94 & 94 & 92.9 & 97.3 & 97.3 \\ 
     &  & (-0.05,-0.25,-0.05) & (0.08,0.008) & 95.1 & 95.2 & 96.6 & 93.7 & 96.6 & 96.9 \\ 
     &  &  & (0.05,0.05) & 94 & 94.1 & 94.3 & 93.3 & 97 & 97.2 \\ 
     &  & (-0.05,0.25,-0.05) & (0.08,0.008) & 94.9 & 94.9 & 96.2 & 93.6 & 96.2 & 96.5 \\ 
     &  &  & (0.05,0.05) & 94.2 & 94.4 & 94.3 & 93.2 & 97.3 & 97.4 \\ 
     & high & (-0.05,-0.5,-0.05) & (0.08,0.008) & 94.5 & 94.6 & 95.5 & 93.2 & 96.4 & 96.7 \\ 
     &  &  & (0.05,0.05) & 94.7 & 94.8 & 94.8 & 93.5 & 97.4 & 98 \\ 
     &  & (-0.05,-0.25,-0.05) & (0.08,0.008) & 94.4 & 94.4 & 95.5 & 93.4 & 96.5 & 96.8 \\ 
     &  &  & (0.05,0.05) & 94.3 & 94.5 & 94.4 & 93 & 97.2 & 97.6 \\ 
     &  & (-0.05,0.25,-0.05) & (0.08,0.008) & 94 & 94.3 & 95.1 & 93.3 & 96.5 & 96.9 \\ 
     &  &  & (0.05,0.05) & 94.8 & 95.2 & 95.1 & 93.8 & 97.3 & 97.7 \\ 
   \hline
\end{tabular}
\caption{\textit{Simulated coverage probabilities (in \%) of various 95\% confidence bands for the cumulative incidence function given a 45 years old female individual without pneumonia at time of hospital admission (trivariate) for centered $\text{Pois}(1)$ multiplier distribution.}} \label{tab:Poi_tri_0}
\end{table}

\begin{table}[ht]
\centering
\begin{tabular}{llllllllll}
  \hline
  \hline
  &  &  &  & $q^*_0$ & $q^*_1$ & $q^*_2$ & $\tilde{q}^*_0$ & $\tilde{q}^*_1$ & $\tilde{q}^*_2$ \\ 
     &  &  &  &  &  &  &  &  &  \\ 
     n & cens. & ${\beta}_0$ & $(\alpha_{010},\alpha_{020})$ &  &  & &  &  &  \\ 
  100 & low & (-0.05,-0.5,-0.05) & (0.08,0.008) & 95.6 & 95.3 & 96.8 & 95.1 & 96.2 & 96.6 \\ 
     &  &  & (0.05,0.05) & 98.7 & 98.5 & 98.2 & 96.4 & 97.4 & 97.8 \\ 
     &  & (-0.05,-0.25,-0.05) & (0.08,0.008) & 94.1 & 95.1 & 96.4 & 94.2 & 95.2 & 95.8 \\ 
     &  &  & (0.05,0.05) & 97.3 & 96.9 & 97.2 & 95.9 & 96.8 & 97.1 \\ 
     &  & (-0.05,0.25,-0.05) & (0.08,0.008) & 94.5 & 95.6 & 95.3 & 93.7 & 94.8 & 95.2 \\ 
     &  &  & (0.05,0.05) & 95.5 & 95.6 & 97.7 & 95.2 & 96.5 & 97 \\ 
     & high & (-0.05,-0.5,-0.05) & (0.08,0.008) & 96.5 & 96.9 & 96.1 & 95.9 & 96.3 & 96.7 \\ 
     &  &  & (0.05,0.05) & 98.3 & 99 & 98.5 & 95.7 & 96.8 & 97.5 \\ 
     &  & (-0.05,-0.25,-0.05) & (0.08,0.008) & 95.6 & 95.8 & 97.1 & 95.4 & 96.4 & 96.9 \\ 
     &  &  & (0.05,0.05) & 98.1 & 98.2 & 98.3 & 96.5 & 97.5 & 98 \\ 
     &  & (-0.05,0.25,-0.05) & (0.08,0.008) & 94.3 & 95.4 & 96.7 & 94.6 & 95.7 & 96.1 \\ 
     &  &  & (0.05,0.05) & 96.3 & 96.2 & 97.9 & 95.7 & 96.8 & 97.7 \\ 
    200 & low & (-0.05,-0.5,-0.05) & (0.08,0.008) & 94.2 & 94.7 & 95.4 & 94.8 & 95.2 & 95.7 \\ 
     &  &  & (0.05,0.05) & 97.7 & 95.6 & 98.2 & 95.6 & 96.5 & 96.8 \\ 
     &  & (-0.05,-0.25,-0.05) & (0.08,0.008) & 93.3 & 94 & 93.4 & 93.5 & 94.4 & 94.8 \\ 
     &  &  & (0.05,0.05) & 96.2 & 95.1 & 97.2 & 94.7 & 95.7 & 96 \\ 
     &  & (-0.05,0.25,-0.05) & (0.08,0.008) & 95 & 95.7 & 94.8 & 93.5 & 94.3 & 94.6 \\ 
     &  &  & (0.05,0.05) & 93.9 & 94.7 & 95.4 & 93.9 & 95.1 & 95.7 \\ 
     & high & (-0.05,-0.5,-0.05) & (0.08,0.008) & 94.9 & 94.6 & 96 & 94.3 & 94.9 & 95.5 \\ 
     &  &  & (0.05,0.05) & 98.6 & 97.3 & 98.6 & 96 & 96.7 & 97 \\ 
     &  & (-0.05,-0.25,-0.05) & (0.08,0.008) & 93.9 & 94.5 & 95.2 & 93.8 & 94.3 & 95.1 \\ 
     &  &  & (0.05,0.05) & 97.4 & 95.5 & 97.9 & 95.9 & 96.7 & 97 \\ 
     &  & (-0.05,0.25,-0.05) & (0.08,0.008) & 94.5 & 95 & 94.4 & 94.2 & 94.7 & 95.6 \\ 
     &  &  & (0.05,0.05) & 94.3 & 94.4 & 95.8 & 94.3 & 95.4 & 96 \\ 
    300 & low & (-0.05,-0.5,-0.05) & (0.08,0.008) & 94.2 & 94.8 & 94.5 & 93.9 & 94.4 & 94.9 \\ 
     &  &  & (0.05,0.05) & 96 & 94.4 & 98.2 & 95.1 & 95.8 & 96.3 \\ 
     &  & (-0.05,-0.25,-0.05) & (0.08,0.008) & 93.9 & 94.4 & 93.5 & 93.7 & 94.4 & 94.8 \\ 
     &  &  & (0.05,0.05) & 94.8 & 94.5 & 97 & 94.2 & 95 & 95.6 \\ 
     &  & (-0.05,0.25,-0.05) & (0.08,0.008) & 94.4 & 95 & 94.4 & 94 & 94.6 & 95 \\ 
     &  &  & (0.05,0.05) & 94.3 & 94.9 & 94.4 & 93.8 & 94.9 & 95.3 \\ 
     & high & (-0.05,-0.5,-0.05) & (0.08,0.008) & 94.1 & 94.3 & 95 & 93.7 & 94.2 & 94.9 \\ 
     &  &  & (0.05,0.05) & 97.5 & 95.1 & 98.5 & 95.5 & 96 & 96.5 \\ 
     &  & (-0.05,-0.25,-0.05) & (0.08,0.008) & 93.4 & 94.1 & 93.8 & 93.8 & 94.2 & 94.9 \\ 
     &  &  & (0.05,0.05) & 95.8 & 94.6 & 97.8 & 94.8 & 95.6 & 96.3 \\ 
     &  & (-0.05,0.25,-0.05) & (0.08,0.008) & 94.7 & 95.3 & 94.2 & 94.2 & 94.6 & 95.3 \\ 
     &  &  & (0.05,0.05) & 93.7 & 94.3 & 94.9 & 94.2 & 95 & 95.5 \\ 
   \hline
\end{tabular}
\caption{\textit{Simulated coverage probabilities (in \%) of various 95\% confidence bands for the cumulative incidence function given a 70 years old male individual with pneumonia at time of hospital admission (trivariate) for $\mathcal{N}(0,1)$ multiplier distribution.}} \label{tab:Norm_tri_1}
\end{table}

\begin{table}[ht]
\centering
\begin{tabular}{llllllllll}
  \hline
  \hline
  &  &  &  & $q^*_0$ & $q^*_1$ & $q^*_2$ & $\tilde{q}^*_0$ & $\tilde{q}^*_1$ & $\tilde{q}^*_2$ \\ 
     &  &  &  &  &  &  &  &  &  \\ 
     n & cens. & ${\beta}_0$ & $(\alpha_{010},\alpha_{020})$ &  &  & &  &  &  \\ 
   100 & low & (-0.05,-0.5,-0.05) & (0.08,0.008) & 95.4 & 97.9 & 97 & 93.3 & 94.8 & 95.8 \\ 
     &  &  & (0.05,0.05) & 99 & 99.2 & 97.8 & 95.6 & 96.8 & 97.5 \\ 
     &  & (-0.05,-0.25,-0.05) & (0.08,0.008) & 95 & 98 & 95 & 92 & 94 & 95.2 \\ 
     &  &  & (0.05,0.05) & 98.2 & 98.1 & 96.6 & 94.6 & 96 & 97.1 \\ 
     &  & (-0.05,0.25,-0.05) & (0.08,0.008) & 96.5 & 98.5 & 94.8 & 92 & 93.7 & 95 \\ 
     &  &  & (0.05,0.05) & 95.5 & 97.3 & 97.2 & 93.4 & 95.8 & 96.9 \\ 
     & high & (-0.05,-0.5,-0.05) & (0.08,0.008) & 97.7 & 98.5 & 96.1 & 94.2 & 95.7 & 96.2 \\ 
     &  &  & (0.05,0.05) & 98.8 & 99.4 & 98.4 & 94.7 & 96.4 & 97.6 \\ 
     &  & (-0.05,-0.25,-0.05) & (0.08,0.008) & 96.1 & 97.9 & 97.3 & 93.7 & 95.2 & 96.3 \\ 
     &  &  & (0.05,0.05) & 98.6 & 99 & 98.1 & 95.4 & 97.1 & 98.1 \\ 
     &  & (-0.05,0.25,-0.05) & (0.08,0.008) & 95 & 97.8 & 94.9 & 92.9 & 94.6 & 95.6 \\ 
     &  &  & (0.05,0.05) & 96.9 & 97.4 & 97.7 & 94.4 & 96.1 & 97.7 \\ 
    200 & low & (-0.05,-0.5,-0.05) & (0.08,0.008) & 96.1 & 98.6 & 92.3 & 94.4 & 95 & 95.8 \\ 
     &  &  & (0.05,0.05) & 95.7 & 97.7 & 95.8 & 94.1 & 95.2 & 96.3 \\ 
     &  & (-0.05,-0.25,-0.05) & (0.08,0.008) & 96.6 & 98.6 & 92.9 & 93.5 & 94 & 95.1 \\ 
     &  &  & (0.05,0.05) & 94.2 & 97.6 & 95.1 & 93.1 & 94.2 & 95.6 \\ 
     &  & (-0.05,0.25,-0.05) & (0.08,0.008) & 97.7 & 98.1 & 96.9 & 93 & 93.8 & 95 \\ 
     &  &  & (0.05,0.05) & 94.8 & 97.6 & 92.8 & 92.6 & 93.9 & 95.5 \\ 
     & high & (-0.05,-0.5,-0.05) & (0.08,0.008) & 94.1 & 98.1 & 93.6 & 93.1 & 93.7 & 94.9 \\ 
     &  &  & (0.05,0.05) & 98.2 & 98.4 & 97.8 & 94.9 & 95.9 & 96.6 \\ 
     &  & (-0.05,-0.25,-0.05) & (0.08,0.008) & 95.2 & 98 & 92.8 & 93 & 93.7 & 94.7 \\ 
     &  &  & (0.05,0.05) & 96 & 97.6 & 96.4 & 94.4 & 95.6 & 96.4 \\ 
     &  & (-0.05,0.25,-0.05) & (0.08,0.008) & 96.8 & 98.2 & 94 & 93.6 & 94.2 & 95.1 \\ 
     &  &  & (0.05,0.05) & 94 & 97.4 & 94.2 & 92.7 & 93.9 & 95.4 \\ 
    300 & low & (-0.05,-0.5,-0.05) & (0.08,0.008) & 97.2 & 98.5 & 93.9 & 94.2 & 94.6 & 95.4 \\ 
     &  &  & (0.05,0.05) & 93.4 & 98.1 & 93.3 & 94.1 & 94.6 & 95.7 \\ 
     &  & (-0.05,-0.25,-0.05) & (0.08,0.008) & 97.2 & 98.1 & 95 & 93.9 & 94.3 & 95.4 \\ 
     &  &  & (0.05,0.05) & 94.5 & 98.5 & 92.7 & 93.7 & 94.4 & 95.5 \\ 
     &  & (-0.05,0.25,-0.05) & (0.08,0.008) & 97.2 & 97.6 & 97 & 94 & 94.6 & 95.3 \\ 
     &  &  & (0.05,0.05) & 95.9 & 97.9 & 92.8 & 93.5 & 94.4 & 95.5 \\ 
     & high & (-0.05,-0.5,-0.05) & (0.08,0.008) & 95.7 & 98.5 & 90.9 & 94 & 94.2 & 94.9 \\ 
     &  &  & (0.05,0.05) & 94.3 & 97.7 & 95.7 & 94.2 & 94.8 & 95.5 \\ 
     &  & (-0.05,-0.25,-0.05) & (0.08,0.008) & 95.9 & 98.1 & 91.9 & 93.6 & 94 & 94.8 \\ 
     &  &  & (0.05,0.05) & 93.3 & 97.8 & 93.8 & 93.7 & 94.4 & 95.2 \\ 
     &  & (-0.05,0.25,-0.05) & (0.08,0.008) & 97 & 97.8 & 95.5 & 94.1 & 94.4 & 95.1 \\ 
     &  &  & (0.05,0.05) & 94.7 & 97.7 & 92.4 & 93.2 & 94.2 & 95.1 \\ 
   \hline
\end{tabular}
\caption{\textit{Simulated coverage probabilities (in \%) of various 95\% confidence bands for the cumulative incidence function given a 70 years old male individual with pneumonia at time of hospital admission (trivariate) for centered $\text{Exp}(1)$ multiplier distribution.}} \label{tab:Exp_tri_1}
\end{table}

\begin{table}[ht]
\centering
\begin{tabular}{llllllllll}
  \hline
  \hline
  &  &  &  & $q^*_0$ & $q^*_1$ & $q^*_2$ & $\tilde{q}^*_0$ & $\tilde{q}^*_1$ & $\tilde{q}^*_2$ \\ 
     &  &  &  &  &  &  &  &  &  \\ 
    n & cens. & ${\beta}_0$ & $(\alpha_{010},\alpha_{020})$ &  &  & &  &  &  \\ 
   100 & low & (-0.05,-0.5,-0.05) & (0.08,0.008) & 95.6 & 96.3 & 96.4 & 94.5 & 95.7 & 96.8 \\ 
     &  &  & (0.05,0.05) & 98.6 & 99.2 & 97.7 & 96.3 & 97.1 & 97.9 \\ 
     &  & (-0.05,-0.25,-0.05) & (0.08,0.008) & 94.1 & 95.9 & 95.2 & 93.2 & 94.6 & 96.1 \\ 
     &  &  & (0.05,0.05) & 97.6 & 97.9 & 96.6 & 95.3 & 96.6 & 97.4 \\ 
     &  & (-0.05,0.25,-0.05) & (0.08,0.008) & 95 & 96.6 & 94.4 & 93 & 94.3 & 95.6 \\ 
     &  &  & (0.05,0.05) & 95.5 & 96.2 & 97.1 & 94.7 & 96.2 & 97.3 \\ 
     & high & (-0.05,-0.5,-0.05) & (0.08,0.008) & 97.1 & 97.4 & 95.7 & 95.1 & 96 & 96.7 \\ 
     &  &  & (0.05,0.05) & 98.3 & 99.2 & 98.2 & 95.1 & 96.5 & 97.7 \\ 
     &  & (-0.05,-0.25,-0.05) & (0.08,0.008) & 95.9 & 96.5 & 96.9 & 95 & 96 & 96.9 \\ 
     &  &  & (0.05,0.05) & 98.2 & 98.9 & 98 & 96.1 & 97.2 & 98.2 \\ 
     &  & (-0.05,0.25,-0.05) & (0.08,0.008) & 93.8 & 96 & 95.3 & 93.8 & 95.4 & 96.4 \\ 
     &  &  & (0.05,0.05) & 96.6 & 97 & 97.4 & 95.3 & 96.6 & 97.9 \\ 
    200 & low & (-0.05,-0.5,-0.05) & (0.08,0.008) & 94.3 & 96.5 & 92.6 & 94.2 & 95 & 96 \\ 
     &  &  & (0.05,0.05) & 96.8 & 96.1 & 96.9 & 94.9 & 96.1 & 96.9 \\ 
     &  & (-0.05,-0.25,-0.05) & (0.08,0.008) & 94 & 96.1 & 92 & 93.1 & 94.1 & 95.5 \\ 
     &  &  & (0.05,0.05) & 94.8 & 96 & 96 & 93.9 & 95 & 96.4 \\ 
     &  & (-0.05,0.25,-0.05) & (0.08,0.008) & 96.1 & 96.8 & 95.5 & 93.3 & 94 & 95 \\ 
     &  &  & (0.05,0.05) & 93 & 95.8 & 93.1 & 93.2 & 94.5 & 96 \\ 
     & high & (-0.05,-0.5,-0.05) & (0.08,0.008) & 93.3 & 95.8 & 94.5 & 93.9 & 94.5 & 95.7 \\ 
     &  &  & (0.05,0.05) & 98.4 & 98.1 & 98.1 & 95.6 & 96.4 & 97.1 \\ 
     &  & (-0.05,-0.25,-0.05) & (0.08,0.008) & 93.3 & 96 & 93 & 93.5 & 94.2 & 95.4 \\ 
     &  &  & (0.05,0.05) & 96.9 & 96.1 & 96.9 & 95.5 & 96.3 & 97.1 \\ 
     &  & (-0.05,0.25,-0.05) & (0.08,0.008) & 95.2 & 96.5 & 93.2 & 93.9 & 94.5 & 95.7 \\ 
     &  &  & (0.05,0.05) & 93.4 & 95.4 & 94.6 & 93.6 & 94.8 & 96.2 \\ 
    300 & low & (-0.05,-0.5,-0.05) & (0.08,0.008) & 95 & 96.4 & 92.7 & 93.9 & 94.3 & 95.2 \\ 
     &  &  & (0.05,0.05) & 93.4 & 95.5 & 95.3 & 94.6 & 95.1 & 96.2 \\ 
     &  & (-0.05,-0.25,-0.05) & (0.08,0.008) & 94.7 & 96 & 93.4 & 93.8 & 94.3 & 95.3 \\ 
     &  &  & (0.05,0.05) & 93.2 & 96.2 & 93.7 & 94 & 94.5 & 95.8 \\ 
     &  & (-0.05,0.25,-0.05) & (0.08,0.008) & 95.5 & 95.9 & 95.4 & 93.8 & 94.5 & 95.3 \\ 
     &  &  & (0.05,0.05) & 94.3 & 96 & 92.8 & 93.8 & 94.7 & 95.7 \\ 
     & high & (-0.05,-0.5,-0.05) & (0.08,0.008) & 93.3 & 96.1 & 91.5 & 93.9 & 94.1 & 95.2 \\ 
     &  &  & (0.05,0.05) & 96.2 & 95.7 & 97.2 & 94.8 & 95.3 & 96.4 \\ 
     &  & (-0.05,-0.25,-0.05) & (0.08,0.008) & 93.8 & 95.7 & 91.7 & 93.7 & 94.3 & 95 \\ 
     &  &  & (0.05,0.05) & 93.3 & 95.7 & 95.8 & 94.2 & 94.8 & 96.2 \\ 
     &  & (-0.05,0.25,-0.05) & (0.08,0.008) & 95.4 & 96.1 & 94.2 & 94.1 & 94.6 & 95.3 \\ 
     &  &  & (0.05,0.05) & 93.1 & 95.8 & 92.8 & 93.7 & 94.5 & 95.8 \\ 
   \hline
\end{tabular}
\caption{\textit{Simulated coverage probabilities (in \%) of various 95\% confidence bands for the cumulative incidence function given a 70 years old male individual with pneumonia at time of hospital admission (trivariate) for centered $\text{Pois}(1)$ multiplier distribution.}} 
\label{tab:Poi_tri_1}
\end{table}

%
%%
%%%
%%%%
%%%%%
%%%%%%
%%%%%%%
\section*{Appendix B: Proofs and Remarks}
%\appendix
%%%%%%%
%%%%%%
%%%%%
%%%%
%%%
%%
%
Throughout the appendix, we will use a simplified version of the notation introduced in \cref{sec:Fine-Gray}. In particular, we will use the following notation:
\begin{itemize}
    \item $\textbf{C}_n = \textbf{C}_n^{(1)}$ and $\textbf{C}^*_n = \textbf{C}_n^{*(1)}$;
    \item $\textbf{D}_{n,g} = \textbf{D}_{n,g}^{(1)}$ and $\textbf{D}^*_{n,g} = \textbf{D}_{n,g}^{*(1)}$ with $g_{n,i} = g_{n,i}^{(1)}$;
    \item $\textbf{B}_n = \textbf{B}_n^{(2)}$ and $\textbf{B}^*_n = \textbf{B}_n^{*(2)}$;
    \item ${D}_{n,k} = {D}_{n,k}^{(2)}$ and ${D}^*_{n,k} = {D}_{n,k}^{*(2)}$ with $k_{n,i} = k_{n,i}^{(2)}$.
\end{itemize}

%%%%%%%%%%%%%%%%%%%%
%%%%%%%%%%%%%%%%%%%%
\subsection*{B.1 Proofs and Remarks of \cref{subsec:F-G-estimators}}

%Put your short appendix here.  Remember, longer appendices are possible when presented as Supplementary Web Material.  Please review and follow the journal policy for this material, available under Instructions for Authors at \texttt{http://www.biometrics.tibs.org}.
\textbf{Proof of \cref{ass:A1}}.\\
\medskip
\textbf{Proof of {\cref{ass:A1}\ref{ass:A1_1}}}: 
First, we show that \cref{assump2}~\ref{assump2_1}~-~\ref{assump2_4_new} imply parts~(i),~(ii),~and~(iii) of  \cref{assump_general} for $\textbf h_{n,i}(t,\bs{\beta}) = \big(k_{n}(t,\bs{\beta}),\textbf g_{n,i}(t,\bs{\beta})^{\top} \big)^\top = \big(J_n(t)S_n^{(0)}(t,\bs{\beta}),(\textbf Z_i - \textbf E_n(t,{\bs{\beta}}))^\top \big)^\top $ and analogously for its limit in probability $\tilde{\textbf{h}}_{i}(t,\bs{\beta}) = $ $\big( \tilde{k}(t,\bs{\beta}),\tilde{\textbf{g}}_i(t,\bs{\beta})^\top\big)^\top = \big( s^{(0)}(t,\bs{\beta})^{-1}, (\textbf Z_i - \textbf e(t,{\bs{\beta}}))^\top \big)^\top $,  $t \in \mathcal{T}$ and $\bs{\beta} \in \mathcal{B}$. 
Let $\check{\bs{\beta}}_n$ be a consistent estimator of ${\bs{{\beta}}_0}$. Because
\begin{align*}
  &\sup_{t\in\mathcal{T},i\in\{1,\ldots,n\}}\lVert \big( k_n(t,\check{\bs{\beta}}_n),\textbf g_{n,i}(t,\check{\bs{\beta}}_n)^{\top}\big)^\top - \big( \tilde{k}(t,{\bs{{\beta}}_0}), \tilde{\textbf{g}}_i(t,\bs{{\beta}}_0)^\top \big)^\top \rVert_\infty\\
  &\leq \sup_{t\in\mathcal{T}}\lVert k_n(t,\check{\bs{\beta}}_n) - \tilde{k}(t,{\bs{{\beta}}_0})\rVert_\infty + \sup_{t\in\mathcal{T},i\in\{1,\ldots,n\}}\lVert \textbf g_{n,i}(t,\check{\bs{\beta}}_n) - \tilde{\textbf{g}}_i(t,\bs{{\beta}}_0) \rVert_\infty,  
\end{align*}
it suffices for proving part (i)  of  \cref{assump_general} of Part~I to consider the convergence of each of the two terms separately. Obviously, for proving the parts (ii) and (iii) we can also treat the two components of $h_{n,i}(t,\bs{\beta})$ separately. Let us consider $k_n(t,\bs{\beta})$ first.
We have 
\begin{align}\label{remA1:k_tilde_k}
\begin{split}
    &\sup_{t\in\mathcal{T}}\lvert k_n(t,\check{\bs{\beta}}_n) - \tilde{k}(t,{\bs{{\beta}}_0})\rvert \\
    & = \sup_{t\in\mathcal{T}}\lvert J_n(t)S_n^{(0)}(t,\check{\bs{\beta}}_n)^{-1} - s^{(0)}(t,\bs{{\beta}}_0)^{-1} \rvert\\ 
    & = \sup_{t\in\mathcal{T}}\lvert (J_n(t) -1 +1 ) \cdot (\frac{s^{(0)}(t,\bs{{\beta}}_0)}{S_n^{(0)}(t,\check{\bs{\beta}}_n)} -1 +1 )\cdot s^{(0)}(t,\bs{{\beta}}_0)^{-1} - s^{(0)}(t,\bs{{\beta}}_0)^{-1} \rvert\\
    &= \sup_{t\in\mathcal{T}}\lvert \big [(J_n(t) -1)\cdot (\frac{s^{(0)}(t,\bs{{\beta}}_0)}{S_n^{(0)}(t,\check{\bs{\beta}}_n)} -1 ) +  (J_n(t) -1) + (\frac{s^{(0)}(t,\bs{{\beta}}_0)}{S_n^{(0)}(t,\check{\bs{\beta}}_n)} -1 ) + 1 \big ] \\
    & \qquad  \cdot s^{(0)}(t,\bs{{\beta}}_0)^{-1}-s^{(0)}(t,\bs{{\beta}}_0)^{-1} \rvert\\
    &= \sup_{t\in\mathcal{T}}\lvert \big [(J_n(t) -1)\cdot (\frac{s^{(0)}(t,\bs{{\beta}}_0)}{S_n^{(0)}(t,\check{\bs{\beta}}_n)} -1 ) +  (J_n(t) -1) + (\frac{s^{(0)}(t,\bs{{\beta}}_0)}{S_n^{(0)}(t,\check{\bs{\beta}}_n)} -1 ) \big ]\\
    &\qquad \cdot s^{(0)}(t,\bs{{\beta}}_0)^{-1}\rvert.
\end{split}
\end{align}
Moreover, we know that 
\begin{align*}
&\sup_{t\in\mathcal{T}} \lvert \big( \frac{S_n^{(0)}(t,\check{\bs{\beta}}_n)}{s^{(0)}(t,\bs{{\beta}}_0)} -1 \big ) \frac{s^{(0)}(t,\bs{{\beta}}_0)}{s^{(0)}(t,\bs{{\beta}}_0)} \rvert \\
    &= \sup_{t\in\mathcal{T}} \lvert \big (S_n^{(0)}(t,\check{\bs{\beta}}_n) - s^{(0)}(t,\bs{{\beta}}_0) \big ) s^{(0)}(t,\bs{{\beta}}_0)^{-1} \rvert\\
    &= \sup_{t\in\mathcal{T}} \big\{ \lvert [S_n^{(0)}(t,\check{\bs{\beta}}_n) - s^{(0)}(t,\check{\bs{\beta}}_n) + s^{(0)}(t,\check{\bs{\beta}}_n)- S_n^{(0)}(t,{\bs{\beta}}_0) + S_n^{(0)}(t,{\bs{\beta}}_0) - s^{(0)}(t,\bs{{\beta}}_0) ]\\
    & \quad \cdot s^{(0)}(t,\bs{{\beta}}_0)^{-1}\rvert \big\} \\
    &\leq \sup_{t\in\mathcal{T}} \big\{ \lvert S_n^{(0)}(t,\check{\bs{\beta}}_n) - s^{(0)}(t,\check{\bs{\beta}}_n) \rvert + \lvert S_n^{(0)}(t,{\bs{\beta}}_0) - s^{(0)}(t,{\bs{{\beta}}_0}) + s^{(0)}(t,{\bs{\beta}}_0) - s^{(0)}(t,\check{\bs{\beta}}_n) \rvert\\
    & \quad +  \lvert S_n^{(0)}(t,{\bs{\beta}}_0) - s^{(0)}(t,\bs{{\beta}}_0) \rvert\big\} \cdot \sup_{t\in\mathcal{T}} \lvert s^{(0)}(t,\bs{{\beta}}_0)^{-1}\rvert\\
    &\stackrel{\mathbb P}{\longrightarrow} 0, \text { as } n\rightarrow\infty.
\end{align*}
The above convergence in probability to zero, as $n\rightarrow\infty$, holds for any consistent estimator $\check{\bs{\beta}}_n\in\mathcal{B}$ of $\bs{{\beta}}_0$ due to \cref{assump2}~\ref{assump2_1}, the continuity of $s^{(0)}(t,\cdot)$ in $\bs{\beta}\in\mathcal{B}$ (\cref{assump2}~\ref{assump2_2}), and the boundedness of $s^{(0)}(\cdot,\bs{{\beta}}_0)^{-1}$ for all $t\in\mathcal{T}$ according to \cref{assump2}~\ref{assump2_3}~\&~\ref{assump2_4_new}, see \eqref{eq:form_of_s}. Hence, it follows from the continuous mapping theorem that 
\begin{align}\label{remA1:S_s}
    \sup_{t\in\mathcal{T}} \lvert  \frac{s^{(0)}(t,\bs{{\beta}}_0)}{S_n^{(0)}(t,\check{\bs{\beta}}_n)} -1   \rvert \stackrel{\mathbb P}{\longrightarrow} 0, \text { as } n\rightarrow\infty,
\end{align}
for any consistent estimator $\check{\bs{\beta}}_n\in\mathcal{B}$ of $\bs{{\beta}}_0$. Additionally, it holds that 
\begin{align}\label{remA1:J_1}
    \sup_{t\in\mathcal{T}} \lvert J_n(t) -1 \rvert\stackrel{\mathbb{ P}}{\longrightarrow} 0 ,\text{ as } n\rightarrow\infty,
\end{align}
according to \cref{assump2}~\ref{assump2_3}. Based on \eqref{remA1:S_s}, \eqref{remA1:J_1} and the boundedness of $s^{(0)}(\cdot,\bs{{\beta}}_0)^{-1}$ for all $t\in\mathcal{T}$ according to \cref{assump2}~\ref{assump2_3}~\&~\ref{assump2_4_new}, the right-hand side of the fourth equation of \eqref{remA1:k_tilde_k} converges to zero in probability, as $n\rightarrow\infty$, i.e.,
\begin{align}\label{eq:J_S_minus_s}
    \sup_{t\in\mathcal{T}}\lvert J_n(t)S_n^{(0)}(t,\check{\bs{\beta}}_n)^{-1} - s^{(0)}(t,\bs{{\beta}}_0)^{-1} \rvert\stackrel{\mathbb{ P}}{\longrightarrow} 0, \text { as } n\rightarrow\infty,
\end{align}
for any consistent estimator $\check{\bs{\beta}}_n\in\mathcal{B}$ of $\bs{{\beta}}_0$. Thus, \cref{assump_general}~\ref{assump_general1} of Part~I is fulfilled for $k_n(t,\bs{\beta})$ under \cref{assump2}~\ref{assump2_1}~-~\ref{assump2_4_new}.
To see that \cref{assump_general}~\ref{assump_general2} of Part~I holds, we note that $\tilde{k}(t,\cdot) = s^{(0)}(t,\cdot )^{-1}$ is a continuous function in $\bs{\beta}\in\mathcal{B}$, because $s^{(0)}(t,\cdot )$ is  a continuous function in $\bs{\beta}\in\mathcal{B}$ according to \cref{assump2}~\ref{assump2_2}, and that the continuity is preserved under the inverse. Additionally, $s^{(0)}(t,\bs{\beta} )^{-1}$ is bounded on $\mathcal{T}\times\mathcal{B}$, since $s^{(0)}(t,\bs{\beta} )$ is bounded away from zero on $\mathcal{T}\times\mathcal{B}$ according to \cref{assump2}~\ref{assump2_3} and \eqref{eq:form_of_s}, which holds due to
\cref{assump2}~\ref{assump2_4_new}. Hence, \cref{assump2}~\ref{assump2_2}~-~\ref{assump2_4_new} imply \cref{assump_general}~\ref{assump_general2} of Part~I for $k_n(t,\bs{\beta})$.
With respect to part (iii) of \cref{assump_general} of Part~I, we remark that the couples $(\tilde{k}(t,\bs{{\beta}}_0),\lambda_i(t,\bs{{\beta}}_0))$, $i=1,\dots,n$, are pairwise independent and identically distributed for all $t\in\mathcal{T}$, because $\tilde{k}(t,\bs{{\beta}}_0)=s^{(0)}(t,\bs{{\beta}}_0)^{-1}$ is a deterministic function in $t\in\mathcal{T}$ (see \eqref{eq:form_of_s}) and $\lambda_1(t,\bs{{\beta}}_0),\ldots , \lambda_n(t,\bs{{\beta}}_0)$ with $\lambda_i(t,\bs{{\beta}}_0) = Y_i(t)\exp(\textbf Z_i^\top\bs{{\beta}}_0)\alpha_{1;0}(t)$ are pairwise independent and identically distributed for all $t\in\mathcal{T}$ according to \cref{assump2}~\ref{assump2_4_new}. In conclusion, \cref{assump_general} of Part~I is fulfilled for $k_n(t,\bs{\beta})$ under \cref{assump2}~\ref{assump2_1}~-~\ref{assump2_4_new}.

Let us now consider $\textbf g_{n,i}(t,\bs{\beta})$.  We first show under which conditions of \cref{assump2}  \cref{assump_general}~\ref{assump_general1} of Part~I follows for $\textbf g_{n,i}(t,\bs{\beta})$, i.e., we have to prove that for any consistent estimator $\check{\bs{\beta}}_n$ of $\bs{{\beta}}_0$.
\begin{align}\label{remA1:conv_g}
    \sup_{t\in\mathcal{T},i\in\{1,\ldots,n\}}\lVert \textbf g_{n,i}(t,\check{\bs{\beta}}_n) - \tilde{\textbf{g}}_i(t,\bs{{\beta}}_0) \rVert\stackrel{\mathbb{ P}}{\longrightarrow} 0, \text{ as } n \rightarrow \infty.
\end{align}
 Recall that we have
\begin{align*}
    \sup_{t\in\mathcal{T},i\in\{1,\ldots,n\}}\lVert \textbf g_{n,i}(t,\check{\bs{\beta}}_n) - \tilde{\textbf{g}}_i(t,\bs{{\beta}}_0) \rVert &= \sup_{t\in\mathcal{T},i\in\{1,\ldots,n\}}\lVert \textbf Z_i - \textbf E_n(t,\check{\bs{\beta}}_n) - (\textbf Z_i - \textbf e(t,{\bs{{\beta}}_0})) \rVert\\
    &= \sup_{t\in\mathcal{T}}\big\lVert  \frac{\textbf S^{(1)}_n(t,\check{\bs{\beta}}_n)}{S^{(0)}_n(t,\check{\bs{\beta}}_n)} - \frac{\textbf s^{(1)}(t,{\bs{{\beta}}_0})}{s^{(0)}(t,{\bs{{\beta}}_0})}  \big\rVert.
\end{align*}
It is straightforward to show that the term on the right-hand side of the second equation above converges to zero in probability as $n\rightarrow\infty$  for any consistent estimator $\check{\bs{\beta}}_n\in\mathcal{B}$ of $\bs{{\beta}}_0$ according to \cref{assump2}~\ref{assump2_1}~-~\ref{assump2_4_new}. In order to see this one may rewrite $\displaystyle{\sup_{t\in\mathcal{T}}\lVert\frac{\textbf S^{(1)}_n(t,\check{\bs{\beta}}_n)}{S^{(0)}_n(t,\check{\bs{\beta}}_n)}\rVert}$ as 
\begin{align*}
    &\sup_{t\in\mathcal{T}}\lVert \big(\textbf S^{(1)}_n(t,\check{\bs{\beta}}_n) - \textbf s^{(1)}(t,{\bs{{\beta}}_0}) + \textbf s^{(1)}(t,{\bs{{\beta}}_0})\big)\big( \frac{s^{(0)}(t,{\bs{{\beta}}_0})}{S^{(0)}_n(t,\check{\bs{\beta}}_n)} -1 +1 \big)\cdot \frac{1}{s^{(0)}(t,{\bs{{\beta}}_0})}\rVert\\
    &\leq \sup_{t\in\mathcal{T}}\big\{\big[\lVert \textbf S^{(1)}_n(t,\check{\bs{\beta}}_n) - \textbf s^{(1)}(t,{\bs{{\beta}}_0})\big)\rVert \cdot \lvert \frac{s^{(0)}(t,{\bs{{\beta}}_0})}{S^{(0)}_n(t,\check{\bs{\beta}}_n)} -1 \rvert
    + \lVert\textbf S^{(1)}_n(t,\check{\bs{\beta}}_n) - \textbf s^{(1)}(t,{\bs{{\beta}}_0})\rVert\\
    &\quad +\lVert\textbf s^{(1)}(t,{\bs{{\beta}}_0})\rVert\cdot \lvert \frac{s^{(0)}(t,{\bs{{\beta}}_0})}{S^{(0)}_n(t,\check{\bs{\beta}}_n)} -1 \rvert\big ]\cdot\lvert  s^{(0)}(t,{\bs{{\beta}}_0})^{-1}\rvert
    + \lVert\frac{\textbf s^{(1)}(t,{\bs{{\beta}}_0})}{s^{(0)}(t,{\bs{{\beta}}_0})}\rVert\big\}.
\end{align*}
Here, the term in squared brackets converges in probability to zero as $n\rightarrow\infty$ for any consistent estimator $\check{\bs{\beta}}_n\in\mathcal{B}$ of $\bs{{\beta}}_0$ according to \cref{assump2}~\ref{assump2_1}, \eqref{remA1:S_s}, which holds under \cref{assump2}~\ref{assump2_1}~-~\ref{assump2_4_new}, and the boundedness of $s^{(1)}(\cdot,{\bs{{\beta}}_0})$ for all $t\in\mathcal{T}$ according to \cref{assump2}~\ref{assump2_2}. Then, due to the boundedness of $s^{(0)}(\cdot,{\bs{{\beta}}_0})^{-1}$ for all $t\in\mathcal{T}$ according to \cref{assump2}~\ref{assump2_3}~\&~\ref{assump2_4_new}, it holds that $\displaystyle{\sup_{t\in\mathcal{T}}\lVert\frac{\textbf S^{(1)}_n(t,\check{\bs{\beta}}_n)}{S^{(0)}_n(t,\check{\bs{\beta}}_n)}\rVert}$ is asymptotically equivalent to $\displaystyle{\sup_{t\in\mathcal{T}}\lVert \frac{\textbf s^{(1)}(t,{\bs{{\beta}}_0})}{s^{(0)}(t,{\bs{{\beta}}_0})}\rVert}$ under \cref{assump2}~\ref{assump2_1}~-~\ref{assump2_4_new}. Hence,
\begin{align}\label{eq:E_minus_e}
     \sup_{t\in\mathcal{T}}\big\lVert  \frac{\textbf S^{(1)}_n(t,\check{\bs{\beta}}_n)}{S^{(0)}_n(t,\check{\bs{\beta}}_n)} - \frac{\textbf s^{(1)}(t,{\bs{{\beta}}_0})}{s^{(0)}(t,{\bs{{\beta}}_0})}  \big\rVert \stackrel{\mathbb P}{\longrightarrow} 0,\text{ as } n\rightarrow\infty,
\end{align}
for any consistent estimator $\check{\bs{\beta}}_n\in\mathcal{B}$ of $\bs{{\beta}}_0$. From this, \eqref{remA1:conv_g} immediately follows and  \cref{assump_general}~\ref{assump_general1} of Part~I holds for $\textbf g_{n,i}(t,\bs{\beta})$.
Furthermore, $\tilde{g}_i(t,\cdot) = \big(\textbf Z_i - \displaystyle{\frac{\textbf s^{(1)}(t,\cdot)}{s^{(0)}(t,\cdot)}}\big)  $ is a continuous function in $\bs{\beta}\in\mathcal{B}$, because $\textbf s^{(1)}(t,\cdot)$ is a continuous function in $\bs{\beta}\in\mathcal{B}$ according to \cref{assump2}~\ref{assump2_2}, and $s^{(0)}(t,\cdot)^{-1}$ is a continuous function in $\bs{\beta}\in\mathcal{B}$ according to \cref{assump2}~\ref{assump2_2} as argued in the context of $\tilde k(t,\cdot)$. 
Additionally, $\tilde{g}_i$ is bounded on $\mathcal{T}\times\mathcal{B}$ for all $i\in\mathbb N$, since $\textbf Z_i$ is assumed to be bounded for $i\in\mathbb N$ and $\textbf e = \displaystyle{\frac{\textbf s^{(1)}}{s^{(0)}} }$ is bounded on $\mathcal{T}\times\mathcal{B}$, because $\textbf s^{(1)}$ is bounded on $\mathcal{T}\times\mathcal{B}$ according to \cref{assump2}~\ref{assump2_2} and $s^{(0)}(\cdot)^{-1}$ is bounded on $\mathcal{T}\times\mathcal{B}$ according to \cref{assump2}~\ref{assump2_3}~\&~\ref{assump2_4_new} as argued in the context of $\tilde k(t,\cdot)$. Thus we conclude that under \cref{assump2}~\ref{assump2_2}~-~\ref{assump2_4_new},  \cref{assump_general}~\ref{assump_general2} of Part~I holds for $g_{n,i}(t,\bs{\beta})$.
Finally, with respect to part (iii) of \cref{assump_general} of Part~I we note that  the couples $(\tilde{g}_i(t,\bs{{\beta}}_0),\lambda_i(t,\bs{{\beta}}_0))$, $i=1,\dots,n$, with $\lambda_i(t,\bs{{\beta}}_0) = Y_i(t)\exp(\textbf Z_i^\top\bs{{\beta}}_0)\alpha_{1;0}(t)$ are pairwise independent and identically distributed for all $t\in\mathcal{T}$, because $\textbf e(t,\bs{{\beta}}_0)$ is a deterministic function in $t\in\mathcal{T}$, and $(Y_i,N_i,\textbf Z_i)$, $i=1,\ldots,n$, are pairwise independent and identically distributed according to \cref{assump2}~\ref{assump2_4_new}. In conclusion, \cref{assump_general} of Part~I is fulfilled for $g_{n,i}(t,\bs{\beta})$ under \cref{assump2}~\ref{assump2_1}~-~\ref{assump2_4_new}. Combining this with our results for $k_{n}(t,\bs{\beta})$ above, it follows that under \cref{assump2}~\ref{assump2_1}~-~\ref{assump2_4_new} that \cref{assump_general} of Part~I holds for $\textbf h_{n,i}(t,\bs{\beta}) = \big(k_{n}(t,\bs{\beta}),\textbf g_{n,i}(t,\bs{\beta})^{\top} \big)^\top $.

Next, we derive from which conditions of \cref{assump2}  \cref{ass:Bn_Cn} of Part~I can be inferred. We start by considering \cref{ass:Bn_Cn}~\ref{item:ass_Bn_1}, i.e.,
\begin{align}\label{eq:Dk_tilde_K}
    \sup_{t\in\mathcal{T}} \lVert \nabla k_n(t,\check{\bs{\beta}}_n) - \tilde{\textbf K} (t,\bs{{\beta}}_0)\rVert \stackrel{\mathbb P}{\longrightarrow} 0, \text { as } n\rightarrow\infty,
\end{align}
for any consistent estimator $\check{\bs{\beta}}_n$ of $\bs{{\beta}}_0$. According to \cref{subsec:F-G-estimators} the gradient $\nabla k_n$ of $k_n$ with respect to $\bs{\beta}$ at $\bs{\beta}= \check{\bs{\beta}}_n$ is given by $\nabla k_n(t,\check{\bs{\beta}}_n) = -J_n(u)\cdot \textbf E_n(t,\check{\bs{\beta}}_n)^\top \cdot S^{(0)}_n(t,\check{\bs{\beta}}_n)^{-1} $. We claim that \eqref{eq:Dk_tilde_K} holds for $\tilde{\textbf K} (t,\bs{{\beta}}_0) = \textbf e(t,\bs{{\beta}}_0)^\top \cdot s^{(0)}(t,{\bs{{\beta}}_0})^{-1}$. For this $\tilde{\textbf K}$ we have
\begin{align*}
&\sup_{t\in\mathcal{T}} \lVert \nabla k_n(t,\check{\bs{\beta}}_n) - \tilde{\textbf K} (t,\bs{{\beta}}_0)\rVert\\
&=\sup_{t\in\mathcal{T}} \lVert -J_n(u)\cdot \textbf E_n(t,\check{\bs{\beta}}_n)^\top \cdot S^{(0)}_n(t,\check{\bs{\beta}}_n)^{-1} - \textbf e(t,\bs{{\beta}}_0)^\top \cdot s^{(0)}(t,{\bs{{\beta}}_0})^{-1}\rVert\\
    &=\sup_{t\in\mathcal{T}}\big\{ \lVert -J_n(u)\cdot S^{(0)}_n(t,\check{\bs{\beta}}_n)^{-1}\cdot\big( \textbf E_n(t,\check{\bs{\beta}}_n)^\top - \textbf e(t,\bs{{\beta}}_0)^\top + \textbf e(t,\bs{{\beta}}_0)^\top\big)  \\
    &  \quad - \textbf e(t,\bs{{\beta}}_0)^\top \cdot s^{(0)}(t,{\bs{{\beta}}_0})^{-1}\rVert \big\}\\
    &\leq \sup_{t\in\mathcal{T}} \big\{  \lvert J_n(u)\cdot S^{(0)}_n(t,\check{\bs{\beta}}_n)^{-1} - s^{(0)}(t,{\bs{{\beta}}_0})^{-1} \rvert\cdot \lVert \textbf e(t,\bs{{\beta}}_0)^\top \rVert \\
    &\quad + \lVert J_n(u)\cdot S^{(0)}_n(t,\check{\bs{\beta}}_n)^{-1} - s^{(0)}(t,{\bs{{\beta}}_0})^{-1} + s^{(0)}(t,{\bs{{\beta}}_0})^{-1} \rVert\cdot \lVert \textbf E_n(t,\check{\bs{\beta}}_n)^\top - \textbf e(t,\bs{{\beta}}_0)^\top \rVert \big\}\\
    &\leq \sup_{t\in\mathcal{T}} \big\{  \lVert J_n(u)\cdot S^{(0)}_n(t,\check{\bs{\beta}}_n)^{-1} - s^{(0)}(t,{\bs{{\beta}}_0})^{-1} \rVert\cdot \lVert \textbf e(t,\bs{{\beta}}_0)^\top \rVert \\
    &\quad + \big[\lVert J_n(u)\cdot S^{(0)}_n(t,\check{\bs{\beta}}_n)^{-1} - s^{(0)}(t,{\bs{{\beta}}_0})^{-1} \rVert + \lVert s^{(0)}(t,{\bs{{\beta}}_0})^{-1} \rVert\big]\cdot \lVert \textbf E_n(t,\check{\bs{\beta}}_n)^\top - \textbf e(t,\bs{{\beta}}_0)^\top \rVert \big\}.
\end{align*}
Hence, \eqref{eq:Dk_tilde_K} holds due to \eqref{eq:J_S_minus_s}, \eqref{eq:E_minus_e}, which hold under \cref{assump2}~\ref{assump2_1}~-~\ref{assump2_3}, and the boundedness of $\textbf e(t,\bs{{\beta}}_0)$ and $s^{(0)}(t,{\bs{{\beta}}_0})^{-1}$ on $\mathcal{T}$ according to \cref{assump2}~\ref{assump2_2}~-~\ref{assump2_4_new}. We conclude that \cref{ass:Bn_Cn}~\ref{item:ass_Bn_1} of Part~I holds under \cref{assump2}~\ref{assump2_1}~-~\ref{assump2_4_new}. Moreover, because in view of \eqref{eq:form_of_s}, $ \textbf e(t,\bs{{\beta}}_0)^\top $ and $ s^{(0)}(t,{\bs{{\beta}}_0})^{-1}$ are deterministic functions and thus, predictable with respect to $\mathcal{F}_1$, we have that \cref{ass:Bn_Cn}~\ref{item:ass_Bn_2} of Part~I clearly is satisfied due to \cref{assump2}~\ref{assump2_2}~-~\ref{assump2_4_new}. Additionally, since $\textbf e(t,\bs{{\beta}}_0)$ respectively $s^{(0)}(t,{\bs{{\beta}}_0})^{-1}$ are bounded on $\mathcal{T}$ under \cref{assump2}~\ref{assump2_2}~-~\ref{assump2_4_new} (see above), $\tilde{\textbf K} (t,\bs{{\beta}}_0) = \textbf e(t,\bs{{\beta}}_0)^\top \cdot s^{(0)}(t,{\bs{{\beta}}_0})^{-1}$ is bounded on $\mathcal{T}$. Furthermore, $(\tilde{\textbf K}(t,\bs{{\beta}}_0),\lambda_i(t,\bs{{\beta}}_0))$, $i=1,\dots,n$, are pairwise independent and identically distributed for all $t\in\mathcal{T}$, because $\tilde{\textbf K}(t,\bs{{\beta}}_0)$ is a deterministic function in $t\in\mathcal{T}$, and $(Y_i,N_i,\textbf Z_i)$, $i=1,\ldots,n$, are pairwise independent and identically distributed according to \cref{assump2}~\ref{assump2_4_new}. Thus, \cref{ass:Bn_Cn}~\ref{item:ass_Bn_3} of Part~I is fulfilled under \cref{assump2}~\ref{assump2_4_new}.
To sum up, \cref{ass:Bn_Cn} of Part~I holds under \cref{assump2}~\ref{assump2_1}~-~\ref{assump2_4_new}, and \cref{assump_general} and \cref{ass:Bn_Cn} of Part~I are valid under \cref{assump2}~\ref{assump2_1}~-~\ref{assump2_4_new}.

\noindent
\textbf{Proof of {\cref{ass:A1}\ref{ass:A1_3}}}: 
We derive the limit in probability of $\textbf{C}_n $, as $n\rightarrow\infty$. Note that
$\langle \textbf D_{n,g}\rangle (t) = \frac{1}{n}\sum_{i=1}^n\int_0^t \big(\textbf Z_i - \textbf E_n(u,\bs{{\beta}}_0)\big)^{\otimes 2}d\Lambda_i(u,\bs{{\beta}}_0)
=\frac{1}{n}\sum_{i=1}^n\int_0^t \textbf R_n(u,\bs{{\beta}}_0) d\Lambda_i(u,\bs{{\beta}}_0)$. 
Hence, we have
$$\frac{1}{n}\textbf I_n(t,\bs{{\beta}}_0) - \langle \textbf D_{n,g}\rangle (t)  = \frac{1}{n} \sum_{i=1}^n \int_0^t \textbf R_n(u,\bs{{\beta}}_0) dM_i(u),$$
where the right-hand side of the equation above is  a local square integrable martingale, according to Proposition II.4.1 of \cite{Andersen}. Following the notation introduced in Part~I, we denote this martingale by $\frac{1}{\sqrt{n}}\textbf D_{n,R}(t)$. Under \cref{assump2}~\ref{assump2_1}-\ref{assump2_4_new} it follows from \cref{lem:Dn} of Part~I that $\langle \textbf D_{n,R}\rangle (t) \stackrel{\mathbb P}{\longrightarrow} \langle \textbf D_{r}\rangle (t) $ for all $t\in\mathcal{T}$, as $n\rightarrow\infty$, where $\langle \textbf D_{r}\rangle (t)$ is some covariance function bounded for all $t\in\mathcal{T}$. Thus, $\frac{1}{n}\langle \textbf D_{n,R}\rangle (\tau)$ and likewise the corresponding martingale $\frac{1}{\sqrt{n}}\textbf D_{n,R}$ converge to zero in probability, as $n\rightarrow\infty$, according to Lenglart's Inequality. In other words, $\frac{1}{n}\textbf I_n(\tau,\bs{{\beta}}_0)$ and $\langle \textbf D_{n,g}\rangle (\tau)$ are asymptotically equivalent and we get
\begin{align*}
    \frac{1}{n}\textbf I_n(\tau,\bs{{\beta}}_0) = \langle \textbf D_{n,g}\rangle (\tau) + o_p(1) \stackrel{\mathbb P}{\longrightarrow} \langle \textbf D_{\tilde g}\rangle (\tau) = \textbf V_{\tilde g}(\tau), \text{ as } n\rightarrow\infty,
\end{align*}
with $\textbf V_{\tilde g}(t) =  \int_0^t \textbf r(u,\boldsymbol{\beta}_0) s^{(0)}(u,\boldsymbol{\beta}_0)dA_{1;0}(u)$.
By the continuous mapping theorem and because $\frac{1}{n}\textbf I_n(\tau ,\boldsymbol{\beta}_0)$ is asymptotically invertible under \cref{assump2}~\ref{assump2_4}, it follows from \cref{assump2} that 
\begin{align}\label{eq:In_pred-cov-Dg}
     \textbf C_n = \big(\frac{1}{n}\textbf I_n(\tau ,\boldsymbol{\beta}_0)\big)^{-1} \stackrel{\mathbb{P}}{\longrightarrow}   \textbf V_{\tilde g}(\tau)^{-1} = \textbf C, \text{ as } n\rightarrow\infty.
\end{align}
Hence, \cref{ass:Cn-C} of Part~I is satisfied under \cref{assump2}.

Recall that the wild bootstrap counterpart $\textbf C^*_n $ of $\textbf C_n =  \big(\frac{1}{n}\textbf I_n(\tau ,\boldsymbol{\beta}_0)\big)^{-1}$ is defined through the optional covariation process $[ \textbf D^*_{n,g} ] (\tau) $ of $\textbf D^*_{n,g} $, in this case as
\begin{align}
\label{eq:C_n^*-part2-def}
\textbf C^*_n =\big( [\textbf D^*_{n,g}](\tau)\big)^{-1} = \big(\frac{1}{n}\sum_{i=1}^n \int_0^\tau (\bs Z_{i}-\bs E_n(u,\hat{\boldsymbol{\beta}}_n))^{\otimes 2}G_i^2dN_i(u) \big)^{-1};
\end{align}
cf.\ \cref{lemma:mgale} of Part~I.
The particular choice of $\textbf C^*_n$ is motivated by the fact that, under \cref{assump2}~\ref{assump2_1}-\ref{assump2_4_new} and conditionally on $\mathcal{F}_2(0)$, we have $[ \textbf D^*_{n,g} ] (t) \stackrel{\mathbb P}{\longrightarrow} [ \textbf D_{\tilde g} ] (t) = \textbf V_{\tilde g}(t) $ for all $t\in\mathcal{T}$ as $n\rightarrow\infty$, according to \cref{cor:optCov_D*} of Part~I. Hence, from the continuous mapping theorem and because of the asymptotic invertibility of $ \textbf V_{\tilde g}(\tau)$ according to \cref{assump2}~\ref{assump2_4} it follows under \cref{assump2} that
\begin{align}\label{eq:C_n^*-part2}
    \textbf C^*_n %= \big( [\textbf D^*_{n,g}](\tau)\big)^{-1}
    \stackrel{\mathbb P }{\longrightarrow} \textbf V_{\tilde g}(\tau)^{-1} = \textbf C, \text{ as } n\rightarrow\infty.
\end{align}
From \eqref{eq:In_pred-cov-Dg} and \eqref{eq:C_n^*-part2} we conclude that 
%$\textbf C^*_n$ and $\textbf C_n$ share the same limit in probability, %i.e., 
$$\lVert \textbf C_n^* - \textbf C_n \rVert \stackrel{\mathbb P}{\longrightarrow} 0, \quad n \rightarrow\infty,$$
which is why \cref{ass:Cn_star-Cn} of Part~I is fulfilled under \cref{assump2}.
In conclusion, under \cref{assump2} both \cref{ass:Cn-C} and \cref{ass:Cn_star-Cn} of Part~I are satisfied.

\noindent
\textbf{Proof of {\cref{ass:A1}\ref{ass:A1_2}}}: 
We need to prove that under \cref{assump2}, Condition VII.2.1 of \cite{Andersen} holds. It is easy to see that \cref{assump2}~\ref{assump2_1}~-~\ref{assump2_3} and \cref{assump2}~\ref{assump2_4} are identical to  Condition VII.2.1~(a)~-~(c)
and Condition VII.2.1~(e), respectively. Thus, it is only left to show that \cref{assump2}~\ref{assump2_4_new} implies Condition VII.2.1~(d). In particular, we need to prove that under \cref{assump2}~\ref{assump2_4_new} the following holds:
\begin{align}
\begin{split}
    \label{eq:cond_d}
    \frac{\partial}{\partial \bs{\beta}}  s^{(0)}(t,\boldsymbol{\beta}) = \textbf s^{(1)}(t,\boldsymbol{\beta}),\quad
    \frac{\partial^2}{\partial \bs{\beta}^2}  s^{(0)}(t,\boldsymbol{\beta}) = \textbf s^{(2)}(t,\boldsymbol{\beta}), \text{ for } \bs{\beta} \in\mathcal{B},  t\in\mathcal{T}.
\end{split}
\end{align} 
For this we recall \eqref{eq:form_of_s}, this is, under \cref{assump2}~\ref{assump2_4_new} we have
\[\textbf s^{(m)}(t,\boldsymbol{\beta}) = \mathbb{E}(Y_1(t)\textbf Z_1^{\otimes m}\exp(\textbf Z_1^\top\bs{\beta})),\]
for all fixed $t\in\mathcal{T}$, $m \in \{0,1,2\}$ (in non-bold-type for $m=0$), and $\bs{\beta}\in\mathcal{B}$.
Furthermore, we have
%the first-order partial derivative of $Y_1(t)\exp(\textbf Z_1^\top\bs{\beta})$ with respect to $\bs{\beta}_j$  
$$|\frac{\partial}{\partial \bs{\beta}_j} Y_1(t)\exp(\textbf Z_1^\top\bs{\beta})|
=|Y_1(t) Z_{1j}\exp(\textbf Z_1^\top\bs{\beta})|
\leq \lvert Z_{1j} \rvert \exp(K)$$
and 
%the second-order partial derivative of $Y_1(t)\exp(\textbf Z_1^\top\bs{\beta})$ with respect to $\bs{\beta}_l$ equals 
$$ |\frac{\partial^2}{\partial \bs{\beta}_j\partial \bs{\beta}_l}Y_1(t)\exp(\textbf Z_1^\top\bs{\beta})| 
=|Y_1(t) Z_{1j}Z_{1l}\exp(\textbf Z_1^\top\bs{\beta})|
\leq \lvert Z_{1j}Z_{1l} \rvert \exp(K),$$ 
where $Z_{1j}$ is the $j$-th component of $\textbf Z_1$, $j,l = 1,\ldots , q$. 
%Additionally, the absolute value of these derivatives are bounded from above by the integrable random variables $\lvert Z_{1j} \rvert \exp(K)$ and $\lvert Z_{1j}Z_{1l} \rvert \exp(K)$, respectively, where $K= \sup_{\bs{\beta}\in\mathcal{B}}\lVert \bs{\beta}\rVert_\infty \cdot  q\lVert \textbf Z_1\rVert_\infty$. 
Note that $K$ is bounded due to the boundedness of the covariates and the boundedness of $\mathcal{B}$, so that the bounds on the right-hand side of the two formulas above are integrable random variables. 
According to Theorem~12.5 of \cite{schilling_measure}, it then follows that the integral and the differential operator can be interchanged, which yields
\begin{align*}
    \frac{\partial}{\partial \bs{\beta}_j} s^{(0)}(t,\boldsymbol{\beta}) &= \mathbb{E}( \frac{\partial}{\partial \bs{\beta}_j} Y_1(t)\exp(\textbf Z_1^\top\bs{\beta}))\\
    &= \mathbb{E}( Y_1(t) Z_{1j} \exp(\textbf Z_1^\top\bs{\beta})),
\end{align*}
and
\begin{align*}
    \frac{\partial^2}{\partial \bs{\beta}_j \partial \bs{\beta}_l} s^{(0)}(t,\boldsymbol{\beta}) &= \mathbb{E}( \frac{\partial^2}{\partial \bs{\beta}_j \partial \bs{\beta}_l} Y_1(t)\exp(\textbf Z_1^\top\bs{\beta}))\\
    &= \mathbb{E}( Y_1(t) Z_{1j} Z_{1l} \exp(\textbf Z_1^\top\bs{\beta})),
\end{align*}
for $j,l = 1,\ldots , q$. Hence, the gradient and the Hessian matrix of $s^{(0)}(t,\boldsymbol{\beta})$ are given by 
$$\textbf s^{(1)}(t,\boldsymbol{\beta}) = \mathbb{E}(Y_1(t)\textbf Z_1\exp(\textbf Z_1^\top\bs{\beta})) , \quad \textbf s^{(2)}(t,\boldsymbol{\beta}) = \mathbb{E}(Y_1(t)\textbf Z_1^{\otimes 2}\exp(\textbf Z_1^\top\bs{\beta})),$$ for all fixed $t\in\mathcal{T}$ and $\bs{\beta}\in\mathcal{B}$, respectively, so that \eqref{eq:cond_d} holds under \cref{assump2}~\ref{assump2_4_new}. Hence, Condition VII.2.1 of \cite{Andersen} follows from \cref{assump2}. This completes the proof of \cref{ass:A1}.\hfill\qedsymbol

\begin{rem}\label{rem:towards_asyRep_beta}
As explained in \cref{rem:F-G_Cox} and mentioned in \cite{Fine-Gray}, the structures related to the Fine-Gray model coincide with those under the Cox model. In particular, this holds for the log Cox partial likelihood and the log partial likelihood under the Fine-Gray model. Thus, by means of \cref{ass:A1}~\ref{ass:A1_2} we resort to Theorem VII.2.1 of \cite{Andersen} for the Cox model in which it is shown via the log Cox partial likelihood that $\hat{\bs{\beta}}_n$ is unique with probability converging to 1 and that $\hat{\bs{\beta}}_n$ is a consistent estimator for $\bs{{\beta}}_0$.
\end{rem}

\begin{rem}\label{rem:U_a_martingale}
The score statistic $\bs U_{n} (t,\bs{{\beta}}_0)$ is a local square integrable martingale in $t\in\mathcal{T}$. 
In order to see this, we point out the following two observations
\begin{align*}
    \sum_{i=1}^n \int_0^t \textbf Z_i d\Lambda_i(u,\bs{{\beta}}_0) &= n\int_0^t \frac{1}{n} \sum_{i=1}^n \textbf Z_i Y_i(u) \exp (\textbf Z_i^\top \bs{{\beta}}_0) dA_{1;0}(u) \\
&= n\int_0^t \textbf S^{(1)}_n(u,\bs{{\beta}}_0)dA_{1;0}(u)
\end{align*}
and
\begin{align*}
    \sum_{i=1}^n \int_0^t \textbf E_n(u,\bs{{\beta}}_0)  d\Lambda_i(u,\bs{{\beta}}_0) &=  \int_0^t \frac{\textbf S^{(1)}_n(u,\bs{{\beta}}_0)}{S^{(0)}_n(u,\bs{{\beta}}_0)} n S^{(0)}_n(u,\bs{{\beta}}_0)  dA_{1;0}(u)\\
    &= n \int_0^t \textbf S^{(1)}_n(u,\bs{{\beta}}_0)  dA_{1;0}(u).
\end{align*}
Thus, $\bs U_{n} (\cdot,\bs{{\beta}}_0)$ can be expressed as integrals with respect to counting process martingales, i.e.,
\begin{align*}
\bs U_{n} (t,\bs{{\beta}}_0) &= \sum_{i=1}^n \int_0^t \big( \bs Z_i - \bs E_n(u,\bs{{\beta}}_0)\big )(dM_i(u)+d\Lambda_i(u,\bs{{\beta}}_0))\\
&=\sum_{i=1}^n \int_0^t \big( \bs Z_i - \bs E_n(u,\bs{{\beta}}_0)\big) dM_i(u)    
\end{align*}
with predictable and locally bounded integrands $\bs Z_i - \bs E_n(u,\bs{{\beta}}_0)$, $i=1,\ldots ,n$. It follows with Proposition II.4.1 of \cite{Andersen} that $\bs U_{n} (\cdot,\bs{{\beta}}_0)$ is a local square integrable martingale with respect to $\mathcal{F}_1$.
\end{rem}

%%%%%%%%%%%%%%%%%%%%%%%
%%%%%%%%%%%%%%%%%%%%%%%
\subsection*{B.2 Proofs and Remarks of \cref{subsec:Fine-Gray_WB}}

\begin{rem}\label{rem:towards_asyRep_A}
According to the facts below, all assumptions necessary for the asymptotic representation  \eqref{eq:Xn-X_inD} of Part~I to hold are satisfied for $ X_n^{(2)} = \hat A_{1;0,n}(\cdot,\hat{\bs{\beta}}_n)$ and $ X^{(2)} = A_{1;0} $.
\begin{itemize}
    \item The integrand $k_{n}(t,\bs{\beta}) = J_n(t)S^{(0)}_n(t,\bs{\beta})^{-1}$ of $X_n^{(2)}$ is almost surely continuously differentiable in $\bs{\beta}$ by definition of $J_n(t)$ and $S^{(0)}_n(t,\bs{\beta})$.
    \item The regularity assumption  \eqref{eq:intLambda-X=o_p} of Part~I holds, since
\begin{align*}
\sqrt{n}\big(\frac{1}{n} \sum_{i=1}^n \int_0^t k_{n}(u, \bs{{\beta}}_0) d \Lambda_i(u, \bs {\beta}_0) - A_{1;0}(t)\big)= \sqrt{n}\big(\int_0^t (J_n(u) - 1) dA_{1;0}\big)\stackrel{\mathbb P}{\longrightarrow} 0, 
\end{align*}
as $n\rightarrow\infty$. Here we have used $\frac{1}{n} \sum_{i=1}^n d\Lambda_i(t,\bs{{\beta}}_0) = S^{(0)}_n(t,{\bs{{\beta}}_0})dA_{1;0}$, $\sup_{t\in\mathcal{T}}\sqrt{n}\lvert J_n(t) - 1\rvert = o_p(1)$, and $A_{1;0}(\tau)<\infty$.
\item The asymptotic representation \eqref{eq:beta_asy_lin} of Part~I is fulfilled because of \eqref{eq:asy_rep_beta_D_2}, which has been derived under \cref{assump2} by means of \cref{ass:A1}~\ref{ass:A1_2} and Theorem VII.2.1 of \cite{Andersen}.
\item The consistency assumption \eqref{eq:beta-Op} of Part~I, i.e., $\hat {\bs{\beta}}_n - \bs{\beta}_0 = O_p(n^{-1/2})$ holds under \cref{assump2} according to \cref{lem:results_beta}.
\end{itemize}
\end{rem}

\begin{rem}\label{rem:assump_4_I_18}
From the following facts we have that all assumptions necessary for \eqref{eq:X*-Xn_4} of Part~I to hold are satisfied for $ X_n^{*(2)} = \hat A^*_{1;0,n}(\cdot,\hat{\bs{\beta}}^*_n)$ and $ X_n^{(2)} = \hat{A}_{1;0,n}(\cdot,\hat{\bs{\beta}}_n) $.
\begin{itemize}
    \item The integrand $k_{n}(t,\bs{\beta}) = J_n(t)S^{(0)}_n(t,\bs{\beta})^{-1}$ of $X_n^{(2)}$ is almost surely continuously differentiable in $\bs{\beta}$ by definition of $J_n(t)$ and $S^{(0)}_n(t,\bs{\beta})$.
    \item We use the same wild bootstrap representation for $\sqrt{n}(\hat{\boldsymbol{\beta}}^*_n-\hat{\boldsymbol{\beta}})$ as  in \eqref{eq:beta*} of Part~I, cf.\ \eqref{eq:beta_star}.
    \item $\hat {\bs{\beta}}^*_n - \hat{\bs{\beta}}_n = O_p(n^{-1/2})$ holds under \cref{assump2} according to \cref{lem:results_beta_star}.
    \item The wild bootstrap estimator $\hat A^*_{1;0,n}(\cdot,\hat{\bs{\beta}}^*_n)$ has been obtained by applying \cref{WB_replacement} of Part~I to $\hat{A}_{1;0,n}(\cdot,\hat{\bs{\beta}}_n)$, just like $ X_n^{*(2)}$ has been obtained based on $X_n^{(2)}$.
\end{itemize}

\end{rem}

%\medskip
\noindent 
\textbf{Proof of \cref{thm:F-G_asyEquivalence}}\\ 
We write
\begin{align}\label{eq:theta_hat-theta}
\begin{split}
    \sqrt{n}(\hat{\bs\theta}_n - \bs{\theta}_0)(\cdot) &= \sqrt{n}(\hat{\bs{\beta}}_n^\top - \bs{{\beta}}_0^\top,\hat{A}_{1;0,n}(\cdot,\hat{\bs{\beta}}_n) - A_{1;0}(\cdot))^\top\\
    &= 
    \begin{pmatrix}
    \bs 0_{q\times 1} & +\textbf I_{q\times q} \cdot \textbf C_n \cdot \textbf D_{n,g}(\tau) &+ o_p(1)\\
    D_{n,k}(\cdot ) & + \textbf B_n(\cdot)\cdot \textbf C_n \cdot  \textbf D_{n,g}(\tau) &+ o_p(1)
    \end{pmatrix}\\
    &= {\textbf D}_{n,\check k}(\cdot) + \check{\textbf B}_n(\cdot)\cdot \check{\textbf C}_n \cdot {\textbf D}_{n,\check g}(\tau) + o_p(1),    
\end{split}
\end{align}
where ${\textbf D}_{n,\check k}(t) = \frac{1}{\sqrt{n}}\sum_{i=1}^n \int_0^t \check{\textbf k}_{n}(u,\bs{{\beta}}_0) dM_i(u)$ and ${\textbf D}_{n,\check g}(t) = \frac{1}{\sqrt{n}}\sum_{i=1}^n \int_0^t \check{\textbf g}_{n,i}(u,\bs{{\beta}}_0) dM_i(u)$, $t\in\mathcal{T} $, with
\[\check{\textbf k}_n (t,\bs{{\beta}}_0)=
\begin{pmatrix}
    \bs 0_{q\times 1} \\
    k_n(t,\bs{{\beta}}_0)
 \end{pmatrix}
\quad \text{and}\quad \check{\textbf g}_{n,i} (t,\bs{{\beta}}_0)=
\begin{pmatrix}
    {\textbf g}_{n,i} (t,\bs{{\beta}}_0) \\
    {\textbf g}_{n,i} (t,\bs{{\beta}}_0)
 \end{pmatrix},
 \]
and
\[\check{\textbf B}_n(t) =
\begin{pmatrix}
    \textbf I_{q\times q} & \bs 0_{q\times q} \\
    \bs 0_{1\times q}  & \textbf B_n(t)
 \end{pmatrix}
\quad \text{and}\quad  \check{\textbf C}_n =
\begin{pmatrix}
    \textbf C_n & \bs 0_{q\times q} \\
     \bs 0_{q\times q}  & \textbf C_n
 \end{pmatrix},
 \]
$t\in\mathcal{T} $, where $ \bs 0_{q\times 1}$ denotes the $q$-dimensional vector of zeros, $\bs 0_{q\times q}$ denotes the $q\times q$-dimensional matrix of zeros, $\textbf I_{q\times q}$ denotes the $q\times q$-dimensional identity matrix, and ${\textbf B}_n$ and ${\textbf C}_n$ as given in \eqref{eq:B_n-part2} and \eqref{eq:C_n-part2}, respectively. 
The main consequence of \eqref{eq:theta_hat-theta} is that the particular structure of the asymptotic representation of $\sqrt{n}(\hat{\bs{\beta}}_n - \bs{{\beta}}_0) $ and $\sqrt{n}(\hat{A}_{1;0,n}(\cdot,\hat{\bs{\beta}}_n) - A_{1;0}(\cdot))$ carries over to the structure of the asymptotic representation of $ \sqrt{n}(\hat{\bs\theta}_n - \bs{\theta}_0)$. Additionally, the components $ {\textbf D}_{n,\check k}$, $ {\textbf D}_{n,\check g}$, $\check{\textbf B}_n$ and $ \check{\textbf C}_n  $ have the same properties as $ {\textbf D}_{n,k}$, $ {\textbf D}_{n, g}$, ${\textbf B}_n$ and $ {\textbf C}_n  $. Especially, $ {\textbf D}_{n,\check k}$ and ${\textbf D}_{n,\check g}$ are square integrable martingales with respect to $\mathcal{F}_1$, respectively, and under \cref{assump2}~\ref{assump2_1}-\ref{assump2_4_new} $ ({\textbf D}_{n,\check k},{\textbf D}_{n,\check g})$ converges in law, as $n\rightarrow\infty$, to the zero-mean Gaussian vector martingale $({\textbf D}_{\tilde{\check{ k}}},{\textbf D}_{\tilde{\check {g}}})$ with covariance function 
\[
\textbf V_{(\tilde{\check{ k}},\tilde{\check{ g}})}
=
\begin{pmatrix}
    \textbf V_{\tilde{\check{ k}}} & \textbf V_{\tilde{\check{ k}} , \tilde{\check{ g}}} \\
    \textbf V_{\tilde{\check{ g}} , \tilde{\check{ k}}} & \textbf V_{\tilde{\check{ g}}}
 \end{pmatrix},
\]
where
\begin{align*}
    \textbf V_{\tilde{\check{ k}}}(t) &=  \langle \textbf D_{\tilde{\check{ k}}} \rangle (t) =\int_0^t \mathbb E( \tilde{\check{\textbf k}}(u,\boldsymbol{\beta}_0)^{\otimes 2} \lambda_1(u,\bs{{\beta}}_0))du = \int_0^t  \tilde{\check{\textbf k}}(u,\boldsymbol{\beta}_0)^{\otimes 2} s^{(0)}(u,\boldsymbol{\beta}_0)dA_{1;0}(u),\\
    \textbf V_{\tilde{\check{ g}}}(t)  &= \langle \textbf D_{\tilde{\check{ g}}} \rangle (t) = \int_0^t \mathbb E( \tilde{\check{\textbf g}}_1(u,\boldsymbol{\beta}_0)^{\otimes 2} \lambda_1(u,\bs{{\beta}}_0))du= \int_0^t  \tilde{\check{\textbf g}}_1(u,\boldsymbol{\beta}_0)^{\otimes 2} s^{(0)}(u,\boldsymbol{\beta}_0)dA_{1;0}(u),
\end{align*}
with $\tilde{\check{\textbf k}}(t,\boldsymbol{\beta}_0) = (\bs 0_{q\times 1}^\top, \tilde k(t,\boldsymbol{\beta}_0))^\top$, $\tilde{\check{\textbf g}}_1(t,\boldsymbol{\beta}_0) = (\tilde{{\textbf g}}_1(t,\boldsymbol{\beta}_0)^\top, \tilde{{\textbf g}}_1(t,\boldsymbol{\beta}_0)^\top)^\top$,
and 
\begin{align*}
\textbf V_{\tilde{\check{ k}},\tilde{\check{ g}}}(t)^\top = \textbf V_{\tilde{\check{ g}},\tilde{\check{ k}}}(t) 
&= \langle {\textbf D}_{\tilde{\check{ g}}}, {\textbf D}_{\tilde{\check{ k}}} \rangle      \int_0^t \mathbb{E}( \tilde{\check{\textbf g}}(u,\boldsymbol{\beta}_0)\cdot \tilde{\check{\textbf k}}(u,\boldsymbol{\beta}_0)^\top \lambda_1(u,\bs{{\beta}}_0))du \\
     &= \int_0^t \mathbb{E}(
    \begin{pmatrix}
    \bs 0_{q\times q} & \tilde{{\textbf g}}_1(u,\boldsymbol{\beta}_0)\cdot \tilde{{ k}}(u,\boldsymbol{\beta}_0)  \\
    \bs 0_{q\times q}  & \tilde{{\textbf g}}_1(u,\boldsymbol{\beta}_0)\cdot \tilde{{ k}}(u,\boldsymbol{\beta}_0)
    \end{pmatrix}
    \lambda_1(u,\bs{{\beta}}_0))du\\
     &=\bs 0_{2q\times (q+1)},
\end{align*}
as $\textbf V_{\tilde g,\tilde k}(t) =
%\int_0^t \mathbb{E}( \tilde{{\textbf g}}_1(u,\boldsymbol{\beta}_0)\cdot \tilde{{ k}}(u,\boldsymbol{\beta}_0)  \lambda_1(u,\bs{{\beta}}_0))du=
\bs 0_{q\times 1}$ by \eqref{eq:covar_k_g_tilde}. In particular, the orthogonality of the Gaussian martingales $D_{\tilde{{ k}}}$ and $ \textbf D_{\tilde{{ g}}} $ carries over to $\textbf D_{\tilde{\check{ k}}}$ and $ \textbf D_{\tilde{\check{ g}}} $. 
Moreover, under \cref{assump2}, the limits in probability of $\check{\textbf B}_n$ and $\check{\textbf C}_n$ are given by
\[\check{\textbf B}(t) =
\begin{pmatrix}
    \textbf I_{q\times q} & \bs 0_{q\times q} \\
    \bs 0_{1\times q}  & \textbf B (t)
 \end{pmatrix}
\quad \text{and}\quad  \check{\textbf C} =
\begin{pmatrix}
    \textbf C & \bs 0_{q\times q} \\
     \bs 0_{q\times q}  & \textbf C
 \end{pmatrix},
 \]
$t\in\mathcal{T} $, because from $\sup_{t\in\mathcal{T}}\lVert \textbf B_n(t) - \textbf B(t)\rVert = o_p(1) $ and $\lVert \textbf C_n - \textbf C \rVert = o_p(1)$, it follows that $\sup_{t\in\mathcal{T}}\lVert  \check{\textbf B}_n(t) - \check{\textbf B}(t)\rVert = o_p(1) $ and $\lVert \check{\textbf C}_n - \check{\textbf C} \rVert = o_p(1)$, respectively. %Recall from \cref{subsec:F-G-estimators} that $\textbf B = \int_0^\cdot -e(u,\bs{{\beta}}_0)dA_{1;0}(u) $ and $\textbf C = \textbf V_{\tilde g}(\tau )^{-1}$. 
Finally, under \cref{assump2} and due to \eqref{eq:theta_hat-theta} it follows with \cref{thm:Xn-X_convergence} of Part~I that
\begin{align}\label{eq:kvgz_theta-hat}
    \sqrt{n}(\hat{\bs\theta}_n - \bs{\theta}_0) \stackrel{\mathcal{L}}{\longrightarrow}{\textbf D}_{\tilde{\check{ k}}} + \check{\textbf B}\cdot \check{\textbf C} \cdot {\textbf D}_{\tilde{\check {g}}}(\tau ),\text{ in } (D(\mathcal{T}))^{(q+1)},
\end{align}
as $n\rightarrow\infty$. Furthermore, the covariance function of ${\textbf D}_{\tilde{\check{ k}}} + \check{\textbf B}\cdot \check{\textbf C} \cdot {\textbf D}_{\tilde{\check {g}}}(\tau)$ is given by
\[
   t \mapsto \textbf V_{\tilde{\check{ k}}}(t) + \check{\textbf B}(t) \cdot \check{\textbf C}\cdot\textbf V_{\tilde{\check{ g}}}(\tau)\cdot \check{\textbf C}^\top \cdot\check{\textbf B}(t)^\top  ,
\]
as $\textbf V_{\tilde{\check{ k}},\tilde{\check{ g}}}(t)^\top =  \textbf V_{\tilde{\check{ g}}, \tilde{\check{ k}}}(t)=\bs 0_{2q\times (q+1)}.$

For the wild bootstrap counterpart $\sqrt{n}(\hat{\bs\theta}_n^* - \hat{\bs\theta}_n)$ of $\sqrt{n}(\hat{\bs\theta}_n - \bs{\theta}_0) $ we have
\begin{align}\label{eq:theta*-theta_hat}
\begin{split}
    \sqrt{n}(\hat{\bs\theta}_n^* - \hat{\bs\theta}_n)(\cdot) &= \sqrt{n}(\hat{\bs{\beta}}_n^{*\top} - \hat{\bs{\beta}}_n^\top,\hat{A}^*_{1;0,n}(\cdot,\hat{\bs{\beta}}_n^*) - \hat{A}_{1;0,n}(\cdot,\hat{\bs{\beta}}_n))^\top\\
    &= 
    \begin{pmatrix}
    \bs 0_{q\times 1} & +\textbf I_{q\times q} \cdot \textbf C^*_n \cdot \textbf D^*_{n,g}(\tau) + o_p(1)\\
    D^*_{n,k}(\cdot) & + \textbf B^*_n(\cdot)\cdot \textbf C^*_n \cdot  \textbf D^*_{n,g}(\tau) + o_p(1)
    \end{pmatrix}\\
    &= {\textbf D}^*_{n,\check k}(\cdot) + \check{\textbf B}^*_n(\cdot)\cdot \check{\textbf C}^*_n \cdot {\textbf D}^*_{n,\check g}(\tau) + o_p(1),    
\end{split}
\end{align}
where ${\textbf D}^*_{n,\check k}(t) = \frac{1}{\sqrt{n}}\sum_{i=1}^n \int_0^t \check{\textbf k}_{n}(u,\hat{\bs{\beta}}_n) G_idN_i(u)$, ${\textbf D}^*_{n,\check g}(t) = \frac{1}{\sqrt{n}}\sum_{i=1}^n \int_0^t \check{\textbf g}_{n,i}(u,\hat{\bs{\beta}}_n) G_i dN_i(u)$, $t\in\mathcal{T} $, with
\[\check{\textbf k}_n (t,\hat{\bs{\beta}}_n)=
\begin{pmatrix}
    \bs 0_{q\times 1} \\
    k_n(t,\hat{\bs{\beta}}_n)
 \end{pmatrix}
\quad \text{and}\quad \check{\textbf g}_{n,i} (t,\hat{\bs{\beta}}_n)=
\begin{pmatrix}
    {\textbf g}_{n,i} (t,\hat{\bs{\beta}}_n) \\
    {\textbf g}_{n,i} (t,\hat{\bs{\beta}}_n)
 \end{pmatrix}.
 \]
Additionally,
\[\check{\textbf B}^*_n(t) =
\begin{pmatrix}
    \textbf I_{q\times q} & \bs 0_{q\times q} \\
    \bs 0_{q\times 1}^\top  & \textbf B^*_n(t)
 \end{pmatrix}
\quad \text{and}\quad  \check{\textbf C}^*_n =
\begin{pmatrix}
    \textbf C^*_n & \bs 0_{q\times q} \\
     \bs 0_{q\times q}  & \textbf C^*_n
 \end{pmatrix},
 \]
$t\in\mathcal{T} $, where ${\textbf B}^*_n $ and ${\textbf C}^*_n $ are defined in \eqref{eq:B_n^*-part2} and \eqref{eq:C_n^*-part2-def}, respectively.
Note that the structure of the asymptotic representation of $\sqrt{n}(\hat{\bs\theta}_n^* - \hat{\bs\theta}_n)(\cdot ) = \sqrt{n}(\hat{\bs{\beta}}_n^{*\top} - \hat{\bs{\beta}}_n^\top,\hat{A}^*_{1;0,n}(\cdot,\hat{\bs{\beta}}_n^*) - \hat{A}_{1;0,n}(\cdot,\hat{\bs{\beta}}_n))^\top $ resembles the structure of the asymptotic representations of its components $\sqrt{n}(\hat{\bs{\beta}}_n^{*\top} - \hat{\bs{\beta}}_n^\top) $ and $\sqrt{n}(\hat{A}^*_{1;0,n}(\cdot,\hat{\bs{\beta}}_n^*) - \hat{A}_{1;0,n}(\cdot,\hat{\bs{\beta}}_n))$.
Moreover, just like for ${\textbf D}^*_{n,k} $ and $ {\textbf D}^*_{n, g}$, it holds that ${\textbf D}^*_{n,\check k} $ and $ {\textbf D}^*_{n,\check g}$ are square integrable martingales with respect to $\mathcal{F}_2$.
Additionally, under \cref{assump2}~\ref{assump2_1}-\ref{assump2_4_new} and conditionally on $\mathcal{F}_2(0)$, it follows with \cref{lem:D*->D} of Part~I
that $({\textbf D}^{*\top}_{n,\check k} ,{\textbf D}^{*\top}_{n,\check g})^\top$ converge in law to $({\textbf D}_{\tilde{\check k}}^\top , {\textbf D}_{\tilde{\check g}}^\top)^\top$, as $n\rightarrow\infty$. Furthermore, under \cref{assump2}, we have
\[\sup_{t\in\mathcal{T}}\lVert  \check{\textbf B}^*_n(t) - \check{\textbf B}(t)\rVert\stackrel{\mathbb P}{\longrightarrow}0 \quad \text{ and } \quad \lVert \check{\textbf C}^*_n - \check{\textbf C} \rVert \stackrel{\mathbb P}{\longrightarrow}0,  \text{ as }   n\rightarrow\infty ,\]
because $\sup_{t\in\mathcal{T}}\lVert \textbf B^*_n(t) - \textbf B(t)\rVert = o_p(1) $ and $\lVert \textbf C^*_n - \textbf C \rVert = o_p(1)$. From \cref{assump2} and \eqref{eq:theta*-theta_hat} we conclude by means of \cref{thm:asyEquivalence} of Part~I that, conditionally on $\mathcal{F}_2(0)$,
\begin{align}\label{eq:kvgz_theta*}
    \sqrt{n}(\hat{\bs\theta}_n^* - \hat{\bs\theta}_n) \stackrel{\mathcal{L}}{\longrightarrow}{\textbf D}_{\tilde{\check{ k}}} + \check{\textbf B}\cdot \check{\textbf C} \cdot {\textbf D}_{\tilde{\check {g}}}(\tau ),\text{ in } (D(\mathcal{T}))^{(q+1)}, 
\end{align}
in probability as $n\rightarrow\infty$. Comparison of \eqref{eq:kvgz_theta-hat} with \eqref{eq:kvgz_theta*} leads to the final conclusion that the (conditional) distributions of $\sqrt{n}(\hat{\bs\theta}_n^* - \hat{\bs\theta}_n)$ and $\sqrt{n}(\hat{\bs\theta}_n - {\bs\theta}_0)$ are asymptotically equivalent, as $n\rightarrow\infty$. This completes the proof of \cref{thm:F-G_asyEquivalence}.
\hfill\hfill\qedsymbol

%%%%%%%%%%%%%%%%%%%%%%%
%%%%%%%%%%%%%%%%%%%%%%%
\subsection*{B.3 Proofs of \cref{subsec:Fine-Gray_CIF}}

\noindent
\textbf{Proof of \cref{lem:Hadamard}}\\
In order to derive the Hadamard derivative, we consider $\Gamma$ as the composition of the following three functionals 
\begin{align*}
    & \varphi_{Z}: (\bs x^\top,y)^\top(t) \mapsto (\exp(\textbf Z^\top \bs x),y(t))^\top;\\
    & \zeta: (x,y)(t) \mapsto x\cdot y(t);\\
    & \psi: x(t) \mapsto 1-\exp(-x(t)).
\end{align*}
This yields $$\Gamma (\tilde{\bs\theta}^{(j)})(t)  = 1-\exp\big\{-\exp(\textbf Z^\top\tilde{\bs{\beta}}^{(j)})\tilde{A}_{1;0}^{(j)}(t)\big\}= (\psi\circ\zeta\circ\varphi_{Z})(\tilde{\bs\theta}^{(j)})(t), \qquad
j=0,1,2,$$ 
where 
$\tilde{\bs\theta}^{(j)}(t) = (\tilde{\bs{\beta}}^{(j)\top},\tilde{A}_{1;0}^{(j)}(t))^\top$ 
with $\tilde{\bs{\beta}}^{(0)} = \bs{{\beta}}_0$, $\tilde{\bs{\beta}}^{(1)} = \hat{\bs{\beta}}_n$, $\tilde{\bs{\beta}}^{(2)} = \hat{\bs{\beta}}^*_n$ 
and $\tilde{A}_{1;0}^{(0)}(t) = A_{1;0}(t)$, $\tilde{A}_{1;0}^{(1)}(t) = \hat{A}_{1;0,n}(t,\hat{\bs{\beta}}_n)$, $\tilde{A}_{1;0}^{(2)}(t) = \hat{A}^*_{1;0,n}(t,\hat{\bs{\beta}}^*_n)$. 
Furthermore, with the chain rule, we obtain for $j=1,2$, 
\begin{align}
\label{eq:dgam}
\begin{split}
&\text{d}\Gamma(\tilde{\bs\theta}^{(j-1)})\cdot\sqrt{n}(\tilde{\bs\theta}^{(j)}-\tilde{\bs\theta}^{(j-1)})(t)\\
    &=\text{d}(\psi\circ\zeta\circ\varphi_{Z})({\tilde{\bs\theta}^{(j-1)}})\cdot\sqrt{n}(\tilde{\bs\theta}^{(j)}-\tilde{\bs\theta}^{(j-1)})(t)\\
    & = \text{d}\psi({\zeta(\varphi_{Z}(\tilde{\bs\theta}^{(j-1)}})))\cdot \text{d}\zeta({\varphi_{Z}(\tilde{\bs\theta}^{(j-1)}}))\cdot \text{d}\varphi_{Z}(\tilde{\bs\theta}^{(j-1)})\cdot  \sqrt{n}(\tilde{\bs\theta}^{(j)}-\tilde{\bs\theta}^{(j-1)})(t). 
\end{split}
\end{align}
Evaluating the last expression in \eqref{eq:dgam} step by step, we first get
\begin{align}
\label{eq:dphi}
\begin{split}
  &\text{d}\varphi_{Z}({\bs\theta}) \cdot (\textbf x^\top, y)^\top (t) = ( \exp(\textbf Z^\top{\bs\theta}_1)\textbf Z^\top\textbf x, y(t) )^\top\\
  &= ( \exp(\textbf Z^\top\tilde{\bs{\beta}}^{(j-1)\top})\textbf Z^\top\sqrt{n}(\tilde{\bs{\beta}}^{(j)} - \tilde{\bs{\beta}}^{(j-1)}),\sqrt{n}(\tilde{A}_{1;0}^{(j)}(t) - \tilde{A}_{1;0}^{(j-1)}(t)) )^\top
\end{split}
\end{align}
with $\bs\theta = (\bs\theta_1^\top,\theta_2)^\top =  (\tilde{\bs{\beta}}^{(j-1)\top},\tilde{A}_{1;0}^{(j-1)}(t))^\top$, $ \textbf x=\sqrt{n}(\tilde{\bs{\beta}}^{(j)} - \tilde{\bs{\beta}}^{(j-1)})$, $y(t) = \sqrt{n}(\tilde{A}_{1;0}^{(j)}(t) - \tilde{A}_{1;0}^{(j-1)}(t))$.
Then, with \eqref{eq:dphi} we find
\begin{align}
\label{eq:dzeta}
\begin{split}
&\text{d}\zeta(\bs\theta) \cdot (x,y)^\top (t) = \theta_2(t)\cdot x + \theta_1\cdot y(t)\\ 
&= \tilde{A}_{1;0}^{(j-1)}(t)\cdot \exp(\textbf Z^\top\tilde{\bs{\beta}}^{(j-1)})\textbf Z^\top\sqrt{n}(\tilde{\bs{\beta}}^{(j)} - \tilde{\bs{\beta}}^{(j-1)}) \\
&\quad + \exp(\textbf Z^\top \tilde{\bs{\beta}}^{(j-1)})\cdot \sqrt{n}(\tilde{A}_{1;0}^{(j)}(t) - \tilde{A}_{1;0}^{(j-1)}(t))
\end{split}
\end{align}
with $\bs\theta = (\theta_1,\theta_2)^\top = {\varphi_{Z}(\tilde{\bs\theta}^{(j-1)}}) =(\exp(\textbf Z^\top \tilde{\bs{\beta}}^{(j-1)}),\tilde{A}_{1;0}^{(j-1)}(t))^\top$, $(x,y)^\top(t)= \text{d}\varphi_{Z}(\tilde{\bs\theta}^{(j-1)})\cdot  \sqrt{n}(\tilde{\bs\theta}^{(j)}-\tilde{\bs\theta}^{(j-1)})(t)$. 
Finally, with \eqref{eq:dzeta} we obtain
\begin{align}
\label{eq:dpsi}
\begin{split}
    &\text{d}\psi (\theta) \cdot x(t) = \exp(-\theta(t))\cdot x(t)\\
    &=\exp\{-\exp(\textbf Z^\top \tilde{\bs{\beta}}^{(j-1)})\cdot \tilde{A}_{1;0}^{(j-1)}(t)\} \exp(\textbf Z^\top\tilde{\bs{\beta}}^{(j-1)}) \\
    & \quad \cdot \big [\tilde{A}_{1;0}^{(j-1)}(t)\cdot \textbf Z^\top\sqrt{n}(\tilde{\bs{\beta}}^{(j)} - \tilde{\bs{\beta}}^{(j-1)}) +  \sqrt{n}(\tilde{A}_{1;0}^{(j)}(t) - \tilde{A}_{1;0}^{(j-1)}(t))\big]
   \end{split}
\end{align}
with $ \theta = \zeta(\varphi_{Z}(\tilde{\bs\theta}^{(j-1)}) ) = \exp(\textbf Z^\top \tilde{\bs{\beta}}^{(j-1)})\cdot \tilde{A}_{1;0}^{(j-1)}(t)$, $x(t) = \text{d}\zeta({\varphi_{Z}(\tilde{\bs\theta}^{(j-1)}}))\cdot \text{d}\varphi_{Z}(\tilde{\bs\theta}^{(j-1)})\cdot  \sqrt{n}(\tilde{\bs\theta}^{(j)}-\tilde{\bs\theta}^{(j-1)})(t) $.
Combining \eqref{eq:dgam} and \eqref{eq:dpsi} yields \cref{lem:Hadamard}.
\hfill\hfill\qedsymbol

\bigskip
%\noindent
For the proof of \cref{thm:asyEquiv_CIF} we will use, like in Part~I, that the probability space can be modelled as a product space $ (\Omega, \mathcal A, \mathbb P) = (\Omega_1 \times \Omega_2, \mathcal A_1 \otimes \mathcal A_2, \mathbb P_1 \otimes \mathbb P_2) = (\Omega_1, \mathcal A_1, \mathbb P_1) \otimes (\Omega_2, \mathcal A_2, \mathbb P_2)$. Where necessary, we will distinguish between the probability space $(\Omega_1, \mathcal A_1, \mathbb P_1)$ underlying the data sets $\{\mathbbm{1}\{C_i \geq t\}, N_i(t), Y_i(t), \textbf{Z}_i, t\in\mathcal{T}, i=1,\ldots , n\}$, and the probability space $(\Omega_2, \mathcal A_2, \mathbb P_2)$ underlying the multipliers $G_1, \dots, G_n$. Additionally, we denote by $\stackrel{\mathcal{L}_{\mathbb P_2}}{\longrightarrow}$ the convergence in law w.r.t. the probability measure $\mathbb P_2$. Moreover, for some stochastic quantity $\textbf H_n$, we denote $\textbf H_n$ given the data as $\textbf H_n|\mathcal{F}_2(0)(\omega)$, $\omega\in\Omega_1$.

\bigskip
\noindent
\textbf{Proof of \cref{thm:asyEquiv_CIF}}\\
We wish to show that the conditional limiting distribution of $\sqrt{n}(\Gamma (\hat{\bs\theta}_n^* ) - \Gamma (\hat{\bs\theta}_n ))$ is asymptotically equivalent to the limiting distribution of $\sqrt{n}(\Gamma (\hat{\bs\theta}_n ) - \Gamma ({\bs\theta}_0 ))$. For this we recall the asymptotic representation  \eqref{eq:Gamma-hadamard} of $\sqrt{n}(\Gamma (\tilde{\bs\theta}^{(j)} ) - \Gamma (\tilde{\bs\theta}^{(j-1)} ))(t)$.
%that with the functional $\delta$-method as  in Theorem II.8.1. of \cite{Andersen} we get for $j=1,2$:
%\begin{align}\label{eq:Hadamard_proof}
%    \sqrt{n}(\Gamma (\tilde{\bs\theta}^{(j)} ) - \Gamma (\tilde{\bs\theta}^{(j-1)} ))(t) &=\text{d}\Gamma (\tilde{\bs\theta}^{(j-1)} )\cdot \sqrt{n}(\tilde{\bs\theta}^{(j)} - \tilde{\bs\theta}^{(j-1)})(t) + o_p(1),
%\end{align}
%where $\tilde{\bs\theta}^{(0)} = \bs{\theta}_0= (\bs{{\beta}}_0^\top,A_{1;0}(\cdot))^\top$, $\tilde{\bs\theta}^{(1)} = \hat{\bs\theta}_n=(\hat{\bs{\beta}}_n^\top,\hat{A}_{1;0,n}(\cdot,\hat{\bs{\beta}}_n))^\top$, and $\tilde{\bs\theta}^{(2)} = \hat{\bs\theta}^*_n=(\hat{\bs{\beta}}^{*\top}_n,\hat{A}^*_{1;0,n}(\cdot,\hat{\bs{\beta}}^*_n))^\top$. 
In the proof of \cref{lem:Hadamard} we have introduced the functional $\Gamma$ as a composition of the three functionals $\varphi_{Z}$, $\zeta$ and $\psi$. For the present proof it is useful to consider the Hadamard derivatives $\text{d}\varphi_{Z}(\tilde{\bs\theta}^{(j-1)})$, $\text{d}\zeta(\varphi_{Z}(\tilde{\bs\theta}^{(j-1)}))$ and $\text{d}\psi (\zeta(\varphi_{Z}(\tilde{\bs\theta}^{(j-1)})))$ without directly multiplying them by $\sqrt{n}(\tilde{\bs\theta}^{(j)}-\tilde{\bs\theta}^{(j-1)})$ as we did in \eqref{eq:dgam} . In particular, we now identify the Hadamard-derivatives with
\begin{align*}
    \text{d}\varphi_{Z}(\tilde{\bs\theta}^{(j-1)})&=
    \begin{pmatrix}
    \exp(\textbf Z^\top\tilde{\bs{\beta}}^{(j-1)})\textbf Z^\top & 0 \\
     \bs 0_{1\times q}  &  1
 \end{pmatrix},\\
 \text{d}\zeta(\varphi_{Z}(\tilde{\bs\theta}^{(j-1)}))&=
    \Big(
    \varphi_{Z}(\tilde{\bs\theta}^{(j-1)})_2 , \varphi_{Z}(\tilde{\bs\theta}^{(j-1)})_1  \Big)\\
 &=
 \Big(
    \tilde A^{(j-1)}_{1;0}(\cdot),  \exp (\textbf Z^\top \tilde{\bs{\beta}}^{(j-1)}) \Big),\\
  \text{d}\psi (\zeta(\varphi_{Z}(\tilde{\bs\theta}^{(j-1)})))&=\exp \{-\zeta(\varphi_{Z}(\tilde{\bs\theta}^{(j-1)})\}\\
  &= \exp\{ - \exp (\textbf Z^\top \tilde{\bs{\beta}}^{(j-1)})\cdot \tilde A^{(j-1)}_{1;0}(\cdot)\}.
\end{align*}
In the above,  $\varphi_{Z}(\cdot )_i$ denotes the i-th component of $\varphi_{Z}$, and $\tilde{\bs\theta}^{(j-1)} = (\tilde{\bs\theta}_1^{(j-1)\top},\tilde{\theta}_2^{(j-1)})^\top = (\tilde{\bs{\beta}}^{(j-1)\top},\tilde{A}_{1;0}^{(j-1)}(\cdot))^\top$ with $\tilde{\bs{\beta}}^{(0)} = \bs{{\beta}}_0$, $\tilde{\bs{\beta}}^{(1)} = \hat{\bs{\beta}}_n$, $\tilde{A}_{1;0}^{(0)}(\cdot) = A_{1;0}(\cdot)$ and $\tilde{A}_{1;0}^{(1)}(\cdot) = \hat{A}_{1;0,n}(\cdot,\hat{\bs{\beta}}_n)$. 
With the chain rule, we can express the Hadamard derivative $\text{d}\Gamma$ of $\Gamma$ as follows:
\begin{align}\label{eq:hadamard_3func}
  \text{d}\Gamma (\tilde{\bs\theta}^{(j-1)} ) = \text{d}\psi({\zeta(\varphi_{Z}(\tilde{\bs\theta}^{(j-1)}})))\cdot \text{d}\zeta({\varphi_{Z}(\tilde{\bs\theta}^{(j-1)}}))\cdot \text{d}\varphi_{Z}(\tilde{\bs\theta}^{(j-1)}).
\end{align}

We first consider the case $j=1$. In this case, $\tilde{\bs\theta}^{(j-1)} =\tilde{\bs\theta}^{(0)} = \bs{\theta}_0$ is a constant point in the space $\mathbb R^q\times \mathcal C[0,\tau] $, where $\mathcal C[0,\tau]^x$ is the set of all continuous functions mapping from $[0,\tau]$ to $\mathbb R^x$, $x\in\mathbb N$. Thus, $\big(\text{vec}(\text{d}\varphi_{Z}({\bs\theta}_0))^\top,\text{d}\zeta(\varphi_{Z}({\bs\theta}_0)),\text{d}\psi(\zeta(\varphi_{Z}({\bs\theta}_0)))\big)$ is a constant in the space $\mathbb R^{2q+2}\times \mathcal{C}[0,\tau]\times \mathbb R\times \mathcal{C}[0,\tau]\subset  \mathcal{C}[0,\tau]^{2q+5} $. We now turn to the second term of the expression on the right-hand side of \eqref{eq:Gamma-hadamard}. For $j=1$ we have $\sqrt{n}(\tilde{\bs\theta}^{(1)} - \tilde{\bs\theta}^{(0)}) = \sqrt{n}(\hat{\bs\theta}_n - \bs{\theta}_0)$ and as formulated in the proof of \cref{thm:F-G_asyEquivalence} it holds that 
\begin{align}\label{eq:theta-hat_equals_linearTerm}
    \sqrt{n}(\hat{\bs\theta}_n - \bs{\theta}_0)={\textbf D}_{n,\check k} + \check{\textbf B}_n\cdot \check{\textbf C}_n \cdot {\textbf D}_{n,\check g}(\tau) + o_p(1).
\end{align}
From the proof of \cref{thm:Xn-X_convergence} it follows that the convergence in distribution of this term is based on the joint convergence in distribution of $\big({\textbf D}_{n,\check k}^\top,{\textbf D}_{n,\check g}^\top,\text{vec}(\check{\textbf B}_n)^\top,\text{vec}(\check{\textbf C}_n)\big)$ to $\big({\textbf D}_{\tilde{\check k}}^\top,{\textbf D}_{\tilde{\check g}}^\top,\text{vec}(\check{\textbf B})^\top,\text{vec}(\check{\textbf C})^\top\big)$, as $n\rightarrow\infty$, with $\big({\textbf D}_{\tilde{\check k}}^\top,{\textbf D}_{\tilde{\check g}}^\top,\text{vec}(\check{\textbf B})^\top,\text{vec}(\check{\textbf C})^\top\big) \in \mathcal{C}[0,\tau]^{10q+2}$. 
From the continuous mapping theorem and the maps $f_1, f_2$, and $f_3$ defined in the proof of \cref{thm:Xn-X_convergence} it follows that 
\[{\textbf D}_{n,\check k} + \check{\textbf B}_n\cdot \check{\textbf C}_n \cdot {\textbf D}_{n,\check g}(\tau)\stackrel{\mathcal{L}}{\longrightarrow} {\textbf D}_{\tilde{\check k}} + \check{\textbf B}\cdot \check{\textbf C} \cdot {\textbf D}_{\tilde{\check g}}(\tau), \text{ in } \mathcal{D}[0,\tau]^{(q+1)}, \text{ as } n\rightarrow\infty.\]
In order to derive the convergence in distribution of $\text{d}\Gamma ({\bs\theta}_0 )\cdot \sqrt{n}(\hat{\bs\theta}_n - {\bs\theta}_0)(t)$, we enlarge 
$$\big({\textbf D}_{n,\check k}^\top,{\textbf D}_{n,\check g}^\top,\text{vec}(\check{\textbf B}_n)^\top,\text{vec}(\check{\textbf C}_n)^\top\big)$$ by $\big(\text{vec}(\text{d}\varphi_{Z}({\bs\theta}_0))^\top,\text{d}\zeta(\varphi_{Z}({\bs\theta}_0)),\text{d}\psi(\zeta(\varphi_{Z}({\bs\theta}_0)))\big)$. As the first vector converges in distribution to a limit that is continuous and thus separable, and the latter vector is a constant of the space $\mathcal{C}[0,\tau]^{2q+5}, $
it holds according to Example 1.4.7 of \cite{Vaart_Wellner} that
\begin{align}\label{eq:extended_vec_hat}
\begin{split}
    &\big({\textbf D}_{n,\check k}^\top,{\textbf D}_{n,\check g}^\top,\text{vec}(\check{\textbf B}_n)^\top,\text{vec}(\check{\textbf C}_n)^\top,\text{vec}(\text{d}\varphi_{Z}({\bs\theta}_0))^\top,\text{d}\zeta(\varphi_{Z}({\bs\theta}_0)),\text{d}\psi(\zeta(\varphi_{Z}({\bs\theta}_0)))\big)\\
    &\stackrel{\mathcal{L}}{\longrightarrow} \big({\textbf D}_{\tilde{\check k}}^\top,{\textbf D}_{\tilde{\check g}}^\top,\text{vec}(\check{\textbf B})^\top,\text{vec}(\check{\textbf C})^\top,\text{vec}(\text{d}\varphi_{Z}({\bs\theta}_0))^\top,\text{d}\zeta(\varphi_{Z}({\bs\theta}_0)),\text{d}\psi(\zeta(\varphi_{Z}({\bs\theta}_0)))\big),
\end{split}
\end{align}
in $\mathcal{D}[0,\tau]^{12q + 7}$, as $n\rightarrow\infty$. Next, we make use of the continuous mapping theorem. For this we consider the following map
\begin{alignat*}{2}
%\begin{array}{r@{}l}
& f_4:\; &&\big([{\textbf D}_{n,\check k} + \check{\textbf B}_n\cdot \check{\textbf C}_n \cdot {\textbf D}_{n,\check g}(\tau) ]^\top,\text{vec}(\text{d}\varphi_{Z}({\bs\theta}_0))^\top,\text{d}\zeta(\varphi_{Z}({\bs\theta}_0)),\text{d}\psi(\zeta(\varphi_{Z}({\bs\theta}_0)))\big) \\
& &&\mapsto \big(\text{d}\psi(\zeta(\varphi_{Z}({\bs\theta}_0)))\cdot\text{d}\zeta(\varphi_{Z}({\bs\theta}_0))\cdot \text{d}\varphi_{Z}({\bs\theta}_0)\cdot[\textbf D_{n,k} + \textbf B_n\cdot \textbf C_n\cdot  \textbf D_{n,g}(\tau)]\big)
%\big((\text{d}\varphi_{Z}({\bs\theta}_0)\cdot[\textbf D_{n,k} + \textbf B_n\cdot \textbf C_n\cdot  \textbf D_{n,g}(\tau)])^\top,\text{d}\zeta(\varphi_{Z}({\bs\theta}_0)),\text{d}\psi(\zeta(\varphi_{Z}({\bs\theta}_0)))\big)\\
%& f_5: &&\big((\text{d}\varphi_{Z}({\bs\theta}_0)\cdot[\textbf D_{n,k} + \textbf B_n\cdot \textbf C_n\cdot  \textbf D_{n,g}(\tau)])^\top,\text{d}\zeta(\varphi_{Z}({\bs\theta}_0)),\text{d}\psi(\zeta(\varphi_{Z}({\bs\theta}_0)))\big) \\
%&  &&\mapsto \big(\text{d}\zeta(\varphi_{Z}({\bs\theta}_0))\cdot \text{d}\varphi_{Z}({\bs\theta}_0)\cdot[\textbf D_{n,k} + \textbf B_n\cdot \textbf C_n\cdot  \textbf D_{n,g}(\tau)],\text{d}\psi(\zeta(\varphi_{Z}({\bs\theta}_0)))\big)\\
%& f_6: &&\big(\text{d}\zeta(\varphi_{Z}({\bs\theta}_0))\cdot \text{d}\varphi_{Z}({\bs\theta}_0)\cdot[\textbf D_{n,k} + \textbf B_n\cdot \textbf C_n\cdot  \textbf D_{n,g}(\tau)],\text{d}\psi(\zeta(\varphi_{Z}({\bs\theta}_0)))\big) \\
%&  &&\mapsto \big(\text{d}\psi(\zeta(\varphi_{Z}({\bs\theta}_0)))\cdot\text{d}\zeta(\varphi_{Z}({\bs\theta}_0))\cdot \text{d}\varphi_{Z}({\bs\theta}_0)\cdot[\textbf D_{n,k} + \textbf B_n\cdot \textbf C_n\cdot  \textbf D_{n,g}(\tau)]\big)
%&f_2: (\textbf D_{n,k}^\top, \textbf D_{n,g}(\tau)^\top, \text{vec}(\textbf B_n\cdot \textbf C_n)^\top) \mapsto (\textbf D_{n,k}^\top, (\textbf B_n\cdot \textbf C_n\cdot  \textbf D_{n,g}(\tau))^\top)\\
%&f_3: (\textbf D_{n,k}^\top, (\textbf B_n\cdot \textbf C_n\cdot  \textbf D_{n,g}(\tau))^\top) \mapsto \big(\textbf D_{n,k} + \textbf B_n\cdot \textbf C_n\cdot  \textbf D_{n,g}(\tau)\big).\\
%\end{array}
\end{alignat*}
Since 
\begin{align}
    \label{eq:12q+7}
    \begin{split}
\big({\textbf D}_{\tilde{\check k}}^\top,{\textbf D}_{\tilde{\check g}}^\top,&\text{vec}(\check{\textbf B})^\top,\text{vec}(\check{\textbf C})^\top,\text{vec}(\text{d}\varphi_{Z}({\bs\theta}_0))^\top,\text{d}\zeta(\varphi_{Z}({\bs\theta}_0)),\text{d}\psi(\zeta(\varphi_{Z}({\bs\theta}_0)))\big)\\
&\in \mathcal{C}[0,\tau]^{12q+7},
\end{split}
\end{align}
it follows successively with the continuous mapping theorem and the maps $f_1$, $f_2$, $f_3$, and $f_4$ applied to \eqref{eq:extended_vec_hat} that
\begin{align}\label{eq:CMT_applied2Hadamard}
\begin{split}
    &\text{d}\psi(\zeta(\varphi_{Z}({\bs\theta}_0)))\cdot\text{d}\zeta(\varphi_{Z}({\bs\theta}_0))\cdot \text{d}\varphi_{Z}({\bs\theta}_0)\cdot[\textbf D_{n,k} + \textbf B_n\cdot \textbf C_n\cdot  \textbf D_{n,g}(\tau)]\\
    &\stackrel{\mathcal{L}}{\longrightarrow}\text{d}\psi(\zeta(\varphi_{Z}({\bs\theta}_0)))\cdot\text{d}\zeta(\varphi_{Z}({\bs\theta}_0))\cdot \text{d}\varphi_{Z}({\bs\theta}_0)\cdot [\textbf D_{\tilde {\check k}} + \check{\textbf B }\cdot \check{\textbf C}\cdot \textbf D_{\tilde{\check g}}(\tau)],
\end{split}
\end{align}
in $  D[0,\tau]^{q+1}$, as $n\rightarrow\infty$. In conclusion, \eqref{eq:Gamma-hadamard}, \eqref{eq:hadamard_3func}, \eqref{eq:theta-hat_equals_linearTerm}, and \eqref{eq:CMT_applied2Hadamard} combined yield
\begin{align}\label{eq:CIF-hat}
\sqrt{n}(\Gamma (\hat{\bs\theta}_n ) - \Gamma ({\bs\theta}_0))\stackrel{\mathcal{L}}{\longrightarrow}\text{d}\Gamma ({\bs\theta}_0 )\cdot [\textbf D_{\tilde {\check k}} + \check{\textbf B }\cdot \check{\textbf C}\cdot \textbf D_{\tilde{\check g}}(\tau)],\text{ in }  \mathcal D[0,\tau]^{q+1}, \text{ as } n\rightarrow\infty.
\end{align}
This completes the proof for the case $j=1$.

For the case $j=2$, we have $\tilde{\bs\theta}^{(j-1)} =\tilde{\bs\theta}^{(1)} = \hat{\bs\theta}_n$ and
\[ \hat{\bs\theta}_n\stackrel{\mathbb P}{\longrightarrow} \bs{\theta}_0, \text{ as } n\rightarrow\infty, \] 
follows from \cref{thm:F-G_asyEquivalence}. Recall that $\bs{\theta}_0\in\mathbb R^q\times \mathcal C[0,\tau]$ holds. Thus, $\hat{\bs\theta}_n$ is asymptotically degenerate. Furthermore, $\text{d}\varphi_{Z}(\cdot )$ is continuous at every point of the set $\mathbb R^q\times \mathcal C[0,\tau]  $.
%, where $\cdot$ indicates that the space to which the second argument $\hat{A}_{1;0,n}(t,\hat{\bs{\beta}}_n)$ of $\hat{\bs\theta}_n$ belongs is arbitrary.
Hence, with the continuous mapping theorem as in, e.g., Theorem 1.3.6 of \cite{Vaart_Wellner} we get
\[ \text{d}\varphi_{Z}(\hat{\bs\theta}_n)\stackrel{\mathbb P}{\longrightarrow} \text{d}\varphi_{Z}(\bs{\theta}_0), \text{ as } n\rightarrow\infty. \] 
Moreover, $\varphi_{Z}(\cdot) $ is continuous at all points of the space $\mathbb R^q\times \mathcal C[0,\tau] $ mapping the space $\mathbb R^q\times \mathcal C[0,\tau] $ to $\mathbb R\times \mathcal C[0,\tau] $. Thus, by means of the continuous mapping theorem we have
\[ \varphi_{Z}(\hat{\bs\theta}_n)\stackrel{\mathbb P}{\longrightarrow} \varphi_{Z}(\bs{\theta}_0), \text{ as } n\rightarrow\infty. \] 
Furthermore, $\text{d}\zeta(\cdot )$ is a continuous at all points of the space $\mathbb R\times \mathcal C[0,\tau] $. Hence, it follows again with the continuous mapping theorem that
\[ \text{d}\zeta(\varphi_{Z}(\hat{\bs\theta}_n))\stackrel{\mathbb P}{\longrightarrow} \text{d}\zeta(\varphi_{Z}(\bs{\theta}_0)), \text{ as } n\rightarrow\infty. \] 
Additionally, $\zeta(\cdot )$ is continuous at all points of the set $\mathbb R\times \mathcal C[0,\tau] $ and maps the space $\mathbb R\times \mathcal C[0,\tau] $ to $\mathcal C[0,\tau]$. This yields
\[ \zeta(\varphi_{Z}(\hat{\bs\theta}_n))\stackrel{\mathbb P}{\longrightarrow} \zeta(\varphi_{Z}(\bs{\theta}_0)), \text{ as } n\rightarrow\infty,\] 
according to the continuous mapping theorem. Finally, $\text{d}\psi(\cdot )$ is continuous at all points of the set $\mathcal C[0,\tau]$. Hence, with the continuous mapping theorem we get 
\[ \text{d}\psi(\zeta(\varphi_{Z}(\hat{\bs\theta}_n)))\stackrel{\mathbb P}{\longrightarrow} \text{d}\psi(\zeta(\varphi_{Z}(\bs{\theta}_0))), \text{ as } n\rightarrow\infty.\] 
In conclusion, $\text{d}\varphi_{Z}(\hat{\bs\theta}_n)$, $\text{d}\zeta(\varphi_{Z}(\hat{\bs\theta}_n))$, and $\text{d}\psi(\zeta(\varphi_{Z}(\hat{\bs\theta}_n)))$ are asymptotically degenerate. It immediately follows that 
\begin{align}\label{eq:Hadamard_vec_cond}
\begin{split}
    &\big( \text{vec}(\text{d}\varphi_{Z}(\hat{\bs\theta}_{n}))^\top,\text{d}\zeta(\varphi_{Z}(\hat{\bs\theta}_{n})),\text{d}\psi(\zeta(\varphi_{Z}(\hat{\bs\theta}_{n}))) \big)\\
    &\stackrel{\mathbb P}{\longrightarrow}\big( \text{vec}(\text{d}\varphi_{Z}({\bs\theta}_0))^\top,\text{d}\zeta(\varphi_{Z}({\bs\theta}_0)),\text{d}\psi(\zeta(\varphi_{Z}({\bs\theta}_0)))\big),\text{ as } n\rightarrow\infty.
    %&\text{d}\varphi_{Z}(\hat{\bs\theta}_n)\stackrel{\mathbb P}{\longrightarrow}\text{d}\varphi_{Z}({\bs\theta}_0), \text{ as } n\rightarrow\infty;\\
    %&\text{d}\zeta(\varphi_{Z}(\hat{\bs\theta}_n))\stackrel{\mathbb P}{\longrightarrow}\text{d}\zeta(\varphi_{Z}({\bs\theta}_0)), \text{ as } n\rightarrow\infty;\\
    %&\text{d}\psi(\zeta(\varphi_{Z}(\hat{\bs\theta}_n)))\stackrel{\mathbb P}{\longrightarrow}\text{d}\psi(\zeta(\varphi_{Z}({\bs\theta}_0))), \text{ as } n\rightarrow\infty.
\end{split} 
\end{align}
By means of the notation introduced just outside the proof of \cref{thm:asyEquiv_CIF}, by Fact 1 of the supplement of \cite{dobler19}, which states that convergence in probability is equivalent to convergence in conditional probability, and by the subsequence principle, we can infer from \eqref{eq:Hadamard_vec_cond} that for every subsequence $n_1$ of $n$ there exists a further subsequence $n_2$ such that 
\begin{align}\label{eq:Hadamard_conditional}
\begin{split}
    &\big( \text{vec}(\text{d}\varphi_{Z}(\hat{\bs\theta}_{n_2}))^\top,\text{d}\zeta(\varphi_{Z}(\hat{\bs\theta}_{n_2})),\text{d}\psi(\zeta(\varphi_{Z}(\hat{\bs\theta}_{n_2}))) \big)|\mathcal{F}_2(0)(\omega)\\
    &{\longrightarrow}\big( \text{vec}(\text{d}\varphi_{Z}({\bs\theta}_0))^\top,\text{d}\zeta(\varphi_{Z}({\bs\theta}_0)),\text{d}\psi(\zeta(\varphi_{Z}({\bs\theta}_0)))\big),\text{ as } n\rightarrow\infty,
    %&\text{d}\varphi_{Z}(\hat{\bs\theta}_{n_2})|\mathcal{F}_2(0)(\omega)\stackrel{\sout{\mathbb P_2}}{\longrightarrow}\text{d}\varphi_{Z}({\bs\theta}_0), \text{ as } n\rightarrow\infty;\\
    %&\text{d}\zeta(\varphi_{Z}(\hat{\bs\theta}_{n_2}))|\mathcal{F}_2(0)(\omega)\stackrel{\sout{\mathbb P_2}}{\longrightarrow}\text{d}\zeta(\varphi_{Z}({\bs\theta}_0)), \text{ as } n\rightarrow\infty;\\
    %&\text{d}\psi(\zeta(\varphi_{Z}(\hat{\bs\theta}_{n_2})))|\mathcal{F}_2(0)(\omega)\stackrel{\sout{\mathbb P_2}}{\longrightarrow}\text{d}\psi(\zeta(\varphi_{Z}({\bs\theta}_0))), \text{ as } n\rightarrow\infty
    \end{split}
\end{align}
for $\mathbb P_1$-almost all $\omega\in\Omega_1$. Moreover, for $j=2$, we have $\sqrt{n}(\tilde{\bs\theta}^{(2)} - \tilde{\bs\theta}^{(1)}) = \sqrt{n}(\hat{\bs\theta}_n^* - \hat{\bs\theta}_n) $ for which it follows according to the proof of \cref{thm:F-G_asyEquivalence} that 
\begin{align}\label{eq:term_BCD_WB}
    \sqrt{n}(\hat{\bs\theta}_n^* - \hat{\bs\theta}_n) = {\textbf D}^*_{n,\check k} + \check{\textbf B}^*_n\cdot \check{\textbf C}^*_n \cdot {\textbf D}^*_{n,\check g}(\tau) + o_p(1). 
\end{align}
According to the proof of \cref{thm:asyEquivalence}, we know that $({\textbf D}^*_{n_6,\check k},{\textbf D}^*_{n_6,\check g},\check{\textbf B}^*_{n_6},\check{\textbf C}^*_{n_6})|\mathcal{F}_2(0)(\omega)$ converges in $\mathbb P_2$-law to $({\textbf D}_{\tilde{\check k}},{\textbf D}_{\tilde{\check g}},\check{\textbf B},\check{\textbf C})$ for $\mathbb P_1$-almost all $\omega\in\Omega_1$, as $n\rightarrow\infty$. Additionally, by means of the continuous mapping theorem and the maps $f_1 $, $f_2$, and $ f_3$, which are defined in that proof, it follows that 
\begin{align}\label{eq:term_BCD_conditional_2}
    {\textbf D}^*_{n_6,\check k} + \check{\textbf B}^*_{n_6}\cdot \check{\textbf C}^*_{n_6} \cdot {\textbf D}^*_{n_6,\check g}(\tau)|\mathcal{F}_2(0)(\omega)\stackrel{\mathcal{L}_{\mathbb P_2}}{\longrightarrow} {\textbf D}_{\tilde{\check k}} + \check{\textbf B}\cdot \check{\textbf C} \cdot {\textbf D}_{\tilde{\check g}}(\tau), \text{ in } \mathcal{D}[0,\tau]^{(q+1)}, 
\end{align}
as $n\rightarrow\infty$ and for $\mathbb P_1$-almost all $\omega\in\Omega_1$. Clearly, the convergence in \eqref{eq:Hadamard_conditional} and \eqref{eq:term_BCD_conditional_2} holds along  a joint subsequence $n_8$ as well. We also have that the limit in law with respect to $\mathbb P_2$ of $({\textbf D}^{*\top}_{n_8,\check k},{\textbf D}^{*\top}_{n_8,\check g},\text{vec}(\check{\textbf B}^*_{n_8})^\top,\text{vec}(\check{\textbf C}^*_{n_8})^\top)|\mathcal{F}_2(0)(\omega)$ is separable for $\mathbb P_1$-almost all $\omega\in\Omega$ and $\big(\text{vec}(\text{d}\varphi_{Z}(\hat{\bs\theta}_{n_8}))^\top,\text{d}\zeta(\varphi_{Z}(\hat{\bs\theta}_{n_8})),\text{d}\psi(\zeta(\varphi_{Z}(\hat{\bs\theta}_{n_8})))\big)|\mathcal{F}_2(0)(\omega)$ is asymptotically degenerate. Therefore, we can conclude based on Example 1.4.7 of \cite{Vaart_Wellner} that, conditionally on 
%the data, 
$\mathcal{F}_2(0)(\omega)$,
\begin{align}\label{eq:extended_vec_hat_2}
\begin{split}
    &\big({\textbf D}^{*\top}_{n_8,\check k},{\textbf D}^{*\top}_{n_8,\check g},\text{vec}(\check{\textbf B}^{*}_{n_8})^\top,\text{vec}(\check{\textbf C}^{*}_{n_8})^\top,\text{vec}(\text{d}\varphi_{Z}(\hat{\bs\theta}_{n_8}))^\top,\text{d}\zeta(\varphi_{Z}(\hat{\bs\theta}_{n_8})),\text{d}\psi(\zeta(\varphi_{Z}(\hat{\bs\theta}_{n_8})))\big)\\
    &\stackrel{\mathcal{L}_{\mathbb P_2}}{\longrightarrow} \big({\textbf D}_{\tilde{\check k}}^\top,{\textbf D}_{\tilde{\check g}}^\top,\text{vec}(\check{\textbf B})^\top,\text{vec}(\check{\textbf C})^\top,\text{vec}(\text{d}\varphi_{Z}({\bs\theta}_0))^\top,\text{d}\zeta(\varphi_{Z}({\bs\theta}_0)),\text{d}\psi(\zeta(\varphi_{Z}({\bs\theta}_0)))\big),
\end{split}
\end{align}
in $\mathcal{D}[0,\tau]^{12q+7}$, as $n\rightarrow\infty$ and for $\mathbb P_1$-almost all $\omega\in\Omega_1$. 
%Recall, that we have $\big({\textbf D}_{\tilde{\check k}}^\top,{\textbf D}_{\tilde{\check g}}^\top,\text{vec}(\check{\textbf B})^\top,\text{vec}(\check{\textbf C})^\top,\text{vec}(\text{d}\varphi_{Z}({\bs\theta}_0))^\top,\text{d}\zeta(\varphi_{Z}({\bs\theta}_0)),\text{d}\psi(\zeta(\varphi_{Z}({\bs\theta}_0)))\big)\in \mathcal{C}[0,\tau]^{12q+7}$. 
From \eqref{eq:12q+7},  the continuous mapping theorem, and application of the maps  $f_1 $, $f_2$, $ f_3$, and $f_4$ to \eqref{eq:extended_vec_hat_2} it follows that 
\begin{align}\label{eq:CMT_applied2Hadamard_2}
\begin{split}
    &\text{d}\psi(\zeta(\varphi_{Z}(\hat{\bs\theta}_{n_8})))\cdot\text{d}\zeta(\varphi_{Z}(\hat{\bs\theta}_{n_8}))\cdot \text{d}\varphi_{Z}(\hat{\bs\theta}_{n_8})\cdot[\textbf D^*_{n_8,k} + \textbf B^*_{n_8}\cdot \textbf C^*_{n_8}\cdot  \textbf D^*_{n_8,g}(\tau)]|\mathcal{F}_2(0)(\omega)\\
    &\stackrel{\mathcal{L}_{\mathbb P_2}}{\longrightarrow}\text{d}\psi(\zeta(\varphi_{Z}({\bs\theta}_0)))\cdot\text{d}\zeta(\varphi_{Z}({\bs\theta}_0))\cdot \text{d}\varphi_{Z}({\bs\theta}_0)\cdot [\textbf D_{\tilde {\check k}} + \check{\textbf B }\cdot \check{\textbf C}\cdot \textbf D_{\tilde{\check g}}(\tau)],
\end{split}
\end{align}
in $\mathcal{D}[0,\tau]^{(q+1)}$, as $n\rightarrow\infty$ and for $\mathbb P_1$-almost all $\omega\in\Omega_1$. Eventually, by invoking the subsequence principle again and combining \eqref{eq:Gamma-hadamard}, \eqref{eq:hadamard_3func}, \eqref{eq:term_BCD_WB}, and \eqref{eq:CMT_applied2Hadamard_2}, we find that, conditionally on $\mathcal{F}_2(0)$,
\begin{align}\label{eq:CIF-star}
    \sqrt{n}(\Gamma (\hat{\bs\theta}^*_n ) - \Gamma (\hat{\bs\theta}_n ))\stackrel{\mathcal{L}_{\mathbb P_2}}{\longrightarrow}\text{d}\Gamma ({\bs\theta}_0 )\cdot [\textbf D_{\tilde {\check k}} + \check{\textbf B }\cdot \check{\textbf C}\cdot \textbf D_{\tilde{\check g}}(\tau)],\text{ in } \mathcal D[0,\tau]^{q+1}, \text{ as } n\rightarrow \infty,
\end{align}
in $\mathbb P_1$-probability. This completes the proof for the case $j=2$.

As the (conditional) limits in distribution of $\sqrt{n}(\Gamma (\hat{\bs\theta}_n ) - \Gamma ({\bs\theta}_0 ))$ and $\sqrt{n}(\Gamma (\hat{\bs\theta}^*_n ) - \Gamma (\hat{\bs\theta}_n ))$  in \eqref{eq:CIF-hat} and \eqref{eq:CIF-star}, respectively, are the same, we have proved
\cref{thm:asyEquiv_CIF}.
%, i.e.,
%\[d[\mathcal{L}_{\mathbb P_2}(\sqrt{n}(\Gamma (\hat{\bs\theta}_n^*) - \Gamma( \hat{\bs\theta}_n))|\mathcal{F}_2(0)),\mathcal{L}_{\mathbb P_1}(\sqrt{n}(\Gamma (\hat{\bs\theta}_n) - \Gamma ({\bs{\theta}_0})))]\stackrel{\mathbb P_1}{\longrightarrow} 0, \text{ as } n\rightarrow \infty.\]
%This completes the proof of \cref{thm:asyEquiv_CIF}. 
\hfill\hfill\qedsymbol
\label{lastpage}

%%%%%%%%%%%%%%%%%%%%%%%
%%%%%%%%%%%%%%%%%%%%%%%
\subsection*{B.4 Proofs of \cref{sec:CBs}}

\noindent
\textbf{Proof of \cref{lem:sigma_hat_sigma_star}}\\
We first show that under \cref{assump2},  $\hat{\sigma}^2_n(t)$ defined in \eqref{eq:cov_sigma} is a consistent estimator of the variance of $W_{n,\phi,1}(t)$ for $t\in\mathcal{T}$. For this, we point out that
%for  $n\rightarrow \infty$, 
\begin{align}
\begin{split}
\label{asymW}
W_{n,\phi,1}(t)
& = \sqrt{n}(\textbf Z^\top(\hat{\bs{\beta}}_n-\bs{{\beta}}_0) + \log(\hat{A}_{1;0,n}(t,\hat{\bs{\beta}}_n)) - \log(A_{1;0}(t)))\\ 
& = \sqrt{n}(\textbf Z^\top(\hat{\bs{\beta}}_n-\bs{{\beta}}_0) + \log^\prime (A_{1;0}(t) ) (\hat{A}_{1;0,n}(t,\hat{\bs{\beta}}_n) - A_{1;0}(t))) + o_p(1).  
\end{split}
\end{align}
%on which $\hat{\sigma}^2_n$ will be based. 
From  \cref{lem:results_beta} and \eqref{eq:In_pred-cov-Dg} we see that under \cref{assump2},  the asymptotic  covariance matrix of  $\sqrt{n}(\hat{\bs{\beta}}_n-\bs{{\beta}}_0)$ equals  $\textbf C$. 
%According to \eqref{eq:In_pred-cov-Dg} it holds under \cref{assump2} that
%$$\textbf C_n = \big(\frac{1}{n}\textbf I_n(\tau ,\boldsymbol{\beta}_0)\big)^{-1}\stackrel{\mathbb{P}}{\longrightarrow} \textbf V_{\tilde g}(\tau)^{-1} =  \textbf C,\text{ as }n\rightarrow\infty.$$ 
Moreover, in view of \eqref{eq:In_pred-cov-Dg} we have that  $\lVert \textbf I_n(\tau ,\hat{\boldsymbol{\beta}}_n) - \textbf I_n(\tau ,\boldsymbol{\beta}_0) \rVert = o_p(1)$, since $\hat{\boldsymbol{\beta}}_n$ is a consistent estimator of $\boldsymbol{\beta}_0$. Hence,  $\big(\frac{1}{n}\bs I_n(\tau,\hat{\bs{\beta}}_n)\big)^{-1}$ is a  consistent estimator of the covariance of $\sqrt{n}(\hat{\bs{\beta}}_n-\bs{{\beta}}_0)$, cf.\ Corollary VII.2.4 of \cite{Andersen}.
Next, 
%we derive the uniformly consistent estimator of the variance function of $\sqrt{n}(\hat{A}_{1;0,n}(t,\hat{\bs{\beta}}_n) - A_{1;0}(t))$. For this, we recall the covariance function of the limiting distribution $  D_{\tilde k} + \textbf B  \cdot \textbf C\cdot \textbf D_{\tilde g}(\tau)$ of $\sqrt{n}(\hat{A}_{1;0,n}(t,\hat{\bs{\beta}}_n) - A_{1;0}(t))$ stated 
from \cref{lem:results_A} 
%under \cref{assump2}, 
%i.e.,
%$$t \mapsto  V_{\tilde k}(t) + \textbf B(t) \cdot \textbf C \cdot \textbf B(t)^\top,$$
%with $V_{\tilde k}(t) =  \int_0^t  s^{(0)}(u,\bs{{\beta}}_0)^{ -1}dA_{1;0}(u)$, and $\textbf B(t) = \int_0^t - e(u,\bs{{\beta}}_0)^\top dA_{1;0}(u)$, $t\in\mathcal{T}$. Thus, 
it is easy to see that under \cref{assump2},
\begin{align*}
&\int_0^{t} S^{(0)}_n(u,\hat{\bs{\beta}}_n)^{-1}d\hat{A}_{1;0,n}(u,\hat{\bs{\beta}}_n) \\
&+ \int_0^t\textbf E_n(u,\hat{\bs{\beta}}_n)^\top d \hat A_{1;0,n}(u,\hat{\bs{\beta}}_n)  \big(\frac{1}{n}\bs I_n(\tau,\hat{\bs{\beta}}_n)\big)^{-1} \int_0^t \textbf E_n(u,\hat{\bs{\beta}}_n) d \hat A_{1;0,n}(u,\hat{\bs{\beta}}_n),     
\end{align*}
is a uniformly consistent estimator of the variance function of $\sqrt{n}(\hat{A}_{1;0,n}(t,\hat{\bs{\beta}}_n) - A_{1;0}(t))$, $t\in\mathcal{T}$.
%cf. Corollary VII.2.4 of \cite{Andersen}. 
As according  to \cref{lem:results_beta} and \cref{lem:results_A}, and due to the asymptotic orthogonality of $D_{\tilde k}$ and $\textbf D_{\tilde g}$, it holds that under \cref{assump2}, the covariance function of $ \textbf C\cdot \textbf D_{\tilde g}(\tau)$ and $  D_{\tilde k} + \textbf B  \cdot \textbf C\cdot \textbf D_{\tilde g}(\tau)$ equals $\textbf C\cdot \textbf{B}^\top $. Hence, we find that 
%\begin{align*}
    $-\Big(\frac{1}{n}\bs I_n(\tau,\hat{\bs{\beta}}_n)\Big)^{-1} \int_0^t \textbf E_n(u,\hat{\bs{\beta}}_n) d \hat A_{1;0,n}(u,\hat{\bs{\beta}}_n),$ $t\in\mathcal{T},$
%\end{align*}  
is a uniformly consistent estimator of the  covariance of $\sqrt{n}(\hat{\bs{\beta}}_n-\bs{{\beta}}_0)$ and $\sqrt{n}(\hat{A}_{1;0,n}(t,\hat{\bs{\beta}}_n) - A_{1;0}(t))$.
%with the following uniformly consistent estimator
%\begin{align*}
%    -\Big(\frac{1}{n}\bs I_n(\tau,\hat{\bs{\beta}}_n)\Big)^{-1} \int_0^t \textbf E_n(u,\hat{\bs{\beta}}_n) d \hat A_{1;0,n}(u,\hat{\bs{\beta}}_n),\quad t\in\mathcal{T};
%\end{align*} 
%cf. Corollary VII.2.5 of \cite{Andersen}. 
Combining this with \eqref{asymW}, we see that under \cref{assump2}, $\hat{\sigma}^2_n(t)$ defined in  \eqref{eq:cov_sigma} is a consistent estimator of the variance of $W_{n,\phi,1}(t)$, $t\in\mathcal{T}$.
%the following consistent estimator for the variance of $W_{n,\phi , 1}$
%\begin{align}\label{eq:cov_sigma}
%\begin{split}
%\hat{\sigma}^2_n(t) = \hat{A}_{1;0,n}(t,\hat{\bs{\beta}}_n)^{-2} &\Big[ %\int_0^{t} S^{(0)}_n(u,\hat{\bs{\beta}}_n)^{-1}d\hat{A}_{1;0,n}(u,\hat{\bs%{\beta}}_n)\\
%   & +\int_0^t (\textbf Z - \textbf E_n(u,\hat{\bs{\beta}}_n))^\top d\hat{A}_{1;0,n}(u,\hat{\bs{\beta}}_n) \Big(\frac{1}{n}\bs I_n(\tau,\hat{\bs{\beta}}_n)\Big)^{-1}\\
%    & \cdot \int_0^t (\textbf Z - \textbf E_n(u,\hat{\bs{\beta}}_n)) d\hat{A}_{1;0,n}(u,\hat{\bs{\beta}}_n)  \Big].
%\end{split}
%\end{align}

We now consider the wild bootstrapped variance estimator $\hat{\sigma}^{*2}_n(t)$, $t\in\mathcal{T}$, from \eqref{eq:WB_variance_estimator}. According to \cref{thm:F-G_asyEquivalence}, under \cref{assump2} the (conditional) covariance functions of  $\sqrt{n}(\hat{\bs\theta}^*_n(t)-\hat{\bs\theta}_n(t))$ and $\sqrt{n}(\hat{\bs\theta}_n(t) - \bs{\theta}_0(t))$ coincide asymptotically. 
Thus, we use the same general structure for $\hat{\sigma}^{*2}_n(t)$ as given in \eqref{eq:cov_sigma} for $\hat{\sigma}^{2}_n(t)$.
Yet, we replace the basic estimator $\big(\frac{1}{n}\bs I_n(\tau,\hat{\bs{\beta}}_n)\big)^{-1}$ by the wild bootstrap counterpart $\big(\frac{1}{n}\bs I^*_n(\tau,\hat{\bs{\beta}}^*_n)\big)^{-1}$ with 
$$\bs I_n^*(\tau,\hat{\boldsymbol{\beta}}^*_n) = \sum_{i=1}^n \int_0^\tau (\bs Z_{i}-\bs E_n(u,\hat{\boldsymbol{\beta}}^*_n))^{\otimes 2}G_i^2dN_i(u),$$ 
which is the optional covariation process 
$$[\textbf D^*_{n,g}] (\tau ) = \sum_{i=1}^n \int_0^\tau (\bs Z_{i}-\bs E_n(u,\hat{\boldsymbol{\beta}}_n))^{\otimes 2}G_i^2dN_i(u)$$
of $\textbf D^*_{n,g}(t)$ at $t=\tau$ with $\hat{\bs{\beta}}_n$ replaced by $\hat{\bs{\beta}}_n^*$.
We also replace the basic estimator 
$$\int_0^{t} S^{(0)}_n(u,\hat{\bs{\beta}}_n)^{-1}d\hat{A}_{1;0,n}(u,\hat{\bs{\beta}}_n)$$
by the wild bootstrap counterpart
$$\frac{1}{n}\sum_{i=1}^n \int_0^{t} S^{(0)}_n(u,\hat{\bs{\beta}}^*_n)^{-2}G_i^2 dN_i(u),\quad t\in\mathcal{T},$$
which originates from the optional covariation process 
$$[D_{n,k}^*](t) = \frac{1}{n}\sum_{i=1}^n \int_0^{t} S^{(0)}_n(u,\hat{\bs{\beta}}_n)^{-2}G_i^2 dN_i(u)$$
of $D_{n,k}^*(t)$, $t\in\mathcal{T}$, with again $\hat{\bs{\beta}}_n$ replaced by $\hat{\bs{\beta}}_n^*$. 
Note that according to \cref{ass:A1}~\ref{ass:A1_1} in combination with \cref{cor:optCov_D*} of Part~I, under \cref{assump2} the optional covariation processes of $\textbf D^*_{n,g}$ and $ D^*_{n,k}$ converge in probability to $\textbf V_{\tilde g}$ and $V_{\tilde k}$, respectively. Therefore, the corresponding wild bootstrap estimators are consistent estimators. For the particular form of the respective optional covariation process we refer to \cref{lemma:mgale} of Part~I. Additionally, we substitute $\hat{A}_{1;0,n}(t,\hat{\bs{\beta}}_n)$ and $\textbf E_n(u,\hat{\bs{\beta}}_n)$ in \eqref{eq:cov_sigma} by $\hat{A}^*_{1;0,n}(t,\hat{\bs{\beta}}^*_n)$ and $\textbf E_n(u,\hat{\bs{\beta}}^*_n)$, respectively.
All in all, we have that under \cref{assump2},  $\hat{\sigma}^{*2}_n(t)$ defined in \eqref{eq:WB_variance_estimator}
%\begin{align}\label{eq:WB_variance_estimator}
%\begin{split}
% \hat{\sigma}^{*2}_n(t) =  \hat{A}^*_{1;0,n}(t,\hat{\bs{\beta}}^*_n))^{-2} %&\big[ \frac{1}{n}\sum_{i=1}^n \int_0^{t} %S^{(0)}_n(u,\hat{\bs{\beta}}^*_n)^{-2}G_i^2 dN_i(u)\\
%& +\int_0^t (\textbf Z - \textbf E_n(u,\hat{\bs{\beta}}^*_n))^\top d\hat{A}^*_{1;0,n}(u,\hat{\bs{\beta}}^*_n) (\frac{1}{n}\bs I^*_n(\tau,\hat{\bs{\beta}}^*_n))^{-1}\\
%& \cdot\int_0^t (\textbf Z - \textbf E_n(u,\hat{\bs{\beta}}^*_n)) d\hat{A}^*_{1;0,n}(u,\hat{\bs{\beta}}^*_n)  \big],
%\end{split}
%\end{align} 
is a consistent wild bootstrap estimator for the variance of $W_{n,\phi , 1}(t)$, $t\in\mathcal{T}$.
This completes the proof of \cref{lem:sigma_hat_sigma_star}.
\hfill\hfill\qedsymbol